\documentclass[a4paper,12pt,times,numbered,print,index]{Classes/PhDThesisPSnPDF}

\input{Preamble/preamble}

\title{Beyond special relativity and the notion of spacetime}

\author{José Javier Relancio Martínez}

\dept{Departamento de Física Teórica}

\university{Universidad de  Zaragoza}

\collegeshield{\includegraphics[width=0.5\textwidth]{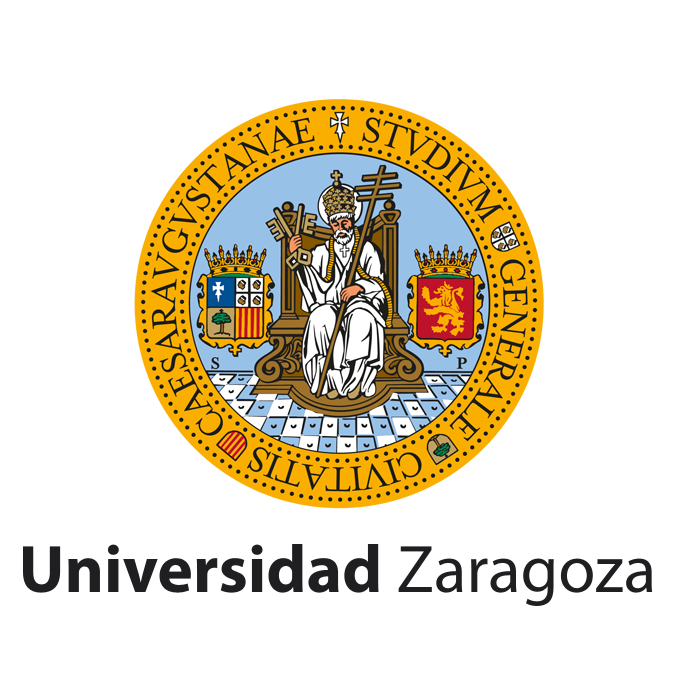}}

\supervisor{Prof. José Manuel Carmona Martínez\newline
Prof. José Luis Cortés Azcoiti}

\degreetitle{Doctor in Physics}

\graphicspath{ {./Figs/} }
\ifdefineAbstract
\fi

\ifdefineChapter
\fi

\begin{document}

\frontmatter

\maketitle


\begin{declaration}

I hereby declare that except where specific reference is made to the work of 
others, the contents of this dissertation are original and have not been 
submitted in whole or in part for consideration for any other degree or 
qualification in this, or any other university. This dissertation is my own 
work and contains nothing which is the outcome of work done in collaboration 
with others, except as specified in the text and Acknowledgements. 


\end{declaration}


\begin{acknowledgements}

Firstly, I would like to express my gratitude to my supervisors, Professor Carmona and Professor Cortés, for their patient guide, encouragement, and fruitful criticisms during this thesis. Their support goes beyond the academical sphere and helped me a lot not only in the research work but in everything surrounding my Ph.D. 

I would also thank Professor Asensio and Professor Clemente for their advice and useful discussions in the geometric part of the work, which would not be carried out without their aid. Besides, conversations with Professor Liberati were a guidance during my visit in Trieste, helping me with the last part of the thesis. Moreover, the help of Raúl Carballo-Rubio during that time was also indispensable.

Finally, I would like to express my profound gratitude to my parents for their unfailing support and encouragement during all these years. Also to my best friends, Eduardo, Adri\'{a}n and Miguel, for all the good moments we spent.    

\end{acknowledgements}

 Esta tesis ha sido llevada a cabo gracias al contrato predoctoral para la formación de personal
investigador concedido por el Ministerio de Economía, Industría y Competitividad (España) con referencia BES-2016-077005. Asimismo, la investigación aquí presentada ha sido parcialmente financiada por los proyectos de investigación FPA2015-65745-P y PGC2018-095328-B-I00 del Ministerio de  Economía, Industría y Competitividad (España).


\begin{abstract} 
One of the challenges of theoretical physics nowadays is the unification of general relativity and quantum field theory, or equivalently, the formulation of a quantum gravity theory. Both theories, well checked experimentally in the past century, present fundamental incompatibilities which have their origin in the role that spacetime plays in them (it is a dynamical variable in general relativity, and a static frame in quantum field theory).
There have been many attempts to formulate a quantum field theory of gravity, such as string theory, quantum loop gravity, set causal theory, etc. In some of these frameworks, the spacetime acquires a fundamental and characteristic structure, very much different from the notion of a continuous spacetime in special relativity. However, neither the dynamics of these theories are fully understood, nor they are easily contrastable with experimental observations.
At the beginning of the century a new theory started to be developed and it is still germinating, doubly special relativity. The starting point is very different from the other perspectives: it is not a fundamental theory, but it is considered a low energy limit of a quantum gravity theory that tries to study its possible residual elements. In particular, in doubly special relativity the Einstenian relativity principle is generalized, adding to the speed of light $c$ another relativistic invariant, the Planck length $l_P$. This idea can really have possible experimental observations, giving place to what it is known as quantum gravity phenomenology. On the other hand, doubly special relativity implies the existence of deformed composition laws  for energy and momentum, which leads to a spacetime with nonlocal ingredients, an element that also appears in other approaches of quantum gravity.

In this thesis, after exposing the motivations to consider deformations of special relativity, we will study the role that the changes of momentum variables play in a deformed relativistic kinematics, observing that there is a simple way to define such a deformation just by using a change of variables. We will see that one of the most studied kinematics in doubly special relativity models, $\kappa$-Poincaré, can be obtained through this method order by order. This leads to too many deformed composition laws, bringing us to think that a mathematical or physical criteria might be needed in order to restrict the possible kinematics. 

In many works in the literature, a connection between the $\kappa$-Poincaré model and a curved momentum space has been explored. We will see that considering a maximally symmetric momentum space one can construct a deformed relativistic kinematics, and that the $\kappa$-Poincaré model is obtained as a particular case when the curvature of the space is positive. 
 
This deformed composition law changes the notion of spacetime. As we will see, in  the doubly special relativity framework, there is a loss of locality of interactions due to the deformed composition law for the momenta. We will study how this loss arises and how a new spacetime, which turns out to be noncommutative, can be considered in order to make the interactions local. We will also see that there is a relation between the locality and geometry frameworks.

After that, we will consider two different phenomenological studies. In the first one, we will analyze the possible time delay in the flight of photons as a consequence of a deformed kinematics. This will be done considering that observables are defined either on a commutative or on a noncommutative spacetime. We will find that, while in the first case a time delay could exist, depending on the choice of momentum variables one is working with, in the last scheme one obtains that there is no time delay, independently of the choice of variables. 
Since photon time delays measurements may be the only phenomenological test of doubly special relativity for small energies compared with the Planck scale, the absence of time delays would
imply that the constraints on the high energy scale that characterizes doubly special relativity might be orders of magnitude smaller than the Planck energy. With this observation in mind, 
we will do some computations in quantum field theory with a simple ansatz for the modified Feynman rules corresponding to a particle process, concluding that a deformed kinematics with an energy scale of some TeV's is compatible with the observational data. 

However, the previous study of time delays is carried out in flat spacetime, which is not the correct way to consider the propagation of a photon in an expanding universe. In the last part of the thesis, we will study the generalization of the geometrical approach for a curved spacetime. We will develop the construction of a metric in the cotangent bundle that takes into account the deformed relativistic kinematics in the presence of a nontrivial geometry in spacetime. With a generalization of the usual procedures of general relativity, we will study the phenomenological consequences of a momentum dependent metric in the cotangent bundle for an expanding universe and a stationary static black hole.  
\end{abstract}

\clearpage 
\begin{center}
\hfill \break\vspace{3cm}
\section*{Resumen}
\end{center}

\justify
Uno de los desafíos de la física teórica hoy en día es la unificación de la relatividad general y la teoría cuántica de campos, o equivalentemente, la formulación de una teoría de gravedad cuántica. Ambas teorías, bien comprobadas experimentalmente durante el siglo pasado, presentan incompatibilidades fundamentales que tienen su origen en el papel que el espacio-tiempo juega en ellas (es una variable dinámica en relatividad general, y un marco estático en teoría cuántica de campos). Ha habido numerosos intentos de formular una teoría de gravedad cuántica, como la teoría de cuerdas, teoría cuántica de bucles, teoría de conjuntos causales, etc. En algunos de estos marcos, el espacio-tiempo adquiere una estructura fundamental y característica, muy diferente de la noción de espacio-tiempo continuo de relatividad especial. Sin embargo, ni se entiende completamente la dinámica de estas teorías, ni son fácilmente contrastables con observaciones experimentales. Al principio de este siglo ha empezado a desarrollarse una teoría que todavía está germinando, la relatividad doblemente especial. El punto de partida de esta teoría es completamente distinto al de las otras perspectivas: no es una teoría fundamental, sino que es considerada un límite de bajas energías de una teoría de gravedad cuántica que intenta estudiar sus posibles elementos residuales. En particular, en relatividad doblemente especial se generaliza el principio de relatividad de Einstein, añadiendo a la velocidad de la luz $c$ otro invariante relativista, la longitud de Planck $l_p$. Esta idea puede tener evidencias experimentales, dando lugar a lo que se conoce como fenomenología de gravedad cuántica. Por otro lado, la relatividad doblemente especial implica la existencia de una ley de composición deformada para la energía y el momento, lo que lleva a un espacio-tiempo con ingredientes no locales, un elemento que también aparece en otras aproximaciones de gravedad cuántica.    

En esta tesis, tras mostrar las motivaciones para considerar deformaciones de la relatividad especial, estudiaremos el papel que  juegan los cambios en las variables momento en una cinemática relativista deformada, observando que hay una forma simple de definir una deformación usando simplemente un cambio de variables. Veremos que una de las cinemáticas más estudiadas en los modelos de relatividad doblemente especial, $\kappa$-Poincaré, puede obtenerse a través de este método orden a orden. Esto conduce a demasiadas leyes de composición deformadas, llevándonos a pensar que podría ser necesario un criterio matemático o físico para restringir las cinemáticas posibles.

En muchos trabajos de la literatura se ha explorado una conexión entre el modelo de $\kappa$-Poincaré y un espacio de momentos curvo. Veremos que considerando un espacio de momentos maximalmente simétrico se puede construir una cinemática relativista deformada, y que entre las posibles cinemáticas se obtiene $\kappa$-Poincaré como un caso particular cuando la curvatura del espacio de momentos es positiva. 

Esta ley de composición deformada altera el comportamiento del espacio-tiempo. Como veremos, en el marco de la relatividad doblemente especial, hay una pérdida de la noción de localidad de interacciones debido a la ley de composición deformada para los momentos. Estudiaremos cómo aparece esta pérdida y cómo un nuevo espacio-tiempo, que es no conmutativo, puede considerarse para hacer que las interacciones sean locales. Veremos también que hay una relación entre los marcos de localidad y geometría.

Después, consideraremos dos estudios fenomenológicos. En el primero, analizaremos el posible retraso en tiempo de vuelo para fotones como consecuencia de una cinemática deformada. Esto se hará considerando que los observables están definidos en un espacio-tiempo conmutativo o no conmutativo. Encontraremos que, mientras que en el primer caso podría existir un retraso en el tiempo, dependiendo de la elección de variables momento con las que uno trabaja, en este último esquema se obtiene que no hay retraso en el tiempo, independientemente de la elección de variables. Ya que las medidas de retrasos en tiempos de vuelo podrían ser el único test fenomenológico de relatividad doblemente especial para pequeñas energías comparadas con la escala de Planck, la ausencia de retrasos en tiempos de vuelo implicaría que las restricciones en la escala de alta energía que caracteriza relatividad doblemente especial podría ser órdenes de magnitud menor que la energía de Planck. Con esta observación en mente, haremos algunos cálculos en teoría cuántica de campos con una suposición simple para las reglas de Feynman modificadas correspondientes a procesos de partículas, concluyendo que una cinemática deformada con una escala de energía de unos pocos TeV's es compatible con los datos experimentales.

Sin embargo, los estudios anteriores de tiempos de retrasos se llevan a cabo en espacio-tiempo llano, que no es la forma correcta de considerar la propagación de un fotón en un universo en expansión. En la última parte de la tesis, estudiaremos la generalización del enfoque geométrico para un espacio-tiempo curvo. Desarrollaremos la construcción de una métrica en el fibrado cotangente que tiene en cuenta la cinemática relativista deformada en presencia de una geometría no trivial en el espacio-tiempo. Con una generalización de los procedimientos usuales de relatividad general, estudiaremos las consecuencias fenomenoló-\\gicas de una métrica dependiente del momento en el fibrado cotangente para un universo en expansión y para un agujero negro estacionario.


\tableofcontents



\printnomenclature

\mainmatter


\chapter{Introduction}  
\label{chapter_intro}
\ifpdf
    \graphicspath{{Chapter1/Figs/Raster/}{Chapter1/Figs/PDF/}{Chapter1/Figs/}}
\else
    \graphicspath{{Chapter1/Figs/Vector/}{Chapter1/Figs/}}
\fi

\epigraph{I study myself more than any other subject. That is my metaphysics, that is my physics.}{Michel de Montaigne}

\nomenclature[z-DSR]{DSR}{Doubly Special Relativity}%
\nomenclature[z-PDG]{PDG}{Particle Data Group}%
\nomenclature[z-BSR]{BSR}{Beyond Special Relativity}%
\nomenclature[z-UHECR]{UHECR}{ultra-high  Energy Cosmic Rays}%
\nomenclature[z-CMB]{CMB}{Cosmic Microwave Background}%
\nomenclature[z-pp]{pp}{proton-proton}%
\nomenclature[z-SR]{SR}{Special Relativity}%
\nomenclature[z-QFT]{QFT}{Quantum Field Theory}%
\nomenclature[z-LEP]{LEP}{Large Electron-Positron collider}%
\nomenclature[z-GR]{GR}{General Relativity}%
\nomenclature[z-QT]{QT}{Quantum Theory}%
\nomenclature[z-LIV]{LIV}{Lorentz Invariance Violation}%
\nomenclature[z-LI]{LI}{Lorentz Invariance}%
\nomenclature[z-EFT]{EFT}{Effective Field Theory}%
\nomenclature[z-SME]{SME}{Standard Model Extension}%
\nomenclature[z-SM]{SM}{Standard Model}%
\nomenclature[z-mSME]{mSME}{minimal Standard Model Extension}%
\nomenclature[z-GRB]{GRB}{Gamma Ray Burst}%
\nomenclature[z-AGN]{AGN}{Active Galactic Nuclei}%
\nomenclature[z-DSR]{DSR}{Doubly Special Relativity}%
\nomenclature[z-DSR]{DSR}{Doubly Special Relativity}%
\nomenclature[z-DSR]{DSR}{Doubly Special Relativity}%
\nomenclature[z-DCL]{DCL}{Deformed Composition Law}%
\nomenclature[z-DDR]{DDR}{Deformed Dispersion Relation}%
\nomenclature[z-DRK]{DRK}{Deformed  Relativistic Kinematics}%
\nomenclature[z-DLT]{DLT}{Deformed Lorentz Transformations}%
\nomenclature[g-lambda]{$\Lambda$}{High energy scale}%
\nomenclature[g-varphi]{$\varphi$}{Tetrad in momentum space/ function characterizing the noncommutative spacetime}%
\nomenclature[g-vvarphi]{$\bar{\varphi}$}{Inverse function of $\varphi$}%
\nomenclature[x-oplus]{$\oplus$}{Composition law}%
\nomenclature[r-hat]{$\hat{}$}{Antipode}%
\nomenclature[x-n]{$n^\mu$}{Unitary timelike vector $(1,0,0,0)$}%

Since the dawn of humanity, human beings have tried to explain all observed phenomena. With scientific research new theories have been developed, so more phenomena have been explained. Thanks to the technology provided by this research, smaller scales have been studied, leading to new processes that had to be understood. The problem arises when one knows that 
there are some missing parts or inconsistencies in the theory, while no new experimental observations are available to guide its development.
This undesirable fact emerges in theoretical physics nowadays. 

One source of inconsistencies appears when one tries to unify general relativity (GR) and quantum theory (QT). One of the possible issues that impedes the unification of these two theories is the role that spacetime plays in them. While in quantum field theory (QFT) spacetime is given
from the very beginning, 
as a framework in which the processes of interactions can be described, in GR the spacetime is understood as the deformation of a flat 4-dimensional space modeled by matter and radiation. Of course, one can consider a quantum theory of gravitation where the mediation of the interaction is carried out by the graviton, a spin-2 particle, leading to Einstein’s equations~\cite{Feynman:1996kb}. The problem of this approach is that this theory is not renormalizable, and then it gives well-defined  predictions only for energies below the Planck scale.

With the huge machinery that these two theories provide we can describe, on the one hand, the massive objects (GR), and on the other one, the lightest particles (QFT), so one could naively say that a theory that contains both, a quantum gravity theory (QGT), would be completely unnecessary. But this is not the case if one wants to study the propagation and interaction of very tiny and energetic particles: the kinematics of the processes should take into account the quantum and gravitational effects together, something unthinkable if QFT and GR cannot be studied in the same framework. This kind of interactions did take place at the beginning of the universe, where a huge amount of matter was concentrated in a minute region of space. So in order to describe the first instants of the universe, a complete understanding of a QGT should be indispensable. 

Besides this, we do not know what happens inside a black hole, which is a source of contradiction between GR and QT~\cite{Hawking:1976ra}. What happens with the information when it crosses the event horizon? If one considers that the information is lost, one is going against what QT says. If on  the other hand, the information remains encrypted in the horizon surface, the evaporation of the black hole~\cite{Hawking:1974sw} would lead to a contradiction between pure and mixed states. In fact, one of the possible solutions to the information paradox, named firewall~\cite{Almheiri:2012rt} because it proposes that due to the existence of mixed states there would be particles ``burning'' an observer in free fall into the black hole, violates the equivalence principle, which states that one should not feel anything while crossing the event horizon.
Another question is: what happens when one comes to the singularity? To answer all these questions, we need a QGT.

Another problem that one finds is that in QT one assumes that spacetime is given and studies with all detail the properties and movement of particles in it, including matter and radiation. In GR, and specially in cosmology, one takes the opposite way: the properties of matter and radiation are given (through state equations) and one describes the resultant spacetime. 

There is also a difficulty in defining spacetime. Einstein thought about being able to describe the space-time coordinates through the exchange of light signals~\cite{Einstein1905}, but when one uses this procedure, one neglects all information about the energy of the photons and assumes that the same spacetime is rebuilt by exchange of light signals of different frequencies. However, what would happen if the speed of light depends on the energy of the photon, as it happens in many theoretical frameworks which try to unify GR with QT? In this case, the energy of the photon would affect the own structure of spacetime. Also, this procedure of identifying points of spacetime assumes that interactions are local events, happening at the same point of spacetime. This is no longer valid when one has a deformed relativistic kinematics~\cite{AmelinoCamelia:2011bm,AmelinoCamelia:2011pe}, as we will see. 

Presumably, all these paradoxes and inconsistencies could be avoided if a QGT was known. Despite our ignorance about the possible consequences and implications of a complete QGT, we can pose the main properties that such a theory should have, and the characteristic phenomenological impacts that may result. 

\section{Towards a QGT: main ideas and ingredients}
\label{sec:QGT}
In the last 60 years, numerous theories have tried to avoid the inconsistencies that appear when one tries to put in the same scheme GR and QFT: string theory~\cite{Mukhi:2011zz,Aharony:1999ks,Dienes:1996du}, loop quantum gravity~\cite{Sahlmann:2010zf,Dupuis:2012yw}, supergravity~\cite{VanNieuwenhuizen:1981ae,Taylor:1983su}, or causal set theory~\cite{Wallden:2010sh,Wallden:2013kka,Henson:2006kf}. The main problem is the lack of experimental observations that could tell us which is the correct theory corresponding to a QGT~\cite{LectNotes702}. 

In most of these theories, as they are trying to consider a generalization of the classical version of spacetime, a minimum length appears~\cite{Gross:1987ar,Amati:1988tn,Garay1995}, which is usually considered to be the Planck one, and then the Planck energy is taken as a characteristic energy. In order to obtain the Planck length $l_p$, time $t_p$, mass $M_p$ and energy $E_p$, one only needs to use the physical constants of quantum mechanics ~$\hbar$, relativity $c$ and gravitation $G$,

\begin{eqnarray}
l_p\,&=&\, \sqrt{\frac{\hbar G}{c^3}}\,=\, 1.6 \times \,10^{-35}\,\text{m}\,, \nonumber \\
t_p\,&=&\, \sqrt{\frac{\hbar G}{c^5}}\,=\, 5.4 \times\,10^{-44}\,\text{s}\,, \nonumber \\
\frac{E_p}{c^2}\,&=&\,M_p\,=\, \sqrt{\frac{\hbar c}{G}}\,=\, 2.2 \times\,10^{-8}\,\text{kg}\,=\, 1.2\times\,10^{19} \,\text{GeV}/c^2\,.
\end{eqnarray}

The consideration of a minimum length provokes that the spacetime acquires some very particular features that must be studied in order to know what kind of theories we are facing. 

Along this thesis (although, exceptionally, not in this chapter), we will be using natural units, in which ~$\hbar$, $c$ and $G$ are 1.

\subsection{Minimum length scenario}
There are numerous consequences of having a minimal length (see Ref.~\cite{Schiller:1996fw} for more information). We can enumerate some of them: 
\begin{itemize}
	\item The concept of a space-time manifold disappears. SR, QFT and GR are developed under the idea that time is a continuous concept (it admits a description in terms of real numbers). But due to the presence of a minimum length and time, we have an uncertainty in the measure of distances and times that impedes us to synchronize two clocks with more precision than the Planck time. Due to this impossibility of synchronizing clocks in a precise way, the idea of a univocal time coordinate for a reference frame is only approximated, and it cannot be maintained in an accurate description of nature. We do not have a way to sort events for times smaller than the Planck one either. One then is forced to forget about the idea of time as a unique ``point''. For example, at Planckian scales the concept of proper time disappears. 
			
	\item 		In this way one has a quantized spacetime, in the sense that it is discrete and non-continuous. Due to this quantization, the concepts of a point in space and of an instant of time is lost, as a consequence of the impossibility of measuring with a greater resolution than the Planck scale. Also, this gives place to a modification of the commutation rules (as we will see below), because the measure of space and time leads to  non-vanishing uncertainties for position and time, $\Delta x \, \Delta t \geq l_p \, t_p$. 
			
 	\item Since one cannot determine the metric at these scales, the sense of curvature is lost. That is, the impossibility of measuring lengths is exactly equivalent to curvature fluctuations. It is possible then to imagine that spacetime is like a foam ~\cite{Wheeler:1955zz,Ng:2011rn} at very small scales. Particles would notice these effects due to the quantum fluctuations of spacetime, being more and more relevant for higher energies.
	
		\item 
		Due to this imprecision of measuring at Planckian scales, the concepts of spatial order, translational invariance, vacuum isotropy and global coordinates systems, lose all experimental support at these dimensions. Moreover, spacetime is neither invariant under Lorentz transformations, nor diffeomorphisms or dilation transformations, so all fundamental symmetries of SR and GR are only valid approximations for scales larger than the Planck one. 
			
			\item At the Planck scale we lose the naive sense of dimensions. The number of dimensions of a space can be obtained by determining how many points can be chosen such the distance between them are equal. Then, if one can find $n$ points, the space has $n-1$ dimensions. For example, in 1D one has two points, in 2D three points, and so on. The lack of 
			precise measurements makes impossible to determine the number of dimensions at Planckian scales with this method. With all this, we see that the physical spacetime cannot be a set of mathematical points. We also are not able to distinguish at small scales if a distance is timelike or spacelike. At Planckian scales, space and time cannot be distinguished. Summarizing, spacetime at these scales is neither continuous, nor ordered, nor metric gifted, nor four-dimensional, nor made of points.
				
		\item Since space and time are not continuous, observables do not vary continuously, either. This means that at Planckian scales, observables cannot be described with real numbers with (potentially) infinity precision. Nor the physical fields can be described as continuous functions.
		
		\item Also the concept of point particle disappears. In fact, it is completely senseless. Of course, the existence of a minimum length, for empty space so for objects, is related with this fact. If the term of point is senseless, also the concept of point particle is lost. 
		\item If one takes as valid that the size of an elementary particle is always smaller than its Compton wavelength and always bigger than the Planck one, one can prove that the mass of particles must be less than the Planck mass. In QFT we know that the difference between a real or virtual particle is if it is on-shell or off-shell. Due to these uncertainties in measurements, at Planckian scales one cannot know if a particle is real or virtual. As antimatter can be described as matter moving backwards in time, and since the difference between backwards and forwards cannot be determined at Planckian scales, one cannot distinguish between matter and antimatter at these ranges. Since we do not have well-defined  rotations, the spin of a particle cannot be properly defined at Planck scales, and then, we cannot distinguish between bosons and fermions, or, in other words, we cannot distinguish matter and radiation at these scales. 
			\item Finally, let us think about the inertial mass of a tiny object. In order to determine it, we must push it, that is, elaborate a scattering experiment. In order to determine the inertial mass inside a region of size $R$, a wavelength smaller than $R$ must be used, so one needs high energies. That means that the particle will feel attraction due to gravity interaction to the probe (as we will see in the next subsection). Then, at Planckian scales, inertial and gravitational mass cannot be distinguished. To determine the mass in a Planck volume, a wavelength of Planck size has to be used. But,  as the minimal error in the wavelength is also the Planck length, the error in the mass becomes so big as the Planck energy is. 
			In this way, one cannot differentiate between matter and vacuum, and then, when a particle with Planck energy is traveling through spacetime, it can be scattered by the own fluctuations of spacetime, making impossible to say if it has been scattered by vacuum or matter.
\end{itemize}
With all these examples, we see that physics at Planck scales is completely different to what we are used to and to what we can even imagine. 
In the following section we will consider some gedanken experiments in order to shed some light about how new physics effects may arise.

\subsection{GUP: generalized uncertainty principle}
\label{sec:gup}
We have seen many consequences of having a minimum length, but we do not have explored through a physical intuition how this minimum length could appear. In this subsection we study a thought experiment in which the Planck scale arises. 

First of all, we consider the gedanken Heisenberg microscope experiment in QT. According to classical optics, the wavelength of a photon with momentum ``$\omega$'' establishes a limit in the possible resolution $\Delta x$ in the position of the particle which interacts with the photon 
\be
\Delta x \gtrsim \frac{1}{2 \pi \omega \sin{\epsilon}}\,,
\label{eq:delta_x}
\ee
where $\epsilon$ is the aperture angle of the microscope lens. But the photon used to measure the position of the particle has a recoiling when it is scattered and it transfers momentum to the particle. As one does not know the direction of the photon with more resolution than $\epsilon$, this leads to an uncertainty in the momentum of the particle in the $x$ direction 
\be
\Delta p_x  \gtrsim \omega \sin{\epsilon}\,.
\ee
Taking all this together, one obtains the uncertainty 
\be
\Delta x\, \Delta p_x  \gtrsim \frac{1}{2 \pi}\,.
\ee
This is a fundamental property of the quantum nature of matter.

We can recreate the mental Heisenberg microscope experiment including the gravitational attraction between the particle whose position one wants to know and the probe used for that aim~\cite{Hossenfelder:2012jw,Garay1995}. As we have seen, the interaction of the photon with the particle does not take place in a well-defined  point, but in a region of size $R$. For the interaction to take place and the measurement to be possible, the time passed between the interaction and the measurement has to be at least of the order $\tau \gtrsim R$. The photon carries an energy that, even though small, it exerts a gravitational attraction over the particle whose position we want to measure. The gravitational acceleration acting over the particle is at least of the order of 
\be
a \approx \frac{G \omega}{R^2}\,,
\ee 
and, assuming that the particle is non-relativistic and much slower than the photon, the acceleration acts approximately along the time the photon is in the region of the interaction, so the particle acquires a speed 
\be
v\approx a\,R\,=\, \frac{G \omega}{R}\,.
\ee
So, in a time $R$ the acquired velocity allows the particle to travel a distance 
\be
L\approx G\omega\,.
\ee 
However, since the direction of the photon is unknown with a width of angle $\epsilon$, the direction of the acceleration and the movement of the particle are also unknown. The projection over the $x$ axis gives and additional uncertainty of 
\be
\Delta x \gtrsim G\omega\,\sin{\epsilon}\,.
\label{eq:delta_x2}
\ee 
Combining Eq.~\eqref{eq:delta_x} and Eq.~\eqref{eq:delta_x2}  we see that 
\be
\Delta x \gtrsim \sqrt{G}\,=\,l_p\,.
\ee 
One can refine this argument taking into account that, strictly speaking, during the experiment the photon momentum is increased by 
\be
\frac{Gm\omega}{R}\,,
\ee
where $m$ is the mass of the particle. This increases the uncertainty of the momentum of the particle
\be
\Delta p_x \gtrsim \omega\left(1+\frac{G m}{R}\right)\,\sin{\epsilon}\,,
\ee
and during the time in which the photon is in the interaction region, it is translated 
\be
\Delta x \approx \frac{R \,\Delta p_x}{m}\,, \qquad \text{so} \qquad \Delta x \gtrsim \omega\left(G+\frac{R}{m}\right)\,\sin{\epsilon}\,,
\ee
which is bigger than the previous uncertainty and then the limit in which one is not considering gravity is still satisfied. 

Assuming that the regular uncertainty and the gravitational one add linearly, one gets
\be
\Delta x\gtrsim \frac{1}{\Delta p_x}+G\,\Delta p_x\,.
\ee
This result is also obtained in string theory through completely different assumptions~\cite{Kato:1990bd,Susskind:1993ki}. 

With this thought experiment, we see that when one adds the gravitational interaction to the usual Heisenberg microscope we obtain a generalized uncertainty principle, which leads to a modification of the commutation rules. This could be then considered as an ingredient that a QGT should have. From the fact we have different commutation rules, one can guess that it is necessary a completely different notion of spacetime, and then the symmetries acting on it should also be different.  

\subsection{Spacetime and symmetries in a QGT}

As we have seen in the previous subsection, the introduction of a minimum length leads to nontrivial commutation rules, which could be considered as a way to parametrize a quantum nature of spacetime. The idea of a quantum spacetime was firstly proposed by Heisenberg and Ivanenko as an attempt to avoid the ultraviolet divergences of QFT. This idea passed from Heisenberg to Peierls and to Robert Oppenheimer, and finally to Snyder, who published the first concrete example in 1947~\cite{Snyder:1946qz}. This is a Lorentz covariant model in which the commutator of two coordinates is proportional to the Lorentz generator
\begin{equation}
\left[x^\mu,x^\nu\right]\,=\,i\frac{J^{\mu \nu}}{\Lambda^2}  \,,
\end{equation}
where $\Lambda$ has dimensions of energy by dimensional arguments\footnote{Remember that we are using natural units, making that the inverse of length is an energy.}. But this model, originally proposed to try to avoid the ultraviolet divergences in QFT,  was forgotten when renormalization appeared as a systematic way to avoid the divergences at the level of the relations between observables. Recently, the model has been reconsidered when noncommutativity was seen as a way to go towards a QGT.  

Another widely studied model is the canonical noncommutativity~\cite{Szabo:2001kg,Douglas:2001ba}, 
\begin{equation}
\left[x^\mu,x^\nu\right]\,=\, i \Theta^{\mu \nu} \,,
\end{equation}
where $\Theta^{\mu \nu}$ is a constant matrix with dimensions of length squared. In this particular simple case of noncommutativity it has been possible to study a QFT with the standard perturbative approach.

The last model we mention here, named $\kappa$-Minkowski\footnote{We will study it in more detail in Sec.\ref{sec:DSR} and Sec.~\ref{sec:examples} as it is included in the scheme of  $\kappa$-Poincar\'{e}.}~\cite{Smolinski1994}, has the following non-vanishing commutation rules 
\begin{equation}
\left[x^0,x^i\right]\,=\,- i \frac{x^i}{\Lambda}  \,,
\end{equation}
where $\Lambda$ has also dimensions of energy.

Snyder noncommutative spacetime is very peculiar from the point of view of symmetries since the usual Lorentz transformations used in SR are still valid. But in general, in the other models of noncommutativity, linear Lorentz invariance is not a symmetry of the new spacetime, which is in agreement with what we have seen previously: the classical concept of a continuum spacetime has to be replaced somehow for Planckian scales, where new effects due to the quantum nature of gravity (for example, creation and evaporation of virtual black holes~\cite{Kallosh:1995hi}) should appear. So, while SR postulates Lorentz invariance as an exact symmetry of Nature (every experimental test up to date is in accordance with it~\cite{Kostelecky:2008ts,Long:2014swa,Kostelecky:2016pyx,Kostelecky:2016kkn}; see also the papers in~Ref.~\cite{LectNotes702}), a QGT is expected to modify  someway this symmetry. Many theories which try to describe a QGT include a modification of Lorentz invariance in a form or another (for a review, see Ref.~\cite{AmelinoCamelia:2008qg}), and the possible experimental observations that confirms or refutes this hypothesis would be very important in order to constrain these possible theories. A way to go beyond the Lorentz invariance is to consider that this symmetry would be violated for energies comparable with the high energy scale. This is precisely what is studied in the so-called Lorentz-invariance violation theories (LIV). In this way, the SR symmetries are only low energy approximations of the true symmetries of spacetime. We will study in the next subsection the usual theoretical framework in which these kind of theories are formulated and the main experiments where a LIV effect could be manifest.

\subsection{Lorentz Invariance Violation}

As we have previously mentioned, the symmetries of the ``classical'' spacetime have to be broken of deformed at high energies due to the possible new effects of the quantum spacetime. LIV theories consider that Lorentz symmetry is violated at high energies, establishing that there is a preferred frame of reference (normally an observer aligned with the cosmic microwave background (CMB), in such a way that this radiation is isotropic). A conservative way to consider this theory is to assume the validity of the field theory framework. Then, all the terms that violate Lorentz invariance (LI) are added to the standard model (SM), leading to an effective field theory (EFT) known as the standard model extension (SME)~\cite{Colladay:1998fq} (in the simplest model one considers only operators of dimension 4 or less, known as the minimal SME, or mSME) with the condition that they do not change the field content and that the gauge symmetry is not violated.

Historically, in the middle of the past century researchers realized that LIV could have some phenomenological observations~\cite{Dirac:1951:TA,Bjorken:1963vg,Phillips:1966zzc,Pavlopoulos:1967dm,Redei:1967zz},   and in the seventies and eighties theoretical bases were settled pointing how LI could be established for low energies without being an exact symmetry at all scales~\cite{Nielsen:1978is,Ellis:1980jm,Zee:1981sy,Nielsen:1982kx,Chadha:1982qq,Nielsen:1982sz}. However, this possible way to go beyond SR did not draw much attention since it was thought that effects of new physics would only appear for energies comparable to the Planck mass. It seems impossible to talk about phenomenology of such a theory being the Planck energy of the order of $10^{19}$ GeV and having only access to energies of $10^4$  GeV from particle accelerators and  $10^{11}$ GeV from particles coming from cosmic rays. But over the past few years people have realized that there could be some effects at low energy that could find out evidences of a LIV due to amplification processes~\cite{Mattingly:2005re}. These effects were baptized as ``Windows on Quantum Gravity''. A partial list of these {\em windows on QG} includes (see Refs.~\cite{Mattingly:2005re,Liberati2013} for a review):
\begin{itemize}
\item \textbf{Change in the results of the experiment as the laboratory moves}

Due to the existence of a preferred frame of reference, there is a change in the measurement since the laboratory moves (due to the rotation and translation of the Earth), and then different results should be obtained depending on the spatial location where the experiment takes place and the time when it is done. To carry out the experiment, two ``clocks'', i.e., two atomic transition frequencies of different materials or with different orientations, are positioned in the same point of the space. During the movement of the ``clocks'', they would take different components of the tensors appearing in the deformed Lorentz violating EFT of the mSME. This supposed difference of time between the clock frequencies should be measured over a long time in order to be appreciable, and its absence puts constraints in the parameters of the model (generally for protons and neutrons~\cite{Kostelecky:1999mr}). 

\item \textbf{Cumulative effects}

There are two important effects trying to be measured. On the one hand, if there is a deformed dispersion relation (DDR) with Lorentz invariance violating terms, the velocity of particles (and particularly photons), would depend on their energy  (this effect was considered for the first time in Ref.~\cite{Amelino-Camelia1998}). This could be measured for photons coming from a gamma-ray burst (GRB), pulsars, or active galactic nuclei (AGN), due to the long distance they travel, amplifying the possible effect\footnote{This effect will be seen in more detail in Sec.\ref{sec:DSR}}. On the other hand, some terms in the mSME would produce a time delay for photons due to a helicity dependence of the velocity, phenomenon baptized as birefringence~\cite{Maccione:2008tq}.

\item \textbf{Threshold of allowed (SR forbidden) reactions} 

Due to the existence of a preferred reference frame, some reactions forbidden in SR are now allowed starting at some threshold energy. For example, photon splitting $\gamma \rightarrow e^+e^- $ is not allowed in usual QFT because of kinematics and charge-parity (CP) conservation, but it could be possible in a LIV scenario from some threshold energy~\cite{Jacobson:2002hd}. 

\item  \textbf{Shifting of existing threshold reactions} 

The GZK cutoff~\cite{Greisen:1966jv,Zatsepin:1966jv} is a theoretical limit on the energy of the ultra-high  energy cosmic rays (UHECR) that come to our galaxy due to interactions with CMB photons, i.e. interactions like $\gamma_{CMB}+p \rightarrow \Delta^{+} \rightarrow p + \pi^0$ or $\gamma_{CMB}+p \rightarrow \Delta^{+} \rightarrow n + \pi^{+}$. In the LI case these interactions have a threshold energy of $5\times 10^{19}$ eV. Experimentally, the suppression of the UHECR flux was confirmed only recently~\cite{Roth:2007in,Thomson:2006mm} in the Auger and HiRes experiments and also in the AGASA collaboration~\cite{Takeda:1998ps}. Even though this cutoff could be expected due to the finite acceleration power of UHECR sources, the fact that the maximum energies of UHECR coincide with the proposed GZK cutoff  makes plausible its explanation as due to the interaction with the CMB photons. However, if a LIV scenario is present this cutoff could be modified. The GZK cutoff is a good arena for constraining LIV since the threshold of the interaction of high energy protons with CMB photons is very sensitive to a LIV in the kinematics~\cite{Jacobson:2002hd}.

\end{itemize}

Despite the efforts of the scientific community, until now there is no clear evidences of LIV. Current experiments have only been able to put constraints in the SME parameters~\cite{Kostelecky:2008ts}. 

In this thesis, the main field of research is a different way to go beyond SR. In this framework there is also a high energy scale parameterizing departures from SR, but preserving a relativity principle. 

\section{DSR: Doubly Special Relativity}
\label{sec:DSR}

In the previous subsection we briefly summarized the most important features of LIV. Now we can wonder if there is another option instead of violating Lorentz symmetry for going beyond SR (BSR). One could consider that Lorentz symmetry is not violated at Planckian scales but deformed. This is nothing new in physics; some symmetries have been deformed when another, more complete theory which encompasses the previous one, is considered. For example, Poincar\'{e} transformations, that are the symmetries of SR, are a deformation of the Galilean transformations in classical mechanics. In this deformation, a new invariant parameter appears, the speed of light. Similarly, in a theory beyond SR (thinking on some approximation of a QGT), one could have a Poincar\'{e} deformed symmetry with a new parameter. This is what doubly special relativity (DSR) considers (see Ref.~\cite{AmelinoCamelia:2008qg} for a review).

In this theory, the Einstein relativity principle is generalized adding a new relativistic invariant to the speed of light $c$, the Planck length $l_P$. This is why this theory is also-called Doubly Special Relativity. The Planck length is normally considered as a minimum length. Of course, it is assumed that in the limit in which $l_P$ tends to 0, DSR becomes the standard SR.

\subsection{Introduction of the theory}
We start this subsection by summarizing the first papers~\cite{Amelino-Camelia2001,Amelino-Camelia2002a}, in which DSR was formulated as a low energy limit of a QGT that could have some experimental consequences. These papers formulate DSR as the result of the introduction of a new invariant scale in SR, the Planck length $l_P$, in a parallel way as the speed of light $c$ is introduced as a fundamental scale to obtain SR from the Galilean relativity principle.

One starts with the relativity principle (R.P.) introduced by Galileo:
\begin{itemize}
	\item (R.P.): The laws of physics take the same form in all inertial frame, i.e. these laws are the same for all inertial frames.
\end{itemize} 
In this approach there is not any fundamental scale. Using this postulate, one can derive the composition law of velocities $v^\prime=v_0+v$ that describes the velocity of a projectile measured by an observer, when a second observer,  moving with velocity $v_0$ with respect to the former, sees it with a velocity $v$. One just impose that $v^\prime=f(v_0,v)$ where $f$ must satisfy $f(0, v) = v$, $f(v_0, 0) = v_0$,  $f(v, v_0) = f(v_0, v)$,  $f(-v_0,-v) =- f(v_0, v)$, and by dimensional analysis the known law is obtained.

In SR, Einstein introduced a fundamental velocity scale in such a way that it is consistent with the relativity principle. Then in SR, every observer agrees that the speed of light is $c$. The new relativity principle (Einstein laws, E.L.) can be written as  
\begin{itemize}
	\item (E.L.): The laws of physics involve a fundamental velocity scale $c$, corresponding with the speed of light measured by each inertial observer.
\end{itemize} 
Form (R.P.) and (E.L.) one can read the new expression for the composition of velocities, being $v^\prime=f(v_0,v;c)=(v_0+v)/(1+v_0 v /c^2)$. 

If one wants to include a new invariant scale one can proceed as before:
\begin{itemize}
	\item (L.1.): The laws of physics involve a fundamental velocity scale $c$, and a fundamental length scale $l_P$. 
	\item (L.1.b):  The value of the fundamental velocity scale $c$ can be measured by each
inertial observer as the speed of light with wavelength $\lambda$ much larger than $l_P$
(more rigorously, $c$ is obtained as the $\lambda/ l_P \rightarrow \infty $  limit of the speed of light of wavelength $\lambda$).
\end{itemize}

(L.1.b) appears since the addition of a new length scale would introduce in principle a dependence of the speed of light as a function of the energy of the photon, or equivalently, the speed of the photon would depend on the quotient  $\lambda/ l_P $. So due to this scenario, a new addendum to the relativity principle can be written:
\begin{itemize}
	\item (L.1.c): Each inertial observer can establish the value of $l_p$ (same value for all
inertial observers) by determining the dispersion relation for photons, which takes
the form $E^2- c^2p^2 + f(E, p; l_p) = 0$, where the function $f$ is the same for all
inertial observers. In particular, all inertial observers agree on the leading $l_p$
dependence of $f$: $ f(E, p; l_p) \simeq \eta l_p c \vec{p}^2 E$.
\end{itemize}
Here $\eta$ is the dimensionless constant coefficient of the first term of an infinite series expansion. This expression was used in  Refs.\cite{Amelino-Camelia2001,Amelino-Camelia2002a} only as an example in order to study the possible effects of this new deformed relativity principle.

\subsection{Deformed relativistic kinematics}
\label{sub_DRK}
Since the dispersion relation has changed, the usual Lorentz transformations are no longer valid, and in order to save the relativity principle, one has to consider deformed transformation rules assuring that every inertial observer uses the same dispersion relation\footnote{This is a crucial difference between LIV and DSR. In LIV scenarios there is a deformed dispersion relation, different for each observer, while in DSR the dispersion relation is the same for every observer, so deformed transformation rules are needed.}. 

Normally, it is considered that the isotropy of the space remains unaltered (so rotations are not deformed), but there is a modification of the differential operators 
\begin{equation}
B_i\,=\, i c p_i \frac{\partial}{\partial E} +i [E/c- \eta l_p ( E^2/c^2-\vec{p}^2)] \frac{\partial}{\partial p_i}-i\eta l_p p_i p_j   \frac{\partial}{\partial p_j}\,,
\label{eq:boosts}
\end{equation} 
that represent the generators of ``boosts'' acting on momentum space. The quotation marks are added in order to remark that these transformations are no longer the SR boosts. 

Now we have seen how the kinematics is changed in the one-particle sector, we can wonder what happens while considering a simple scattering process, $a+b\rightarrow c+d$. The conservation law has to be consistent with the deformed transformation rules in order to be valid in every inertial frame (in particular, all observers must agree on whether or not a certain process is allowed). In SR the conservation law is the sum, but in the case we are considering, the deformed composition law is 
\begin{equation}
\begin{split}
E_a\oplus E_b\,&=\,E_a+E_b+l_p c p_a p_b\,,\\ \nonumber
p_a\oplus p_b\,&=\,p_a+p_b+l_p (E_a p_b+p_a E_b)/c\,.
\end{split}  
\end{equation}
One can check that this composition rule is compatible with the deformed transformations~\eqref{eq:boosts}~\footnote{Note that in fact this example can be obtained from the SR kinematics through a change of momentum basis (in Ch.~\ref{chapter_second_order} we will study this in detail).}. 

We see that the main ingredients of a DSR relativistic kinematics are: a deformed dispersion relation, a deformed composition law for the momenta, and nonlinear Lorentz transformations making compatible the two previous pieces with a relativity principle \footnote{There are DSR models where there is no modification in the dispersion relation, and the Lorentz transformations in the one-particle system are linear~\cite{Borowiec2010}.}.

\subsection{Thought experiments in DSR}
\label{sec:thought_experiments}
We have seen in Sec.~\ref{sec:QGT} that one of the ingredients that a QGT should have is a minimum length, and this implies an uncertainty in the measurement of time and position. If DSR considers a minimum length, somehow these uncertainties should appear.

As in Sec.~\ref{sec:QGT}, in order to understand how a minimum length, deformed commutation rules and uncertainties in measurements emerge when the gravitational interaction plays a role in the Heisenberg microscope, in this subsection we will study some thought experiments and see what are the consequences of having a new fundamental length scale introduced in the dispersion relation as before, leading to a momentum dependent velocity for photons~\cite{Amelino-Camelia2002,Amelino-Camelia2002a,AmelinoCamelia:2010pd}.  

\subsubsection{Minimum length uncertainty}

Starting from the dispersion relation for photons $E^2\simeq c^2 p^2+\eta l_p c E p^2$ one can obtain the velocity as 
\begin{equation}
\frac{d E}{d p}\,=\,v_\gamma (p)\,\simeq \,c \left(1+\eta l_p \frac{|p|}{2} \right)\,.
\end{equation}
Since the velocity is momentum dependent, any uncertainty on the momentum of a photon used as a probe for length measurements would induce an uncertainty on its speed.  The first uncertainty comes from the position of the photon, which is related to the momentum uncertainty through $\Delta x_1 \geq 1/\Delta p$ (since we are using units in which $\,\hbar=1$). As there is a velocity uncertainty $\Delta v_\gamma\sim |\eta| \Delta p l_p c$ and the distance traveled during its flight is $L = v_\gamma T$, the uncertainty of the distance is  $\Delta x_2\sim |\eta| \Delta p l_p  L$. As $\Delta x_1$ decreases with $\Delta p$ while $\Delta x_2$ increases with it, one easily finds that $\Delta L\geq \sqrt{\eta L\, l_p}$. If $|\eta|\simeq 1$, this procedure is only meaningful for $L>l_p$, finding $\Delta L>l_p$. This result is obviously observer independent by construction. One can see that the fact of postulating a deformed dispersion relation including a length scale leads to the interpretation of this scale as a minimum length.

\subsubsection{Minimum length and time}

In SR, the Lorentz contraction implies that, given a length measured by an observer, it is always possible to find another observer for which the measured length is arbitrarily small. In DSR, this can be no longer valid, so in order to understand better what would happen in this frame, let us consider a simple thought experiment.

Let us imagine two observers with their own spaceships moving in the same direction with different velocities, i.e. one is at rest and the other is moving with respect to the first one with a velocity $V$. In order to measure the distance between $A$ and $B$, two points on the ship at rest, there is a mirror at $B$ and the distance is measured as half of the time needed by a photon with momentum $p_0$ emitted at $A$ to come back to the initial position after reflection by the mirror. Timing is provided by a digital light clock with the same system used before: a mirror placed at $C$ (at the same rest ship, in a cross direction to $AB$) and a photon with the same energy emitted from $A$ would measure the distance between these two points. The observer at rest measures the distance between $A$ and $B$ obtaining $AB= v_\gamma (p_0) N \tau_0/2$, where $N$ is the number of ticks done by the digital light clock during the journey of the photon traveling from $A\rightarrow B \rightarrow A$ and $\tau_0$ is the interval of time corresponding to each tick of the clock $(\tau_0=2\,AC/v_\gamma (p_0))$. The observer on the second ship moving with velocity $V$ with respect to the one at rest will see that the elapsed time for the photon going from $A\rightarrow C \rightarrow A$ is given by 
\begin{equation}
\tau\,=\,\frac{v_\gamma(p_0)}{\sqrt{v^2_\gamma (p^\prime)-V^2}}\tau_0\,,
\label{eq:tau_int}
\end{equation}
where $p^\prime$ is related to $p_0$ through the formula for boosts in a direction orthogonal to the one of motion of the photon. On the other hand, the observer who sees the ship moving measures that the time in which the photon goes and come back from A to B is given by
\begin{equation}
N \tau\,=\,\frac{AB'}{v_\gamma(p)-V}+\frac{AB'}{v_\gamma(p)+V}\,=\,\frac{2 AB'v_\gamma(p) }{v^2_\gamma(p)-V^2}\,,
\end{equation}
and then,
\begin{equation}
AB'\,=\,\frac{v^2_\gamma(p)-V^2}{v_\gamma (p)}N\frac{\tau}{2}\,,
\label{eq:AB}
\end{equation}
where $p$ is given by the action of a finite boost Eq.~\eqref{eq:boosts} over $p_0$. Combining Eq.\eqref{eq:AB} and Eq.\eqref{eq:tau_int} one obtains 
\begin{equation}
AB'\,=\,\frac{\left(v^2_\gamma(p)-V^2\right)v_\gamma(p_0)}{v_\gamma (p) \sqrt{v^2_\gamma (p^\prime)-V^2}}N\frac{\tau_0}{2}\,=\,\frac{v^2_\gamma(p)-V^2}{v_\gamma (p) \sqrt{v^2_\gamma (p^\prime)-V^2}}AB\,.
\end{equation}
This derivation is analogous to the SR one, where a momentum velocity of the photon is taken into account. The implications of this formula are very easy when one considers the small $V$ and small momentum limit, since one recovers the result of SR. For large $V$, $AB'$ has two important contributions 
\begin{equation}
AB'\,> \frac{\sqrt{c^2-V^2}}{c}\,AB+\frac{\eta c l_p \,AB}{ \sqrt{c^2-V^2}\,AB'}\,,
\end{equation}
where $|p|>|\Delta p|>1/AB'$ is imposed (the probe wavelength must be shorter than the distance being measured). The first one is the usual Lorentz contribution and the last one makes that, for $\eta$ positive and of order 1, $AB'>l_p$ for all values of $V$. This study is only taking into account leading order corrections, so when $V$ is large enough, the correction term is actually bigger than the 0th order contribution to $AB'$. But we see that there is a modification of the Lorentz contraction in such a way that a minimum length appears, unlike in the SR case, where for photons ($V=c$) the measured length is zero. 

\subsubsection{Spacetime fuzziness for classical particles}
The necessity of imposing nonlinear boosts in order to keep the relativity principle has important consequences in the propagation of particles and in the identification of intervals of time. 

Let us consider an observer $O$ who sees two different particles of masses $m_1$ and $m_2$ moving with the same speed and following the same trajectory. Another observer $O'$ boosted with respect to $O$ would see that these particles are ``near'' only for a limited amount of time: the particles would become more and more separated as they move. According to this, the concept of ``trajectory'' should be removed from this picture.

A similar effect occurs considering Eqs.\eqref{eq:boosts} and \eqref{eq:tau_int}. For simplicity, let us consider two photons with energies $E_2$ and $E_1$ such that $v(E_2)=2 v(E_1)$, i.e. the difference in energy is large enough to induce a doubling of the speed, making the time lapsed for the first photon to describe the same trajectory twice that of the second, $\tau_1=2 \tau_2$. But when one considers another observer moving with a speed $V$ with respect to the first one
\begin{equation}
\tau_1^\prime \,=\,\frac{v(E_1)}{\sqrt{v(E_1^\prime)^2-V^2}}\tau_1\,\neq\, 2\tau_2^\prime\,=\,2 \tau_2 \frac{v(E_2)}{\sqrt{v(E_2^\prime)^2-V^2}}\,.
\end{equation}

These implications on spacetime that DSR kinematics is provoking would lead to think that the usual concept of spacetime would have to be modified in this scheme, leading to a new spacetime where these ``paradoxes'' do not appear.

\subsection{Relation with Hopf algebras}
\label{sec:Hopf_algebras}
The example set in (L.1.c) is considering only the first order modification in $l_p$ of the dispersion relation. In order to construct a relativistic kinematics at all order with a fundamental length scale, one needs a new ingredient, a mathematical tool. In this context, the use of Hopf algebras is introduced~\cite{Majid:1995qg}, and a particular example is considered, the deformation of Poincar\'{e} symmetries through quantum algebras known as $\kappa$-Poincar\'{e}~\cite{Lukierski:1992dt,Lukierski:1993df,Majid1994,Lukierski1995}.  

We have previously seen that there are modifications in the kinematics when a minimum length is considered. For the one-particle sector, one has a deformed dispersion relation and a deformed Lorentz transformation. For the two-particle system, a deformed composition law (DCL) for the momenta appears.  In $\kappa$-Poincar\'{e},  there is a modification of the dispersion relation, of the Lorentz symmetries in the one-particle system, and a coproduct of momenta and Lorentz transformations in the two-particle system\footnote{The coproduct of the boost does not appear in the example considered in  Refs.\cite{Amelino-Camelia2001,Amelino-Camelia2002a} due to the fact that the composition law is just the sum expressed in other variables.}. The coproduct of momenta is considered as a deformed composition law and the coproduct of the boosts tell us how one momentum changes under Lorentz transformations in presence of another momentum. One of the most studied bases in $\kappa$-Poincar\'{e} is the bicrossproduct basis\footnote{For this and other bases see Refs.~\cite{KowalskiGlikman:2002we,Lukierski:1991pn}.}. All the ingredients of this basis are
\begin{equation}
\begin{split}
m^2\,&=\,\left(2 \kappa \sinh{\left(\frac{p_0}{2 \kappa}\right)} \right)^2-\vec{p}^2 e^{p_0/\kappa}\,,\\ 
\left[N_i,p_j\right]\,&=\, i \delta_{ij}\left(\frac{\kappa}{2}\left(1-e^{-2p_0/\kappa}\right)+\frac{\vec{p}^2}{2\kappa}\right) -i \frac{p_i p_j}{\kappa}\,, \qquad \left[N_i,p_0\right]\,=\, i p_i\,,\\ 
\Delta\left(M_i\right)\,&=\,M_i \otimes \mathbb{I}+ \mathbb{I}\otimes M_i\,, \qquad \Delta\left(N_i\right)\,=\,N_i \otimes \mathbb{I}+  e^{-p_0/\kappa}\otimes M_i +\frac{1}{\kappa}\epsilon_{i\,j\,k} p_j\otimes M_k\,, \\ 
\Delta\left(p_0\right)\,&=\,p_0 \otimes \mathbb{I}+ \mathbb{I}\otimes p_0\,, \qquad 
\Delta\left(p_1\right)\,=\,p_1 \otimes \mathbb{I}+ \mathbb{I}\otimes e^{-p_0/\kappa}\,. 
\end{split}
\label{eq:coproducts}
\end{equation}

Besides the deformed relativistic kinematics (DRK), as we have seen previously in this section through thought experiments, a minimum length appears, and we know from Sec.\ref{sec:QGT} this fact is related to a noncommutativity of the space-time coordinates. Hopf algebras also gives the commutators of phase space coordinates and in particular, in the bicrossproduct basis of $\kappa$-Poincar\'{e}, the commutators are
\begin{equation}
\begin{split}
\left[x^0,x^i\right]\,&=\,-i\,\frac{x^i}{\kappa} \,,\qquad \left[x^0,p_0\right]\,=\,-i\,\\ 
\left[x^0,p_i\right]\,&=\,i\,\frac{p_i}{\kappa} \,,\qquad \left[x^0,p_i\right]\,=\,-\delta^i_j \,,\qquad \left[x^i,p_0\right]\,=\,0\,. 
\end{split}
\label{eq:pairing_intro}
\end{equation}

We see that this gives all the ingredients that a deformed relativistic kinematics should have and also gives nontrivial commutators in phase space, making Hopf algebras an attractive way for studying DSR theories.

\subsection{Phenomenology}
\label{sec_phenomenology_DSR}
We have said at the subsection dedicated to LIV that there are numerous ways to look for possible LIV effects. But in the context of DSR, the phenomenology is completely different. In LIV, there is not an equivalence of inertial frames, so in order to observe an effect on the threshold of a reaction, the particles involved in the process must have enough energy. The first order correction for the threshold of a reaction is $E^3/m^2 \Lambda$, where $E$ is the energy of a particle involved in the process measured by our Earth-based laboratory frame, and $m$ is a mass that controls the corresponding SR threshold, so the energy has to be high enough in order to have a non-negligible correction. In contrast, in DSR there is a relativity principle, so the threshold of a reaction cannot depend on the observer; there is no new threshold for particle decays at a certain energy of the decaying particle: the energy of the initial particle is not relativistic invariant, so the threshold of such reaction cannot depend on it. Moreover, as a consequence of having a relativity principle, cancellations of effects in the deformed dispersion relation and the conservation law appear~\cite{Carmona:2010ze,Carmona:2014aba}, so many of the effects that can be observed in the LIV case  are completely invisible in this context. 

Then, in principle, the only experiment that can report some observations are time delay of astroparticles\footnote{This is true if the parameter that characterizes the deformation is of the order of the Planck energy, but as we will see in Ch.~\ref{chapter_twin}, if one leaves out this restriction, there could appear other possible observations in the next generation of particle accelerators. This in principle seems absurd because the time delay experiments put high restrictions a to first order deviation to SR~\cite{Vasileiou:2015wja,Abdalla:2019krx,Ellis:2018lca}, but in Refs.~\cite{Carmona:2017oit,Carmona:2018xwm,Carmona:2019oph} it is shown that such a modification does not necessarily imply a time delay. We will study in more detail  how this possibility appears in Ch.~\ref{chapter_time_delay}.}, so many models of emission, propagation and detection of photons and neutrinos have been studied~\cite{AmelinoCamelia:2011cv,Freidel:2011}. In those works, a time delay for photons could appear due to a DDR, which leads to a velocity depending on the energy. For energies much smaller than the Planck one, the DDR can be written in a power series
\be
E^2-\vec{p}^2-m^2\,\approx\,\zeta_n\, E^2\left(\frac{E}{\Lambda}\right)^n\,,
\ee 
where the coefficients $\zeta_n$ are the $n$-th order to the modification of the dispersion relation. Considering the speed as 
\be
v\,=\,\frac{d E}{d p}\,,
\ee
one can check that this leads to a flight time delay 
\be
\Delta t \sim \frac{d}{c} \zeta_n \left(\frac{E}{\Lambda}\right)^n\,,
\ee
where $d$ is the distance between the emission and detection. This time delay could be measured, as in the LIV case, for photons with different energies that come from a very distant source. But unlike the LIV scenario, the time delay depends on the model for the propagation and detection, making important the choice of momentum space-time variables (see Refs.~\cite{Carmona:2017oit,Carmona:2018xwm}). In Ch.~\ref{chapter_time_delay} we will study this phenomenology in more detail.

\subsection{Curved momentum space}
In the 30's, Born considered a duality between spacetime and momentum space, leading to a curved momentum space~\cite{Born:1938} (this idea was discussed also by Snyder some years after~\cite{Snyder:1946qz}). This proposal was postulated as an attempt to avoid the ultraviolet divergences in QFT, and until some years ago, it was not considered as a way to go beyond SR. 

In this work, it was shown that a ``reciprocity'' (name chosen from the lattice theory of crystals) between space-time and momentum variables appears in physics. For example, in the description of a free particle in quantum theory through a plane wave 
\be
\psi (x)\,=\,  e^{i p_\mu \,x^\mu}\,,
\ee
the role played by the $x$'s or the $p$'s are completely identical. $p_\mu$ can be seen as usual as the translation generator of the space-time coordinates, $-i \partial /\partial x^\mu$, or, conversely, $x ^\mu$ can be treated as the translations in momentum space, being $-i \partial /\partial p_\mu$. As in GR, the line element is written as
\be
ds^2\,=\, dx^\mu \,g_{\mu \nu} \,dx^\nu\,.
\ee
Now, thanks to this duality, a line element in momentum space can be considered:
\be
ds^2\,=\, dp_\mu\, \gamma^{\,\mu \nu}\, dp_\nu\,.
\ee
Also, it is proposed that this new metric in momentum space, that has not to be the same that the space-time metric, can be also  a Riemannian metric, leading then to the Einstein's equations
\be
P^{\mu\nu}-\left(\frac{1}{2} \,P+\lambda^\prime\right)\gamma^{\,\mu\nu}\,=\, -\kappa^\prime T^{\mu\nu}\,,
\ee
where $P^{\mu\nu}$ is the Ricci tensor, $P$ the scalar of curvature, and $\lambda^\prime$ and  $\kappa^\prime$ play the dual role of the cosmological constant and $8\pi G$ of GR. We can understand the $T^{\mu\nu}$ term if one keeps in mind that the integrals $\int{T_{\mu 0}dx dy dz}$ are the four momentum of the system considered, so  $\int{T^{\mu 0}dp_x dp_y dp_z}$ would be the space-time coordinates. The interpretation of the $\lambda^\prime$ and  $\kappa^\prime$ and their possible connection with their dual of the space-time counterpart was not clear for Born. 

An anti-de Sitter momentum space is considered in such a way that there is a maximum value for the momentum (note that the aim of this proposal is to get rid of the ultraviolet divergences in QFT, and a cutoff would be necessary). With this assumption, the author sees that a lattice structure for the spacetime appears, an ingredient that, as we have seen, a QGT should have.

In Refs.~\cite{AmelinoCamelia:2008qg,AmelinoCamelia:2011nt,Lobo:2016blj} it was proposed a way to establish a connection between a geometry in momentum space and a deformed relativistic kinematics. In Ch.~\ref{chapter_curved_momentum_space} we will explain this work in depth and will make another proposal to relate a curved momentum space and DSR, in such a way that a noncommutative spacetime appears in a natural way (this will be studied in Ch.~\ref{chapter_locality}), leading to a similar conclusion to the lattice structure obtained in  Ref.~\cite{Born:1938}.  

\subsection{About momentum variables in DSR}
Since the first papers  laying the foundations of the theory, DSR was criticized because it was considered to be just SR in a complicated choice of coordinates~\cite{Jafari:2006rr}. The point is that the deformed dispersion relation and the deformed composition law of momenta proposed in Sec.~\ref{sub_DRK} can be obtained through a change of momentum basis. But in general, a DRK cannot be obtained in such a way, like for example $\kappa$-Poincar\'{e}. Any deformed kinematics with a non-symmetric composition law cannot be SR, because there is no change of momentum variables that reproduce it. In such a way, DSR is safe.

In a collision of particles in DSR, the energy and momentum of the initial and final state particles do not fix the total initial and final momenta, since there are different channels for the reactions due to the non-symmetric  DCL (see for example Ref.~\cite{Albalate:2018kcf}), i.e different total momentum states would be characterized by different orderings of  momenta in the DCL. This will be studied in detail in Ch.~\ref{chapter_twin}. 

Also, there is a controversy about what the physical momentum variables are~\cite{AmelinoCamelia:2010pd}. In SR we use the variables where the conservation law for momenta is the sum and where the dispersion relation is quadratic in momentum. We could wonder why no other coordinates are used. It seems a silly question in the sense that every study in SR  is easier in the usual coordinates, and the use of another (more complicated) ones would be a mess and a waste of time. But in the DSR scheme, this naive and useful argument is no longer valid. There are a lot of representations of $\kappa$-Poincaré~\cite{KowalskiGlikman:2002we}, and in some of them, the dispersion relation is the usual one, but the DCL takes a non-simple  form (the so-called ``classical basis'' is an example); however, in other bases, the DCL is a simple expression but, conversely, the DDR is not trivial (the ``bicrossproduct'' basis). So the criteria used in SR to choose the physical variables cannot be used in these schemes. From the point of view of the algebra, any basis is completely equivalent, but from the point of view of physics, only one should be the nature choice (supposing $\kappa$-Poincar\'{e} is the correct deformation of SR).  Ideally, one could use any momentum variable if it were possible to identify the momentum variable from a certain signal in the detector. The problem resides in the fact that the physics involved in the detection is too complicated to be able to take into account the effect of a change in momentum variables in relation with the detector signal. Maybe some physical criteria could identify the physical momentum variables. 

We have discussed in this section the fact that there are many ways to represent the kinematics of $\kappa$-Poincaré in different momentum variables. But in addition to this particular model, there are also a lot of them characterizing a DRK (this will be studied in Ch.~\ref{chapter_curved_momentum_space}). In the next chapter, we will study how to construct a generic DRK from a simple trick, and how $\kappa$-Poincaré is contained as a particular example.


\chapter{Beyond Special Relativity kinematics}
\label{chapter_second_order}
\ifpdf
    \graphicspath{{Chapter2/Figs/Raster/}{Chapter2/Figs/PDF/}{Chapter2/Figs/}}
\else
    \graphicspath{{Chapter2/Figs/Vector/}{Chapter2/Figs/}}
\fi

\epigraph{Science, my lad, is made up of mistakes, but they are mistakes which it is useful to make, because they lead little by little to the truth.}{Jules Verne, A Journey to the Center of the Earth}


As we have discussed at the end of Sec~\ref{sec:DSR}, there are many ways to consider a relativistic kinematics that goes beyond the usual SR framework. In order to satisfy a relativity principle, all the ingredients that integrate the kinematics must be related, i.e. given a DCL and a DDR, some deformed Lorentz transformations (DLT) making them compatible have to exist.

In Ref.~\cite{AmelinoCamelia:2011yi}, a generic DRK with only first order terms in the high energy scale was studied, analyzing how the ingredients of the kinematics must be related in order to have a relativity principle. In that work, a particular simple process (a photon decay to an electron-positron pair) was used in order to obtain what are called ``golden rules'', relationships between the coefficients of the DDR and DCL. Due to the considered simplifications (the particular choice of the process), only one such a rule was obtained. In Ref.~\cite{Carmona:2012un}, a generalization of the previous work was carried out, without restricting to a particular process, and including a new term proportional to the Levi-Civita symbol in the DCL. In that work, the most general DRK is obtained at first order and it is shown that in fact, there are two golden rules.  

In the previous work there was also a discussion about the choice of momentum variables. It was found that the one-particle sector, i.e. the DDR and the DLT for the one-particle system, can be reduced to SR kinematics just through a nonlinear change of momentum basis. But when the two-particle sector is considered, one cannot carry out this simplification in a general case. Only when the composition law is symmetric, one can find such a change of basis that indicates the considered kinematics is just SR in a fancy choice of momentum variables. The discussion whether different bases reproduce different physics will be treated along this thesis. We consider here that the main ingredient that characterizes if a kinematics is just SR in other variables, is the fact that one can reproduce the SR kinematics through a change of basis\footnote{We will see strong motivations for such assumption in Chs.~\ref{chapter_curved_momentum_space}-\ref{chapter_locality}.}. Obviously, with this prescription, any non-symmetric composition law is a kinematics which goes beyond SR. 

In Ref.~\cite{Carmona:2016obd}, we developed a systematic way to consider different kinematics compatible with the relativity principle. Starting from the SR kinematics, one can obtain the most general DRK at first order studied in~\cite{Carmona:2012un}. With this method, a DRK up to second order is worked out, in such a way that (by construction) it is relativistic invariant, i.e. the appropriate relationship between the ingredients of the kinematics holds. A DRK with only second order additional terms had been previously considered in the DSR literature in relation to the Snyder noncommutativity~\cite{Battisti:2010sr}, but it had not been studied with such a detail before this work. Obviously, possible new physics effects at second order will be of less relevance than first order effects if both corrections are present, which is what is usually considered in the literature for near-future experiments~\cite{AmelinoCamelia:2008qg}. The motivation to go beyond first order DRK is that there are strong limits (constraining the high energy scale to values larger than the Planck one) to this possible modification imposed by experiments that try to measure photon time delays coming from astrophysical sources, like gamma-ray bursts and blazars~\cite{Vasileiou:2015wja,Abdalla:2019krx,Ellis:2018lca}. Also, in Ref.~\cite{Stecker:2014oxa}, a possible explanation for the apparent cutoff in the energy spectrum of neutrinos observed by IceCube is offered assuming second order Planckian physics instead of a first order modification (see Ref.~\cite{Carmona:2019xxp} for an analytic version of the same study). From a theoretical point of view, as we have seen in Sec.~\ref{sec:gup}, a generalized uncertainty principle of the Heisenberg microscope incorporates corrections at second order in the characteristic length. Moreover, in the supersymmetry framework, $d=6$ Lorentz-violating operators (terms proportional to $\Lambda^{-2}$) can suppress Lorentz violation effects at low energies generated through radiative corrections, while these effects appear for $d=5$ Lorentz-violating operators ($\Lambda^{-1}$ corrections to SR)~\cite{Collins2004,Bolokhov2005,Kislat2015}.

In this chapter, we will start by summarizing the results obtained in Ref.~\cite{Carmona:2012un}, and how the results from that work can be reproduced by a change of variables. After that, we will compute the second order DRK with the same procedure, and we will see that the particular example of $\kappa$-Poincar\'e kinematics in the classical basis~\cite{Borowiec2010} is obtained through this method. 

\section{Beyond SR at first order}
\label{sec:firstorder}

\subsection{A summary of previous results}
\label{sec:summary}

We will start by reviewing the results obtained in Ref.~\cite{Carmona:2012un}. The most general expression for a first-order (in a series expansion in the inverse of the high energy scale) DDR compatible with rotational invariance as a function of the components of the momentum is parametrized by two adimensional coefficients $\alpha_1, \alpha_2$:
\begin{equation}
C(p)\,=\,p_0^2-\vec{p}^2+\frac{\alpha_1}{\Lambda}p_0^3+\frac{\alpha_2}{\Lambda}p_0\vec{p}^2=m^2\,,
\label{eq:DDR}
\end{equation}
while the DCL is parametrized by five adimensional coefficients $\beta_1, \beta_2, \gamma_1, \gamma_2, \gamma_3$,
\begin{equation}
\left[p\oplus q\right]_0 \,=\, p_0 + q_0 + \frac{\beta_1}{\Lambda} \, p_0 q_0 + \frac{\beta_2}{\Lambda} \, \vec{p}\cdot\vec{q}\,, \,\,\,\,\,  \left[p \oplus q\right]_i \,=\, p_i + q_i + \frac{\gamma_1}{\Lambda} \, p_0 q_i + \frac{\gamma_2}{\Lambda} \, p_i q_0
+ \frac{\gamma_3}{\Lambda} \, \epsilon_{ijk} p_j q_k \,,
\label{eq:DCL}
\end{equation}
where $\epsilon_{ijk}$ is the Levi-Civita symbol. By definition, a DCL has to satisfy the following conditions
\begin{equation}
(p \oplus q)|_{q=0} \,=\, p\,, \quad \quad  \quad \quad (p \oplus q)|_{p=0} \,=\, q\,,
\label{eq:cl0}
\end{equation}
leaving room only for these five parameters when one assumes a linear implementation of rotational invariance.

The most general form of the Lorentz transformations in the one-particle system is 
\begin{eqnarray}
\left[T(p)\right]_0 &\,=\,& p_0 + (\vec{p} \cdot \vec{\xi}) + \frac{\lambda_1}{\Lambda} \, p_0 (\vec{p} \cdot \vec{\xi})\,, \nonumber \\
\left[T(p)\right]_i &\,=\,& p_i + p_0 \xi_i + \frac{\lambda_2}{\Lambda} \, p_0^2 \xi_i + \frac{\lambda_3}{\Lambda} \, {\vec p}^{\,2} \xi_i + \frac{(\lambda_1 + 2\lambda_2 + 2\lambda_3)}{\Lambda} \, p_i ({\vec p} \cdot {\vec \xi}) \,,
\label{T-one}
\end{eqnarray}
where $\vec{\xi}$ is the vector parameter of the boost, and the $\lambda_i$ are dimensionless coefficients. These expressions are obtained after imposing that these transformations must satisfy the Lorentz algebra, i.e.  the commutator of two boosts corresponds to a rotation. 

The invariance of the dispersion relation under this transformation, $C(T(p))=C(p)$, requires the coefficients of the DDR to be a function of those of the boosts
\begin{equation}
\alpha_1 \,=-\,2(\lambda_1+\lambda_2+2\lambda_3)\,, \quad\quad \alpha_2\,=\,2(\lambda_1+2\lambda_2+3\lambda_3)\,.
\label{alphalambda}
\end{equation}

As we have mentioned previously, a modification in the transformations of the two-particle system is needed in order to have a relativity principle, making the DLT to depend on both momenta. Then, we are looking for a DLT such that $(p,q) \to (T^L_q(p),T^R_p(q))$, where
\begin{equation}
T^L_q(p) \,=\, T(p) + {\bar T}^L_q(p)\,, {\hskip 1cm} T^R_p(q) \,=\, T(q) + {\bar T}^R_p(q) \,.
\label{eq:boost2}
\end{equation}
When one considers the most general transformation in the two-particle system and imposes that they are Lorentz transformations and that they leave the DDR invariant, the final form for the DLT in the two-particle system is obtained:
\begin{eqnarray}
\left[{\bar T}^L_q(p)\right]_0 &\,=\,& \frac{\eta_1^L}{\Lambda} \, q_0 ({\vec p} \cdot {\vec \xi})  + \frac{\eta_2^L}{\Lambda} \, ({\vec p} \wedge {\vec q}) \cdot {\vec \xi}\,, \nonumber \\
\left[{\bar T}^L_q(p)\right]_i &\,=\,&  \frac{\eta_1^L}{\Lambda} \, p_0 q_0 \xi_i + \frac{\eta_2^L}{\Lambda}\left( \, q_0 \epsilon_{ijk} p_j \xi_k - p_0 \epsilon_{ijk} q_j \xi_k\right)+\frac{\eta_1^L}{\Lambda}\left(q_i ({\vec p} \cdot {\vec \xi})-({\vec p} \cdot {\vec q}) \xi_i \right) \,,
 \nonumber \\
\left[{\bar T}^R_p(q)\right]_0 &\,=\,& \frac{\eta_1^R}{\Lambda} \, p_0 ({\vec q} \cdot {\vec \xi}) + \frac{\eta_2^R}{\Lambda} \, ({\vec q} \wedge {\vec p}) \cdot {\vec \xi}\,, \nonumber \\
\left[{\bar T}^R_p(q)\right]_i &\,=\,& \frac{\eta_1^R}{\Lambda} \, q_0 p_0 \xi_i-\frac{\eta_2^R}{\Lambda} \left( p_0 \epsilon_{ijk} q_j \xi_k-q_0 \epsilon_{ijk} p_j \xi_k\right)+\frac{\eta_1^R}{\Lambda} \left(p_i ({\vec q} \cdot {\vec \xi}) - ({\vec q} \cdot {\vec p}) \xi_i\right) \,.\nonumber \\
\label{eq:gen2pboost}
\end{eqnarray}

The last step is to find the relationship between the boosts and the DCL coefficients. In order to do so, we have to impose the relativity principle for a simple process, a particle with momentum $(p \oplus q)$ decays into two particles of momenta $p$ and $q$. The relativity principle imposes that the conservation law for the momenta must be satisfied for every inertial observer, that is,
\begin{equation}
T(p\oplus q)\,=\,T_q^L(p)\oplus T_p^R(q) \,.
\label{eq:RP-1}
\end{equation}
This imposes the following relations between the coefficients of the DLT and DCL:
\begin{alignat}{3}
\beta_1 &\,=\, 2 \,(\lambda_1 + \lambda_2 + 2\lambda_3)\,, \quad\quad & 
\beta_2 &\,=\, -2 \lambda_3 - \eta_1^L - \eta_1^R\,, \quad\quad & \label{betalambda}\\
\gamma_1 &\,=\, \lambda_1 + 2\lambda_2 + 2\lambda_3 - \eta_1^L\,, \quad\quad & 
\gamma_2 &\,=\, \lambda_1 + 2\lambda_2 + 2\lambda_3 - \eta_1^R\,, \quad\quad & \gamma_3 \,=\, \eta_2^L - \eta_2^R \,.
\label{gammalambda}
\end{alignat}
Now one can establish a relationship between the DDR and DCL through the DLT coefficients, i.e.  the two conditions imposed by the relativity principle, the ``golden rules'', over the adimensional coefficients of the DDR and DCL:
\begin{equation}
\alpha_1\,=\,-\beta_1\,, \quad \quad \quad \alpha_2\,=\,\gamma_1+\gamma_2-\beta_2\,.
\label{eq:GR}
\end{equation}

\subsection{Change of variables and change of basis}
\label{sec:change}

In Ref.~\cite{Carmona:2016obd} we introduced a new mathematical trick in order to avoid the previous tedious computation that leads to Eqs.~\eqref{alphalambda}, \eqref{betalambda}, \eqref{gammalambda}. The main idea is that one can construct the same DRK in an easy way through a change of variables in the two-particle system. We will distinguish here between two different changes of momentum variables. The first one is what we will call a \emph{change of basis}  $(p_0,\vec{p}) \to (P_0,\vec{P})$, following Ref.~\cite{Carmona:2012un}, where the new momentum variables are just a function of the old ones (preserving rotational invariance),
\begin{equation}
\begin{split}
p_0& \,=\,P_0+\frac{\delta_1}{\Lambda}P_0^2+\frac{\delta_2}{\Lambda}\vec{P}^2\equiv \mathcal{B}_0(P_0,\vec{P})\equiv \mathcal{B}_0(P)\,, \\
p_i&=P_i+\frac{\delta_3}{\Lambda} P_0 P_i \equiv \mathcal{B}_i(P)\,.
\end{split}
\label{eq:ch-base}
\end{equation} 
This change of basis is the same for all particles involved in a process. The name of change of basis is taken form the Hopf algebra context, where different coproducts (DCL) in the $\kappa$-Poincar\'e deformation are related through this kind of transformation. From a geometrical point of view, different bases would correspond to different momentum coordinates on a curved momentum space. So from the algebraical and geometrical perspectives, a change of basis has no content. However, this is not the case from the point of view of DSR, where different bases are supposed to be physically nonequivalent\footnote{This will be treated from a theoretical point of view in Ch.~\ref{chapter_locality}, and the possible phenomenological consequences of this non-equivalence in Ch.~\ref{chapter_time_delay}.}. 

Let us consider that $(P_0,\vec{P})$ are the SR momentum variables with a linear conservation law, i.e the sum. Let us see what is the new kinematics one gets with the change of basis~\eqref{eq:ch-base} (this was done systematically in Ref.~\cite{Carmona:2012un}). In order to compute the DCL in the new variables one has to use the inverse of~\eqref{eq:ch-base}
\begin{equation}
\begin{split}
P_0& \,=\,p_0-\frac{\delta_1}{\Lambda}p_0^2-\frac{\delta_2}{\Lambda}\vec{p}^2\equiv \mathcal{B}^{-1}_0(p)\,, \\
P_i&\,=\,p_i-\frac{\delta_3}{\Lambda} p_0 p_i \equiv \mathcal{B}^{-1}_i(p)\,,
\end{split}
\label{eq:invch-base}
\end{equation}
and then
\begin{equation}
(P+Q)_0\,=\,P_0+Q_0=\mathcal{B}^{-1}_0(p)+\mathcal{B}^{-1}_0(q)\,=\,p_0+q_0-\frac{\delta_1}{\Lambda} (p_0^2+q_0^2)-\frac{\delta_2}{\Lambda}(\vec{p}^2+\vec{q}^2)\,.
\label{eq:DCL-1}
\end{equation}
We see that this cannot define a DCL since it does not satisfy~\eqref{eq:cl0}. However, this condition can be implemented as 
\begin{equation}
(p\oplus q)_\mu \,\equiv\, \mathcal{B}_\mu\left(\mathcal{B}^{-1}(p)+\mathcal{B}^{-1}(q)\right)\,.
\label{eq:DCLdef}
\end{equation}
This procedure is used in the DSR literature in order to describe the ``physical variables'' from the ``auxiliary variables'' (the SR variables which compose and transform linearly)~\cite{Judes:2002bw}. One then  gets
\begin{align}
(p\oplus q)_0&\, =\,p_0+q_0+\frac{2\delta_1}{\Lambda}p_0 q_0+\frac{2\delta_2}{\Lambda}\vec{p}\cdot\vec{q}\,,\\
(p\oplus q)_i& \,=\,p_i+q_i+\frac{\delta_3}{\Lambda}p_0 q_i+\frac{\delta_3}{\Lambda}q_0 p_i\,.
\label{eq:DCL-2}
\end{align}
We see that there is a relationship between the $\delta$'s and $\beta$'s and $\gamma$'s, in such a way that only the symmetric part of the composition law is reproduced. In particular, one sees that $\beta_1=2\delta_1$, $\beta_2=2\delta_2$, $\gamma_1=\gamma_2=\delta_3$, and $\gamma_3=0$. 

Besides the modification in the conservation law, a change of basis produces a modification in the dispersion relation and in the Lorentz transformation in the one-particle system. In particular, if $(P_0,\vec{P})$ transform as in SR under a Lorentz boost,  $P'_0=P_0+\vec{P}\cdot\vec{\xi}\,$, $\vec{P}'=\vec{P}+P_0\vec{\xi}\,$, the new boosts for the new momentum coordinates are 
\begin{align}
[T(p)]_0 \equiv \mathcal{B}_0(P')&\,=\,P_0+\vec{P}\cdot\vec{\xi}+\frac{\delta_1}{\Lambda}(p_0+\vec{p}\cdot\vec{\xi})^2+\frac{\delta_2}{\Lambda}(\vec{p}+p_0\vec{\xi})^2\nonumber \\
&=\,p_0+\frac{(2\delta_1+2\delta_2-\delta_3)}{\Lambda}p_0\vec{p}\cdot\vec{\xi}\,, \\
[T(p)]_i \equiv \mathcal{B}_i(P')&\,=\,P_i+P_0\xi_i+\frac{\delta_3}{\Lambda}(p_0+\vec{p}\cdot\vec{\xi})(p_i+p_0\xi_i)  \nonumber \\
&=\,p_i+p_0 \xi_i+\frac{(\delta_3-\delta_1)}{\Lambda}p_0^2 \xi_i-\frac{\delta_2}{\Lambda}\vec{p}^2\xi_i+\frac{\delta_3}{\Lambda}p_i \vec{p}\cdot\vec{\xi} \, .
\label{eq:ch-LT}
\end{align}
Again, we can compare these results with Eq.~\eqref{T-one},
\begin{equation}
\lambda_1\,=\,2\delta_1+2\delta_2-\delta_3\,,\qquad \lambda_2\,=\,\delta_3-\delta_1 \,,\qquad \lambda_3\,=-\,\delta_2\,.
\label{eq:lambdafromdelta}
\end{equation}

For the DDR, one obtains 
\begin{equation}
C(p)\equiv P_0^2-\vec{P}^2\,=\,p_0^2-\vec{p}^2-\frac{2\delta_1}{\Lambda}p_0^3+\frac{2(\delta_3-\delta_2)}{\Lambda}p_0\vec{p}^2\,=\,m^2,
\label{eq:ch-DR}
\end{equation}
and, comparing with the results of Eq.~\eqref{eq:DDR},
\begin{equation}
\alpha_1\,=\,-2\delta_1=-\beta_1\,, \quad \quad \alpha_2\,=\,2(\delta_3-\delta_2)=\gamma_1+\gamma_2-\beta_2\,.
\label{eq:alphafromdelta}
\end{equation}
By construction, all the results agree with Eqs.~\eqref{eq:lambdafromdelta} and~\eqref{alphalambda}, and with the golden rules~\eqref{eq:GR}. 

The other kind of transformations will be denoted \emph{change of variables}, which is in fact a change of variables in the two-particle system, where
\be
\begin{split}
(P,Q) \to (p,q)=(\mathcal{F}^L(P,Q),\mathcal{F}^R(P,Q)) \,\text{ such that } \mathcal{F}^L(P,0)\,=\,P \,,\, \mathcal{F}^L(0,Q)\,=\,0\,, \\ 
\text{ and } \mathcal{F}^R(0,Q)\,=\,Q \,,\, \mathcal{F}^R(P,0)\,=\,0\,.
\label{eq:restrictedch}
\end{split}
\ee
While a change of basis has a clear interpretation in the algebraic or geometric approaches, this is not the case for this kind of transformation. It is only used as a mathematical trick in order to compute a DRK from SR kinematics in a simple way without doing all the tedious work showed in Ref.~\cite{Carmona:2012un} for the first order case. In fact, we are going to see that this trick reproduces the most general DRK at first order. 

Let us consider again that $(P,Q)$ are the SR variables with linear Lorentz transformations and conservation law. Instead of considering a change of basis as in Eq.~\eqref{eq:ch-base}, we are going to see what DRK is obtained from the change of variables of Eq.~\eqref{eq:restrictedch}, which will be compatible with a relativity principle by construction. Since we are only considering a change of variables and not a change of basis, we will reproduce the DRK and their relation with the $\eta$'s coefficients obtained in Eqs.~\eqref{betalambda}-\eqref{gammalambda} without the $\lambda$'s, as they only appear when a change of basis is explored. We will denote this kind of basis where there are no modification of the Lorentz transformations in the one-particle system the  \emph{classical basis}, as it is usually the name appearing in the Hopf algebras scheme (see Sec.~\ref{sec:Hopf}). So through a change of variables, we can only construct a DRK in the classical basis.

In order to check that the kinematics obtained through a change of variables is compatible with the relativity principle, we start by defining
\begin{equation}
p\oplus q \equiv P+Q\,,
\label{eq:DCL-ch}
\end{equation}
which satisfies the condition~\eqref{eq:cl0},
\begin{equation}
(p \oplus q)|_{q=0} \,=\,\left[\left(\mathcal{F}^{-1}\right)^L (p,q)+ \left(\mathcal{F}^{-1}\right)^R (p,q)\right]_{q=0}\,=\,p+0\,=\,p\,,
\label{eq:demo1}
\end{equation}
being then a good definition of a DCL. Here we have used the inverse of the change of variables, $P=\left(\mathcal{F}^{-1}\right)^L (p,q)$, $Q=\left(\mathcal{F}^{-1}\right)^R (p,q)$, and the features of  Eq.~\eqref{eq:restrictedch}.  As we are not changing the basis, the total momentum also transforms linearly, as a single momentum does.

The transformations of $(p,q)$ are given by
\begin{equation}
T_q^L(p)\equiv\mathcal{F}^L(P',Q')\,,  \quad \quad T_p^R(q)\equiv\mathcal{F}^R(P',Q')\,,
\label{eq:MTL-ch}
\end{equation}
where $P'$, $Q'$, are the linear Lorentz transformed momenta of $P$, $Q$. Then,
\begin{equation}
T_q^L(p) \oplus T_p^R(q)\,= \,\mathcal{F}^L(P',Q') \oplus \mathcal{F}^R(P',Q') \,= \,P'+Q' \,= \,(P+Q)'\, =\, T(p\oplus q)\, ,
\label{eq:demo2}
\end{equation}
where we have used Eq.~\eqref{eq:MTL-ch}, that $P$ and $Q$ transform linearly, and the definition appearing in Eq.~\eqref{eq:DCL-ch}. We can see that condition~\eqref{eq:RP-1} is automatically satisfied when a change of variables is made.

Since we have not deformed the one-particle transformations, the Casimir is the SR one, $C(p)=p_0^2-\vec{p}^2=m^2$ (this is a particular characteristic of the classical basis, being the Casimir different for any other basis). 

One cannot take the most general expression of the change of variables compatible with rotational invariance because, as it was explained in Sec.~\ref{sec:summary}, the Casimir of the two-particle system must be $P_0^2-\vec{P}^2=p_0^2-\vec{p}^2$, $Q_0^2-\vec{Q}^2=q_0^2-\vec{q}^2$. Once this condition is implemented at first order in $1/\Lambda$, one obtains
\begin{equation}
\begin{split}
P_{0}\,=\,p_{0}+\frac{v_{1}^{L}}{\Lambda}\vec{p}.\vec{q}\,,\qquad &  P_{i}\,=\,p_{i}+\frac{v_{1}^{L}}{\Lambda}p_{0}q_{i}+\frac{v_{2}^{L}}{\Lambda}\epsilon_{ijk}p_{j}q_{k}\,,
\\
Q_{0}\,=\,q_{0}+\frac{v_{1}^{R}}{\Lambda}\vec{p}.\vec{q}\,,\qquad & Q_{i}\,=\,q_{i}+\frac{v_{1}^{R}}{\Lambda}q_{0}p_{i}+\frac{v_{2}^{R}}{\Lambda}\epsilon_{ijk}q_{j}p_{k}\,.
\label{ch-var}
\end{split}
\end{equation}
Using  definition~\eqref{eq:DCL-ch}, one gets
\begin{equation}
\begin{split}
\left[p\oplus q\right]_0&\,=\,P_0+Q_0=p_{0}+q_{0}+\frac{v_{1}^{L}+v_{1}^{R}}{\Lambda}\vec{p}.\vec{q}\,,
\\
\left[p\oplus q\right]_i&\,=\,P_i+Q_i=p_{i}+q_{i}+\frac{v_{1}^{L}}{\Lambda}p_{0}q_{i}+\frac{v_{1}^{R}}{\Lambda}q_{0}p_{i}+\frac{v_{2}^{L}-v_{2}^{R}}{\Lambda}\epsilon_{ijk}p_{j}q_{k}\,,
\end{split}
\label{chvar-cl-1st}
\end{equation}
so we can establish a correspondence with Eq.~\eqref{eq:DCL}, obtaining
\begin{align}
\beta_{1}\,=\,0\,,{\hskip1cm}\beta_{2}\,=\,v_{1}^{L}+v_{1}^{R}\,,
{\hskip1cm}\gamma_{1}\,=\,v_{1}^{L}\,,{\hskip1cm}\gamma_{2}\,=\,v_{1}^{R}\,,{\hskip1cm}\gamma_{3}\,=\,v_{2}^{L}-v_{2}^{R}\,.
\label{DCLpar-ch}
\end{align}
We see that we have obtained the general solution of the golden rules appearing in Eq.~\eqref{eq:GR} when $\alpha_1=\alpha_2=0$ (note that this is the main property of the classical basis). 

One can obtain from Eq.~\eqref{eq:MTL-ch} the transformation law in the two-particle system:
\begin{align}
\eta_{1}^{L,R}\,=\,-v_{1}^{L,R}\,,\qquad \eta_{2}^{L,R}\,=\,v_{2}^{L,R}\,.
\label{eta-v}
\end{align}
In this way, we obtain Eqs.~\eqref{betalambda}-\eqref{gammalambda} in the particular case when all $\lambda$'s vanish. Then, one can combine a change of basis and change of variables in order to obtain the most general DRK at first order in $1/\Lambda$, as there is a one-to-one correspondence between the parameters of the kinematics and those of the change of basis~\eqref{eq:ch-base} and the change of variables~\eqref{ch-var}. We will now assume that this property is true at higher orders, so that through this mathematical trick we can obtain the most general relativistic kinematics at any order from what we will denote a covariant composition law.\footnote{There is not a proof of this assumption, but even if this is not true, this is a way to produce examples of DRKs without tedious calculations.}
 
\subsection{Covariant notation}
\label{sec:covariant}

Before studying the kinematics at second order, it is very convenient to use a covariant notation in order to simplify the calculations. We will study now how this can be done for the change of variables satisfying again
\be
(P, 0) \to (P, 0)\,, {\hskip 1cm} (0, Q) \to (0, Q) \,,{\hskip 1cm} P^2 \,=\, p^2\,, {\hskip 1cm} Q^2\, =\, q^2 \,.
\ee
In order for the Casimir in the two-particle system to be the same as the Casimir that appears in the one-particle system, the change of variables must be such that the terms proportional to  $(1/\Lambda)$ appearing in $P$ must be orthogonal to the momentum $p$, and also to $q$. We will introduce a fixed vector $n$ and the Levi-Civita tensor with $\epsilon_{0123}=-1$ in order to rewrite the deformed kinematics. The most general change of variables with this requirement written in covariant notation is
\begin{align}
P_\mu &\,=\, p_\mu + \frac{v_1^L}{\Lambda} \left[q_\mu (n\cdot p) - n_\mu (p\cdot q)\right] + \frac{v_2^L}{\Lambda} \epsilon_{\mu\nu\rho\sigma} p^\nu q^\rho n^\sigma\,, \\
Q_\mu &\,= \,q_\mu + \frac{v_1^R}{\Lambda} \left[p_\mu (n\cdot q) - n_\mu (p\cdot q)\right] + \frac{v_2^R}{\Lambda} \epsilon_{\mu\nu\rho\sigma} q^\nu p^\rho n^\sigma \,.
\label{p,q}
\end{align}
One obtains the previous results of~\eqref{ch-var} when $n_\mu=(1,0,0,0)$.~\footnote{The expression obtained is not really covariant because $n$ does not transform under Lorentz transformations like a vector (because it is a fixed vector).} 

As before, we suppose that $P$ variables compose additively, and then we obtain the composition law in the $p$ variables
\be
\begin{split}
\left(p\oplus q\right)_\mu \equiv  \left[P \bigoplus Q\right]_\mu \,=\, P_\mu + Q_\mu & \,= \,p_\mu + q_\mu + \frac{v_1^L}{\Lambda} (n\cdot p) q_\mu + \frac{v_1^R}{\Lambda} (n\cdot q) p_\mu \\ & - \frac{(v_1^L+v_1^R)}{\Lambda} n_\mu (p\cdot q) +
\frac{(v_2^L-v_2^R)}{\Lambda} \epsilon_{\mu\nu\rho\sigma} p^\nu q^\rho n^\sigma \,.
\end{split}
\label{cl1}
\ee
Taking $n_\mu = (1, 0, 0, 0)$ in this composition law we obtain
\be
\begin{split}
\left[p\oplus q\right]_0 &\,= \,p_0 + q_0 + \frac{(v_1^L + v_1^R)}{\Lambda} \vec{p}\cdot \vec{q}\,, \\
\left[p\oplus q\right]_i &\,= \,p_i + q_i + \frac{v_1^L}{\Lambda} p_0 q_i + \frac{v_1^R}{\Lambda} q_0 p_i + \frac{(v_2^L - v_2^R)}{\Lambda} \epsilon_{ijk} p_j q_k \,,
\end{split}
\ee
giving the same result appearing in Eq.~\eqref{chvar-cl-1st}.  

In order to simplify the notation for the Lorentz transformations of Secs.~\ref{sec:summary}-\ref{sec:change}, we will denote by $(p',q')$ or $(P',Q')$ the transformed momenta of $(p,q)$ or $(P,Q)$, instead of the previous convention of $T_q^L(p)$or $T_p^R(q)$. Also, we will use the notation $\tilde{X}_\mu \equiv \Lambda_\mu^{\:\nu} X_\nu$, where the $\Lambda_\mu^{\:\nu}$ are the usual Lorentz transformation matrices. Then, we see that
\be
\begin{split}
&p'_\mu + \frac{v_1^L}{\Lambda} \left[q'_\mu (n\cdot p') - n_\mu (p'.q')\right] + \frac{v_2^L}{\Lambda} \epsilon_{\mu\nu\rho\sigma} p^{\prime\,\nu} q^{\prime\,\rho} n^\sigma \equiv P'_\mu \,, \\
& =\, \Lambda_\mu^{\:\nu} P_\nu \,= \,\tilde{p}_\mu + \frac{v_1^L}{\Lambda} \left[\tilde{q}_\mu (\tilde{n}\cdot\tilde{p}) - \tilde{n}_\mu (\tilde{p}\cdot\tilde{q})\right] + \frac{v_2^L}{\Lambda} \epsilon_{\mu\nu\rho\sigma} \tilde{p}^\nu \tilde{q}^\rho \tilde{n}^\sigma \,.
\end{split}
\ee
We can realize that $p'$ is equal to $\tilde{p}$ when one neglects terms proportional to $(1/\Lambda)$, which is obvious since the Lorentz transformations of the new momentum variables are a consequence of the change of variables. So at first order we have
\be
p'_\mu \,= \,\tilde{p}_\mu + \frac{v_1^L}{\Lambda} \left[\tilde{q}_\mu \left((\tilde{n}-n)\cdot\tilde{p}\right) - \left(\tilde{n}_\mu - n_\mu\right) (\tilde{p}\cdot\tilde{q})\right] + \frac{v_2^L}{\Lambda} \epsilon_{\mu\nu\rho\sigma} \tilde{p}^\nu \tilde{q}^\rho (\tilde{n}^\sigma - n^\sigma) \,.
\ee
For an infinitesimal Lorentz transformation, we have
\be
\tilde{X}_\mu \,= \,X_\mu + \omega^{\alpha\beta} \eta_{\mu\alpha} X_\beta \,,
\ee
where $\omega^{\alpha\beta}=-\omega^{\beta\alpha}$ are the infinitesimal transformation parameters, and then
\be
p'_\mu \,= \,\tilde{p}_\mu + \omega^{\alpha\beta} n_\beta \left[\frac{v_1^L}{\Lambda} \left(q_\mu p_\alpha - \eta_{\mu\alpha} (p\cdot q)\right) + \frac{v_2^L}{\Lambda} \epsilon_{\mu\alpha\nu\rho} p^\nu q^\rho\right] \,.
\label{p'1}
\ee
A similar procedure leads to the transformation for the second variable $q$ 
\be
q'_\mu \,=\, \tilde{q}_\mu + \omega^{\alpha\beta} n_\beta \left[\frac{v_1^R}{\Lambda} \left(p_\mu q_\alpha - \eta_{\mu\alpha} (p\cdot q)\right) + \frac{v_2^R}{\Lambda} \epsilon_{\mu\alpha\nu\rho} q^\nu p^\rho\right]\,.
\label{q'1}
\ee
Eqs.~(\ref{p'1})-(\ref{q'1}) are the new Lorentz transformations of the variables $(p, q)$ at first order.   
For $n_\mu = (1, 0, 0, 0)$, one obtains
\be
\omega^{0\beta} n_\beta \,= \,0\,, {\hskip 1cm} \omega^{i\beta} n_\beta \,= \,\omega^{i0}=\xi^i = -\xi_i \,,
\ee
and hence, (\ref{p'1})-(\ref{q'1}) become
\be
\begin{split}
p'_0 &\,= \,p_0 + (\vec{p} \cdot \vec{\xi}) - \frac{v_1^L}{\Lambda} q_0 (\vec{p} \cdot \vec{\xi}) + \frac{v_2^L}{\Lambda} ({\vec p} \wedge {\vec q}) \cdot {\vec \xi}\,, \\ 
p'_i &\,= \,p_i + p_0 \xi_i - \frac{v_1^L}{\Lambda} \left(q_i (\vec{p} \cdot \vec{\xi}) - (p\cdot q) \xi_i\right) - \frac{v_2^L}{\Lambda} \left(q_0 (\vec{p} \wedge \vec {\xi})_i - p_0 (\vec{q} \wedge \vec {\xi})_i\right)\,, \\
q'_0 &\,= \,q_0 + (\vec{q} \cdot \vec{\xi}) - \frac{v_1^R}{\Lambda} p_0 (\vec{q} \cdot \vec{\xi}) + \frac{v_2^R}{\Lambda} ({\vec q} \wedge {\vec p}) \cdot {\vec \xi}\,, \\ 
q'_i &\,= \,q_i + q_0 \xi_i - \frac{v_1^R}{\Lambda} \left(p_i (\vec{q} \cdot \vec{\xi}) - (p\cdot q) \xi_i\right) - \frac{v_2^R}{\Lambda} \left(p_0 (\vec{q} \wedge \vec {\xi})_i - q_0 (\vec{p} \wedge \vec {\xi})_i\right) \,,
\end{split}
\label{DLT-1st}
\ee
leading to the same result of Eq.~\eqref{eta-v}.

With this, we conclude the discussion of how a DRK at first order can be obtained from a change of variables in covariant notation. Now we can consider a change of basis written in covariant notation in order to reproduce the previous results. The most general expression of a change of basis has also three terms,
\be
X_\mu \,=\, \hat{X}_\mu + \frac{b_1}{\Lambda} \hat{X}_\mu (n\cdot\hat{X}) + \frac{b_2}{\Lambda} n_\mu \hat{X}^2 + \frac{b_3}{\Lambda} n_\mu (n\cdot\hat{X})^2
\label{eq:basiscov}
\ee
(where $X$ stands for $p$ or $q$), which leads to the  dispersion relation
\be
m^2 \,=\, p^2 \,= \,\hat{p}^2 + \frac{2(b_1+b_2)}{\Lambda} \hat{p}^2 (n\cdot\hat{p}) + \frac{2 b_3}{\Lambda} (n\cdot\hat{p})^3 \,,
\label{drhat1}
\ee
and also to a new composition law at first order $\hat{p}\ohat\hat{q}$ obtained from Eqs.~\eqref{eq:DCLdef} and~\eqref{eq:basiscov}
\be
\left[p\oplus q\right]_\mu \,=\, \left[\hat{p}\ohat\hat{q}\right]_\mu + \frac{b_1}{\Lambda} (\hat{p}+\hat{q})_\mu \left(n\cdot(\hat{p}+\hat{q})\right) + \frac{b_2}{\Lambda} n_\mu (\hat{p}+\hat{q})^2 + \frac{b_3}{\Lambda} n_\mu \left(n\cdot(\hat{p}+\hat{q})\right)^2 \,.
\ee 
In the hat variables, the DCL finally reads
\be
\begin{split}
\left[\hat{p}\ohat \hat{q}\right]_\mu & \,=\, \hat{p}_\mu + \hat{q}_\mu + \frac{v_1^L - b_1}{\Lambda} (n\cdot\hat{p}) \hat{q}_\mu + \frac{v_1^R - b_1}{\Lambda} (n\cdot\hat{q}) \hat{p}_\mu \\ & - \frac{(v_1^L+v_1^R) + 2 b_2}{\Lambda} n_\mu (\hat{p}\cdot\hat{q}) - \frac{2 b_3}{\Lambda} n_\mu (n\cdot\hat{p}) (n\cdot\hat{q}) + \frac{(v_2^L-v_2^R)}{\Lambda} \epsilon_{\mu\nu\rho\sigma} \hat{p}^\nu \hat{q}^\rho n^\sigma \,.
\end{split}
\label{clhat1}
\ee   
Taking $n_\mu=(1, 0, 0, 0)$ in Eqs.~\eqref{drhat1} and~\eqref{clhat1}, we obtain the new DDR for these variables
\be
\begin{split}
m^2 \,&=\, \hat{p}^2 + \frac{2(b_1+b_2)}{\Lambda} \hat{p}^2 \hat{p}_0 + \frac{2 b_3}{\Lambda} \left(\hat{p}_0\right)^3\\
\,&=\, 
\hat{p}_0^2 - {\vec{\hat{p}}}^2 + \frac{2(b_1+b_2+b_3)}{\Lambda} \left(\hat{p}_0\right)^3 - \frac{2(b_1+b_2)}{\Lambda}\hat{p}_0 \vec{\hat{p}}^2 \,,
\end{split}
\ee
to be compared with Eq.~\eqref{eq:DDR}, and
\be
\begin{split}
& \left[\hat{p}\ohat\hat{q}\right]_0 \,=\, \hat{p}_0 + \hat{q}_0 - \frac{2(b_1+b_2+b_3)}{\Lambda} \hat{p}_0 \hat{q}_0 + \frac{(v_1^L + v_1^R) + 2 b_2}{\Lambda}\, \vec{\hat{p}}\cdot \vec{\hat{q}}\,, \\
& \left[\hat{p}\ohat \hat{q}\right]_i \,=\, \hat{p}_i + \hat{q}_i
+ \frac{v_1^L - b_1}{\Lambda} \hat{p}_0 \hat{q}_i + \frac{v_1^R - b_1}{\Lambda} \hat{q}_0 \hat{p}_i + \frac{(v_2^L - v_2^R)}{\Lambda} \epsilon_{ijk} \hat{p}_j \hat{q}_k \,,
\end{split}
\ee
to be compared with Eq.~\eqref{eq:DCL}. As expected,  the golden rules~\eqref{eq:GR} are satisfied, obtaining the same results appearing in Sec.~\ref{sec:summary} for a DRK at first order (DDR and DCL), compatible with rotational invariance.

\section{Beyond SR at second order}
\label{sec:second}

In this section we will obtain a deformed kinematics at second order by performing a change of variables from momentum variables which transform linearly. As we will see, this does not imply that the composition law of the original variables is just the sum. We could get a general kinematics by making a change of basis over the obtained kinematics, but since we want to compare our results with those of the literature, and in particular with the kinematics derived from Hopf algebras, this is not mandatory. We will compare our kinematics with the one obtained in the Hopf algebra framework in the classical basis, where the one-particle momentum variable transforms linearly.  

\subsection{Change of variables up to second order}

We proceed as in Sec.~\ref{sec:covariant} finding the most general expression of a change of variables at second order $(P, Q) \to (p, q)$ compatible with $p^2=P^2$, $q^2=Q^2$. The complete calculation is developed in Appendix~\ref{appendix_second_order_a}. Here, we only summarize the main results and procedures. As in the first order case,  we first start by obtaining the most general change of variables compatible with $p^2=P^2$ up to second order, obtaining Eqs.~\eqref{P->p}-\eqref{Q->q}, which have a total of 14 parameters $(v_1^L,\ldots,v_7^L;v_1^R,\ldots,v_7^R)$. 

As we are applying a change of variables to momenta which transform linearly, the starting composition law (which we will call \textit{covariant} composition law, also noted in Ref.~\cite{Ivetic:2016qtz}) will be a sum of terms covariant under linear Lorentz transformations:
\be
\left[P\bigoplus Q\right]_\mu \,= \,P_\mu + Q_\mu + \frac{c_1}{\Lambda^2} P_\mu Q^2 + \frac{c_2}{\Lambda^2} Q_\mu P^2 + \frac{c_3}{\Lambda^2} P_\mu (P\cdot Q) + \frac{c_4}{\Lambda^2} Q_\mu (P\cdot Q) \,.
\label{ccl2}
\ee
Then, we can obtain a generic DCL and DLT in the two-particle system by applying a generic change of variables to this covariant composition law.

As it is showed in Appendix~\ref{appendix_second_order_a}, a generic DCL obtained through a change of variables up to second order has coefficients depending on 16 parameters: the four parameters appearing in the covariant composition law~\eqref{ccl2}, and 12 combinations of the 14 parameters of the change of variables \eqref{P->p}-\eqref{Q->q}. For the case $n_\mu=(1, 0, 0, 0)$, the composition law reads
\be
\begin{split}
&\left[p\oplus q\right]_0 \,=\, p_0 + q_0 + \frac{(v_1^L + v_1^R)}{\Lambda} \vec{p}\cdot\vec{q} + \frac{(2 c_1 - v_1^L v_1^L- 2v_3^R)}{2 \Lambda^2} p_0 q^2 + \frac{(2 c_2 - v_1^R v_1^R - 2 v_3^L)}{2 \Lambda^2} q_0 p^2 \\ &  + 
\frac{(2 c_3 + v_1^R v_1^R - v_2^R v_2^R - 2 v_4^L - 2 v_5^R)}{2 \Lambda^2} p_0 (p\cdot q) + \frac{(2 c_4 + v_1^L v_1^L - v_2^L v_2^L - 2 v_5^L - 2 v_4^R)}{2 \Lambda^2} q_0 (p\cdot q) \\ & + \frac{(v_2^R v_2^R + 2 v_3^L+ 2 v_4^L + 2 v_5^R)}{2 \Lambda^2} p_0^2 q_0 + \frac{(v_2^L v_2^L + 2 v_3^R + 2 v_5^L + 2 v_4^R)}{2 \Lambda^2} p_0 q_0^2\,, \\
&\left[p\oplus q\right]_i\,= \,p_i + q_i + \frac{v_1^L}{\Lambda} p_0 q_i + \frac{v_1^R}{\Lambda} q_0 p_i + \frac{(v_2^L - v_2^R)}{\Lambda} \epsilon_{ijk}p_{j} q_{k} + \frac{(2c_1 - v_2^L v_2^L)}{2 \Lambda^2} p_i q^2 + \\ &  \frac{(2 c_2 - v_2^R v_2^R)}{2 \Lambda^2} q_i p^2 + \frac{(2c_3 - v_1^R v_1^R + v_2^R v_2^R)}{2 \Lambda^2} p_i (p\cdot q) + \frac{(2c_4 - v_1^L v_1^L + v_2^L v_2^L)}{2 \Lambda^2} q_i (p\cdot q)  + \\ & \frac{(v_2^R v_2^R + 2 v_4^L)}{2 \Lambda^2} p_0^2 q_i + \frac{(v_2^L v_2^L + 2 v_4^R)}{2 \Lambda^2} p_i q_0^2  + \frac{(v_3^L+ v_5^R)}{\Lambda^2} p_i p_0 q_0 + \frac{(v_3^R + v_5^L)}{\Lambda^2} q_i p_0 q_0 \\ &  + \frac{(v_6^L-v_7^R)}{\Lambda^2} p_0\, \epsilon_{ijk}p_{j} q_{k} + \frac{(v_7^L-v_6^R)}{\Lambda^2} q_0\, \epsilon_{ijk}p_{j} q_{k} \,,
\end{split}
\label{generalCL}
\ee
where we can identify, following the same notation used at first order, the dimensionless coefficients:
{\footnotesize
\begin{align}
\beta_1 & = 0 &  \beta_2 & = v_1^L + v_1^R & 2\beta_3 &= 2 c_1 - v_1^L v_1^L- 2v_3^R \nonumber \\
2\beta_4 &= 2 c_2 - v_1^R v_1^R - 2 v_3^L  & 2\beta_5 &=  2 c_3 + v_1^R v_1^R - v_2^R v_2^R - 2 v_4^L - 2 v_5^R & 2\beta_6 &= 2 c_4 + v_1^L v_1^L - v_2^L v_2^L - 2 v_5^L - 2 v_4^R  \nonumber \\
2\beta_7 &= v_2^R v_2^R + 2 v_3^L+ 2 v_4^L + 2 v_5^R & 2\beta_8 & =v_2^L v_2^L + 2 v_3^R + 2 v_5^L + 2 v_4^R & \gamma_1 & = v_1^L \nonumber \\
\gamma_2 & =v_1^R & \gamma_3 &=v_2^L - v_2^R & 2\gamma_4 &= 2c_1 - v_2^L v_2^L \nonumber \\
2\gamma_5 &= 2 c_2 - v_2^R v_2^R & 2\gamma_6 &=2c_3 - v_1^R v_1^R + v_2^R v_2^R & 2\gamma_7 &=2c_4 - v_1^L v_1^L + v_2^L v_2^L\nonumber \\
2\gamma_8 &=v_2^R v_2^R + 2 v_4^L & 2\gamma_9 &=v_2^L v_2^L + 2 v_4^R & \gamma_{10} & =v_3^L+ v_5^R \nonumber \\
\gamma_{11} &=v_3^R + v_5^L & \gamma_{12} &=v_6^L-v_7^R & \gamma_{13} &=v_7^L-v_6^R\,.
\label{DCLpar-ch-2nd}
\end{align}}
\normalsize 
These are the generalization of Eq.~\eqref{DCLpar-ch} at second order. As it was shown in the first-order case, we can obtain the golden rules at second order by using the relations in Eq.~\eqref{DCLpar-ch-2nd}
\be
\begin{split}
\beta_1\,=\,\beta_2 - \gamma_1 - \gamma_2 &\,=\,0\\
\beta_3 + \beta_6 - \gamma_4 -\gamma_7 + \gamma_9 +\gamma_{11} -\frac{\gamma_1^2}{2}&\,=\,0\\
\beta_4 + \beta_5 - \gamma_5 -\gamma_6 + \gamma_8 +\gamma_{10} -\frac{\gamma_2^2}{2}&=0\\
\beta_7 - \gamma_8 -\gamma_{10}=\beta_8 - \gamma_9 -\gamma_{11} &\,=\,0\,.
\end{split}
\label{gr-upto-2nd}
\ee

Also, in Appendix~\ref{appendix_second_order_a}, we obtain the DLT in the two-particle system, $(p,q) \to (p',q')$, using the same procedure used to obtain Eqs.~\eqref{p'1}-\eqref{q'1} in Sec.~\ref{sec:covariant}. Their coefficients depend on the 14 parameters that characterize a generic change of variables in the two-particle system. The 4 parameters $c_i$ of the covariant composition law do not appear, since a covariant DCL is compatible with linear LT. Note also that only at first order the coefficients of the DCL are determined by the ones of the DLT. This is no longer true at second order due to the Lorentz covariant terms that can be present in the DCL. 

The previous kinematics can be generalized by means of a change of basis that will modify the DDR, making it invariant at second order. In the following subsection we will consider a simplified case in which the corrections to SR start directly at second order.  

\subsection{Change of variables and change of basis starting at second order}
\label{sec:second order}

As we saw at the beginning of the chapter, there are some phenomenological indications, and also theoretical arguments, that seem to suggest that the corrections in a deformed kinematics could start at second order. In this subsection, we will study this case, finding the most general DRK in the same way we did in Sec.~\ref{sec:covariant} for the first order case.

The DRK obtained  from a change of variables can be easily obtained by making  $v_1^L, v_1^R , v_2^L, v_2^R$ equal to zero in Eqs.~\eqref{cl2}-\eqref{q'2}.

The change of basis starting at second order is
\be
X_\mu \,= \,\hat{X}_\mu + \frac{b_4}{\Lambda^2} n_\mu \hat{X}^2 (n\cdot\hat{X}) + \frac{b_5}{\Lambda^2} \hat{X}_\mu (n\cdot\hat{X})^2 + \frac{b_6}{\Lambda^2} n_\mu (n\cdot\hat{X})^3 \,,
\label{eq:basiscov2}
\ee
that leads to the DDR
\be
m^2 \,= \,p^2\, =\, \hat{p}^2 + \frac{2(b_4+b_5)}{\Lambda^2} \hat{p}^2 (n\cdot\hat{p})^2 + \frac{2 b_6}{\Lambda^2} (n\cdot\hat{p})^4 \,.\
\label{drhat2}
\ee

Choosing $n_\mu=(1, 0, 0, 0)$ in Eq.~\eqref{drhat2}, we obtain 
\be
m^{2}=\hat{p}_{0}^{2}-{\vec{\hat{p}}}^{2}+\frac{\alpha_{3}}{\Lambda^2}\left(\hat{p}_{0}\right)^{4}+\frac{\alpha_{4}}{\Lambda^2}(\hat{p}_{0})^{2}\vec{\hat{p}}^{2} \,,
\ee
with 
\be
\alpha_{3}=2(b_{4}+b_{5}+b_{6})\,,\qquad\alpha_{4}=-2(b_{4}+b_{5})\,,
\ee
which is the DDR that generalizes Eq.~\eqref{eq:DDR} when the corrections to SR start at second order.

The DCL coefficients in this particular case are obtained from Eqs.~\eqref{DCLpar-ch-2nd} are now:
\begin{align}
\beta_3 &= c_1 - v_3^R -b_4 & \beta_4 &= c_2  -  v_3^L -b_4 & \beta_5 &=  c_3 -  v_4^L -  v_5^R-2b_4  \nonumber \\
\beta_6 &=  c_4 -  v_5^L -  v_4^R -2 b_4 & \beta_7 &=  v_3^L+ v_4^L +  v_5^R- 3b_5- 3b_6 & \beta_8 & =  v_3^R +  v_5^L + v_4^R -3b_5 - 3b_6 \nonumber \\
\gamma_4 &= c_1 & \gamma_5 &= c_2  & \gamma_6 &=c_3 \nonumber \\ \gamma_7 &=c_4 & \gamma_8 &= v_4^L-b_5 & \gamma_9 &= v_4^R-b_5 \nonumber \\
\gamma_{10} & =v_3^L+ v_5^R-2b_5 &
\gamma_{11} &=v_3^R + v_5^L-2b_5 & \gamma_{12} &=v_6^L-v_7^R \nonumber \\ \gamma_{13} &=v_7^L-v_6^R\,.
\label{DCLpar-ch-only-2nd}
\end{align}
As we did for the first order case Eq.~\eqref{eq:GR}, we can find the golden rules at second order:
\be
\begin{split}
\beta_3 + \beta_6 - \gamma_4 -\gamma_7 + \gamma_9 +\gamma_{11}\, =\, 
\beta_4 + \beta_5 - \gamma_5 -\gamma_6 + \gamma_8 +\gamma_{10} &\,=\, \frac{3}{2} \, \alpha_4\,,\\
\beta_7 - \gamma_8 -\gamma_{10}=\beta_8 - \gamma_9 -\gamma_{11} &\,=\, -\frac{3}{2} \, (\alpha_3 + \alpha_4)\,.
\end{split}
\label{gr-at-2nd}
\ee

\subsection{Generalized kinematics and the choice of momentum variables}
\label{sec:choice}

In the preceding subsections, we have constructed a DRK at second order in $(1/\Lambda)$ through a change of variables. At the beginning of the chapter, we have mentioned that there is a controversy about the physical meaning of the momentum variables. While there is a clear distinction between kinematics related through a change of variables, those kinematics related through a change of basis are completely equivalent from the algebraic and geometric point of views. However, from a physical point of view, this may not be the case. 


Whatever the situation is, one can wonder whether there are DRKs that cannot be obtained from SR with the procedure proposed in the previous subsections. As we saw in  Sec.~\ref{sec:firstorder}, the most general DRK at first order can be obtained following this prescription. We will see now that this is not the case for a general DRK at second order. The difference lays in the covariant terms of the composition law~\eqref{ccl2}, that cannot be generated by a covariant change of basis.  

In order to do so, we start with the additive composition law in the variables $\left\{ \hat{P}\,,\hat{Q}\right\}$. We can make a covariant change of basis
\be
\tilde{P}_{\mu}\,=\,\hat{P}_{\mu}\left(1+\frac{b}{\Lambda^{2}}\hat{P}^{2}\right)\,,
\ee  
that leaves 
the dispersion relation invariant since $\hat{P}^{2}$ is an invariant. As we did in Eq.~\eqref{eq:DCLdef}, we can find the DCL generated by this change of basis 
\be
\left[\tilde{P}\tilde{\bigoplus} \tilde{Q}\right]_{\mu}\,=\,\tilde{P}_{\mu}+\tilde{Q}_{\mu}-\frac{b}{\Lambda^{2}}\tilde{P}_{\mu}\tilde{Q}^{2}-\frac{b}{\Lambda^{2}}\tilde{Q}_{\mu}\tilde{P}^{2}-\frac{2b}{\Lambda^{2}}\tilde{P}_{\mu}(\tilde{P}\cdot\tilde{Q})-\frac{2b}{\Lambda^{2}}\tilde{Q}_{\mu}(\tilde{P}\cdot\tilde{Q}) \,.
\ee
Moreover, in order to obtain a generic DCL from the procedure proposed in this chapter, we need to make a covariant change of variables in such a way that momentum variables do not mix in the dispersion relations
\be
\tilde{P}_{\mu}\,=\,P_{\mu}+\frac{v^{L}}{\Lambda^{2}}\left(Q_{\mu}P^{2}-P_{\mu}(P\cdot Q)\right)\,,\qquad\tilde{Q}_{\mu}\,=\,Q_{\mu}+\frac{v^{R}}{\Lambda^{2}}\left(P_{\mu}Q^{2}-Q_{\mu}(P\cdot Q)\right) \,.
\ee
We finally obtain
\begin{equation}
\begin{split}
\left[P\bigoplus Q\right]_{\mu}\,=&\,P_{\mu}+Q_{\mu}+\frac{v^{R}-b}{\Lambda^{2}}P_{\mu}Q^{2}+\frac{v^{L}-b}{\Lambda^{2}}Q_{\mu}P^{2}\\&-\frac{v^{L}+2b}{\Lambda^{2}}P_{\mu}(P\cdot Q)-\frac{v^{R}+2b}{\Lambda^{2}}Q_{\mu}(P\cdot Q) \,.
\end{split}
\label{cclcv}
\end{equation}
As one can see comparing Eq.~\eqref{cclcv} with Eq.~\eqref{ccl2}, there are three parameters in the DCL obtained through a change of basis and a change of variables, while in a generic covariant composition law there are four. This shows the impossibility of obtaining the most general covariant composition law with the methods we used for the first order case. 
 
Summarizing, 17 out of the 18 parameters $(v_i^L, v_i^R , c_i)$ can be reproduced by a change of basis and variables. This means that not any DRK can be obtained through this procedure from the linear composition (the sum) but, at least up to second order, it is possible to obtain the most general composition law applying a change of basis and a change of variables to a generic covariant composition. The parameter that cannot be generated is a combination of the coefficients of the covariant composition law $c_i$. 

\section{Relation with the formalism of Hopf algebras}
\label{sec:Hopf}

We can compare now our results of the previous subsections with the kinematics obtained in the formalism of Hopf algebras. Since we have obtained the most general kinematics up to second order with linear Lorentz transformations in the one-particle system, we are able to see if there is a correspondence with the so-called classical basis of  $\kappa$-Poincaré~\cite{Borowiec2010}:
\be
\Delta\left(N_{i}\right)\,=\,N_{i}\otimes \mathbb{1}+\left(\mathbb{1}-\frac{P_{0}}{\Lambda}+\frac{P_{0}^{2}}{2\Lambda^{2}}+\frac{\vec{P}^{2}}{2\Lambda^{2}}\right)\otimes N_{i}-\frac{1}{\Lambda}\epsilon_{ijk}P_{j}\left(\mathbb{1}-\frac{P_{0}}{\Lambda}\right)\otimes J_{k}\,,
\label{co-boost}
\ee
\begin{align}
\Delta\left(P_{0}\right)\,=\,&P_{0}\otimes\left(\mathbb{1}+\frac{P_{0}}{\Lambda}+\frac{P_{0}^{2}}{2\Lambda^{2}}-\frac{\vec{P}^{2}}{2\Lambda^{2}}\right)+\left(\mathbb{1}-\frac{P_{0}}{\Lambda}+\frac{P_{0}^{2}}{2\Lambda^{2}}+\frac{\vec{P}^{2}}{2\Lambda^{2}}\right)\otimes P_{0}\nonumber\\&+\frac{1}{\Lambda}P_{m}\left(\mathbb{1}-\frac{P_{0}}{\Lambda}\right)\otimes P_{m} \,,
\label{co-p0} \\
\Delta\left(P_{i}\right)\,=\,&P_{i}\otimes\left(\mathbb{1}+\frac{P_{0}}{\Lambda}+\frac{P_{0}^{2}}{2\Lambda^{2}}-\frac{\vec{P}^{2}}{2\Lambda^{2}}\right)+\mathbb{1}\otimes P_{i}\, .
\label{co-pi}
\end{align}
One can see that $[\Delta(N_j),C\otimes\mathbb{1}]=[\Delta(N_j),\mathbb{1}\otimes C]=0$, since the Casimir of the algebra $C$, commutes with the $(P_0,P_i,J_i,N_i)$ generators. This shows that the Casimir is trivially extended to the tensor product of the algebras (or in our language of Sec.~\ref{sec:summary}, that the DDR does not mix momentum variables).

In order to find the relation between these algebraic expressions and the kinematical language used in this thesis, we can consider that the generators of the Poincaré algebra act as operators on the basis of the momentum operator, $P_\mu \p=p_\mu \p$. The boost generators $N_j$ in SR satisfy
\be
 \pp \,=\, (\mathbb{1}-i\xi_j N_j+\mathcal{O}(\xi^2)) \p\,,
\ee
where $\pp\equiv \p'$ is the transformed state from $\p$ with a boost. Neglecting terms of order $\mathcal{O}(\xi^2)$, we find
\begin{align}
-i\xi_j [N_j,P_\mu]\p &\,=\, - i\xi_j (N_j P_\mu-P_\mu N_j)\p \,=\, p_\mu (\pp-\p)-p'_\mu\pp+p_\mu\p \nonumber \\ 
& \,=\, (p-p')_\mu\pp\,=\,(p-p')_\mu\p+\mathcal{O}(\xi^2) \,.
\label{deriv}
\end{align}
From here we obtain
\be
p'_\mu \,=\, p_\mu + i\xi_j \fj\,,
\label{relac-1}
\ee
where $\fj$ are the eigenvalues of $[N_j,P_\mu]$, being a function of the $P_\mu$: 
\be
f_j(P_\mu)\p\,\equiv\, [N_j,P_\mu]\p \,= \,\fj \p\,.
\ee

These relations can be extended for the two-particle system. Then, we can define 
\be
(P_\mu \otimes \mathbb{1})\pqp \,=\, p'_\mu\pqp\,, \quad \quad
(\mathbb{1} \otimes P_\mu)\pqp\, =\, q'_\mu\pqp\,,
\ee
and the generators of co-boosts, $\Delta(N_j)$, satisfying
\be
\pqp \,=\, (\mathbb{1}-i\xi_j \Delta(N_j)+\mathcal{O}(\xi^2)) \pq\,.
\ee
So Eq.~\eqref{relac-1} is generalized to
\be
p'_\mu \,=\, p_\mu + i\xi_j \fja\,, \quad \quad q'_\mu\,=\,q_\mu+i\xi_j\fjb\,,
\label{co-transformed}
\ee 
where $\fja$ and $\fjb$ are the eigenvalues of $[\Delta(N_j),P_\mu\otimes \mathbb{1}]$ and $[\Delta(N_j),\mathbb{1}\otimes P_\mu]$, respectively:
\be
[\Delta(N_j),P_\mu\otimes \mathbb{1}] \pq \,=\, \fja \pq\,, \quad \quad [\Delta(N_j),\mathbb{1}\otimes P_\mu]\pq \,=\, \fjb\pq\,.
\label{co-transformed2}
\ee
Finally, the coproduct $\Delta(P_\mu)$ acting in the two-particle system momentum space is
\be
\Delta(P_\mu)\pq\,=\, \cop\pq\,.
\label{coprod-CL}
\ee

With the previous relations, we can now make explicit the correspondence between our language and that of $\kappa$-Poincaré. From Eq.~\eqref{coprod-CL} and Eqs.~\eqref{co-p0} and \eqref{co-pi}, the DCL of $\kappa$-Poincaré in the classical basis is
\be
\begin{split}
(p\oplus q)_0&\,=\,p_0+q_0+\frac{\vec{p}\cdot\vec{q}}{\Lambda}+\frac{p_0}{2\Lambda^2}\left(q_0^2-\vec{q}^2\right) + \frac{q_0}{2\Lambda^2}\left(p_0^2+\vec{p}^2\right) - \frac{p_0}{\Lambda^2}(\vec{p}\cdot \vec{q})\,,\\
(p\oplus q)_i&\,=\,p_i+q_i+\frac{q_0 p_i}{\Lambda} + \frac{p_i}{\Lambda^2}\left(q_0^2-\vec{q}^2\right)\,.
\end{split}
\label{eq:kappa-CL}
\ee
From the coproduct of the boost, Eq.~\eqref{co-boost}, and using Eqs.~\eqref{co-transformed}-\eqref{co-transformed2}, together with the usual commutation relations $[N_i,P_0]=-iP_j$, $[N_i,P_j,]=i\delta_{ij}P_0$, $[J_i,P_j]=i\epsilon_{ijk}P_m$ (we are working in the classical basis, where the Lorentz transformations in the one-particle system are linear), we obtain
\be
\begin{split}
p'_0&\,=\,p_0+\vec{p}\cdot\vec{\xi}\,, \quad \quad \quad \quad p'_i\,=\,p_i+p_0\xi_i \,, \\
q'_0&\,=\,q_{0}+\vec{q}.\vec{\xi}\left(1-\frac{p_{0}}{\Lambda}+\frac{p_{0}^{2}}{2\Lambda^{2}}+\frac{\vec{p}^{2}}{2\Lambda^{2}}\right)\,,
\\
q'_{i}&\,=\,q_{i}+q_{0}\xi_{i}\left(1-\frac{p_{0}}{\Lambda}+\frac{p_{0}^{2}}{2\Lambda^{2}}+\frac{\vec{p}^{2}}{2\Lambda^{2}}\right)+(\vec{p}\cdot\vec{q})\xi_{i}\left(\frac{1}{\Lambda}-\frac{p_{0}}{\Lambda^{2}}\right)+\vec{q}\cdot\vec{\xi}\left(-\frac{p_{i}}{\Lambda}+\frac{p_{0}p_{i}}{\Lambda^{2}}\right) \,.
\end{split}
\label{eq:kappa-transformed}
\ee
Comparing Eq.~\eqref{eq:kappa-CL} with Eq.~\eqref{generalCL}, and Eq.~\eqref{eq:kappa-transformed} with Eqs.~\eqref{generaltr1}-\eqref{generaltr4}, we see that the choice of the coefficients that reproduces $\kappa$-Poincaré in the classical basis is 
\[
v_{1}^{R}\,=\,1\,,\qquad c_{1}\,=\,c_{3}\,=\,\frac{1}{2} \,,
\]
being the rest of the parameters equal to zero. We can see that, as expected, $\kappa$-Poincaré is a particular case of our general framework that includes a DRK beyond SR up to second order in the power expansion of $\kappa$ ($1/\Lambda$). In fact, we can reproduce the covariant terms of $\kappa$-Poincaré kinematics with $b=v^L=-v^R/2=-1/6$. 

We have found a systematic way to obtain all the possible DRKs up to second order, but this work can be generalized order by order.  Then, we have plenty of ways to go beyond SR. A physical criteria is needed in order to constrain the possible kinematics, an additional ingredient that is still not clear. In this sense, to consider a different framework might lead to a better understanding of how a DRK appears and what represents from a physical point of view. This will be the aim of the next chapter, where we will study how a DRK naturally emerges from the geometry properties of a curved momentum space.  

\chapter{Curved momentum space}
\label{chapter_curved_momentum_space}
\ifpdf
    \graphicspath{{Chapter3/Figs/Raster/}{Chapter3/Figs/PDF/}{Chapter3/Figs/}}
\else
    \graphicspath{{Chapter3/Figs/Vector/}{Chapter3/Figs/}}
\fi

\epigraph{Equations are just the boring part of mathematics. I attempt to see things in terms of geometry.}{Stephen Hawking}

As we have mentioned in the previous chapters, Hopf algebras are a mathematical tool which has been used as a way to characterize a DRK, considering the DDR as the Casimir of the Poincar\'{e} algebra in a certain basis, and the DCL as given by the coproduct operation. The description of symmetries in terms of Hopf algebras introduces a noncommutative spacetime~\cite{Majid:1999tc} that can be understood as the dual of a curved momentum space. In the particular case of the deformation of $\kappa$-Poincar\'{e}~\cite{Lukierski:1991pn}, the noncommutative spacetime that arises is $\kappa$-Minkowski, as we have shown in the Introduction, from which one can deduce a momentum geometry corresponding to de Sitter~\cite{KowalskiGlikman:2002ft}. 

In Refs.~\cite{AmelinoCamelia:2011bm,Amelino-Camelia:2013sba,Lobo:2016blj} there are other proposals that try to establish a relation between a geometry in momentum space and a deformed kinematics. In Ref.~\cite{AmelinoCamelia:2011bm}, the DDR is defined as the squared of the distance in momentum space from the origin to a point $p$, and the DCL is associated to a non-metrical connection. The main problem of this work is that there is no mention to Lorentz transformations, and then to a relativity principle, the fundamental ingredient of a DRK. 

Another proposal was presented in Ref.~\cite{Amelino-Camelia:2013sba}, achieving a different path to establish a relation between a DCL and a curved momentum space through a connection, which in this case can be  (but it is not mandatory) affine to the metric that defines the DDR in the same way as before. This link is carried out by parallel transport, implemented by a connection in momentum space, which indicates how momenta must compose. They found a way to implement some DLT implementing the relativity principle; with this procedure any connection could be considered, giving any possible DRK, and then, this would reduce to the study of a generic DRK as we did in the previous chapter.  

In Ref.~\cite{Lobo:2016blj}, a possible correspondence between a DCL and the isometries of a curved momentum space related to translations (transformations that do not leave the origin invariant) is considered. The Lorentz transformations are the homogeneous transformations (leaving the origin invariant), in such a way that a relativity principle holds if the DDR is compatible with the DCL and this one with the DLT. As one would want 10 isometries (6 boosts and 4 translations), one should consider only maximally symmetric spaces. Then, there is only room for three options: Minkowski, de Sitter or anti-de Sitter momentum space.

However, in Ref.~\cite{Lobo:2016blj} there is not a clear way to obtain the DCL, because in fact, there are a lot of isometries that do not leave the origin invariant, so a new ingredient is mandatory. Moreover, the relativity principle argument is not really clear since one needs to talk about the transformed momenta of a set of two particles, as we saw in Ch.~\ref{chapter_second_order}. 

In this chapter, we will first review the geometrical framework proposed in Ref.~\cite{AmelinoCamelia:2011bm}. After that, we will make clear our proposal~\cite{Carmona:2019fwf}. We present a precise way to understand a DCL: it is associated to translations, but in order to find the correct one, we must impose their generators to form a concrete subalgebra inside the algebra of isometries of the momentum space metric. 

We will see how the much studied  $\kappa$-Poincar\'{e} kinematics can be obtained from our proposal. In fact, the method we propose can be used in order to obtain other DRKs, such as Snyder~\cite{Battisti:2010sr} and the so-called hybrid models~\cite{Meljanac:2009ej}.

Finally, we will see the correspondence between our prescription and the one proposed in Refs.~\cite{AmelinoCamelia:2011bm,Amelino-Camelia:2013sba}. 

\section{Momentum space geometry in relative locality}

In Ref.~\cite{AmelinoCamelia:2011nt}, a physical observer who can measure the energies and momenta of particles in her vicinity is considered. 
This observer can define a metric in momentum space by performing measurements in a one-particle system, and a (non-metrical) connection by performing measurements in a multi-particle system.

The one-particle system measurement allows the observer to determine the geometry of momentum space through the dispersion relation, considering it as the square of the geodesic distance from the origin to a point $p$ in momentum space, which corresponds to the momentum of the particle,
\begin{equation}
D^2(p)\,\equiv \,D^2(0,p)\,=\,m^2\,.
\end{equation}

The kinetic energy measurement defines the geodesic distance between two particles of mass $m$: $p$, which is at rest, and another particle $p'$ with kinetic energy $K$, i.e. $D(p)=D(p')=m$, and 
\begin{equation}
D^2(p,p')\,=\,-2 m K\,,
\end{equation}
where the minus sign appears since we are considering a Lorentzian momentum manifold. From both measurements she can reconstruct a metric in momentum space
\begin{equation}
dk^2\,=\, h^{\mu\nu}(k)dk_\mu dk_\nu\,.
\end{equation}
This metric must reduce to the Minkowski space in the limit $\Lambda\rightarrow 0$. Also, they argued this metric must possess 10 isometries  (transformations that leave the form of the metric invariant), 6 related with Lorentz transformations and 4 with translations, so the only possible metrics are those that correspond to a maximally symmetric space, leading to only three options: Minkowski, de Sitter or anti-de Sitter momentum spaces. 

From the measurement of a system of particles, she can deduce the composition law of momenta, which is an operation that joins two momenta, and in order to consider more particles, the total momentum is computed by gathering momenta in pairs. The authors define also a momentum called  \textit{antipode} $\hat{p}$ (which was previously introduced in the context of Hopf algebras~\cite{Majid:1995qg}) in such a way that $\hat{p}\oplus p=0$.

This composition law, which in principle cannot be assumed linear, nor commutative, nor associative, defines the geometry of momentum space related to the algebra of combinations of momentum. The connection at the origin is 
\be
\Gamma^{\tau \lambda}_\nu (0)\,=\,-\left.\frac{\partial^2  (p\oplus q)_\nu}{\partial p_\tau \partial q_\lambda}\right\rvert_{p,q \rightarrow 0}\,,
\ee
and the torsion 
\be
T^{\tau \lambda}_\nu (0)\,=\,-\left.\frac{\partial^2  \left((p\oplus q)-(q\oplus p)\right)_\nu}{\partial p_\tau \partial q_\lambda}\right\rvert_{p,q \rightarrow 0}\,.
\ee
The curvature tensor is determined from the lack of associativity of the composition law
\be
R^{\mu\nu\rho}_\sigma (0)\,=\,2 \left.\frac{\partial^3  \left((p\oplus q)\oplus k-p\oplus (q\oplus k)\right)_\sigma}{\partial p_{[\mu} \partial q_{\nu]} \partial k_\rho}\right\rvert_{p,q,k \rightarrow 0}\,,
\ee
where the bracket denotes the anti-symmetrization. They suggested that the non-associativity of the composition law, giving a non-vanishing curvature tensor in momentum space, could be tested with experiments. 

In order to obtain the connection at any point, they defined a new composition depending on another momentum $k$ 
\be
(p\oplus_k q) \,\doteq\, k\oplus\left((\hat{k}\oplus p)\oplus(\hat{k}\oplus q)\right)\,.
\label{k-DCL}
\ee
Then, they claimed that the connection at a point $k$ can be determined by
\be
\Gamma^{\tau \lambda}_\nu (k)\,=\,-\left.\frac{\partial^2  (p\oplus_{k}q)_\nu}{\partial p_\tau \partial q_\lambda}\right\rvert_{p,q \rightarrow k}\,.
\label{k-connection}
\ee

In principle, the connection is not metrical in the sense that it is not the affine connection given by the metric defining the DDR. This fact is argued from the construction they gave, separating the dispersion relation from the composition law from the very beginning. 

But a DRK is not only composed of a DDR and a DCL. In order to have a relativity principle, a DLT for the one and two-particle systems must make the previous ingredients compatible. The Lorentz transformations of the one-particle system are proposed to be determined by the metric, being directly compatible with the DDR (the explicit expression of the distance is invariant under isometries). However, it is not clear how to implement the two-particle transformations, making all the ingredients of the kinematics compatible with each other. 

In the next section, we present another proposal which tries to avoid these problems and puts all the ingredients of the kinematics in the same framework.

\section{Derivation of a DRK from the momentum space geometry}
\label{sec:derivation}

As we have commented previously, a DRK is composed of a DDR, a DCL and, in order to have a relativity principle, a DLT for the one and two-particle systems, making the previous ingredients compatible. In this section we will explain how we propose to construct a DRK from the geometry of a maximally symmetric momentum space. 

\subsection{Definition of the deformed kinematics}

In a maximally symmetric space, there are 10 isometries. We will denote our momentum space metric as $g_{\mu\nu}(k)$\footnote{There is a particular choice of coordinates in momentum space which leads the metric to take the simple form $g_{\mu\nu}(k)=\eta_{\mu\nu} \pm k_\mu k_\nu/\Lambda^2$, where the de Sitter (anti-de Sitter) space corresponds with the positive (negative) sign.}. By definition, an isometry is a transformation  $k\to k'$ satisfying 
\be
  g_{\mu\nu}(k') \,=\, \frac{\partial k'_\mu}{\partial k_\rho} \frac{\partial k'_\nu}{\partial k_\sigma} g_{\rho\sigma}(k)\, .
\ee

One can always take a system of coordinates in such a way that $g_{\mu\nu}(0)=\eta_{\mu\nu}$, and we write the isometries in the form
\be
k'_\mu \,=\, [T_a(k)]_\mu \,=\, T_\mu(a, k)\,, \quad\quad\quad k'_\mu \,=\, [J_\omega(k)]_\mu \,=\,J_\mu(\omega, k)\,,
\ee
where $a$ is a set of four parameters and $\omega$ of six, and 
\be
T_\mu(a, 0) \,=\, a_\mu\,, \quad\quad\quad J_\mu(\omega, 0) \,=\, 0\,,
\ee
so $J_\mu(\omega, k)$ are the 6 isometries forming a subgroup that leave the origin in momentum space invariant, and $T_\mu(a, k)$ are the other 4 isometries which transform the origin and that one can call translations.

We will identify the isometries $k'_\mu = J_\mu(\omega, k)$ with the DLT of the one-particle system, being $\omega$ the six parameters of a Lorentz transformation. The dispersion relation is defined, rather than as the square of the distance from the origin to a point $k$ (which was the approach taken in the previous section), as any arbitrary function of this distance with the SR limit when the high energy scale tends to infinity\footnote{This disquisition can be avoided with a redefinition of the mass with the same function $f$ that relates the Casimir with the distance $C(k)=f(D(0,k))$.}. Then, under a Lorentz transformation, the equality $C(k)=C(k')$ holds, allowing us to determine the Casimir directly from $J_\mu(\omega, k)$. In this way we avoid the computation of the distance and obtain in a simple way the dependence on $k$ of $C(k)$.

The other 4 isometries $k'_\mu = T_\mu(a, k)$ related with translations define the composition law $p\oplus q$ of two momenta $p$, $q$ through
\be
(p\oplus q)_\mu \doteq T_\mu(p, q)\,.
\label{DCL-translations}
\ee
One can easily see that the DCL is related to the translation composition through 
\be
p\oplus q=T_p(q)=T_p(T_q(0))=(T_p \circ T_q)(0)\,.
\label{T-composition}
\ee
Note that the equation above implies that $T_{(p\oplus q)}$ differs from $(T_p \circ T_q)$ by a Lorentz transformation, since it is a transformation that leaves the origin invariant.

From this perspective, a DRK (in Sec.~\ref{sec:diagram} we will see that with this construction a relativity principle holds) can be obtained by identifying the isometries $T_a$, $J_\omega$ with the composition law and the Lorentz transformations, which fixes the dispersion relation.

Then, starting from a metric, we can deduce the DRK by obtaining $T_a$, $J_\omega$ through
\be
g_{\mu\nu}(T_a(k)) \,=\, \frac{\partial T_\mu(a, k)}{\partial k_\rho} \frac{\partial T_\nu(a, k)}{\partial k_\sigma} g_{\rho\sigma}(k), \quad\quad
g_{\mu\nu}(J_\omega(k)) \,=\, \frac{\partial J_\mu(\omega, k)}{\partial k_\rho} \frac{\partial J_\nu(\omega, k)}{\partial k_\sigma} g_{\rho\sigma}(k)\,.
\label{T,J}
\ee
The previous equations have to be satisfied for any choice of the parameters $a$, $\omega$. From the limit $k\to 0$ in (\ref{T,J})
\be
\begin{split}
g_{\mu\nu}(a) \,=&\, \left[\lim_{k\to 0} \frac{\partial T_\mu(a, k)}{\partial k_\rho}\right] \, 
\left[\lim_{k\to 0} \frac{\partial T_\nu(a, k)}{\partial k_\sigma}\right] \,\eta_{\rho\sigma}\,,  \\
\eta_{\mu\nu} \,=&\, \left[\lim_{k\to 0} \frac{\partial J_\mu(\omega, k)}{\partial k_\rho}\right] \,  
\left[\lim_{k\to 0} \frac{\partial J_\nu(\omega, k)}{\partial k_\sigma}\right] \,\eta_{\rho\sigma}\,,
\end{split}
\ee
one can identify
\be
\lim_{k\to 0} \frac{\partial T_\mu(a, k)}{\partial k_\rho} \,=\, \delta^\rho_\alpha e_\mu^\alpha(a)\,, \quad\quad\quad
\lim_{k\to 0} \frac{\partial J_\mu(\omega, k)}{\partial k_\rho} \,=\, L_\mu^\rho(\omega)\,,
\label{e,L}
\ee
where $e_\mu^\alpha(k)$ is the (inverse of\footnote{Note that the metric $g_{\mu\nu}$ is the inverse of $g^{\mu\nu}$.} the) tetrad of the momentum space, and $L_\mu^\rho(\omega)$ is the standard Lorentz transformation matrix with parameters $\omega$. From Eq.~\eqref{DCL-translations} and Eq.~\eqref{e,L}, one obtains
\be
\lim_{k\to 0} \frac{\partial(a\oplus k)_\mu}{\partial k_\rho} \,=\, \delta^\rho_\alpha e_\mu^\alpha(a)\,,
\label{magicformula}
\ee
which leads to a fundamental relationship between the DCL and the momentum space tetrad.

For infinitesimal transformations, we have
\be
T_\mu(\epsilon, k) = k_\mu + \epsilon_\alpha {\cal T}_\mu^\alpha(k)\,, \quad\quad\quad
J_\mu(\epsilon, k) = k_\mu + \epsilon_{\beta\gamma} {\cal J}^{\beta\gamma}_\mu(k)\,,
\label{infinit_tr}
\ee
and Eq.~(\ref{T,J}) leads to the equations
\be
\frac{\partial g_{\mu\nu}(k)}{\partial k_\rho} {\cal T}^\alpha_\rho(k) \,=\, \frac{\partial{\cal T}^\alpha_\mu(k)}{\partial k_\rho} g_{\rho\nu}(k) +
\frac{\partial{\cal T}^\alpha_\nu(k)}{\partial k_\rho} g_{\mu\rho}(k)\,,
\label{cal(T)}
\ee
\be
\frac{\partial g_{\mu\nu}(k)}{\partial k_\rho} {\cal J}^{\beta\gamma}_\rho(k) \,=\,
\frac{\partial{\cal J}^{\beta\gamma}_\mu(k)}{\partial k_\rho} g_{\rho\nu}(k) +
\frac{\partial{\cal J}^{\beta\gamma}_\nu(k)}{\partial k_\rho} g_{\mu\rho}(k)\,,
\label{cal(J)}
\ee
which allow us to obtain the Killing vectors ${\cal J}^{\beta\gamma}$, but do not completely determine ${\cal T}^\alpha$. This is due to the fact that if ${\cal T}^\alpha$, ${\cal J}^{\beta\gamma}$ are a solution of the Killing equations (\eqref{cal(T)}-\eqref{cal(J)}), then ${\cal T}^{\prime \alpha} = {\cal T}^\alpha + c^\alpha_{\beta\gamma} {\cal J}^{\beta\gamma}$ is also a solution of Eq.~(\ref{cal(T)}) for any arbitrary constants $c^\alpha_{\beta\gamma}$, and then  $T'_\mu(\epsilon, 0)=T_\mu(\epsilon, 0)=\epsilon_\mu$. This observation is completely equivalent to the comment after Eq.~\eqref{T-composition}. In order to eliminate this ambiguity, since we know that the isometry generators close an algebra, we can chose them as
\be
T^\alpha \,=\, x^\mu {\cal T}^\alpha_\mu(k), \quad\quad\quad J^{\alpha\beta} \,=\, x^\mu {\cal J}^{\alpha\beta}_\mu(k)\,,
\label{generators_withx}
\ee
so that their Poisson brackets
\begin{align}
  &\{T^\alpha, T^\beta\} \,=\, x^\rho \left(\frac{\partial{\cal T}^\alpha_\rho(k)}{\partial k_\sigma} {\cal T}^\beta_\sigma(k) - \frac{\partial{\cal T}^\beta_\rho(k)}{\partial k_\sigma} {\cal T}^\alpha_\sigma(k)\right)\,, \\
  &\{T^\alpha, J^{\beta\gamma}\} \,=\, x^\rho \left(\frac{\partial{\cal T}^\alpha_\rho(k)}{\partial k_\sigma} {\cal J}^{\beta\gamma}_\sigma(k) - \frac{\partial{\cal J}^{\beta\gamma}_\rho(k)}{\partial k_\sigma} {\cal T}^\alpha_\sigma(k)\right)\,,
\end{align}
close a particular algebra. Then, we see that this ambiguity in defining the translations is just the ambiguity in the choice of the isometry algebra, i.e., in the basis. Every choice of the translation generators will lead to a different DCL, and then to a different DRK.

\subsection{Relativistic deformed kinematics}
\label{sec:diagram}

In this subsection we will prove that the kinematics obtained as proposed before is in fact a DRK. The proof can be sketched in the next diagram:
\begin{center}
\begin{tikzpicture}
\node (v1) at (-2,1) {$q$};
\node (v4) at (2,1) {$\bar q$};
\node (v2) at (-2,-1) {$p \oplus q$};
\node (v3) at (2,-1) {$(p \oplus q)^\prime$};
\draw [->] (v1) edge (v2);
\draw [->] (v4) edge (v3);
\draw [->] (v2) edge (v3);
\node at (-2.6,0) {$T_p$};
\node at (2.7,0) {$T_{p^\prime}$};
\node at (0,-1.4) {$J_\omega$};
\end{tikzpicture}
\end{center}
where the momentum with prime indicates the transformation through ${\cal J}_\omega$, and $T_p$, $T_{p'}$ are the translations with parameters $p$ and $p'$. One can define $\bar{q}$ as the point that satisfies 
\be
(p\oplus q)' \,=\, (p' \oplus \bar{q})\,.
\label{qbar1}
\ee
One sees that in the case $q=0$, also $\bar{q}=0$, and in any other case with $q\neq 0$, the point $\bar{q}$ is obtained from $q$ by an isometry, which is a composition of  the translation $T_p$, a Lorentz transformation $J_\omega$, and the inverse of the translation $T_{p'}$ (since the isometries are a group of transformations, any composition of isometries is also an isometry). So we have found that there is an isometry  $q\rightarrow \bar{q}$, that leaves the origin invariant, and then
\be
C(q) \,=\, C(\bar{q})\,,
\label{qbar2}
\ee
since they are at the same distance from the origin. Eqs.~\eqref{qbar1}-\eqref{qbar2} imply that the deformed kinematics with ingredients $C$ and $\oplus$ is a DRK if one identifies the momenta $(p', \bar{q})$ as the two-particle Lorentz transformation of $(p, q)$. In particular, Eq.~(\ref{qbar1}) tells us that the DCL is invariant under the previously defined Lorentz transformation and Eq.~(\ref{qbar2}), together with $C(p)=C(p')$, that the DDR of both momenta is also Lorentz invariant. We can see that with this definition of the two-particle Lorentz transformations, one of the particles ($p$) transforms as a single momentum, but the transformation of the other one ($q$) depends of both momenta. This computation will be carried out in the next subsection in the particular example of $\kappa$-Poincaré.

\section{Isotropic relativistic deformed kinematics}
\label{sec:examples}

In this section we derive the construction in detail for two simple isotropic kinematics, $\kappa$-Poincaré and Snyder. Also, we will show how to construct a DRK beyond these two simple cases, the kinematics known as hybrid models. 

If the DRK is isotropic,  the general form of the algebra of the generators of isometries is  
\be
\{T^0, T^i\} \,=\, \frac{c_1}{\Lambda} T^i + \frac{c_2}{\Lambda^2} J^{0i}, \quad\quad\quad \{T^i, T^j\} \,=\, \frac{c_2}{\Lambda^2} J^{ij}\,,
\label{isoRDK}
\ee
where we assume that the generators $J^{\alpha\beta}$ satisfy the standard Lorentz algebra, and due to the fact that isometries are a group, the Poisson brackets of $T^\alpha$ and $J^{\beta\gamma}$ are fixed by Jacobi identities\footnote{The coefficients proportional to the Lorentz generators in Eq.~\eqref{isoRDK} are the same also due to Jacobi identities.}. For each choice of the coefficients $(c_1/\Lambda)$ and $(c_2/\Lambda^2)$ (and then for the algebra) and for each choice of a metric of a maximally symmetric momentum space in  isotropic coordinates, one has to obtain the isometries of such metric so that their generators close the chosen algebra in order to find a DRK. 

\subsection{\texorpdfstring{$\kappa$}{k}-Poincaré relativistic kinematics}
\label{subsection_kappa_desitter}
We can consider the simple case where $c_2=0$ in Eq.~\eqref{isoRDK}, so the generators of translations close a subalgebra\footnote{We have reabsorbed the coefficient $c_1$ in the scale $\Lambda$.}    
\be
\{T^0, T^i\} \,=\, \pm \frac{1}{\Lambda} T^i\,.
\label{Talgebra}
\ee
A well known result of differential geometry (see Ch.6 of Ref.~\cite{Chern:1999jn}) is that if the generators of left-translations $T^\alpha$ transforming $k \to T_a(k) = (a\oplus k)$ form a Lie algebra, the generators of right-translations $\tilde{T}^\alpha$ transforming $k \to (k\oplus a)$, close the same algebra but with a different sign
\be
 \{\tilde{T}^0, \tilde{T}^i\} \,=\, \mp \frac{1}{\Lambda} \tilde{T}^i \,.
\label{Ttildealgebra}
\ee   
We have found the explicit relation between the infinitesimal right-translations and the tetrad of the momentum metric in Eq.~\eqref{magicformula}, which gives
\be
(k\oplus\epsilon)_\mu=k_\mu+\epsilon_\alpha e^\alpha_\mu\equiv \tilde{T}_\mu(k,\epsilon).
\ee
Comparing with Eq.~\eqref{infinit_tr} and Eq.~\eqref{generators_withx}, we see that right-translation generators are given by
\be
\tilde{T}^\alpha \,=\, x^\mu e^\alpha_\mu(k)\,.
\label{Ttilde}
\ee

Since both algebras \eqref{Talgebra}-\eqref{Ttildealgebra} satisfy $\kappa$-Minkowski noncommutativity, the problem to find a tetrad $e^\alpha_\mu(k)$ compatible with the algebra of Eq.~(\ref{Ttildealgebra}) is equivalent to the problem of obtaining a representation of this noncommutativity expressed in terms of canonical coordinates of the phase space. One can easily confirm that the choice of the tetrad
\be
e^0_0(k) \,=\, 1\,, \quad\quad\quad e^0_i(k) \,=\, e^i_0(k) \,=\, 0\,, \quad\quad\quad e^i_j (k) \,=\, \delta^i_j e^{\mp k_0/\Lambda}\,,
\label{bicross-tetrad}
\ee
leads to a representation of $\kappa$-Minkowski noncommutativity. 

In order to obtain the finite translations $T_\mu(a,k)$, which in this case form a group, one can try to generalize  Eq.~\eqref{e,L} to define a transformation that does not change the form of the tetrad:
\be
e_\mu^\alpha(T(a, k)) \,=\, \frac{\partial T_\mu(a, k)}{\partial k_\nu} \,e_\nu^\alpha(k)\,.
\label{T(a,k)}
\ee
Obviously, if $T_\mu(a,k)$ is a solution to the previous equation, it implies that the translation leaves the tetrad invariant, and then the metric, so it is therefore an isometry. Then, one can check that translations form a group since the composition of two transformations leaving the tetrad invariant also leaves the tetrad invariant. Indeed, Eq. \eqref{T(a,k)} can be explicitly solved in order to obtain the finite translations. For the particular choice of the tetrad in Eq.~\eqref{bicross-tetrad}, the translations read (see \ref{appendix_translations})
\be
T_0(a, k) \,=\, a_0 + k_0, \quad\quad\quad T_i(a, k) \,=\, a_i + k_i e^{\mp a_0/\Lambda}\,,
\ee
and then the DCL is 
\be
(p\oplus q)_0 \,=\, T_0(p, q) \,=\, p_0 + q_0\,, \quad\quad\quad
(p\oplus q)_i \,=\, T_i(p, q) \,=\, p_i + q_i e^{\mp p_0/\Lambda}\,,
\label{kappa-DCL}
\ee
which is the one obtained  in the bicrossproduct basis of  $\kappa$-Poincaré kinematics (up to a sign depending on the choice of the initial sign of $\Lambda$ in Eq.~\eqref{bicross-tetrad}).

From the equation
\be
\frac{\partial C(k)}{\partial k_\mu} \,{\cal J}^{\alpha\beta}_\mu(k) \,=\, 0 \,,
\label{eq:casimir_J}
\ee
one can obtain the DDR, where ${\cal J}^{\alpha\beta}$ are the infinitesimal Lorentz transformations satisfying Eq.~\eqref{cal(J)} with the metric $g_{\mu\nu}(k)=e^\alpha_\mu(k)\eta_{\alpha\beta}e^\beta_\nu(k)$ defined by the tetrad~\eqref{bicross-tetrad}:
\be
\begin{split}
&0 \,=\, \frac{{\cal J}^{\alpha\beta}_0(k)}{\partial k_0}\,, \quad
0 \,=\, - \frac{{\cal J}^{\alpha\beta}_0(k)}{\partial k_i} e^{\mp 2k_0/\Lambda} + \frac{{\cal J}^{\alpha\beta}_i(k)}{\partial k_0}\,, \\
&\pm \frac{2}{\Lambda} {\cal J}^{\alpha\beta}_0(k) \delta_{ij} \,=\, - \frac{\partial{\cal J}^{\alpha\beta}_i(k)}{\partial k_j} - \frac{\partial{\cal J}^{\alpha\beta}_j(k)}{\partial k_i}\,.
\end{split}
\ee
One gets finally 
 \be
{\cal J}^{0i}_0(k) \,=\, -k_i\,, \quad \quad \quad {\cal J}^{0i}_j(k)\,=\, \pm \delta^i_j \,\frac{\Lambda}{2} \left[e^{\mp 2 k_0/\Lambda} - 1 - \frac{\vec{k}^2}{\Lambda^2}\right] \pm \,\frac{k_i k_j}{\Lambda}\,,
\label{eq:j_momentum_space}
\ee 
and then 
\be
C(k) \,=\, \Lambda^2 \left(e^{k_0/\Lambda} + e^{-k_0/\Lambda} - 2\right) - e^{\pm k_0/\Lambda} \vec{k}^2  \,,
\label{eq:casimir_momentum_space}
\ee
which is the same function of the momentum which defines the DDR of $\kappa$-Poincaré kinematics in the bicrossproduct basis (up to the sign in $\Lambda$).

The last ingredient we need in order to complete the discussion of the kinematics is the two-particle Lorentz transformations. Using the diagram in Sec.~\ref{sec:diagram}, one has to find $\bar{q}$ so that
\be
(p\oplus q)' \,=\, p'\oplus \bar{q}\,.
\ee
Equating both expressions and taking only the linear terms in $\epsilon_{\alpha\beta}$ (parameters of the infinitesimal Lorentz transformation) one arrives to the equation
\be
\epsilon_{\alpha\beta} {\cal J}^{\alpha\beta}_\mu(p\oplus q) \,=\, \epsilon_{\alpha\beta} \frac{\partial(p\oplus q)_\mu}{\partial p_\nu} {\cal J}^{\alpha\beta}_\nu(p) + \frac{\partial(p\oplus q)_\mu}{\partial q_\nu} (\bar{q}_\nu - q_\nu)\,.
\ee
From the DCL of \eqref{kappa-DCL} with the minus sign, we find
\begin{align}
& \frac{\partial(p\oplus q)_0}{\partial p_0} \,=\, 1\,, \quad\quad
\frac{\partial(p\oplus q)_0}{\partial p_i} \,=\, 0\,, \quad\quad
\frac{\partial(p\oplus q)_i}{\partial p_0} \,=\, - \frac{q_i}{\Lambda} e^{-p_0/\Lambda}\,, \quad\quad
\frac{\partial(p\oplus q)_i}{\partial p_j} \,=\, \delta_i^j\,, \\
& \frac{\partial(p\oplus q)_0}{\partial q_0} \,=\, 1\,, \quad\quad  \frac{\partial(p\oplus q)_0}{\partial q_i} \,=\, 0\,, \quad\quad
\frac{\partial(p\oplus q)_i}{\partial q_0} \,=\, 0\,, \quad\quad  \frac{\partial(p\oplus q)_i}{\partial q_j} \,=\, \delta_i^j e^{-p_0/\Lambda}\,.
\end{align}
Then, we obtain
\be
\begin{split}
\bar{q}_0 \,&=\, q_0 + \epsilon_{\alpha\beta} \left[{\cal J}^{\alpha\beta}_0(p\oplus q) - {\cal J}^{\alpha\beta}_0(p)\right]\,, \\
\bar{q}_i \,&=\, q_i + \epsilon_{\alpha\beta} \, e^{p_0/\Lambda} \, \left[{\cal J}^{\alpha\beta}_i(p\oplus q) - {\cal J}^{\alpha\beta}_i(p) + \frac{q_i}{\Lambda} e^{-p_0/\Lambda} {\cal J}^{\alpha\beta}_0(p)\right]\,,
\end{split}
\label{eq:jr_momentum_space}
\ee
and one can check that this is the Lorentz transformation of the two-particle system of $\kappa$-Poincaré in the bicrossproduct basis~\eqref{eq:coproducts}.

For the choice of the tetrad in Eq.~\eqref{bicross-tetrad}, the metric in momentum space reads~\footnote{This is the de Sitter metric written in the comoving coordinate system used in Ref.~\cite{Gubitosi:2013rna}.}
\be
g_{00}(k) \,=\, 1\,, \quad\quad\quad g_{0i}(k) \,=\, g_{i0}(k) \,=\, 0\,, \quad\quad\quad g_{ij}(k) \,=\, - \delta_{ij} e^{\mp 2k_0/\Lambda}\,.
\label{bicross-metric}
\ee
Computing the Riemann-Christoffel tensor, one can check that it corresponds to a de Sitter momentum space with curvature $(12/\Lambda^2)$.\footnote{In Appendix~\ref{appendix_algebra} it is shown that the way we have constructed the DRK as imposing the invariance of the tetrad cannot be followed for the case of anti-de Sitter space.} 

To summarize, we have found the $\kappa$-Poincaré kinematics in the bicrossproduct basis~\cite{KowalskiGlikman:2002we} from geometric ingredients of a de Sitter momentum space with the choice of the tetrad of Eq.~\eqref{bicross-tetrad}. For different choices of tetrad (in such a way that the generators of Eq.~\eqref{Ttilde} close the algebra Eq.~\eqref{Ttildealgebra}), one will find the $\kappa$-Poincaré kinematics in different bases. Then, the different bases of the deformed kinematics are just different choices of coordinates in de Sitter space. Note that when generators of right-translations constructed from the momentum space tetrad close the algebra of Eq.~\eqref{Ttildealgebra}, the DCL obtained is associative (this can be easily understood since as the generators of translations close an algebra Eq.~\eqref{Talgebra}, translations form a group). 

\subsection{Beyond \texorpdfstring{$\kappa$}{k}-Poincaré relativistic kinematics}

The other simple choice in the algebra of the translation generators is $c_1=0$, leading to the Snyder algebra explained in the introduction. As the generators of translations do not close an algebra, we cannot follow the same procedure we did in the previous case for obtaining the $\kappa$-Poincaré kinematics. But considering the simple covariant form of the de Sitter metric, $g_{\mu\nu}(k) =\eta_{\mu\nu} + k_\mu k_\nu/\Lambda^2$, one can find the DCL just requiring to be covariant 
\be
(p\oplus q)_\mu \,=\, p_\mu f_L\left(p^2/\Lambda^2, p\cdot q/\Lambda^2, q^2/\Lambda^2\right) + q_\mu f_R\left(p^2/\Lambda^2, p\cdot q/\Lambda^2, q^2/\Lambda^2\right)\,,
\label{DCLSnyder-1}
\ee
asking for the following equation to hold:
\be
\eta_{\mu\nu} + \frac{(p\oplus q)_\mu (p\oplus q)_\nu}{\Lambda^2} \,=\, \frac{\partial(p\oplus q)_\mu}{\partial q_\rho} \frac{\partial(p\oplus q)_\nu}{\partial q_\sigma} \left(\eta_{\rho\sigma} + \frac{q_\rho q_\sigma}{\Lambda^2}\right)\,.
\ee
Then, one can solve for the two functions $f_L$, $f_R$ of three variables, obtaining
\be
 \begin{split}
f_L\left(p^2/\Lambda^2, p\cdot q/\Lambda^2, q^2/\Lambda^2\right) \,=&\, \sqrt{1+\frac{q^2}{\Lambda^2}}+\frac{p\cdot q}{\Lambda^2\left(1+\sqrt{1+p^2/\Lambda^2}\right)}\,,\\
f_R\left(p^2/\Lambda^2, p\cdot q/\Lambda^2, q^2/\Lambda^2\right) \,=&\, 1\,,
\label{DCLSnyder-2}
 \end{split}
\ee
which is the DCL of Snyder kinematics in the Maggiore representation previously derived in  Ref.~\cite{Battisti:2010sr} (the first order terms were obtained also in Ref.~\cite{Banburski:2013jfa}).

From the infinitesimal generators of translations
\be
{\cal T}^\mu_\nu(p)\,=\,\left.\frac{\partial \left(k\oplus p\right)_\nu}{\partial k_\mu} \right\rvert_{k \rightarrow 0}\,=\,\delta^\mu_\nu \sqrt{1+\frac{p^2}{\Lambda^2}} \,,
\ee  
one can see that $T^\alpha=x^\nu{\cal T}^\alpha_\nu$ form the Snyder algebra
\be
\{T^\alpha, T^\beta\} \,=\, \frac{1}{\Lambda^2} J^{\alpha\beta}\,.
\ee
From linear Lorentz covariance, one can deduce that the dispersion relation $C(p)$ is just a function of $p^2$, and the Lorentz transformations both in the one and two-particle systems are linear (the same Lorentz transformations used in SR).

Different choices of momentum coordinates making the metric to be expressed in covariant terms will lead to different representations of the Snyder kinematics. For the anti-de Sitter case, the DCL is the one obtained in Eq.~\eqref{DCLSnyder-2} just replacing $(1/\Lambda^2)$ by $-(1/\Lambda^2)$, since the anti-de Sitter metric is the same of de Sitter proposed at the beginning of this subsection interchanging $(1/\Lambda^2)$ by $-(1/\Lambda^2)$.   

When both coefficients $c_1$, $c_2$ are non-zero, one has algebras of the generators of translations  known as hybrid models~\cite{Meljanac:2009ej}. The DCL in these cases can be obtained from a power expansion in $(1/\Lambda)$ asking to be an isometry and that their generators close the desired algebra. With this procedure, one can get the same kinematics found in Ref.~\cite{Meljanac:2009ej}.  

The DCL obtained when the generators of translations  close a subalgebra  (the case of $\kappa$-Poincaré) is the only one which is associative. The other compositions obtained when the algebra is Snyder or any hybrid model do not have this property (see  Eqs.~\eqref{DCLSnyder-1} and~\eqref{DCLSnyder-2}).  This is an important difference between the algebraic and geometric approaches: the only isotropic DRK obtained from the Hopf algebra approach is $\kappa$-Poincaré, since one asks  the generators of translations to close an algebra (and then, one finds an associative composition of momenta), eliminating any other option. With this proposal, identifying a correspondence between translations of a maximally symmetric momentum space whose generators close a certain algebra and a DCL, we open up the possibility to construct more DRK in a simple way. It is clear that associativity is a crucial property for studying processes with a DRK, so somehow the $\kappa$-Poincaré scenario seems special. Note also that the two different perspectives (algebraic and geometrical approaches) has only one common DRK, which might indicate that $\kappa$-Poincaré is a preferred kinematics.

\section{Comparison with previous works}
\label{relative_locality_comparison}

In this section, we will compare the prescription followed in the previous sections with the one proposed in Ref.~\cite{AmelinoCamelia:2011bm}. This comparison can only be carried out for the $\kappa$-Poincaré kinematics, since as we will see, the associativity property of the composition law plays a crucial role. In order to make the comparison, we can derive with respect to  $p_\tau$  the equation of the invariance of the tetrad under translations Eq.~\eqref{T(a,k)}, written in terms of the DCL
\be
\frac{\partial e^\alpha_\nu(p\oplus q)}{\partial p_\tau} \,=\, \frac{\partial  e^\alpha_\nu(p\oplus q)}{\partial (p\oplus q)_\sigma} \frac{\partial(p\oplus q)_\sigma}{\partial p_\tau} \,=\, \frac{\partial^2(p\oplus q)_\nu}{\partial p_\tau \partial q_\rho} e^\alpha_\rho(q)\,.
\ee 
One can find the second derivative of the DCL
\be
\frac{\partial^2(p\oplus q)_\nu}{\partial p_\tau \partial q_\rho} \,=\, e^\rho_\alpha(q) \frac{\partial  e^\alpha_\nu(p\oplus q)}{\partial (p\oplus q)_\sigma} \frac{\partial(p\oplus q)_\sigma}{\partial p_\tau}\,,
\ee
where $e^\nu_\alpha$ is the inverse of $e^\alpha_\nu$, $e^\alpha_\nu e^\mu_\alpha=\delta^\mu_\nu$. 
But also using Eq.~\eqref{T(a,k)}, one has
\be
e^\rho_\alpha(q) \,=\, \frac{\partial(p\oplus q)_\mu}{\partial q_\rho} e^\mu_\alpha(p\oplus q)\,,
\label{magicformula2}
\ee
and then 
\be
\frac{\partial^2(p\oplus q)_\nu}{\partial p_\tau \partial q_\rho} + \Gamma^{\sigma\mu}_\nu(p\oplus q) \,\frac{\partial(p\oplus q)_\sigma}{\partial p_\tau} \,\frac{\partial(p\oplus q)_\mu}{\partial q_\rho} \,=\, 0\,,
\label{geodesic_tetrad}
\ee
where 
\be
\Gamma^{\sigma\mu}_\nu(k) \,\doteq\, - e^\mu_\alpha(k) \, \frac{\partial  e^\alpha_\nu(k)}{\partial k_\sigma}\,.
\label{e-connection}
\ee
It can be checked that the combination of tetrads and derivatives appearing in Eq.~\eqref{e-connection} in fact transforms like a connection under a change of momentum coordinates.

In Ref.~\cite{Amelino-Camelia:2013sba}, it is proposed another way to define a connection and a DCL in momentum space through parallel transport, establishing a link between these two ingredients. It is easy to check that the DCL obtained in this way satisfies Eq.~\eqref{geodesic_tetrad}. This equation only determines the DCL for a given connection if one imposes the associativity property of the composition. Comparing with the previous reference, one then concludes that the DCL obtained from translations that leaves the form of the tetrad invariant is the associative composition law one finds by parallel transport, with the connection constructed from a tetrad and its derivatives as in Eq.\eqref{e-connection}. 

Finally, if the DCL is associative, then Eq.~\eqref{k-DCL} reduces to
\be
(p\oplus_k q) \,=\, p\oplus\hat{k}\oplus q.
\ee
Replacing $q$ by $(\hat{k}\oplus q)$ in Eq.~\eqref{geodesic_tetrad}, which is valid for any momenta ($p, q$), one obtains
\be
\frac{\partial^2  (p \oplus \hat{k} \oplus q)_\nu}{\partial p_\tau \partial(\hat{k} \oplus q)_\rho}+\Gamma^{\sigma \mu}_\nu (p \oplus \hat{k} \oplus q) \frac{\partial (p \oplus \hat{k} \oplus q)_\sigma}{\partial p_\tau}\frac{\partial (p \oplus \hat{k} \oplus q)_\mu}{\partial(\hat{k} \oplus q)_\rho}\,=\,0\,.
\ee
Multiplying by $\partial(\hat{k} \oplus q)_\rho/\partial q_\lambda$, one finds
\be
\frac{\partial^2  (p \oplus \hat{k} \oplus q)_\nu}{\partial p_\tau \partial q_\lambda}+\Gamma^{\sigma \mu}_\nu (p \oplus \hat{k} \oplus q) \frac{\partial (p \oplus \hat{k} \oplus q)_\sigma}{\partial p_\tau}\frac{\partial (p \oplus \hat{k} \oplus q)_\mu}{\partial q_\lambda}\,=\,0\,.
\label{connection_1}
\ee
Taking $p=q=k$ in Eq.~\eqref{connection_1}, one finally gets
\be
\Gamma^{\tau \lambda}_\nu (k)\,=\,-\left.\frac{\partial^2  (p\oplus_{k}q)_\nu}{\partial p_\tau \partial q_\lambda}\right\rvert_{p,q \rightarrow k}\,,
\ee
which is the same expression of Eq.~\eqref{k-connection} proposed in Ref.~\cite{AmelinoCamelia:2011bm}. This concludes that the connection of Eq.~\eqref{e-connection} constructed from the tetrad is the same connection given by the prescription developed in Ref.~\cite{AmelinoCamelia:2011bm} when the DCL is associative. 

\chapter{Spacetime from local interactions}
\label{chapter_locality}
\ifpdf
    \graphicspath{{Chapter4/Figs/Raster/}{Chapter4/Figs/PDF/}{Chapter4/Figs/}}
\else
    \graphicspath{{Chapter4/Figs/Vector/}{Chapter4/Figs/}}
\fi

\epigraph{Like the physical, the psychical is not necessarily in reality what it appears to us to be.}{Sigmund Freud}

In the previous chapter we have seen that a DRK can be understood from the geometry of a curved (maximally symmetric) momentum space. However, we have not discussed the effects that this curvature, and then the kinematics, provoke on spacetime. A possible consequence of a DCL considered in numerous works is a noncommutative spacetime. In particular, in Refs.~\cite{Meljanac:2009ej,Battisti:2010sr}, a composition law is obtained from the product of plane waves for $\kappa$-Minkowski, Snyder and hybrid models noncommutativity. Also, from the Hopf algebra perspective, it is possible to obtain  a modified Heisenberg algebra with $\kappa$-Minkowski spacetime from a DCL through the ``pairing'' operation~\cite{Kosinski_paring}. 

In all these works, however, there is a lack of physical understanding about the relation between a DCL and a noncommutative spacetime. In this chapter, we will try to show how these ingredients are related giving a physical intuition. As we will see in Sec.~\ref{sec:relative_locality_intro}, a DCL produces a loss of locality in canonical spacetime. From an action of free relativistic particles, the authors of Ref.~\cite{AmelinoCamelia:2011bm} derived such effect including an interaction term defined by the energy-momentum conservation, which is determined by the DCL. It is possible to understand this nonlocality from the following argument: since the total momentum can be viewed as the generator of translations in spacetime, a modification of it as a function of all momenta will produce nontrivial translations. This means that only an observer placed where the interaction takes place will see such interaction as local, but not any other related to him by a translation. Along this chapter, we will see that there are different ways to choose  space-time coordinates (depending on momentum) which we call ``physical'' coordinates~\cite{Carmona:2017cry}, in which the interactions are local. We will see that there is a relationship between this approach and the results obtained through Hopf algebras, and also with the momentum space geometry studied in the previous chapter.

\section{Relative Locality}
\label{sec:relative_locality_intro}
In this section we explain the main results of~\cite{AmelinoCamelia:2011nt}. We first start by the following action 
\begin{equation}
S_{\text{total}}\,=\, S_{\text{free}}^{\text{in}}+S_{\text{free}}^{\text{out}} +S_{\text{int}}\,, 
\label{eq:action}
\end{equation}
where the first part describes the free propagation of the $N$ incoming worldlines
\begin{equation}
S_{\text{free}}^{\text{in}}\,=\,\sum_{J=1}^{N}\int^{0}_{-\infty} ds \left(x^{\mu}_J \dot k^{J}_{\mu}+\mathcal{N}_J\left(C(k^{J})-m^2_J\right)\right)\,,
\label{eq:action_in}
\end{equation}
and the outgoing worldlines are given by the second term
\begin{equation}
S_{\text{free}}^{\text{out}}\,=\,\sum_{J=N+1}^{2N}\int_{0}^{\infty} ds \left(x^{\mu}_J \dot k^{J}_{\mu}+\mathcal{N}_J\left(C(k^{J})-m^2_J\right)\right)\,.
\label{eq:action_out}
\end{equation}
In the previous expressions $s$ plays the role of an arbitrary parameter characterizing the worldline of the particle and $\mathcal{N}_J$ is the Lagrange multiplier imposing on all particles the condition to be on mass shell    
\begin{equation}
C(k^J)\,=\,m^2_J\,,
\end{equation}
for a Casimir $C(k^J)$ (which in principle is deformed).

The interaction term appearing in the action is the conservation law of momenta times a Lagrange multiplier
\begin{equation}
S_{\text{int}}\,=\,\left( \bigoplus\limits_{N+1\leq J\leq 2N}  k^J_\nu(0)\,\,- \bigoplus\limits_{1\leq J\leq N} k^J_\nu(0)\right) \xi^\nu\,.
\label{eq:action_int}
\end{equation}
The parametrization $s$ is chosen in such a way that the interaction occurs at $s=0$ for every particle, and $\xi$ can be seen as a Lagrange multiplier imposing the momentum conservation at that point. 
Varying the action and integrating by parts one finds
\begin{equation}
\delta S_{\text{total}}\,=\,\sum_J \int_{s_1}^{s_2}\left(\delta x^{\mu}_J \dot k^J_{\mu} - \delta k^{J}_{\mu}\left[\dot x^{\mu}_J-\mathcal{N}_J\frac{\partial C(k^J)}{\partial k^{J}_{\mu}} \right]\right)+\mathcal{R}\,,
\label{deltaS}
\end{equation}
where the term $\mathcal{R}$ contains the variation of $S_{\text{int}}$ and also the boundary terms appearing after the integration by parts, and $s_{1,2}$ are $-\infty$, 0 or  0, $\infty$ depending on the incoming or outgoing character of the terms. 
One finds
{\small
\be
\begin{split}
\mathcal{R} \,=\,\left(\bigoplus\limits_{N+1\leq J\leq 2N}k^J_\nu(0)\,\,- \bigoplus\limits_{1\leq J\leq N} k^J_\nu(0)\right)& \delta\xi^\nu+ \sum_{J=1}^{N} \left(x^{\mu}_J (0) - \xi^{\nu} \frac{\partial}{\partial k^J_{\mu}} \left[ \bigoplus\limits_{1\leq J\leq N} k^I_\nu\right](0)\right)\delta k^J_{\mu}(0)\\- &
\sum_{J=N+1}^{2N} \left(x^{\mu}_J (0)  -\xi^{\nu} \frac{\partial}{\partial k^J_{\mu}} \left[ \bigoplus\limits_{N+1\leq J\leq 2N} k^I_\nu\right](0)\right) \delta k^J_{\mu}(0)\,,
\end{split}  
\ee}
\normalsize 
where the $x^{\mu}_J (0)$ are the space-time coordinates of the worldline at the initial (final) point for $1\leq J\leq N$ ($N+1\leq J\leq 2N$).
The worldlines of particles must obey the variational principle $\delta S_{\text{total}}=0$ for any variation $\delta\xi^\mu$, $\delta x_J^\mu$, $\delta k^J_\mu$. From the variation of the Lagrange multiplier of the interaction term $\delta \xi^\nu$, one obtains the momentum conservation at the interaction point, and for the variation with respect $\delta k^J_{\mu}(0)$ one finds\footnote{The variation of  $\delta x^\mu_J(s)$ in Eq.(\ref{deltaS}) implies constant momenta along each worldline.}
\begin{equation}
\begin{split}
x^{\mu}_J (0)\,&=\, \xi^{\nu} \frac{\partial}{\partial k^J_{\mu}} \left[\bigoplus\limits_{1\leq J\leq N}  k^I_\nu\right] \, \text{for } J=1,\ldots N
\,,\\
x^{\mu}_J (0)\,&=\, \xi^{\nu} \frac{\partial}{\partial k^J_{\mu}} \left[\bigoplus\limits_{N+1\leq J\leq 2N} k^I_\nu\right] \, \text{for } J=N+1,\ldots 2N \,.
\label{eq:endWL}
\end{split}
\end{equation}

The transformation
\begin{equation}
\begin{split}
\delta \xi^\mu&\,=\,a^\mu, \quad
\delta x^\mu_J\,=\,a^\nu\frac{\partial}{\partial k^J_{\mu}} \left[\bigoplus\limits_{1\leq J\leq N}  k^I_\nu\right] (J=1,\ldots N)\,,\\
 \delta x^\mu_J&\,=\,a^\nu \frac{\partial}{\partial k^J_{\mu}} \left[\bigoplus\limits_{N+1\leq J\leq 2N} k^I_\nu\right] (J=N+1,\ldots 2N)\,,
\quad \delta k^J_\mu\,=\,0\,,
\label{eq:translation}
\end{split}  
\end{equation}
is a symmetry of the action (translational invariance) connecting different solutions from the variational principle. We see from Eq.~\eqref{eq:endWL} that only an observer placed at the interaction point ($\xi^\mu=0$) will see the interaction as local (all $x^\mu_J(0)$ coincide, being zero). One can choose the Lagrange multiplier $\xi^\mu$ so the interaction will be local only for one observer, but any other one will see the interaction as non-local. This shows the loss of absolute locality, effect baptized as relative locality.

In the next sections we will see that we can avoid this nonlocality choosing new space-time coordinates, which, in fact, do not commute.

\subsection{Construction of noncommutative spacetimes}

In the literature, a noncommutative spacetime is usually considered through new space-time coordinates $\tilde{x}$ constructed from canonical phase-space coordinates ($x$, $k$).\,\footnote{See Refs.~\cite{Meljanac:2016jwk},\cite{Loret:2016jrg},\cite{Carmona:2017oit} for recent references where this construction is used.} One can write these coordinates as a linear function of space-time coordinates  and a function $\varphi(k)$ through the combination 
\be
\tilde{x}^\mu \,=\, x^\nu \,\varphi^\mu_\nu(k)\,.
\label{eq:NCspt}
\ee
This set of functions has to reduce to the delta function when the momentum tends to zero (or when the high energy scale tends to infinity) in order to recover the SR result. Using  the usual Poisson brackets \begin{equation} 
\left\lbrace k_{\nu}\,,\,x^{\mu} \right\rbrace \,=\,\delta^{\mu}_{\nu}\,, 
\end{equation} 
the Poisson brackets of these new noncommutative space-time coordinates are  
\be
\begin{split}
\{\tilde{x}^\mu, \tilde{x}^\sigma\} &\,=\, \{ x^\nu \varphi^\mu_\nu(k), x^\rho \varphi^\sigma_\rho(k)\} \,=\, x^\nu \frac{\partial\varphi^\mu_\nu(k)}{\partial k_\rho} \,\varphi^\sigma_\rho(k) \,-\, x^\rho \frac{\partial\varphi^\sigma_\rho(k)}{\partial k_\nu} \,\varphi^\mu_\nu(k) \\
&=\, x^\nu \,\left(\frac{\partial\varphi^\mu_\nu(k)}{\partial k_\rho} \,\varphi^\sigma_\rho(k) \,-\, \frac{\partial\varphi^\sigma_\nu(k)}{\partial k_\rho} \,\varphi^\mu_\rho(k)\right)\,,
\end{split}
\label{eq:commNCspt}
\ee
and in phase space leads to the modified Heisenberg algebra 
\be
\{k_\nu, \tilde{x}^\mu\}\,=\, \varphi^\mu_\nu(k)\,.
\ee

Note that different choices of $\varphi^\mu_\nu(k)$ can lead to the same spacetime noncommutativity, leading to different representations of the same algebra. As it is shown in Appendix~\ref{ncst-rep}, different choices of canonical phase-space coordinates corresponding to different choices of momentum variables give different representations of the same space-time noncommutativity.    

\subsection{Noncommutative spacetime in \texorpdfstring{$\kappa$}{k}-Poincaré Hopf algebra}

The formalism of Hopf algebras gives a DCL, which is referred to as the ``coproduct'', and from it one can obtain, through the mathematical procedure known as ``pairing'', the resulting modified phase space. For the case of  $\kappa$-Poincaré in the bicrossproduct basis, as we have seen in the introduction, the coproduct of momenta $P_\mu$ is 
\be
\Delta(P_0)\,=\,P_0\otimes \mathbb{1}+ \mathbb{1}\otimes P_0 \,,\qquad \Delta(P_i)\,=\,P_i\otimes \mathbb{1}  + e^{-P_0/\Lambda} \otimes P_i\,,
\label{coproduct}
\ee
which leads to the composition law~\eqref{kappa-DCL} obtained in the previous chapter
\be
(p\oplus q)_0\,=\,p_0 + q_0 \,, \qquad (p\oplus q)_i \,=\,p_i  + e^{-p_0/\Lambda} q_i\,.
\label{composition}
\ee
The ``pairing'' operation of Hopf algebras formalism allows us to determine the modified Heisenberg algebra for a given coproduct~\cite{KowalskiGlikman:2002we,KowalskiGlikman:2002jr}. The bracket (pairing) $\langle *,* \rangle$  between momentum and position variables is defined as   
\be
\langle p_\nu , x^\mu\rangle \,=\, \delta^\mu_\nu\,.
\ee
This bracket has the following properties
\be
\langle p, x y\rangle  \,=\, \langle p_{(1)}, x \rangle \langle p_{(2)}, y \rangle\,,\qquad \langle pq, x \rangle  \,=\, \langle p, x_{(1)} \rangle \langle q_{(2)}, x_{(2)}\rangle\,,
\label{eq:pairing_pxx}
\ee
where we have used the notation 
\be
\Delta\, t\,=\,\sum \,t_{(1)}\otimes t_{(2)}\,.
\ee
One can see that by definition
\be
\langle \mathbb{1} , \mathbb{1}\rangle \,=\, \mathbb{1}\,.
\ee
Also, since momenta commute, the position coproduct is 
\be
\Delta\, x^\mu\,=\,\mathbb{1}\otimes x^\mu+ x^\mu \otimes \mathbb{1}\,.
\ee
In order to determine the Poisson brackets between the momentum and position one uses 
\be
\lbrace p,x \rbrace\,=\,x_{(1)}\langle p_{(1)}, x_{(2)}\rangle p_{(2)}- x\,p \,,
\label{eq:parentesis_xpp}
\ee
where $x\,p$ is the usual multiplication. 

For the bicrossproduct basis, one obtains from Eq.~\eqref{eq:pairing_pxx}
\be
\langle k_i, x^0 x^j \rangle\,=\,-\frac{1}{\Lambda}\,,\qquad \lbrace k_i, x^j x^0 \rbrace\,=\,0 \,,
\label{eq:parentesis_pxx}
\ee
and then 
\be
\lbrace x^0 , x^i \rbrace\,=\,-\frac{1}{\Lambda}x^i  \,.
\label{eq:parentesis_xx}
\ee
From Eq.~\eqref{eq:parentesis_xpp} one can deduce the rest of the phase space Poisson brackets (which are the ones showed in~\eqref{eq:pairing_intro}) 
\be
\lbrace\tilde{x}^0, k_0\rbrace \,=\,-1\,,\qquad  \lbrace\tilde{x}^0, k_i\rbrace \,=\,\frac{k_i}{\Lambda}\,,\qquad  \lbrace\tilde{x}^i, k_j\rbrace \,=\,-\delta^i_j\,,
\qquad \lbrace\tilde{x}^i, k_0\rbrace \,=\,0\,.
\label{eq:parentesis_xp}
\ee

In the case of  $\kappa$-Minkowski spacetime, the set of functions $\varphi^\mu_\nu(k)$ that leads to Eqs.~\eqref{eq:parentesis_xx}-\eqref{eq:parentesis_xp} is
\be
\varphi^0_0(k)=1 \,,\qquad \varphi^0_i(k)=-\frac{k_i}{\Lambda} \,,\qquad \varphi^i_j(k)=\delta^i_j \,,\qquad \varphi^i_0(k)=0\,.
\label{eq:phibicross}
\ee
We can also use  a covariant notation as we did in Ch.\ref{chapter_second_order} and rewrite Eq.~\eqref{eq:phibicross} as  
\be
\varphi^\mu_\nu(k)\,=\,\delta^\mu_\nu-\frac{1}{\Lambda} n^\mu k_\nu + \frac{k\cdot n}{\Lambda} n^\mu n_\nu\,,
\label{phi}
\ee
where, as in Ch.\ref{chapter_second_order}, $n^\mu$ is a fixed timelike vector of components $n^\mu=(1,0,0,0)$.

We have seen that, in the context of Hopf algebras, a DCL defines a noncommutative spacetime through the pairing operation. This connection is established from a purely mathematical perspective. As we have shown, to consider an action involving an interacting term with a DCL leads to nonlocal effects. What we propose is that the associated spacetime to a DCL could be defined asking for locality of interactions in such spacetime~\cite{Carmona:2017cry,Carmona:2019vsh}. In the next sections, we will discuss different possibilities to find such spacetime.

\section{First attempt to implement locality}
\label{sec:firstattempt}

For the sake of simplicity, let us consider the process of two particles in the initial state  with momenta $k$, $l$ and a total momentum $k\oplus l$, giving two particles in the final state with momenta $p$, $q$ and total momentum $p\oplus q$, i.e. we are considering the particular case $N=2$ of the relative locality model presented at the beginning of Sec.~\ref{sec:relative_locality_intro}. From Eq.~\eqref{eq:endWL} we find
\be
\begin{split}
w^\mu(0) \,=&\, \xi^\nu \frac{\partial(k\oplus l)_\nu}{\partial k_\mu}\,,\qquad x^\mu(0) \,=\, \xi^\nu \frac{\partial(k\oplus l)_\nu}{\partial l_\mu}\,,\\
y^\mu(0) \,=&\, \xi^\nu \frac{\partial(p\oplus q)_\nu}{\partial p_\mu} \,, \qquad z^\mu(0) \,=\, \xi^\nu \frac{\partial(p\oplus q)_\nu}{\partial q_\mu}\,,
\end{split}
\ee
where $w^\mu(0)$, $x^\mu(0)$ are the space-time coordinates of the end points of the worldlines of the initial state particles with momenta $k$, $l$ and $y^\mu(0)$, $z^\mu(0)$ the coordinates of the starting points of the worldlines of the final state particles with momenta $p$, $q$. 

When the composition law is the sum $p\oplus q = p + q$, which is the case of SR, the interaction is local $w^\mu(0) = x^\mu(0) = y^\mu(0) = z^\mu(0) = \xi^\mu$, so one can define events in spacetime through the interaction of particles. This is no longer possible when the DCL is nonlinear in momenta.   

In order to implement locality, one can introduce new space-time coordinates $\tilde{x}$ as in Eq.~\eqref{eq:NCspt}:
\be
\tilde{x}^\mu \,=\, x^\nu \,\varphi^\mu_\nu(k) ,
\label{eq:firstspt}
\ee
the functions $\varphi^\mu_\nu(k)$ being the same for all particles, so that in these coordinates the end and starting points of the worldlines are
\begin{align}
& \tilde{x}^\mu(0) \,=\, \xi^\nu  \frac{\partial (k\oplus l)_\nu}{\partial k_\rho} \, \varphi^\mu_\rho(k) \,,\qquad \tilde{w}^\mu(0) \,=\, \xi^\nu  \frac{\partial (k\oplus l)_\nu}{\partial l_\rho} \, \varphi^\mu_\rho(l) \,,
\nonumber \\ & \tilde{y}^\mu(0) \,=\, \xi^\nu  \frac{\partial (p\oplus q)_\nu}{\partial p_\rho} \, \varphi^\mu_\rho(p)\,,\qquad \tilde{z}^\mu(0) \,=\, \xi^\nu \frac{\partial (p\oplus q)_\nu}{\partial q_\rho} \, \varphi^\mu_\rho(q) \,.
\end{align}
Therefore, the interaction will be local if one can find for a given DCL a set of functions $\varphi^\mu_\nu$ such that\footnote{Note that the conservation of momenta implies that $k\oplus l = p\oplus q$.}
\be
\frac{\partial(k\oplus l)_\nu}{\partial k_\rho} \, \varphi^\mu_\rho(k) 
\,=\, \frac{\partial(k\oplus l)_\nu}{\partial l_\rho} \, \varphi^\mu_\rho(l) \,=\, \frac{\partial(p\oplus q)_\nu}{\partial p_\rho} \,\varphi^\mu_\rho(p) \,=\, \frac{\partial(p\oplus q)_\nu}{\partial q_\rho} \, \varphi^\mu_\rho(q) \,,
\label{loc0}
\ee
and then having $\tilde{w}^\mu(0)=\tilde{x}^\mu(0)=\tilde{y}^\mu(0)=\tilde{z}^\mu(0)$, making possible the definition of an event in this new spacetime. We can now consider the limit when one of the momenta $l$ goes to zero, using that $\lim_{l\to 0} (k\oplus l) = k, $\footnote{Remember the consistency condition of the DCL Eq.~\eqref{eq:demo1}.} giving place to the conservation law $k=p\oplus q$, and then, Eq.~(\ref{loc0}) implies that
\be
\boxed{\varphi^\mu_\nu(p\oplus q) \,=\, \frac{\partial (p\oplus q)_\nu}{\partial p_\rho} \,\varphi^\mu_\rho(p) \,=\, \frac{\partial (p\oplus q)_\nu}{\partial q_\rho} \, \varphi^\mu_\rho(q)} \,.
\label{loc1}
\ee
Taking the limit $p\to 0$ of the previous equation one finds
\be
\varphi^\mu_\nu(q) \,=\, \lim_{p\to 0} \frac{\partial (p\oplus q)_\nu}{\partial p_\mu} \,=\, \varphi^\mu_\nu(q) \,,
\label{eq:limit1}
\ee
where we have taken into account that $\lim_{p\to 0} \varphi^\mu_\rho(p) = \delta^\mu_\rho$.\footnote{Remember the conditions on the small momentum limit over $\varphi^\mu_\nu(p)$ explained after Eq.~\eqref{eq:NCspt}.} Moreover, taking the limit $q\to 0$ one has
\be
\varphi^\mu_\nu(p) \,=\, \varphi^\mu_\nu(p) \,=\, \lim_{q\to 0} \frac{\partial (p\oplus q)_\nu}{\partial q_\mu} \,.
\label{eq:limit2}
\ee
If we change the labels $p$ and $q$ in Eq.~\eqref{eq:limit2} and compare with Eq.~\eqref{eq:limit1}, we can conclude that
\be
\lim_{p\to 0} \frac{\partial (p\oplus q)_\nu}{\partial p_\mu} \,=\,\lim_{p\to 0} \frac{\partial (q\oplus p)_\nu}{\partial p_\mu}\,.
\label{eq:limitsym}
\ee
This condition is not satisfied by every DCL. In fact, one can see that a symmetric DCL  
\be
p\oplus q = q\oplus p\,,
\ee
satisfies Eq.~\eqref{eq:limitsym}. However, we know that the $\kappa$-Poincaré Hopf algebra composition law does not fulfill this requirement (see Eq.~\eqref{composition}).

Moreover, in Appendix~\ref{append-commut} it is proven that the way of implementing locality of Eq.~\eqref{loc1} gives rise to $\tilde{x}$ coordinates which in fact are commutative, so one can identify new variables $\tilde{p}_\mu=g_\mu(p)$ satisfying $\{\tilde{p}_\nu, \tilde{x}^\mu\} = \delta^\mu_\nu$, which corresponds to a linear DCL, $[\tilde{p}\,\tilde{\oplus}\, \tilde{q}]_\mu = \tilde{p}_\mu + \tilde{q}_\mu$. This is telling us that this implementation is related by a canonical transformation to the SR variables $(\tilde{x},\tilde{p})$ and then, the new spacetime obtained asking for locality is just the spacetime of SR. Since the previous procedure does not let us study a generic (noncommutative) composition law (as is the case $\kappa$-Poincaré), we have to find some way to go beyond Eq.~\eqref{loc1}.  

\section{Second attempt: two different spacetimes in the two-particle system}
\label{section_second_attempt}

We have seen in the previous chapters that, in order to implement a relativity principle in a DRK with a generic composition law, nontrivial Lorentz transformations in the one and two-particle systems are requested. In general, the two-particle system transformations mix both momentum variables. Hence, it is not then strange that the noncommutative coordinates one should introduce for having local interactions for a generic DCL should also mix momenta. One possible way to proceed is to consider the simple case
\be
\tilde{y}^\mu \,=\, y^\nu \,\varphi_{L\,\nu}^{\,\mu}(p, q) \,,\quad
\tilde{z}^\mu \,=\, z^\nu \, \varphi_{R\,\nu}^{\,\mu}(p, q) \,.
\label{y-z-coordinates}
\ee  

The interactions will be local if the following equation holds
\be
\boxed{\varphi^\mu_\nu(p\oplus q) \,=\, \frac{\partial (p\oplus q)_\nu}{\partial p_\rho} \,\varphi_{L\,\rho}^{\,\mu}(p, q) \,=\, \frac{\partial (p\oplus q)_\nu}{\partial q_\rho} \, \varphi_{R\,\rho}^{\,\mu}(p, q)} \,.
\label{loc2}
\ee

We now define the functions $\phi_L$, $\phi_R$ through the composition law $p\oplus q$ as
\be
\phi_{L\,\sigma}^{\:\:\nu}(p, q) \,\frac{\partial(p\oplus q)_\nu}{\partial p_\rho} \,=\, \delta^\rho_\sigma\,,\quad
\phi_{R\,\sigma}^{\:\:\nu}(p, q) \,\frac{\partial(p\oplus q)_\nu}{\partial q_\rho} \,=\, \delta^\rho_\sigma \,.
\ee
These functions allow us to write the spacetime of a two-particle system given the spacetime of a one-particle system (i.e. $\varphi$):
\be
\varphi_{L\,\sigma}^{\:\:\mu}(p, q) \,=\, \phi_{L\,\sigma}^{\:\:\nu}(p, q) \,\, \varphi^\mu_\nu(p\oplus q) \,,\quad
\varphi_{R\,\sigma}^{\:\:\mu}(p, q) \,=\, \phi_{R\,\sigma}^{\:\:\nu}(p, q) \,\, \varphi^\mu_\nu(p\oplus q)\,.
\label{phiL-phiR-phi}
\ee
As $\phi_{L\,\sigma}^{\:\:\nu}(p, 0) = \phi_{R\,\sigma}^{\:\:\nu}(0, q) = \delta^\nu_\sigma$, then 
\be
\varphi_{L\,\sigma}^{\:\:\mu}(p, 0) = \varphi^\mu_\sigma(p)\,, \quad \varphi_{R\,\sigma}^{\:\:\mu}(0, q) = \varphi^\mu_\sigma(q)\,,
\ee
which is the result of taking the limits $q\to 0$, $p\to 0$ in Eq.~(\ref{loc2}).

Then, given a function $\varphi$ and a DCL, without any relation between them, locality can always be implemented. However, if the DCL is constructed by the multiplication of plane waves with a  noncommutative spacetime~\cite{Meljanac:2009ej},\cite{Battisti:2010sr} or in the Hopf algebra framework~\cite{Lukierski:1991pn}, this is not the case: given a specific representation of a particular noncommutativity, one and only one DCL is obtained. Therefore, there is an ambiguity in how to select these two ingredients from the perspective we are considering here. This shows that an additional criteria should be looked for in order to establish such connection.

 A possible way to restrict these two ingredients is to consider the relation given by the geometrical interpretation studied in Ch.~\ref{chapter_curved_momentum_space}. In order to reproduce the result obtained in the previous chapter, and then the relation between DCL and spacetime given by the Hopf algebras formalism, we ask that in the two-particle system the spacetime of one of the particles should not depend on the other momentum,      
\be
\varphi^{\:\:\mu}_{R\rho}(p, q) \,=\, \varphi^{\:\:\mu}_{R\rho}(0, q) \,=\,\varphi^\mu_\rho(q) \,,
\label{eq:restriction_R}
\ee
and Eq.~(\ref{loc2}) implies that
\be
\varphi^\mu_\nu(p\oplus q) \,=\, \frac{\partial (p\oplus q)_\nu}{\partial q_\rho} \,\varphi^\mu_\rho(q)\,,
\label{varphi-oplus}
\ee
which determines the DCL for a given noncommutativity (i.e., for a given function $\varphi$). Taking the limit $p\to 0$ one has
\be
\varphi^\mu_\nu(p)  \,=\, \lim_{q\to 0}  \frac{\partial (p\oplus q)_\nu}{\partial q_\mu}\,,
\label{magic_formula}
\ee
and therefore, it is possible to determine the one-particle spacetime given a certain composition law. Eq.~(\ref{magic_formula}) can be interpreted in a simple way: the infinitesimal change of the momentum variable $p_\mu$  generated by the noncommutative space-time coordinates $\tilde{x}$ with parameters $\epsilon$ is
\be
\delta p_\mu \,=\, \epsilon_\nu \lbrace\tilde{x}^\nu, p_\mu\rbrace \,=\, - \epsilon_\nu \varphi^\nu_\mu(p) \,=\, - \epsilon_\nu \lim_{q\to 0} \frac{\partial(p\oplus q)_\mu}{\partial q_\nu} \,=\, - \left[(p\oplus \epsilon)_\mu - p_\mu\right]  \,.
\label{deltap}
\ee
The noncommutative coordinates can be interpreted as the translation generators in momentum space defined by the DCL. The interpretation of the DCL as the momentum (right-) translation generators is the same one found in the previous chapter, which is obvious since we have imposed the same relation obtained in the geometrical context relating the composition law with the tetrad of the momentum space (observe that in Eqs.~\eqref{varphi-oplus}, \eqref{T(a,k)} the set of functions giving the noncommutativity plays the same role than the tetrad of the momentum space)~\footnote{Note that in fact the $\varphi$ functions transform under a canonical transformation~\eqref{ncst-rep} as a tetrad does under a change of momentum coordinates (see Appendix~\ref{ncst-rep}).}. Then, one can construct the physical coordinates by multiplying the canonical space-time coordinates times the momentum space tetrad. This is restricting the possible noncommutativity if one wants to keep a relativistic kinematics obtained from a geometrical interpretation, since only $\kappa$-Minkowski is allowed (see the discussion of~\ref{subsection_kappa_desitter}). In order to study other possible DRK in the geometrical framework one has to lift the restriction imposed in~\eqref{eq:restriction_R}.     

Similar results would have been obtained if we would have considered the case where it is the spacetime of the particle with momentum $p$ which is chosen to be independent of the particle with momentum $q$. In this case one would have
\be
\varphi^{\:\:\mu}_{L\rho}(p, q) \,=\, \varphi^{\:\:\mu}_{L\rho}(p, 0) \,=\,\varphi^\mu_\rho(p)\,,
\ee
\be
\varphi^\mu_\nu(p\oplus q) \,=\, \frac{\partial (p\oplus q)_\nu}{\partial p_\rho} \,\varphi^\mu_\rho(p) \,,
\label{varphi-oplus2}
\ee
\be
\varphi^\mu_\nu(q)  \,=\, \lim_{p\to 0}  \frac{\partial (p\oplus q)_\nu}{\partial p_\mu}\,,
\label{eq:phi-tau}
\ee
and
\be
\delta p_\mu \,=\,- \left[(\epsilon\oplus p)_\mu - p_\mu\right] \,.
\label{eq:TP}
\ee

Then, for the $\kappa$-Minkowski noncommutativity with the proposed prescription, given a DCL one obtains two possible different noncommutative spacetimes (different representations of the same algebra up to a sign) given by Eqs.~\eqref{magic_formula}, \eqref{eq:phi-tau}\footnote{This ambivalence materializes also in the geometrical framework since the relation between tetrad and translations (composition law) is the same as in the locality framework, causing that the same DCL leads to two different coordinate representations of de Sitter space (and then different tetrads).}. This ambiguity, together with the possible existence of a privileged choice of (physical) momentum variables, are open problems deserving further study.

\subsection{Application to \texorpdfstring{$\kappa$}{k}-Poincaré}

In this subsection we will see how to implement locality in the particular case of $\kappa$-Poincaré kinematics. We start by considering the noncommutativity of $\kappa$-Minkowski 
\be
\lbrace\tilde{x}^\mu, \tilde{x}^\nu\rbrace \,=\, \frac{1}{\Lambda} \,\left( \tilde{x}^\mu n^\nu -  \tilde{x}^\nu n^\mu \right) \,,
\ee
where $\varphi^\mu_\nu(k)$ is such that (using \eqref{eq:NCspt} and \eqref{eq:commNCspt})
\be
\frac{\partial\varphi^\mu_\alpha(k)}{\partial k_\beta} \varphi^\nu_\beta(k) - \frac{\partial\varphi^\nu_\alpha(k)}{\partial k_\beta} \varphi^\mu_\beta(k) \,=\, \frac{1}{\Lambda} \,\left( \varphi^\mu_\alpha(k) n^\nu - \varphi^\nu_\alpha(k) n^\mu \right)\,.
\ee

By virtue of simplicity, we will take $\varphi^\mu_\nu(k)$ to be the one appearing in the bicrossproduct basis, Eq.~\eqref{phi}. Imposing $\varphi_{L\,\nu}^{\:\:\mu}(p, q)\,=\, \varphi^\mu_\nu(p)$, we can find unequivocally the DCL from Eq.~(\ref{varphi-oplus}). The result (see Appendix~\ref{append-bicross}) is the  DCL obtained in that basis~\eqref{composition}. As we saw in  Ch.~\ref{chapter_curved_momentum_space}, if we consider the functions $\varphi^\mu_\nu(k)$ to be the tetrad in momentum space of Eq.~\eqref{bicross-tetrad} and we impose  $\varphi_{R\,\nu}^{\:\:\mu}(p, q)\,=\, \varphi^\mu_\nu(q)$, we obtain exactly the same DCL, the one we understood in that chapter as the translations in a de Sitter momentum space. With all this we see that the framework of Hopf algebras is contained as a particular case in our proposal of implementation of locality.  
  
In  Ch.~\ref{chapter_curved_momentum_space} we have shown how to implement the relativity principle from geometrical considerations. Here we will follow another approach without any mention to geometry, which leads to another implementation of the relativity principle. The DLT in the one-particle system is obtained given the function $\varphi^\mu_\nu(k)$ of \eqref{phi} asking for the noncomutative spacetime to form a ten-dimensional Lie algebra (see Appendix~\ref{append-2pLT}), obtaining~\eqref{eq:j_momentum_space}, and therefore, the Casimir and the Lorentz transformation in the two-particle system are the ones obtained in~\ref{subsection_kappa_desitter}, Eqs.~\eqref{eq:casimir_momentum_space} and~\eqref{eq:jr_momentum_space} respectively.

In order to complete the discussion of  the $\kappa$-Poincaré algebra from the point of view of locality of interactions, one can determine $\varphi_{R\,\nu}^{\:\:\mu}(p, q)$ through Eq.~\eqref{phiL-phiR-phi}:
\be
\varphi_{R\,\nu}^{\:\:\mu}(p, q)\,=\,\delta^\mu_\nu \,e^{pn/\Lambda}+\frac{1}{\Lambda}n^\mu\left(n_\nu(e^{pn/\Lambda}\,pn+qn+(1-e^{pn/\Lambda})\,\Lambda)-e^{pn/\Lambda}\,p_\nu-q_\nu\right) ,
\ee
or, in components,
\be
\varphi_{R\,0}^{\:\:0}(p, q)\,=\,1 \,,\quad \varphi_{R\,0}^{\:\:i}(p, q)\,=\,0 \,,\quad \varphi_{R\,i}^{\:\:0}(p, q)\,=\,-\frac{e^{p_0/\Lambda}p_i+q_i}{\Lambda}\,, \quad \varphi_{R\,j}^{\:\:i}(p, q)\,=\,e^{p_0/\Lambda}\delta^i_j \,.
\ee
As expected,
\be
\varphi_{R\,\nu}^{\:\:\mu}(0, q)\,=\, \varphi^\mu_\nu(q)\,.
\ee
Now, we can compute the two-particle phase-space Poisson brackets that are different from zero:
\be
\begin{split}
\lbrace\tilde{y}^0, \tilde{y}^i\rbrace \,=\,-\frac{\tilde{y}^i}{\Lambda}\,,\quad  \lbrace\tilde{y}^0, p_0\rbrace \,=\,-1\,,\quad  \lbrace\tilde{y}^0, p_i\rbrace \,=\,\frac{p_i}{\Lambda}\,,\quad  \lbrace\tilde{y}^i, p_j\rbrace \,=\,-\delta^i_j\,, \quad \lbrace\tilde{y}^0, \tilde{z}^i\rbrace \,=\,-\frac{\tilde{z}^i}{\Lambda},\\
\lbrace\tilde{z}^0, \tilde{z}^i\rbrace \,=\,-\frac{\tilde{z}^i}{\Lambda}\,,\quad  \lbrace\tilde{z}^0, q_0\rbrace \,=\,-1\,,\quad  \lbrace\tilde{z}^0, q_i\rbrace \,=\,\frac{e^{p_0/\Lambda}p_i+q_i}{\Lambda}\,,\quad  \lbrace\tilde{z}^i, q_j\rbrace \,=\,-e^{p_0/\Lambda}\delta^i_j\,.
\end{split}
\ee
Note that all the Poisson brackets of the two space-time coordinates close an algebra, being independent of momenta. 

The one-particle noncommutative spacetime we get from locality when we impose  $\varphi_{L\,\nu}^{\:\:\mu}(p, q)\,=\, \varphi^\mu_\nu(p)$ is the one obtained through the pairing operation in the Hopf algebra framework. This leads us to interpret that algebraic procedure from the physical criteria of imposing locality of interactions, understanding how a noncommutative spacetime crops up (and thus, a modification of the Poisson brackets of phase-space coordinates) in a natural way when a DCL is considered.    

\section{Third attempt: mixing of space-time coordinates}
\label{sec_st_locality_third}
We have seen in the previous section that, in the way proposed to implement locality, there is no restriction on the noncommutative spacetime, nor on the composition law. Any combination of both ingredients admits the implementation of locality. 

In this section, we pose another way to implement locality, imposing that the noncommutative coordinates are defined as a sum of two terms, each one having only the phase-space coordinates of one of the particles,
\be
\tilde{y}^\alpha \,=\, y^\mu \varphi^\alpha_\mu(p) + z^\mu \varphi^{(2)\alpha}_{(1)\mu}(q)\,, \quad\quad
\tilde{z}^\alpha \,=\, z^\mu \varphi^\alpha_\mu(q) + y^\mu \varphi^{(1)\alpha}_{(2)\mu}(p)\,.
\ee
We impose that $\varphi^{(2)\alpha}_{(1)\mu}(0) = \varphi^{(1)\alpha}_{(2)\mu}(0) = 0$ so, when one of the momenta tends to zero, the one-particle coordinates are $\tilde{x}^\alpha = x^\mu \varphi^\alpha_\mu(k)$. 

Locality in the generalized spacetime requires to find a set of functions $\varphi^\alpha_\mu(k)$, $\varphi^{(2)\alpha}_{(1)\mu}(k)$ and $\varphi^{(1)\alpha}_{(2)\mu}(k)$ satisfying the set of equations
{\small \be
\boxed{\varphi^\alpha_\nu(p\oplus q)\,=\,\frac{\partial(p\oplus q)_\mu}{\partial p_\nu} \varphi^\alpha_\nu(p) + 
\frac{\partial(p\oplus q)_\mu}{\partial q_\nu} \varphi^{(2)\alpha}_{(1)\nu}(q) =\frac{\partial(p\oplus q)_\mu}{\partial q_\nu} \varphi^\alpha_\nu(q) + 
\frac{\partial(p\oplus q)_\mu}{\partial p_\nu} \varphi^{(1)\alpha}_{(2)\nu}(p)}\,.
\label{loc-eq}
\ee}
\normalsize

The set of functions $\varphi^{(2)\alpha}_{(1)\mu}(k)$ and $\varphi^{(1)\alpha}_{(2)\mu}(k)$ can be obtained given $\varphi^\alpha_\mu(k)$ and the DCL taking  the limit $p\to 0$ or $q\to 0$ in \eqref{loc-eq}
\be
\varphi^{(2)\alpha}_{(1)\mu}(q) \,=\, \varphi^\alpha_\mu(q) - \lim_{k\to 0}  \frac{\partial(k\oplus q)_\mu}{\partial k_\alpha}, \quad\quad
\varphi^{(1)\alpha}_{(2)\mu}(p) \,=\, \varphi^\alpha_\mu(p) - \lim_{k\to 0} \frac{\partial(p\oplus k)_\mu}{\partial k_\alpha}\,.
\label{eq:phi12}
\ee

Using the functions $\varphi^{(2)}_{(1)}$, $\varphi^{(1)}_{(2)}$ into the locality equations, we find
\begin{align}
& \frac{\partial(p\oplus q)_\mu}{\partial q_\nu} \, \lim_{l\to 0} \frac{\partial(l\oplus q)_\nu}{\partial l_\alpha} \,=\, \frac{\partial(p\oplus q)_\mu}{\partial p_\nu}  \, \lim_{l\to 0} \frac{\partial(p\oplus l)_\nu}{\partial l_\alpha} \,=\, \nonumber \\ & \:\:\: \frac{\partial(p\oplus q)_\mu}{\partial p_\nu} \varphi^\alpha_\nu(p) + 
\frac{\partial(p\oplus q)_\mu}{\partial q_\nu} \varphi^\alpha_\nu(q) - \varphi^\alpha_\mu(p\oplus q)\,.
\label{loc-oplus-varphi}
\end{align}
The first equality imposes a condition on the DCL in order to be compatible with locality, while the second one is establishing a relation between the functions $\varphi^\alpha_\mu$ and the DCL. 

We can introduce the relative coordinate
\begin{align}
\tilde{x}^\alpha_{(12)} \,\doteq\, \tilde{y}^\alpha - \tilde{z}^\alpha & = y^\mu \left[\varphi^\alpha_\mu(p) - \varphi^{(2)\alpha}_{(1)\mu}(p)\right] - z^\mu 
\left[\varphi^\alpha_\mu(q) - \varphi^{(1)\alpha}_{(2)\mu}(q)\right] \nonumber \\
&= y^\mu \,\lim_{l\to 0} \frac{\partial(p\oplus l)_\mu}{\partial l_\alpha} - z^\mu \,\lim_{l\to 0} \frac{\partial(l\oplus q)_\mu}{\partial l_\alpha}\,.
\label{xtilde}
\end{align}
Now, one might use the total momentum to spot the effect of an infinitesimal translation with parameters $\epsilon^\mu$ on the relative coordinate
\begin{align}
\delta\tilde{x}^\alpha_{(12)} \,&=\, \epsilon^\mu \{\tilde{x}^\alpha_{(12)}, (p\oplus q)_\mu\} \nonumber \\
&=\, \epsilon^\mu \left[- \frac{\partial(p\oplus q)_\mu}{\partial p_\nu} \lim_{l\to 0} \frac{\partial(p\oplus l)_\nu}{\partial l_\alpha} + \frac{\partial(p\oplus q)_\mu}{\partial q_\nu} \lim_{l\to 0} \frac{\partial(l\oplus q)_\nu}{\partial l_\alpha}\right]\,.
\end{align}
The last term of the previous expression is zero as a consequence of the conditions that the DCL must satisfy in order to be possible to implement locality.  This is showing the invariance of the relative coordinate under translations, implying that if one observers sees the interaction as local, it will be local for any other observer translated with respect to the former. 

It is easy to test out that the following identities hold, 
\begin{align}
& \frac{\partial(p\oplus q)_\mu}{\partial q_\nu} \, \lim_{l\to 0} \frac{\partial(l\oplus q)_\nu}{\partial l_\alpha} \,=\, \lim_{l\to 0}  \frac{\partial(p\oplus (l\oplus q))_\mu}{\partial (l\oplus q)_\nu} \, \frac{\partial(l\oplus q)_\nu}{\partial l_\alpha} \,=\, \lim_{l\to 0} \frac{\partial(p\oplus (l\oplus q))_\mu}{\partial l_\alpha}\,, \\ \nonumber
& \frac{\partial(p\oplus q)_\mu}{\partial p_\nu}  \, \lim_{l\to 0} \frac{\partial(p\oplus l)_\nu}{\partial l_\alpha} \,=\,  \lim_{l\to 0} \,\frac{\partial((p \oplus l)\oplus q)_\mu}{\partial (p\oplus l)_\nu}  \, \frac{\partial(p\oplus l)_\nu}{\partial l_\alpha} \,=\, \lim_{l\to 0} \frac{\partial((p\oplus l)\oplus q)_\mu}{\partial l_\alpha}\,,
\end{align}
and then, from the first equality of~\eqref{loc-oplus-varphi} one can find 
\be
\lim_{l\to 0} \frac{\partial(p\oplus (l\oplus q))_\mu}{\partial l_\alpha} \,=\, \lim_{l\to 0} \frac{\partial((p\oplus l)\oplus q)_\mu}{\partial l_\alpha}\,,
\ee
which leads to 
\be
(p\oplus \epsilon)\oplus q \,=\, p\oplus(\epsilon\oplus q)\,.
\label{eq:associativity}
\ee

Now we can check that in fact, any associative composition law is compatible with the implementation of locality. Making the choice  $\varphi^{(2)\alpha}_{(1)\mu}(q)=0$ in Eq.~\eqref{eq:phi12}\footnote{This can be done also for the alternative choice $\varphi^{(1)\alpha}_{(2)\mu}(p)=0$.}, one has
\begin{equation}
\begin{split}
\frac{\partial(p\oplus q)_\mu}{\partial p_\nu} \,\varphi^\alpha_\nu(p) &\,=\, \frac{\partial(p\oplus q)_\mu}{\partial p_\nu} \,\lim_{l\to 0} \frac{\partial(l\oplus p)_\nu}{\partial l_\alpha}\\& =\, \lim_{l\to 0} \left[\frac{\partial((l\oplus p)\oplus q)_\mu}{\partial(l\oplus p)_\nu} \,\frac{\partial(l\oplus p)_\nu}{\partial l_\alpha}\right]
\,=\, \lim_{l\to 0} \frac{\partial((l\oplus p)\oplus q)_\mu}{\partial l_\alpha}\,,  \\
\frac{\partial(p\oplus q)_\mu}{\partial q_\nu} \,\varphi^\alpha_\nu(q) &\,=\, \frac{\partial(p\oplus q)_\mu}{\partial q_\nu} \,\lim_{l\to 0} \frac{\partial(l\oplus q)_\nu}{\partial l_\alpha} \\ & =\, \lim_{l\to 0} \left[\frac{\partial(p\oplus(l\oplus q)_\mu}{\partial(l\oplus q)_\nu} \,\frac{\partial(l\oplus q)_\nu}{\partial l_\alpha}\right]\,=\, \lim_{l\to 0} \frac{\partial(p\oplus(l\oplus q)_\mu}{\partial l_\alpha}\,, \\
\varphi^\alpha_\mu(p\oplus q) &\,=\, \lim_{l\to 0} \frac{\partial(l\oplus(p\oplus q))_\mu}{\partial l_\alpha}\,.
\end{split}
\end{equation}

It is easy to verify that Eqs.~\eqref{loc-oplus-varphi} hold
\be
\begin{split}
\lim_{l\to 0} \frac{\partial(p\oplus (l\oplus q))_\mu}{\partial l_\alpha} \,&=\, \lim_{l\to 0} \frac{\partial((p\oplus l)\oplus q)_\mu}{\partial l_\alpha} \,=\, \lim_{l\to 0} \frac{\partial((l\oplus p)\oplus q)_\mu}{\partial l_\alpha} \\ &\,+\, \lim_{l\to 0} \frac{\partial(p\oplus(l\oplus q))_\mu}{\partial l_\alpha} \,-\,\lim_{l\to 0} \frac{\partial(l\oplus(p\oplus q))_\mu}{\partial l_\alpha}\,,
\end{split}
\label{eq:associativity_locality_proof}
\ee
proving that any associative composition law is locality compatible.

\subsection{First-order deformed composition law of four-momenta (DCL1)}
\label{sec_first_order}

In this subsection, we regard a DCL with only linear terms in the inverse of $\Lambda$ and we see what conditions the implementation of locality enforces. At first order, we can write the most general isotropic composition law (DCL1) in a covariant way  
\be
(p\oplus q)_\mu \,=\, p_\mu + q_\mu + \frac{c_\mu^{\nu\rho}}{\Lambda} p_\nu q_\rho\,,
\ee
where $c_\mu^{\nu\rho}$ is  
\be
c_\mu^{\nu\rho} \,=\, c_1 \,\delta_\mu^\nu n^\rho + c_2 \,\delta_\mu^\rho n^\nu + c_3 \,\eta^{\nu\rho} n_\mu + c_4 \,n_\mu n^\nu n^\rho + c_5 \,\epsilon_\mu^{\:\:\nu\rho\sigma} n_\sigma\,,
\ee
being $n_\mu = (1, 0, 0, 0)$ and $c_i$ arbitrary constants. This leads to the general composition law~\eqref{cl1} studied in Ch.~\ref{chapter_second_order}.

We can wonder now if this composition can satisfy the conditions imposed by locality. We have 
\begin{align}
& \frac{\partial(p\oplus q)_\mu}{\partial q_\nu} \,=\, \delta^\nu_\mu + \frac{c_\mu^{\rho\nu}}{\Lambda} p_\rho\,, \quad\quad \lim_{l\to 0} \frac{\partial(l\oplus q)_\nu}{\partial l_\alpha} \,=\, \delta^\alpha_\nu + \frac{c^{\alpha\sigma}_\nu}{\Lambda} q_\sigma\,, \\ \nonumber 
& \frac{\partial(p\oplus q)_\mu}{\partial p_\nu} \,=\, \delta^\nu_\mu + \frac{c_\mu^{\nu\sigma}}{\Lambda} q_\sigma\,, \quad\quad  \lim_{l\to 0} \frac{\partial(p\oplus l)_\nu}{\partial l_\alpha} \,=\, \delta^\alpha_\nu + \frac{c_\nu^{\rho\alpha}}{\Lambda} p_\rho\,,
\end{align}
and
\begin{align}
& \frac{\partial(p\oplus q)_\mu}{\partial q_\nu} \, \lim_{l\to 0} \frac{\partial(l\oplus q)_\nu}{\partial l_\alpha} \,=\, \delta^\alpha_\mu + \frac{c_\mu^{\rho\alpha}}{\Lambda} p_\rho + \frac{c^{\alpha\sigma}_\mu}{\Lambda} q_\sigma + \frac{c_\mu^{\rho\nu} c^{\alpha\sigma}_\nu}{\Lambda^2} p_\rho q_\sigma\,, \nonumber \\
& \frac{\partial(p\oplus q)_\mu}{\partial p_\nu}  \, \lim_{l\to 0} \frac{\partial(p\oplus l)_\nu}{\partial l_\alpha} \,=\,  \delta^\alpha_\mu + \frac{c_\mu^{\rho\alpha}}{\Lambda} p_\rho + \frac{c^{\alpha\sigma}_\mu}{\Lambda} q_\sigma + \frac{c_\mu^{\nu\sigma} c^{\rho\alpha}_\nu}{\Lambda^2} p_\rho q_\sigma\,.
\end{align}  
Then, a DCL1 will be compatible with locality if the following equality holds 
\be
c_\mu^{\rho\nu} c^{\alpha\sigma}_\nu \,=\, c_\mu^{\nu\sigma} c^{\rho\alpha}_\nu\,.
\ee
This requirement is equivalent to demand to the composition to be associative, which is the condition~\eqref{eq:associativity} when the composition law has only first order terms. We obtain  four possible cases for the DCL1
\be
c_\mu^{\nu\rho} \,=\, \delta_\mu^\rho n^\nu, \quad  c_\mu^{\nu\rho} \,=\, \delta_\mu^\nu n^\rho, \quad c_\mu^{\nu\rho} \,=\, \delta_\mu^\nu n^\rho + \delta_\mu^\rho n^\nu - n_\mu n^\nu n^\rho, \quad  c_\mu^{\nu\rho} \,=\, \eta^{\nu\rho} n_\mu - n_\mu n^\nu n^\rho\,.
\ee
The last two cases are not relevant  because it is easy to check that the compositions are obtained by a change of basis ($k'_\mu = f_\mu(k)$)  from the sum  ($(p'\oplus' q')_\mu \doteq (p\oplus q)'_\mu=p'_\mu + q'_\mu$), and we have SR in knotty variables\footnote{For the first of them the function is $f_0(k)=\Lambda \log(1+k_0/\Lambda)$, $f_i(k)=(1+k_0/\Lambda) k_i$, while for the last one $f_0(k)=k_0+\vec{k}^2/(2\Lambda)$, $f_i(k)=k_i$.}.

In the first two cases, we have a noncommutative composition law (in fact the latter is obtained from the former exchanging the role of the momentum). As the composition is not commutative, it is not possible to find a change of basis in which momenta compose additively. 

The explicit form of the first DCL1 is
\be
(p\oplus q)_0 \,=\, p_0 + q_0 + \epsilon\, \frac{p_0 q_0}{\Lambda} \, ,\quad\quad\quad\quad
(p\oplus q)_i \,=\, p_i + q_i + \epsilon\, \frac{p_0 q_i}{\Lambda}\,,
\label{DCL(1)}
\ee
where $\epsilon = \pm 1$ is an overall sign for the modification in the composition law and an arbitrary constant can be reabsorbed in the definition of the scale $\Lambda$. We will see in Sec.~\ref{sec_comparison} that this composition law corresponds in fact to $\kappa$-Poincaré. 

When $\epsilon=-1$, one has
\be
\left(1-\frac{(p\oplus q)_0}{\Lambda}\right) \,=\, \left(1-\frac{p_0}{\Lambda}\right) \left(1-\frac{q_0}{\Lambda}\right)\,,
\ee
making the scale $\Lambda$ to play the role of a cutoff in the energy, being therefore the choice of sign that reproduces the DCL in the DSR framework, as we will see in Sec.~\ref{sec_comparison}.  With the other choice of sign $\epsilon=+1$, the scale $\Lambda$ is not a maximum energy, going beyond DSR scenarios.

From the explicit form of  the local DCL1~\eqref{DCL(1)} we can obtain the expression for the relative generalized space-time coordinates 
\begin{align}
  & \lim_{l\to 0} \frac{\partial(p\oplus l)_0}{\partial l_0} \,=\, 1 + \epsilon \frac{p_0}{\Lambda}\,,& \quad\quad &\lim_{l\to 0} \frac{\partial(p\oplus l)_0}{\partial l_i} \,=\, 0\,,\nonumber \\
  &\lim_{l\to 0} \frac{\partial(p\oplus l)_i}{\partial l_0} \,=\, 0\,,&\quad\quad &  \lim_{l\to 0} \frac{\partial(p\oplus l)_i}{\partial l_j} \,=\, \delta_i^j \left(1 + \epsilon \frac{p_0}{\Lambda}\right)\,,\nonumber \\
  & \lim_{l\to 0} \frac{\partial(l\oplus q)_0}{\partial l_0} \,=\, 1 + \epsilon \frac{q_0}{\Lambda}\,,& \quad\quad &\lim_{l\to 0} \frac{\partial(l\oplus q)_0}{\partial l_i} \,=\, 0\,, \nonumber \\ 
  &\lim_{l\to 0} \frac{\partial(l\oplus q)_i}{\partial l_0} \,=\, \epsilon \frac{q_i}{\Lambda}\,,&\quad\quad &\lim_{l\to 0} \frac{\partial(l\oplus q)_i}{\partial l_j} \,=\, \delta^j_i\,,
\end{align}
and then
\be
\tilde{x}^0_{(12)} \,=\, y^0 (1+\epsilon p_0/\Lambda) - z^0 (1+\epsilon q_0/\Lambda) - z^j \epsilon q_j/\Lambda\,, \quad
\tilde{x}^i_{(12)} \,=\, y^i (1+\epsilon p_0/\Lambda) - z^i\,.
\ee
Therefore, one can check that the relative space-time coordinates of the two-particle system are in fact the coordinates of a $\kappa$-Minkowski spacetime with $\kappa=\epsilon/\Lambda$
\begin{align}
\{\tilde{x}^i_{(12)}, \tilde{x}^0_{(12)}\} \,=&\,\{y^i (1+\epsilon p_0/\Lambda), y^0 (1+\epsilon p_0/\Lambda)\} + \{z^i, z^j \epsilon q_j/\Lambda\} \nonumber \\ =& (\epsilon/\Lambda) \,\left[y^i (1+\epsilon p_0/\Lambda) - z^i\right] \,=\, (\epsilon/\Lambda) \,\tilde{x}^i_{(12)}\,.
\end{align}

In order to obtain the generalized space-time coordinates of the two-particle system, one has to solve \eqref{loc-oplus-varphi}, using the local DCL1, for the functions $\varphi^\alpha_\mu(k)$. The main issue is that these equations do not completely determine the explicit form of $\varphi^\alpha_\mu(k)$ and therefore, the  generalized space-time coordinates of the one-particle are not completely determined. To do so, we need another requirement.

One can observe that the local DCL1~\eqref{DCL(1)} 
\be
\left(p\oplus q\right)_\mu \,=\, p_{\:\mu} + \left(1 + \epsilon p_0/\Lambda\right) \,q_{\:\mu}\,,
\ee
is a sum of $p_\mu$ (independent of $q$) and a term proportional to $q_\mu$ depending on $p$. Then one can consider an ad hoc prescription in which the generalized space-time coordinates $\tilde{y}^\mu$ depends only on its phase-space coordinates ($y, p$), while $\tilde{z}^\mu$ depends on the phase-space coordinates of both particles ($y, p, z, q$), making 
\be
\varphi^{(2)\alpha}_{(1)\mu}(q) \,=\, 0\,, \quad\quad \rightarrow \quad\quad \varphi^\alpha_\mu(p) \,=\, \lim_{l\to 0} \frac{\partial(l\oplus p)_\mu}{\partial l_\alpha}\,.
\label{simplphi}
\ee
This can be done since, as we have proven in Eq.~\eqref{eq:associativity_locality_proof}, any associative composition law (as is the case for DCL1) is compatible with locality with the choice $\varphi^{(2)\alpha}_{(1)\mu}(q)=0$.
We can obtain the generalized space-time coordinates for the one-particle system through the explicit expression of the DCL1
\begin{align}
\tilde{x}^0 \,=&\, x^\mu \lim_{l\to 0} \frac{\partial(l\oplus k)_\mu}{\partial l_0} \,=\, x^0 (1 + \epsilon k_0/\Lambda) + x^j \epsilon k_j/\Lambda\,, \nonumber \\
\tilde{x}^i \,=&\, x^\mu \lim_{l\to 0} \frac{\partial(l\oplus k)_\mu}{\partial l_i} \,=\, x^i\,,
\label{eq:tilde_1}
\end{align}
and then
\be
\{\tilde{x}^i, \tilde{x}^0\} \,=\, \{x^i, x^j \epsilon k_j/\Lambda\} \,=\, - (\epsilon/\Lambda) x^i \,=\, - (\epsilon/\Lambda) \tilde{x}^i\,.
\label{xtilde-}
\ee
This is obvious from the fact that, as we saw in the previous sections, if the relation between the functions $ \varphi^\alpha_\mu(k)$ and the (associative) composition law is the one given in Eq.~\eqref{simplphi}, the resultant spacetime is $\kappa$-Minkowski.

One can proceed in the same way with the other composition law which allows to implement locality
  \be
  \left(p\oplus q\right)_\mu \,=\, \left(1 + \epsilon q_0/\Lambda\right) \,p_{\:\mu} + q_{\:\mu}\,,
\label{DCL(1')}
  \ee
 considering now that the generalized space-time coordinates $\tilde{z}^\mu$ depend only on the phase-space coordinates ($z, q$), while $\tilde{y}^\mu$ depend on the phase-space coordinates of both particles ($y, p, z, q$). This leads to
\be
\varphi^{(1)\alpha}_{(2)\mu}(p) \,=\, 0\,, \quad\quad \rightarrow \quad\quad \varphi^\alpha_\mu(p) \,=\, \lim_{l\to 0} \frac{\partial(p\oplus l)_\mu}{\partial l_\alpha}\,.
\ee
In this case we have
\begin{align}
\tilde{x}^0 \,=&\, x^\mu \lim_{l\to 0} \frac{\partial(k\oplus l)_\mu}{\partial l_0} \,=\, x^0 (1 + \epsilon k_0/\Lambda) + x^j \epsilon k_j/\Lambda\,, \nonumber \\
\tilde{x}^i \,=&\, x^\mu \lim_{l\to 0} \frac{\partial(k\oplus l)_\mu}{\partial l_i} \,=\, x^i\,.
\label{eq:tilde_2}
\end{align}
We see that, by construction, we obtain the same expressions for the generalized space-time coordinates of the one-particle system.

\subsection{Local DCL1 as a relativistic kinematics}
\label{sec_rel_kinematics}

As we discussed previously, any kinematics has three ingredients: a DCL, a DDR and, in order to have a relativity principle, a DLT in the two-particle system, making the former constituents compatible. This can be done in the same way as we did in the previous section for the second attempt to implement locality, obtaining for the one-particle system (see Appendix~\ref{LT-one-particle}):
\begin{align}
  & {\cal J}^{ij}_0(k) \,=\, 0\,, \quad\quad\quad {\cal J}^{ij}_k(k) \,=\, \delta^j_k \, k_i - \delta^i_k \, k_j\,, \nonumber \\
  & {\cal J}^{0j}_0(k) \,=\, - k_j (1+\epsilon k_0/\Lambda)\,, \quad\quad\quad {\cal J}^{0j}_k \,=\, \delta^j_k \left[-k_0 - \epsilon k_0^2/2\Lambda\right] + (\epsilon/\Lambda) \left[\vec{k}^2/2 - k_j k_k\right]\,,
\label{LT1}
\end{align}  
and for the two-particle system, we impose the condition ${\cal J}_{(1)\mu}^{\,\alpha \beta}(p,q)={\cal J}_{\mu}^{\,\alpha \beta}(p)$ for the first momentum, so the second momentum must transform as 
\begin{align}
{\cal J}_{(2)0}^{\,0i}(p, q)\,=&\left(1+ \epsilon p_0/\Lambda\right) {\cal J}_0^{0i}(q)\,, \nonumber \\
 {\cal J}_{(2)j}^{\,0i}(p, q)\,=&\left(1+\epsilon p_0/\Lambda\right) {\cal J}_j^{0i}(q) + (\epsilon/\Lambda) \,\left(p_j q_i - \delta^i_j \vec{p}^{(1)}\cdot\vec{p}^{(2)}\right)\,, \nonumber\\
{\cal J}_{(2)0}^{\,ij}(p, q)\,=&\, {\cal J}^{ij}_0(q)\,,  \quad\quad\quad
{\cal J}_{(2)k}^{\,ij}(p, q)\,=\, {\cal J}^{ij}_k(q)\,,
\label{calJ(2)}
\end{align}
so that the composition is invariant under the DLT, i.e.
\be
{\cal J}^{\alpha\beta}_\mu(p\oplus q) \,=\, \frac{\partial(p\oplus q)_\mu}{\partial p_\nu} \,  {\cal J}^{\alpha\beta}_{\nu}(p) + \frac{\partial(p\oplus q)_\mu}{\partial q_\nu} \, {\cal J}^{\alpha\beta}_{(2)\nu}(p, q)\,.
\ee

From the DLT of the one-particle system~\eqref{LT1}, we are able to determine the DDR from 
\be
\{C(k), J^{\alpha\beta}\} \,=\, \frac{\partial C(k)}{\partial k_\mu} \, {\cal J}^{\alpha\beta}_\mu(k) \,=\, 0\,,
\label{LI-DDR}
\ee 
obtaining
\be
C(k) \,=\, \frac{k_0^2 - \vec{k}^2}{(1 + \epsilon k_0/\Lambda)}\,.
\ee

In order to have a relativistic kinematics, we need to see that
\be
J^{\alpha\beta}_{(2)} \,=\, y^\mu \,{\cal J}^{\alpha\beta}_\mu(p) + z^\mu \,{\cal J}^{\alpha\beta}_{R\,\mu}(p, q)
\ee
is a representation of the Lorentz algebra and that 
\be
\frac{\partial C(q)}{\partial q_\mu} \,{\cal J}^{\alpha\beta}_{R\,\mu}(p, q) \,=\, 0
\ee
holds. One can check that both statements are true from the expressions~\eqref{LT1}-\eqref{LI-DDR}. So we have shown that one can implement locality and a relativity principle from the composition law~(\ref{DCL(1)}) with $\tilde{y}^\alpha$ depending on the phase-space coordinates ($y, p$) and $\tilde{z}^\alpha$ depending on all the phase-space coordinates. The relativity principle is obtained by making that the Lorentz transformation of the first momentum does not depend on the second one, implying that the Lorentz transformation of the second momentum depends on both momenta. This is a particular (simple) example to implement locality and the relativity principle with the local DCL1 (\ref{DCL(1)}).

\subsection{Local DCL1 and \texorpdfstring{$\kappa$}{k}-Poincaré kinematics}
\label{sec_comparison}

We can wonder about the possible momentum basis in the starting point of the section. This can be analyzed by considering new momentum coordinates $k'_\mu$ related nonlinearly to $k_\nu$, obtaining a new dispersion relation $C'$ and a new deformed composition law $\oplus'$ given by
\be
C(k) \,=\, C'(k')\,, \quad\quad\quad (p'\oplus' q')_\mu \,=\, (p\oplus q)'_\mu\,.
\ee  
Then we have 
\be
\begin{split}
\varphi^{\prime \alpha}_\mu(k') \,=\, \lim_{l'\to 0} \frac{\partial(l'\oplus' k')_\mu}{\partial l'_\alpha} \,&=\, \lim_{l'\to 0} \frac{\partial(l\oplus k)'_\mu}{\partial l'_\alpha} \,=\, \lim_{l\to 0} \frac{\partial l_\beta}{\partial l'_\alpha} \frac{\partial(l\oplus k)'_\mu}{\partial l_\beta}\\
&=\, \lim_{l\to 0} \frac{\partial(l\oplus k)'_\mu}{\partial(l\oplus k)_\nu} \frac{\partial(l\oplus k)_\nu}{\partial l_\alpha} \,=\, \frac{\partial k'_\mu}{\partial k_\nu} \varphi^\alpha_\nu(k)\,,
\end{split}
\label{varphi'-varphi}
\ee
where we have used that $\partial l_\beta/\partial l'_\alpha=\delta_{\beta}^\alpha$ when $l\to 0$.
Moreover, a nonlinear change of momentum basis $k\to k'$ defines a change on canonical phase-space coordinates 
\be
x^{\prime \mu} \,=\, x^\rho \frac{\partial k_\rho}{\partial k'_\mu}\,,
\ee
and then 
\be
x^{\prime \mu} \varphi^{\prime \alpha}_\mu(k') \,=\, x^\rho \frac{\partial k_\rho}{\partial k'_\mu} \varphi^{\prime \alpha}_\mu(k') \,=\, x^\rho \frac{\partial k_\rho}{\partial k'_\mu} \frac{\partial k'_\mu}{\partial k_\nu} \varphi^\alpha_\nu(k) \,=\, x^\nu \varphi^\alpha_\nu(k)\,,
\ee
where we have used Eq.~\eqref{varphi'-varphi} in the second equality.
This falls into the same result obtained in the previous attempts: the non-commutative coordinates are invariant under canonical transformations $\tilde{x}^{\prime \alpha}=\tilde{x}^\alpha$.

For the two-particle system, one has
\be
\varphi^{\prime (2)\alpha}_{(1)\mu}(q') \,=\, \varphi^{\prime \alpha}_\mu(q') - \lim_{l'\to 0}  \frac{\partial(l'\oplus' q')_\mu}{\partial l'_\alpha}\,,
\ee
and the same argument used in (\ref{varphi'-varphi}) leads to 
\be
\lim_{l'\to 0}  \frac{\partial(l'\oplus' q')_\mu}{\partial l'_\alpha} \,=\, \frac{\partial q'_\mu}{\partial q_\nu} \,\lim_{l\to 0}  \frac{\partial(l\oplus q)_\mu}{\partial l_\alpha}\,,
\ee
and then one finds
\be
\varphi^{\prime (2) \alpha}_{(1)\mu}(q') \,=\, \frac{\partial q'_\mu}{\partial q_\nu} \,\varphi^{(2)\alpha}_{(1)\nu}(q)\,.
\ee
The space-time coordinates of the two-particle system change as
\be
y^{\prime  \mu} \,=\, y^\nu \,\frac{\partial p_\nu}{\partial p'_\mu}\,, \quad\quad
z^{\prime  \mu} \,=\, z^\nu \,\frac{\partial q_\nu}{\partial q'_\mu}\,, 
\ee
and as in the one-particle system, we find that the  generalized space-time coordinates of the two-particle system are invariant under canonical transformations
\be
\tilde{y}^{\prime \alpha} \,=\, \tilde{y}^\alpha\,, \quad\quad\quad \tilde{z}^{\prime \alpha} \,=\, \tilde{z}^\alpha\,.
\ee
This implies that all the obtained results for the local DCL1 (\ref{DCL(1)}) (crossing of worldlines, a $\kappa$-Minkowski noncommutative spacetime, and a DRK) do not depend on the phase-space coordinates (momentum basis) one uses.

If we consider the change of momentum basis $k_\mu \to k'_\mu$ 
\be
k_i \,=\, k'_i\,, \quad\quad\quad (1 + \epsilon k_0/\Lambda) \,=\, e^{\epsilon k'_0/\Lambda}\,,
\ee
on the kinematics of the local DCL1, one finds the composition law
\be
(p'\oplus' q')_0 \,=\, p'_0 + q'_0, \quad\quad\quad
(p'\oplus' q')_i \,=\, p'_i + e^{\epsilon p'_0/\Lambda} \,q'_0\,,
\ee
and the  dispersion relation
\be
\frac{k_0^2 - \vec{k}^2}{(1+\epsilon k_0/\Lambda)} \,=\, \Lambda^2 \left(e^{\epsilon k'_0/\Lambda} + e^{-\epsilon k'_0/\Lambda} -2\right) - \vec{k}^{\prime 2} \,e^{-\epsilon k'_0/\Lambda}\,,
\label{C(p)-bcb}
\ee
obtained in Ch.~\ref{chapter_curved_momentum_space}, which is $\kappa$-Poincaré in the bicrossproduct basis (when $\epsilon=-1$). Then we can conclude that the local DCL1 kinematics is the $\kappa$-Poincaré kinematics. 

As we also found in the geometry section, there is a new kinematics corresponding to the case $\epsilon=1$, which cannot allow us to identify $\Lambda$ as a cutoff on the energy. This is a possibility that should be considered and has been overlooked in DSR scenarios.   

In the second attempt of Sec.~\ref{section_second_attempt}, we found that  $\kappa$-Poincar\'e kinematics is compatible with locality. This new way to implement locality allow us to determine the general form of a DCL1 compatible with locality, and $\kappa$-Minkowski as the generalized spacetime of the relative coordinates of the two-particle system.

\subsection{Associativity of the composition law of momenta, locality and relativistic kinematics}
\label{sec:associativity}

In Sec.~\ref{sec_first_order} we saw that a DCL1 must be associative in order to be able to implement locality and then, any kinematics related with it by a change of basis will be associative. Also, at the beginning of Sec.~\ref{sec_st_locality_third}, we have also proved that any associative DCL is locality compatible. Then, this raises the question if associativity will be a necessary condition to have local interactions.

Using the notation 
\be
L^\alpha_\nu(q) \,\doteq\, \lim_{l\to 0} \frac{\partial(l\oplus q)_\nu}{\partial l_\alpha}\,, \quad\quad\quad 
R^\alpha_\nu(p) \,\doteq\, \lim_{l\to 0} \frac{\partial(p\oplus l)_\nu}{\partial l_\alpha}\,,
\ee
we can derive with respect to $p_\rho$ both sides of Eq.~\eqref{loc-oplus-varphi}, finding
\be
\frac{\partial^2(p\oplus q)_\mu}{\partial q_\nu \partial p_\rho} \,L^\alpha_\nu(q) \,=\, 
\frac{\partial^2(p\oplus q)_\mu}{\partial p_\nu \partial q_\rho} R^\alpha_\nu(p) + \frac{\partial(p\oplus q)_\mu}{\partial p_\nu} \frac{\partial R_\nu^\alpha(p)}{\partial p_\rho}\,. 
\ee
Taking the limit $p\to 0$ one has
\be
\frac{\partial L^\rho_\mu(q)}{\partial q_\nu} \, L^\alpha_\nu(q) \,=\, \frac{L^{\alpha\rho}_\mu(q)}{\Lambda} + L^\nu_\mu(q) \, \frac{c^{\rho\alpha}_\nu}{\Lambda}\,,
\label{L}
\ee
being
\be
\frac{L^{\alpha\rho}_\mu(q)}{\Lambda} \,\doteq\, \lim_{p\to 0} \frac{\partial^{2}(p\oplus q)_\mu}{\partial p_\alpha \partial p_\rho}
\ee
the coefficient of the term proportional to $p_\alpha p_\rho$ in $(p\oplus q)_\mu$, and 
\be
\frac{c^{\rho\alpha}_\nu}{\Lambda} \,\doteq\, \lim_{p, q \to 0} \frac{\partial^2(p\oplus q)_\nu}{\partial p_\rho \partial q_\alpha}\,,
\ee
the coefficient of the term proportional to $p_\rho q_\alpha$ in $(p\oplus q)_\nu$. Due to the symmetry under the exchange $\alpha\leftrightarrow \rho$ of $L^{\alpha\rho}_\mu$ in Eq.~\eqref{L}, one finds
\be
\frac{\partial L^\rho_\mu(q)}{\partial q_\nu} \, L^\alpha_\nu(q) - \frac{\partial L^\alpha_\mu(q)}{\partial q_\nu} \, L^\rho_\nu(q) \,=\, \frac{(c^{\rho\alpha}_\nu - c^{\alpha\rho}_\nu)}{\Lambda} \, L^\nu_\mu(q)\,.
\ee
This implies that the generators 
\be 
T_L^\mu \,\doteq\,z^\rho \,L^\mu_\rho(q)\,,
\ee
form a Lie algebra 
\be
\{T_L^\mu, T_L^\nu \} \,=\, \frac{(c^{\mu\nu}_\rho - c^{\nu\mu}_\rho)}{\Lambda} \, T_L^\rho\,.
\ee
Therefore, the infinitesimal transformation of the momentum $q$ with parameter $\epsilon$ is given by
\be
\delta q_\mu \,=\, \epsilon_\nu \{q_\mu, T_L^\nu \} \,=\, \epsilon_\nu L^\nu_\mu(q) \,=\, \epsilon_\nu \lim_{l\to 0} \frac{\partial (l\oplus q)_\mu}{\partial l_\nu} \,=\, (\epsilon\oplus q)_\mu - q_\mu\,.
\ee

If the composition law is associative, this allows us to define the finite transformation starting from the infinitesimal one  generated by the $T_L^\mu$, as
\be
q_\mu \to q'_\mu \,=\, (a \oplus q)_\mu\,,
\ee
for a transformation with parameter $a$.

Proceeding in the same way, we can derive with respect to $q_\rho$ instead of $p_\rho$ the first equality of Eq.\eqref{loc-oplus-varphi}, and taking the limit $q\to 0$ one obtains that
\be
T_R^\mu \doteq y^\nu R^\mu_\nu(p)
\ee
are the generators of a Lie algebra 
\be
\{T_R^\mu, T_R^\nu \} \,=\, - \,\frac{(c^{\mu\nu}_\rho - c^{\nu\mu}_\rho)}{\Lambda} \, T_R^\rho\,,
\ee
which is the same Lie algebra we have found for $T_L$ up to a sign\footnote{This is what we mentioned in Ch.~\ref{chapter_curved_momentum_space}: if the generators of left-translations form a Lie algebra, the generators of right-translations form the same algebra but with a different sign (see Ch.6 of Ref.~\cite{Chern:1999jn}).}. The  infinitesimal transformation of the momentum $p$  with parameter $\epsilon$ is
\be
\delta p_\mu \doteq \epsilon_\nu \{p_\mu, T_R^\nu \} \,=\, \epsilon_\nu R^\nu_\mu(p) \,=\, \epsilon_\nu \,\lim_{l\to 0} \frac{\partial(p\oplus l)_\mu}{\partial l_\nu} \,=\, (p\oplus \epsilon)_\mu - p_\mu\,,
\ee
and this leads to a finite transformation if the composition law is associative
\be
p_\mu \to p'_\mu \,=\, (p\oplus a)_\mu\,.
\ee

In Ch.~\ref{chapter_curved_momentum_space} we saw that $\kappa$-Poincaré kinematics is the only DRK obtained from geometry whose generators of translations form a Lie algebra. This is why the local DCL1 is compatible with locality, since we proved that it is the $\kappa$-Poincaré kinematics in a different basis. 

Any other relativistic kinematics obtained from the geometrical procedure (Snyder and hybrid models)  lead to $T_{L,R}^\mu$ generators  which do not close a Lie algebra, and then do not lead to locality of interactions. So in this scheme, locality  selects $\kappa$-Poincaré kinematics as the exclusive relativistic isotropic kinematics going beyond SR framework and compatible with locality.

In this chapter we have seen a new ingredient, a noncommutative spacetime, that arises from a DRK in a natural way when locality is imposed. This is of vital importance since, as we saw in Sec.~\ref{sec:QGT}, a noncommutativity in space-time coordinates is a main ingredient of a QGT, giving place to a possible minimal length. 

From the implementation of locality, we can solve the apparent paradox we saw in Sec.~\ref{sec:thought_experiments}, where we showed that there is a spacetime fuzziness for classical particles. If one observer $O$ sees two different particles of masses $m_1$ and $m_2$ moving with the same speed and following the same trajectory, another observer $O^\prime$ boosted with respect to $O$ would see that these particles are also following the same trajectory in the physical coordinates, but not in the canonical variables. This is telling us that maybe the physical coordinates are the good arena for considering physical processes and interactions.  

Now that we understand better how a DCL affects the spacetime for all particles involved in an interaction, we can try to study some phenomenological aspects that can be observed in order to test the theory. In Sec.~\ref{sec_phenomenology_DSR}, we saw that time delay of flight of particles is in principle the only experimental observation in the DSR framework for low energies in comparison with the Planck scale. This will be the subject of the next chapter, in which the use of these privileged coordinates will be indispensable.

\chapter{Time delay for photons in the DSR framework}
\label{chapter_time_delay}
\ifpdf
    \graphicspath{{Chapter5/Figs/Raster/}{Chapter5/Figs/PDF/}{Chapter5/Figs/}}
\else
    \graphicspath{{Chapter5/Figs/Vector/}{Chapter5/Figs/}}
\fi

\epigraph{Truth is confirmed by inspection and delay; falsehood by haste and uncertainty.}{Publius Tacitus}

In this chapter we will study the time delay for photons in the DSR context. This is a very important phenomenological study since, as we saw in Sec.~\ref{sec:DSR}, the only window to test DSR theories with a high energy scale of the order of the Planck energy is precisely the time delay of astroparticles.  

There are some studies of time delays in the DSR framework in the literature~\cite{AmelinoCamelia:2011cv,Loret:2014uia,Mignemi:2016ilu}. In the first two works the study is carried out with the noncommutativity of $\kappa$-Minkowski while in the third one, the Snyder spacetime is considered. In the former cases a time delay for photons is found, which differs from the result obtained in the latter, where there is an absence of such effect. Apparently, depending on the noncommutativity of the spacetime in which photons propagate, the results vary. 

Along this chapter, we will consider three different models. In the first one, the results depend on the basis of $\kappa$-Poincar\'{e} one works with, making that the final result of the existence or not of a time delay is basis dependent~\cite{Carmona:2017oit}. Also we will study a generic space-time noncommutativity and see the necessary conditions in order to show a lack of time delay.     

But one could think that something is wrong in the previous analysis since the results depend on the basis we are choosing. This means that the physics in the DSR framework would depend on the choice of coordinates on momentum space, making the results coordinate dependent. This leads us to study another formulation of time delays in such a way that the observables are defined in the physical coordinates of Ch.~\ref{chapter_locality}, and we will see that in this framework, the result of absence of time delay is basis independent in both noncommutative spacetimes, $\kappa$-Minkowski and Snyder~\cite{Carmona:2018xwm}.  

Finally, we will consider another model where the time delay is studied in the framework of interactions, considering the emission and detection of a photon, not only its free propagation~\cite{Carmona:2019oph}. In this context, we will see that one should consider not only the particles involved in the emission and detection processes, but any other related to them, making in principle this model untreatable. This is why a cluster decomposition principle will be suggested as a way to avoid these inconsistencies.       

\section{First approach: relative locality framework}
\label{sec:td_first}
In this section we will study the first model we have mentioned above, previously considered in the literature~\cite{AmelinoCamelia:2011cv,Loret:2014uia}. We will see that the absence or not of a time delay of flight for photons will depend on the realization of the noncommutative spacetime (choice of momentum basis) and also, on the considered noncommutativity~\cite{Carmona:2017oit}. 
\subsection{Determination of time delays}

Let us consider two photons emitted simultaneously from a source at a distance $L$ from our laboratory, and let us suppose a DDR for the high energy photon, as we can neglect this modification for the low energy one. We can consider that the DDR is such that the speed of the high energy photon is lower than $1$, so a detector in our laboratory would measure a time delay $\tilde{T}$ between them. 

But this is not the only contribution to the time delay: there is another correction due to the fact that photons see different (momentum dependent) spacetimes characterized with the function $\varphi^\mu_\nu(k)$, as we saw in the previous chapter. In a noncommutative spacetime translations (given by the DCL) act non trivially, as they depend on the momentum of the particle. This is the main difference between the model in DSR and the corresponding one of LIV. In LIV, the only contribution to time delay is the DDR, but in DSR, as the relativity principle has to be maintained, one needs to include the effect of non-trivial (momentum dependent) translations, whose effect in the one-particle system is depicted by a noncommutative spacetime. 

Due to the effect of translations (we saw in Sec.~\ref{sec:relative_locality_intro} that it leads to non-local effects), we should correct the affirmation at the beginning of the section. When we have said that two photons are emitted simultaneously, we would have to say that they are so only for an observer at the source and then not for us, placed at the laboratory. Hence, in order to study the time delay we need to consider two observers: $A$, which is placed at the source and see the two photons emitted at the same time and at the same point, and $B$, which is at the detection point. 

For simplicity, and without any loss of generality, we can treat the problem in $1+1$ dimensions, so we will write for the photon its energy $E$ and its momentum $k\equiv |\vec{k}|$. Note that we will use the same notation ($k$) for the four-momentum in 3+1 and for the momentum in 1+1. Since we have neglected the modification in the dispersion relation for the low energy photon, we can also consider that it propagates in a commutative spacetime (neglecting the contribution due to the $\varphi^\mu_\nu(k)$), so the low energy photon will behave as in SR, traveling at speed $1$.

We can compute the translations relating the noncommutative coordinates of observers $A$ and $B$ directly from the usual translations of the commutative ones, $x^B=x^A-L$, $t^B=t^A-L$:
\begin{align}
\tilde{t}^B& \,=\,\varphi^0_0 t^B+\varphi^0_1 x^B=\tilde{t}^A-L(\varphi^0_0 + \varphi^0_1) \label{traslaciont} \,,\\
\tilde{x}^B& \,=\,\varphi^1_0 t^B+\varphi^1_1 x^B=\tilde{x}^A-L(\varphi^1_0 + \varphi^1_1) \label{traslacionx} \,.
\end{align}

The worldline of the high energy particle for observer $A$ is
\be
\tilde{x}^A\,=\,\tilde{v}\,\tilde{t}^A\,,
\label{eq:AWL}
\ee
since $\tilde{x}^A=0,\, \tilde{t}^A=0$, are the initial conditions of the worldline, and $\tilde{v}$ is obtained through
\be
\tilde{v}\,=\,\frac{\lbrace{C, \tilde{x}\rbrace}}{\lbrace{C, \tilde{t}\rbrace}}\,=\,\frac{\varphi^1_0 (\partial C/\partial E)-\varphi^1_1(\partial C/\partial k)}{\varphi^0_0 (\partial C/\partial E)-\varphi^0_1(\partial C/\partial k)}\,,
\label{eq:v_tilde}
\ee
where the minus signs appear due to the fact that $k_1=-k^1=-k$, and so $\partial C/\partial k_1=-\partial C/\partial k$. 
We can now compute the  observer $B$ worldline by applying Eqs.~\eqref{traslaciont}-\eqref{traslacionx} to Eq.~\eqref{eq:AWL}:
\be
\tilde{x}^B=\tilde{x}^A-L(\varphi^1_0 + \varphi^1_1)=\tilde{v}\,[\tilde{t}^B+L(\varphi^0_0 + \varphi^0_1)]-L(\varphi^1_0 + \varphi^1_1)\,.
\label{eq:BWL}
\ee
The worldline for observer $B$ ends at $\tilde{x}^B=0$.\footnote{We assume that the detector is at rest, being the spatial location  coincident for the detection of both photons.} The time delay $\tilde{T}\equiv \tilde{t}^B(\tilde{x}^B=0)$ can be obtained from  Eq.~\eqref{eq:BWL}, giving 
\be
\begin{split}
\tilde{T}\,=\,&\tilde{v}^{-1} \,L(\varphi^1_0 + \varphi^1_1)-L(\varphi^0_0 + \varphi^0_1)\\
=\,&L\left[(\varphi^1_0 + \varphi^1_1)\frac{\varphi^0_0 (\partial C/\partial E)-\varphi^0_1(\partial C/\partial k)}{\varphi^1_0 (\partial C/\partial E)-\varphi^1_1(\partial C/\partial k)}-(\varphi^0_0 + \varphi^0_1)\right]\,,
\end{split}
\label{eq:time-delay}
\ee 
since the low energy photon arrives at $\tilde{t}^B=0$. This equation is valid not only for photons but for every relativistic particle. Also, one can check that one obtains the same results of SR for both cases just taking the limit $\Lambda\to \infty$.

\subsection{Momenta as generators of translations in spacetime}

We can write the following Poisson brackets with the functions $\varphi^\mu_\nu$
\be
\{E,\tilde{t}\}\,=\,\varphi^0_0\,, \quad \quad \{E,\tilde{x}\}=\varphi^1_0\,, \quad \quad
\{k,\tilde{t}\}\,=\,-\varphi^0_1\,, \quad  \quad \{k,\tilde{x}\}=-\varphi^1_1\,,
\ee
where again the minus signs appear since $k_1=-k^1=-k$.

Then, we can express Eqs.~\eqref{traslaciont}-\eqref{traslacionx} in the following way
\begin{align}
\tilde{t}^B&\,=\,\tilde{t}^A-L\{E,\tilde{t}\}+L\{k,\tilde{t}\} \,,\\
\tilde{x}^B&\,=\,\tilde{x}^A-L\{E,\tilde{x}\}+L\{k,\tilde{x}\} \,.
\label{eq:transl}
\end{align}
These transformations are the translations generated by the momentum in the noncommutative spacetime, even if the $(\tilde{x},k)$ phase space is non-canonical. This is the procedure used in~\cite{AmelinoCamelia:2011cv,Loret:2014uia,Mignemi:2016ilu}.

Now we can write the time delay formula of Eq.~\eqref{eq:time-delay} in terms of Poisson brackets
\begin{equation}
\tilde{T}\,=\,\left(L\{E,\tilde{x}\}-L\{k,\tilde{x}\}\right) \cdot \left(\frac{(\partial C/\partial E)\{E,\tilde{t}\}+(\partial C/\partial k)\{k,\tilde{t}\}}{(\partial C/\partial E)\{E,\tilde{x}\}+(\partial C/\partial k)\{k,\tilde{x}\}}\right)-L\{E,\tilde{t}\}+L\{k,\tilde{t}\}\,.
\label{eq:genTD}
\end{equation}

For the simple case of a commutative spacetime, as we have $\{E,t\}=1$, $\{E,x\}=0$, $\{k,t\}=0$, $\{k,x\}=-1$, Eq.~\eqref{eq:genTD} gives
\begin{equation}
T=-L\left(1+\frac{\partial C/\partial E}{\partial C/\partial k}\right)\,.
\label{eq:canTD}
\end{equation}
When the dispersion relation is $C(k)=E^2-k^2$, one obtains the result of SR, $T=-L(1-E/p)$, which is zero for photons. 

In order to obtain the first order approximation of Eq.~\eqref{eq:genTD}, keeping the leading terms, one can write the Poisson brackets as their usual value plus an infinitesimal deformation of order $\epsilon$, $\{E,\tilde{t}\}=1+(\{E,\tilde{t}\}-1)=1+\mathcal{O}(\epsilon)$, $\{k,\tilde{x}\}=1+(\{k,\tilde{x}\}-1)=1+\mathcal{O}(\epsilon)$, $\{E,\tilde{x}\}=\mathcal{O}(\epsilon)$, $\{k,\tilde{t}\}=\mathcal{O}(\epsilon)$, and also $(\partial C/\partial E)/(\partial C/\partial k)=-E/k+\mathcal{O}(\epsilon)$, giving
\begin{equation}
\frac{\tilde{T}}{L} \approx -\left(1-\frac{E}{k}\right)-\left(\frac{\partial C/\partial E}{\partial C/\partial k}+\frac{E}{k}\right)-
\left(1-\frac{E}{k}\right)\left(\{E,\tilde{t}\}-1\right) + \left(1-\frac{E}{k}\right)\frac{E}{k} \,\{E,\tilde{x}\}\,.
\label{eq:genTDaprox}
\end{equation}
The first term is the usual time delay in SR, the second one takes into account the effect due to the DDR, and the last two contributions reflect the deformed Heisenberg algebra involving $E$. There are no contributions of the Poisson brackets involving $k$ because they cancel out in the computation.

In the next two subsections, we will use this formula for different bases of $\kappa$-Minkowski and Snyder spacetimes.

\subsection{Photon time delay in \texorpdfstring{$\kappa$}{Lg}-Minkowski spacetime}
As we saw in Sec.~\ref{sec:QGT}, $\kappa$-Minkowski spacetime is defined by
\begin{equation}
[\tilde{x}^0,\tilde{x}^i]\,=\,-\frac{i}{\Lambda}\tilde{x}^i \,, \quad \quad [\tilde{x}^i,\tilde{x}^j]\,=\,0\,,
\label{eq:kM}
\end{equation}
and the non-vanishing  Poisson bracket in $(1+1)$-dimensional spacetime is
\be
\{\tilde{t},\tilde{x}\}\,=\,-\frac{1}{\Lambda}\tilde{x} \,.
\ee

Now we are going to calculate the time delay for three different (well known) choices of momentum coordinates: the bicrossproduct, the classical and the Magueijo-Smolin basis. 

\subsubsection{Bicrossproduct basis}
The DDR in this basis at leading order in $\Lambda^{-1}$ is
\begin{equation}
C(k)\,=\,k_0^2-\vec{k}^2-\frac{1}{\Lambda} k_0 \vec{k}^2\equiv m^2\,,
\label{eq:bicrossCasimir}
\end{equation}
and the Heisenberg algebra in $1+1$ dimensions is given by~\eqref{eq:pairing_intro}
\begin{equation}
\{E,\tilde{t}\}\,=\,1 \,,\quad \quad \{E,\tilde{x}\}\,=\,0\,, \quad \quad \{k,\tilde{t}\}\,=\,-\frac{k}{\Lambda}\,, \quad \quad \{k,\tilde{x}\}\,=\,-1\,.
\end{equation}
Using Eq.~\eqref{eq:bicrossCasimir} one finds
\begin{equation}
\frac{\partial C/\partial E}{\partial C/\partial k}+\frac{E}{k}\,=\,\frac{1}{\Lambda}\left(\frac{E^2}{k}+\frac{k}{2}\right),
\label{eq:bicrossdispterm}
\end{equation}
and then Eq.~\eqref{eq:genTDaprox} gives
\begin{equation}
\frac{\tilde{T}[\text{bicross}]}{L}\,=\,-\left(1-\frac{E}{k}\right)-\frac{1}{\Lambda}\left(\frac{E^2}{k}+\frac{k}{2}\right)\,.
\label{eq:bicrossTD2}
\end{equation}
In the case of photons, Eq.~\eqref{eq:bicrossCasimir} leads to $E=k\,(1+k/2\Lambda)$ at first order, so one obtains a time delay $Lk/\Lambda$ for the high energy photon with respect the low energy one. This result was previously obtained in Refs.~\cite{AmelinoCamelia:2011cv,Loret:2014uia}, making them to conclude that there is an energy dependent time delay for photons in $\kappa$-Minkowski spacetime.

But we are going to see that this result depends on the choice of basis one is working with.

\subsubsection{Classical basis} 

Another choice of basis in $\kappa$-Poincaré is the classical basis (studied in Ch.~\ref{chapter_second_order}), with the same dispersion relation of SR
\begin{equation}
C(k)\,=\,k_0^2-\vec{k}^2\,,
\label{eq:classCasimir}
\end{equation}
and the Poisson brackets in $1+1$ dimensions are at leading order~\cite{KowalskiGlikman:2002jr}
\begin{equation}
\{E,\tilde{t}\}\,=\,1 \,,\quad \quad \{E,\tilde{x}\}\,=\,-\frac{k}{\Lambda} \,,\quad \quad \{k,\tilde{t}\}\,=\,0 \,,\quad \quad \{k,\tilde{x}\}\,=\,-\left(1+\frac{E}{\Lambda}\right)\,.
\label{eq:PP}
\end{equation}
Eq.~\eqref{eq:genTDaprox} gives in this case
\begin{equation}
\frac{\tilde{T}[\text{class}]}{L}\,=\,-\left(1-\frac{E}{k}\right)\left(1+\frac{E}{\Lambda}\right)\,.
\label{eq:classTD}
\end{equation}
Then, for massless particles ($E=k$), there is an absence of time delay in this basis, despite the noncommutativity of the spacetime. 

\subsubsection{Magueijo-Smolin basis}

Another basis described in Ref.~\cite{KowalskiGlikman:2002jr} is the Magueijo-Smolin basis. The DDR at first order in this basis is 
\begin{equation}
C(k)\,=\,k_0^2-\vec{k}^2+\frac{1}{\Lambda}k_0^3-\frac{1}{\Lambda}k_0\vec{k}^2\,,
\label{eq:MG-SCasimir}
\end{equation}
and the Heisenberg algebra in $1+1$ dimensions at leading order is 
\begin{equation}
\{E,\tilde{t}\}\,=\,\left(1-\frac{2E}{\Lambda}\right)\,, \quad \quad \{E,\tilde{x}\}\,=\,-\frac{k}{\Lambda}\,, \quad \quad \{k,\tilde{t}\}\,=\,-\frac{k}{\Lambda}\,, \quad \quad \{k,\tilde{x}\}\,=\,-1\,.
\label{eq:PPM-S}
\end{equation}
From Eq.~\eqref{eq:MG-SCasimir} one can see
\begin{equation}
\frac{\partial C/\partial E}{\partial C/\partial k}+\frac{E}{k}\,=\,\left(1-\frac{E}{k}\right)\frac{E+k}{2\Lambda}\,,
\ee
and then   
\begin{equation}
\begin{split}
\frac{\tilde{T}[\text{M-S}]}{L}&\,=\,-\left(1-\frac{E}{k}\right)-\left(1-\frac{E}{k}\right)\frac{E+k}{2\Lambda}+\left(1-\frac{E}{k}\right)\frac{2E}{\Lambda}-\left(1-\frac{E}{k}\right)\frac{E}{\Lambda}\\
&\,=\,-\left(1-\frac{E}{k}\right)\left[1-\frac{E-k}{2\Lambda}\right]\,.
\label{eq:M-STD2}
\end{split}
\end{equation}
We find that, as in the previous basis,  there is not a time delay for massless particles ($E=k$).

\subsection{Photon time delay in Snyder spacetime}
In Sec.~\ref{sec:QGT} we showed that the noncommutative Snyder spacetime is 
\be
[\tilde{x}_\mu,\tilde{x}_\nu]\,=\,\frac{i}{\Lambda^2} J_{\mu\nu}\,,
\ee
with $J_{\mu\nu}$ the generators of the Lorentz algebra. 

Also in this case, there are different basis (or realizations) of the same algebra in phase space. Here, we will discuss the time delay effect in the representation of Snyder and Maggiore.

\subsubsection{Snyder representation}

In the original representation proposed by Snyder, the Heisenberg algebra in $1+1$ dimensions is
\begin{equation}
\{E,\tilde{t}\}\,=\,\left(1+\frac{E^2}{\Lambda^2}\right)\,, \quad \quad \{E,\tilde{x}\}\,=\,\frac{Ek}{\Lambda^2}\,, \quad \quad \{k,\tilde{t}\}\,=\,\frac{Ek}{\Lambda^2}\,, \quad \quad \{k,\tilde{x}\}\,=\,-\left(1-\frac{k^2}{\Lambda^2}\right),
\label{eq:PPSny}
\end{equation}
and as the Casimir is $(E^2-k^2)$, one finds 
\begin{equation}
\frac{\partial C/\partial E}{\partial C/\partial k}+\frac{E}{k}\,=\,0,
\end{equation}
Then Eq.~\eqref{eq:genTDaprox} gives
\begin{equation}
\frac{\tilde{T}[\text{Snyder}]}{L}\,=\,-\left(1-\frac{E}{k}\right)-\left(1-\frac{E}{k}\right)\frac{E^2}{\Lambda^2}+\left(1-\frac{E}{k}\right)\frac{E}{k}\frac{Ek}{\Lambda^2}\,=\,-\left(1-\frac{E}{k}\right),
\label{eq:TD-Snyder}
\end{equation}
so for the case of photons there is no time delay. 

\subsubsection{Maggiore representation}

The Heisenberg algebra in the Maggiore representation~\cite{Maggiore:1993kv} at leading order is
\begin{equation} 
\{E,\tilde{t}\}\,=\,1+\frac{E^2-k^2}{2\Lambda^2}\,,\quad \quad \{E,\tilde{x}\}\,=\,0\,, \quad \quad \{k,\tilde{t}\}\,=\,0\,, \quad \quad \{k,\tilde{x}\}\,=\,-1-\frac{E^2-k^2}{2\Lambda^2}\,.
\label{eq:PPMagg}
\end{equation}
Also in this representation the Casimir is  $(E^2-k^2)$, so again
\begin{equation}
\frac{\partial C/\partial E}{\partial C/\partial k}+\frac{E}{k}\,=\,0,
\end{equation}
and then from Eq.~\eqref{eq:genTDaprox} one obtains
\begin{equation}
\frac{\tilde{T}[\text{Maggiore}]}{L}\,\,=\,\,-\left(1-\frac{E}{k}\right)-\left(1-\frac{E}{k}\right)\frac{E^2-k^2}{2\Lambda^2}\,\,=\,\,-\left(1-\frac{E}{k}\right)\left[1+\frac{E^2-k^2}{2\Lambda^2}\right] ,
\label{eq:TD-Maggiore}
\end{equation}
so for the Maggiore representation there is not either a time delay for photons.

The result of absence of time delay for photons was claimed in a previous paper~\cite{Mignemi:2016ilu} through a different procedure. These results are particular cases of our general expression of Eq.~\eqref{eq:genTD}.

\subsection{Interpretation of the results for time delays}

One can see that in all cases considered before the time delay is proportional to 
\begin{equation}
L\left[(1+(\partial C/\partial E)/(\partial C/\partial k)\right]\,,
\label{eq:factor}
\end{equation}
i.e. to $(L/v-L)$, where $v$ is the velocity of propagation of the high energy particle in the commutative spacetime,
\be
v\,\,=\,\,-\frac{\partial C/\partial k}{\partial C/\partial E}\,.
\label{velocidadC}
\ee 
This result can be read from Eq.~\eqref{eq:time-delay}:
\be
\begin{split}
\tilde{T}\,\,=\,\,L\left[(\varphi^1_0+\varphi^1_1)\frac{\varphi^0_0+\varphi^0_1 v}{\varphi^1_0+\varphi^1_1 v}-(\varphi^0_0+\varphi^0_1)\right]\,\,=\,&\,\frac{L(\varphi^0_0\varphi^1_1-\varphi^1_0\varphi^0_1)}{\varphi^1_0+\varphi^1_1 v}(1-v)\\
\,=\,&\,\frac{\varphi^0_0\varphi^1_1-\varphi^1_0\varphi^0_1}{\varphi^1_1+\varphi^1_0/v}L\left(\frac{1}{v}-1\right)\,.
\end{split}
\ee
Then, the leading contribution to the time delay is due only to the first terms in the power expansion of the DDR. This is in agreement with what we have found in the previous subsection: the only basis considered here where a time delay is present is the bicrossproduct realization of $\kappa$-Poincar\'{e}, which is the only one with an energy dependent velocity for photons.  

\section{Second approach: locality of interactions}
\label{sec:NC}

We have seen in the previous section that the result of the existence or absence of a time delay is basis dependent, and also depends on the noncommutativity of the considered spacetime. The first dependence is really problematic since one would expect that the same results would be obtained independently of the choice of momentum coordinates. This leads us to consider another model of propagation of particles. 

\subsection{Presentation of the model}

The main ingredient of this model is to consider that all observables are defined in the local (physical) coordinates of Ch.~\ref{chapter_locality}. It means that instead of defining the translations as in the previous section, 
\begin{equation}
\xi^\mu_B\,=\,\xi^\mu_A+a^\mu\,,
\label{eq:xi}
\end{equation}
if we consider the new noncommutative coordinates defined by
\begin{equation}
\zeta^\mu\,=\,\xi^\nu \varphi^\mu_\nu(\mathcal{P}/\Lambda),
\label{eq:zeta-xi}
\end{equation}
where $\mathcal{P}$ is the total momentum of the interaction\footnote{In principle, one should consider all the momenta that intervene in the processes of emission and detection. Nonetheless, we can make a simplification considering that the $\varphi$ function depends only on the momentum of the detected particle. This will be treated in more detail in the next section, in which a cluster decomposition principle will be considered.}, we define the translation in these coordinates: the two observers $A$ and $B$ are connected by a translation with parameter $b$
\begin{equation}
\zeta^\mu_B\,=\,\zeta^\mu_A+b^\mu\,,
\label{eq:zeta}
\end{equation}
where
\begin{equation}
  a^\nu \varphi^\mu_\nu(\mathcal{P}/\Lambda) \,=\, b^\mu\,.
\end{equation}
It is obvious that the results obtained with this relation between observers will be different that those obtained in the previous section. 

\subsection{Computation of the time delay expression}
\label{sec:absence_time_delay}

In this subsection, we will compute the time delay defined by Eq.~\eqref{eq:zeta}. We consider again that both observers are separated  by a distance $L$, but in contrast with the previous model, $L$ is the distance in the noncommutative space. We have then (we are still working in 1+1 dimensions)
\begin{equation}
\zeta^1_B\,=\,\zeta^1_A-L\,.
\label{eq:zetaTD}
\end{equation}

We can now compute the time delay. The detection of the photon for observer $B$ is at $\tilde{x}^1_d=0$ and the emission at $\tilde{x}^1_{e}=-L$, which is at the origin of spatial coordinates for observer $A$, according to Eq.~\eqref{eq:zetaTD}. In fact, we see that since interactions are local, we do not need to consider two different observers. 

Then, the difference in time coordinates from the detection to the emission is
\begin{equation}
\tilde{x}^0_{d}\,=\,\tilde{x}^0_{e}+\frac{L}{\tilde{v}}\,,
\label{eq:TD1}
\end{equation}
where $\tilde{v}$ is the velocity of the photon in the noncommutative spacetime given by  Eq.~\eqref{eq:v_tilde}. So the time delay $\tilde{T}$ is given by
\begin{equation}
\tilde{T}\doteq \tilde{x}^0_d - \tilde{x}^0_e - L \,=\,L\left(\frac{1}{\tilde{v}}-1\right)\,.
\label{eq:TD-2}
\end{equation}

One can check that the velocity defined in the noncommutative spacetime is independent of the choice of basis: 

\begin{equation}
\lbrace{C(k),\tilde{x}^\mu\rbrace}\,=\,\frac{\partial C(k)}{\partial k_\nu}\varphi^\mu_\nu(k)\,=\,\frac{\partial C^{\prime}(k^{\prime})}{\partial k^{\prime}_\sigma}\frac{\partial k^{\prime}_\sigma}{\partial k_\nu}\varphi^\mu_\nu(k)\,=\,\frac{\partial C^{\prime}(k^{\prime})}{\partial k^{\prime}_\sigma}\varphi^{\prime \mu}_\sigma(k^{\prime})\,=\,\lbrace{C^{\prime}(k^{\prime}),\tilde{x}^{\prime \mu}\rbrace}'\,,
\label{eq:velocity_independence}
\end{equation} 
where we have used the relation between $x$ and $x'$ given by the canonical transformation 
\be
k_\mu \,=\, f_\mu(k')\,,\quad x^\mu \,=\, x^{\prime\nu} g^\mu_\nu(k')\,,
\ee
for any set of momentum dependent functions $f_\mu$, with
\be
g^\mu_\rho(k') \frac{\partial f_\nu(k')}{\partial k'_\rho} \,=\, \delta^\mu_\nu \,,
\ee
the  transformation rule of  $\varphi$ (see Appendix~\ref{appendix:locality})
\be
\varphi^{\prime\mu}_\rho(k') \doteq g^\nu_\rho(k') \varphi^\mu_\nu(f(k')) \,=\, \frac{\partial k'_\rho}{\partial k_\nu} \varphi^\mu_\nu(k)\,,
\ee
the fact that $C(k)=C^{\prime}(k^{\prime})$, and we have denoted by
\begin{equation}
\lbrace{A\,,\,B\rbrace}^{\prime}\,=\,\frac{\partial A}{\partial k^{\prime}_\rho}\frac{\partial B}{\partial x^{\prime^\rho}}-\frac{\partial A}{\partial x^{\prime \rho}}\frac{\partial B}{\partial k^{\prime}_\rho}
\end{equation} 
the Poisson brackets in the new canonical coordinates. This reveals that in the physical coordinates, the velocity is the same independently of the canonical coordinates we use. Since the time delay is only a function of $L$ and $\tilde{v}$, this means that whether there is a time delay or not will be independent on the basis in which one makes the computation. 

In particular, one can obtain  $\tilde{v}$ in the bicrossproduct basis with Eq.~\eqref{eq:v_tilde}, finding that $\tilde{v}=1$,\footnote{We will understand why this happens from a geometrical point of view in Ch.~\ref{ch:cotangent}.} and then, there is no time delay in $\kappa$-Poincar\'{e}. This differs from the results of the previous section where the time delay was basis dependent, which support the use of this model against the other one since physics should not depend on the variables one works with.  

Also, one can compute $\tilde{v}$ for the Snyder noncommutativity, obtaining the same result,  $\tilde{v}=1$. The fact that we obtain the same results for these two different cases of noncommutativity can be easily understood since in both models the functions $\varphi^\mu_\nu(k)$, viewed as a tetrad in a de Sitter momentum space as in Ch.~\ref{chapter_curved_momentum_space}, are representing the same curved momentum space. This then leads to the result that there is no observable effect on the propagation of free particles in a flat spacetime due to a de Sitter momentum space. 

\section{Third approach: multi-interaction process}
\label{sec:multi-interaction_td}

In the previous sections, we have considered that particles propagate in the physical spacetime and that there is a simple way to relate observers, Eqs.~\eqref{traslaciont}-\eqref{traslacionx} in Sec.~\ref{sec:td_first} and  Eq.~\eqref{eq:zetaTD} in Sec.~\ref{sec:NC}. This is tricky somehow: the physical coordinates were introduced in order to have local interactions for all observers, but in the previous models we have studied only the propagation, not the interactions leading to the emission and detection processes of the photon. This approximation can be done if one considers a cluster decomposition principle~\cite{Carmona:2019oph}, eliminating the dependence on all other momenta involved in the emission and detection of the particle. 

Moreover, in the model considered in Sec.~\ref{sec:NC}, the translation Eq.~\eqref{eq:zetaTD} is not a symmetry of the action 
\be
S \,=\, \int d\tau \left[\dot{x}^\mu k_\mu - N(\tau) C(k)\right] \,=\, \int d\tau \left[-\tilde{x}^\alpha \varphi^\mu_\alpha(k) \dot{k}_\mu - N(\tau) C(k)\right]\,.
\ee
The translations can be identified with  $x^\mu \to x^\mu + a^\mu$, which leave the action invariant, as it is considered in Sec.~\ref{sec:td_first} (previously studied in Refs.~\cite{AmelinoCamelia:2011cv,Loret:2014uia}). This differs from our second consideration were we used $\tilde{x}^\mu \to \tilde{x}^\mu + a^\mu$ as a translation between observers. This is not a symmetry of the action,  but is a transformation that leaves invariant the set of equations of motion, and then, the set of solutions. 

Another way to study the propagation of particles is to consider a multi-interaction process: one interaction defines the emission of a particle and another one the detection.  This is the subject of this section. 

\subsection{Two interactions}

A model with multi-interactions was proposed in Ref.~\cite{AmelinoCamelia:2011nt} as a way to study a possible time delay.  A first interaction was considered, the  emission of a high energy photon, then its propagation, and finally another interaction defining the detection. It was considered a simplified model where one pion decays into two photons, and one of them (the high energy one) interacts with a particle in the detector producing two particles. 

Here, we will consider the model we proposed in Ref.~\cite{Carmona:2019oph}, a process with two two-particle interactions with three particles in the ingoing state with phase coordinates $(x_{-(i)}, p^{-(i)})$  and another three particles in the outgoing state with phase-space coordinates $(x_{+(j)}, p^{+(j)})$. The two particles participating in the first interaction are labeled with $i=1,2$, and the other two particles involved in the second interaction with $j=2,3$. There is another particle produced in the first interaction with phase-space coordinates $(y, q)$, participating in the second interaction, which will play the role of the detected photon. The action of Eq.~\eqref{eq:action} we used in Sec.~\ref{sec:relative_locality_intro} in order to see how a DRK modifies the nature of spacetime is particularized to
\begin{align}
  S \,=&\, \int_{-\infty}^{\tau_1} d\tau \sum_{i=1,2} \left[x_{-(i)}^\mu(\tau) \dot{p}_\mu^{-(i)}(\tau) + N_{-(i)}(\tau) \left[C(p_{-(i)}(\tau)) - m_{-(i)}^2\right]\right] \nonumber \\
  & + \int_{-\infty}^{\tau_2} \left[x_{-(3)}^\mu(\tau) \dot{p}_\mu^{-(3)}(\tau) + N_{-(3)}(\tau) \left[C(p_{-(3)}(\tau)) - m_{-(3)}^2\right]\right] \nonumber \\   & + \int_{\tau_1}^{\tau_2} d\tau \left[y^\mu(\tau) \dot{q}_\mu(\tau) + N(\tau) \left[C(q(\tau)) - m^2\right]\right] \nonumber \\ 
& + \int_{\tau_1}^{\infty} \left[x_{+(1)}^\mu(\tau) \dot{p}_\mu^{+(1)}(\tau) + N_{+(1)}(\tau) \left[C(p_{+(1)}(\tau)) - m_{+(1)}^2\right]\right] \nonumber \\ & +   \int_{\tau_2}^{\infty} d\tau \sum_{j=2,3} \left[x_{+(j)}^\mu(\tau) \dot{p}_\mu^{+(j)}(\tau) + N_{+(j)}(\tau) \left[C(p_{+(j)}(\tau)) - m_{+(j)}^2\right]\right] \nonumber \\
& + \xi^\mu \left[\left(p^{+(1)}\oplus q\oplus p^{-(3)}\right)_\mu - \left(p^{-(1)}\oplus p^{-(2)}\oplus p^{-(3)}\right)_\mu\right](\tau_1) \nonumber \\ & + \chi^\mu \left[\left(p^{+(1)}\oplus p^{+(2)}\oplus p^{+(3)}\right)_\mu - \left(p^{+(1)}\oplus q\oplus p^{-(3)}\right)_\mu\right](\tau_2),
\end{align}
where we denote by $(k\oplus p\oplus q)$ the total four-momentum of a three-particle system with four-momenta $(k, p, q)$.

The extrema of the action satisfy the set of equations
\be
\dot{p}^{-(i)} \,=\, \dot{p}^{+(j)} \,=\, \dot{q} \,=\, 0, \quad
\frac{\dot{x}_{-(i)}^\mu}{N_{-(i)}} \,=\, \frac{\partial C(p^{-(i)})}{\partial p^{-(i)}_\mu}, \quad \frac{\dot{x}^\mu_{+(j)}}{N_{+(j)}} \,=\, \frac{\partial C(p^{+(j)})}{\partial p_\mu^{+(j)}}, \quad \frac{\dot{y}^\mu}{N} \,=\, \frac{\partial C(q)}{\partial q_\mu}.
\ee
The DRK is given by
\begin{align}
  & C(p^{-(i)}) \,=\, m_{-(i)}^2, \quad\quad C(p^{+(j)}) \,=\, m_{+(j)}^2, \quad\quad C(q^2) \,=\, m^2, \nonumber \\
  & p^{-(1)}\oplus p^{-(2)}\oplus p^{-(3)} \,=\, p^{+(1)}\oplus q\oplus p^{-(3)} \,=\, p^{+(1)}\oplus p^{+(2)} \oplus p^{+(3)},
\label{dk2}
\end{align}
and we also have
\begin{align}
  & x^\mu_{-(i)}(\tau_1) \,=\, \xi^\nu \frac{\partial(p^{-(1)}\oplus p^{-(2)}\oplus p^{-(3)})_\nu}{\partial p^{-(i)}_\mu}, \:\: (i=1,2)\,, \nonumber \\
	&x^\mu_{-(3)}(\tau_2) \,=\, \chi^\nu \frac{\partial(p^{+(1)}\oplus q\oplus p^{-(3)})_\nu}{\partial p^{-(3)}_\mu}\,,\quad  x^\mu_{+(1)}(\tau_1) \,=\, \xi^\nu \frac{\partial(p^{+(1)}\oplus q\oplus p^{-(3)})_\nu}{\partial p^{+(1)}_\mu}\,,\nonumber \\
& x_{+(j)}^\mu(\tau_2) \,=\, \chi^\nu \frac{\partial(p^{+(1)}\oplus p^{+(2)} \oplus p^{+(3)})_\nu}{\partial p^{+(j)}_\mu}, \:\: (j=2,3), \nonumber \\
  & y^\mu(\tau_1) \,=\, \xi^\nu \frac{\partial(p^{+(1)}\oplus q\oplus p^{-(3)})_\nu}{\partial q_\mu}, \quad\quad y^\mu(\tau_2) \,=\, \chi^\nu \frac{\partial(p^{+(1)}\oplus q\oplus p^{-(3)})_\nu}{\partial q_\mu}.
\end{align}
With these equations we can determine the four-momentum $q$ and impose some restrictions between the other momenta. We also have relations between the four-velocities of the particles and their momenta. 

There is a new ingredient due to the presence of two interactions: on the one hand, from the equation for the four-velocity of the photon, one finds
\be
y^\mu(\tau_2) - y^\mu(\tau_1) \,=\, \frac{\partial C(q)}{\partial q_\mu} \,\int_{\tau_1}^{\tau_2} d\tau\, N(\tau),
\ee
and, from the conservation laws of the emission and detection interactions for the photon, one has 
\be
 y^\mu(\tau_2) - y^\mu(\tau_1) \,=\, \left(\chi^\nu - \xi^\nu\right) \,\frac{\partial(p^{+(1)}\oplus q\oplus p^{-(3)})_\nu}{\partial q_\mu}.
\ee
We find, combining both expressions, 
\be
\left(\chi^\nu - \xi^\nu\right) \,\frac{\partial(p^{+(1)}\oplus q\oplus p^{-(3)})_\nu}{\partial q_\mu} \,=\, \frac{\partial C(q)}{\partial q_\mu} \,\int_{\tau_1}^{\tau_2} d\tau\, N(\tau).
\ee
Therefore, the difference of coordinates of the two interaction vertices is fixed and then, we only have a set of solutions depending on four arbitrary constants ($\xi^\mu$) as in the single-interaction process case, which reflects the invariance under translations. 

There is one observer placed at the emission of the photon (for which $\xi^\mu=0$) that sees this process as local, but  not the detection, and another observer at the detection (for which $\chi^\mu=0$) , which is related to the other observer by a translation, seeing the detection as local but not the emission. For any other observer both interactions are not local. 

\subsection{Comparison with the previous models}

This proposal has in common with the first model that the velocity of the photon $v^i \,=\, \dot{y}^i/\dot{y}^0 \,=\, (\partial C(q)/\partial q_i)/(\partial C(q)/\partial q_0)$ has the same momentum dependence, which is determined by the DDR. But in order to determine the time of flight, one has to take into account that the emission and detection points of the photon  ($y^\mu(\tau_1)$ and $y^\mu(\tau_2)$ respectively) and then the trajectory, depend on all momenta involved in both interactions.  So the spectral and timing distribution of photons coming from a short GRB would differ from what one expects in SR,  but in a very complicated and unpredictable way since we do not have access to all the details of the detection and emission interactions. Moreover, we have an inconsistency since, if we consider the emission and propagation of the photon, we should consider also the processes where all the particles involved were produced and so on, so one should know the conditions of every particle in the universe.

If a cluster decomposition principle holds in the DSR framework, one can take the second model as valid since the emission takes place very far from the detection. This is on the other hand the most natural way, since we have seen that in this approach the same velocity holds for all photons, independently on their energy, when one uses the physical coordinates in which the interactions are local. 

We conclude that there are different perspectives of the time delay problem which deserves further investigation.

As we have seen, there are different models in the DSR framework which do not produce a time delay for photons, so the restrictions on the high energy scale that parametrizes DSR based on such experiments are inconclusive. Since this kind of observations are the only possible measurable effects for energies smaller than the high energy scale, this opens up the possibility  that this scale is orders of magnitude lower than expected and then, observable consequences in high energy particle accelerators could be observed. In the next chapter we will consider such possibility, imposing constraints on the scale from the data obtained in them. 

\chapter{Twin Peaks: beyond SR production of resonances}
\label{chapter_twin}
\ifpdf
    \graphicspath{{Chapter6/Figs/Raster/}{Chapter6/Figs/PDF/}{Chapter6/Figs/}}
\else
    \graphicspath{{Chapter6/Figs/Vector/}{Chapter6/Figs/}}
\fi

\epigraph{Harry, I’m going to let you in on a little secret. Every day, once a day, give yourself a present. Don’t plan it, don’t wait for it, just let it happen.}{Dale Cooper, Twin Peaks}

We have seen in the last chapter that there is no time delay of photons with different energies in many models inside the DSR context. Since this is the only phenomenological window to quantum gravity effects due to a deformed relativistic kinematics, the strong constraints based on this kind of experiments may loose they validity, and then, one can consider that the high energy scale that parametrizes a deviation of SR could be orders of magnitude smaller than expected, i.e. the Planck energy. In this chapter, we consider the possibility of  a very low energy (with respect to the Planck energy) scale that characterizes modifications to SR in the framework of DSR, and that this modification could be observed in accelerator physics depending on the value of the scale~\cite{Albalate:2018kcf}. This has been done previously in the canonical noncommutativity we mentioned in the introduction for linear accelerators~\cite{Hewett:2000zp,Hewett:2001im,Mathews:2000we} and for hadron colliders~\cite{Alboteanu:2006hh,Ohl:2010zf,YaserAyazi:2012ni}, obtaining in the latter works a lower bound for the high energy scale of the order of TeV (for a review of canonical noncommutative phenomenology see~\cite{Hinchliffe:2002km}).

In this chapter, we will study the simple process of scattering of two particles, taking $Z$ production at the Large Electron-Positron collider (LEP).  We will obtain a remarkable effect: two correlated peaks, that we have baptized as \emph{twin peaks}, are associated with a single resonance. We study this possible phenomenology using recent experimental data in order to constrain the scale parametrizing the deviation, obtaining a bound of the order of the TeV. Therefore, this effect might be observable in the next very high energy (VHE) proton collider. Also, we will present a more detailed analysis computing the total cross section of the process \mbox{$f_i \overline{f}_i \to X \to f_j \overline{f}_j$} with some prescriptions to include the effects of a particular DCL.  

\section{Twin Peaks}
\label{sec:twin-peaks}

In this section, we start by considering that deviations from SR are characterized by an energy scale $\Lambda$ much smaller than the Planck energy, $\Lambda\ll M_P$. Then, we will try to look for its possible signals in the production of a resonance at a particle accelerator. In fact, we will see a new effect if  the mass of the resonance is of the order of this scale.

We will start by modifying the standard expression of the Breit--Wigner distribution
\be
f(q^2)\,=\,\frac{K}{(q^2-M^2_X)^2+M_X^2\Gamma_X^2}\,,
\label{eq:BW}
\ee
where $q^2$ is the four-momentum squared of the resonance $X$, $M_X$ and $\Gamma_X$ are respectively its mass and decay width, and $K$ is a kinematic factor that can be considered constant in the region $q^2\sim M_X^2$ (i.e. $K$ is a smooth function of $q^2$ near $M_X^2$).

For a resonance produced by the scattering of two particles, and which decays into two particles, $q^2$ is the squared invariant mass of the two particles producing the  resonance or of the two particles into which it decays. In SR, for two particles with four-momenta $p$ and $\overline{p}$, the squared invariant mass is  
\be
\begin{split}
m^2&\,=\,(p+\overline{p})_\mu (p+\overline{p})^\mu\,=\,(p+\overline{p})_0^2-\sum_i(p+\overline{p})_i^2\\
&\,=\,E^2+\overline{E}^2+2E\overline{E}-\sum_i p_i^2-\sum_i \overline{p}_i^2-2p\overline{p}\cos\theta \approx 2E\overline{E}(1-\cos\theta)\,,
\end{split}
\label{eq:s2}
\ee
where $\theta$ is the angle between the directions of the particles and in the last expression we have used that in the ultra-relativistic limit $(E\sim p)$. 

As we have seen in the previous chapters, the main ingredients of a DRK in DSR theories is a DDR and a DCL. In order to study the modification of the production of a resonance, we are going to maintain the usual form of the Breit--Wigner distribution of Eq.~\eqref{eq:BW} but we will modify  the squared invariant mass of the process Eq.~\eqref{eq:s2} using a DCL. This corresponds to  the case in which the dispersion relation is the one of SR. We have seen that in the Hopf algebra frameworks, one can work in the classical basis of $\kappa$ Poincar\'e~\cite{Borowiec2010}, in which the dispersion relation is the usual one. Along this chapter, we will use a simpler case (although it is not included in the Hopf algebras scheme) which satisfies our requirement about the dispersion relation and where the DCL is covariant, corresponding to the composition law for the Snyder algebra~\cite{Battisti:2010sr}. The simplest example is
\begin{equation}\label{eq:DCL_tp}
\mu^2:\,=\,(p\oplus\overline{p})^2\,=\,m^2\left(1+\epsilon\frac{m^2}{\Lambda^2}\right),
\end{equation}
where $\mu^2$ is the new invariant mass squared of the two-particle system, in which a new additional term to the one in SR appears, $\epsilon m^2/\Lambda^2$, where the parameter $\epsilon\,=\,\pm 1$ represents the two possible signs of the correction\footnote{This is related with the curvature sign of the maximally symmetric momentum space, i.e. if de Sitter or anti-de Sitter is considered, as we saw in Ch.~\ref{chapter_curved_momentum_space}.}.

The modification of the expression of the Breit--Wigner distribution of Eq.~\eqref{eq:BW} leads to consider $f_{\text{BSR}}=f(q^2=\mu^2)$ instead of $f_{\text{SR}}=f(q^2=m^2)$ in the production of the resonance~\footnote{This simple modification of the Breit--Wigner distribution is due to the fact that we are considering a modification of SR in which the dispersion relation (and then the propagator) is not modified. Then the only modification appears in the invariant mass and not in the explicit form of the distribution.}. As we do not have a dynamical theory with a DRK, the modification of the Breit--Wigner distribution as a function of $m^2$ is an ansatz, although in Sec.~\ref{sec:cross-section} we will see through a set of prescriptions  a way to compute it. Then, we have 
\begin{equation}
\label{eq:BW-BSR}
f_{\text{BSR}}(m^2)\,=\,\frac {K}{\left[\mu^{2}(m^2)-M_X^{2}\right]^{2}+M_X^{2}\Gamma_X ^{2}}\,.
\end{equation}

In Appendix \ref{appendix:B-W}, we show what are the conditions for a resonance to take place. From them, one can see that the choice $\epsilon\,=\,-1$ leads to a double peak with masses 
\begin{equation}
{m^*_{\pm}}^2\,=\,\frac{\Lambda^2}{2}\left[1\pm\left(1-4\frac{M_X^2}{\Lambda^2}\right)^{1/2} \right]\,,
\end{equation} 
and widths
\begin{equation}
{\Gamma_\pm^*}^2\,=\,\frac{M_X^2\Gamma_X^2}{{m_\pm^*}^2\left(1-4{M_X^2}/{\Lambda^2}\right)}\,.
\end{equation}

From the previous equation, we can find the following relationship between the widths of the two peaks,
\begin{equation}\label{eq:rel-peaks}
\frac{{\Gamma_+^*}^2}{{\Gamma_-^*}^2}\,=\,\frac{{m_-^*}^2}{{m_+^*}^2}\,,
\end{equation}
which will produce a vital distinction between a possible double peak (\emph{twin peaks}) in a BSR scenario and the production of two non-related resonances. 

In the next section, we will study the $Z$-boson production with the previous prescription, which leads us to a lower bound on the high energy scale $\Lambda$, and to consider the possibility to have observable effects in a future VHE hadron collider.

\section{Searches for BSR resonances in colliders}
\label{sec:limits}

\subsection{Bounds on \texorpdfstring{$\Lambda$}{Lg} using LEP data}
\label{sec:limits_1}

The precision obtained in the measurement of the mass and decay width at the LEP collider~\cite{PhysRevD.98.030001}, makes the $Z$ boson a perfect candidate for our study:
\begin{equation}
M_Z^{\text{\text{exp}}}\,=\,91,1876\pm 0.0021\,\mathrm{GeV}\,,\quad\quad \Gamma_Z^{\text{exp}}\,=\,2.4952\pm 0.0023\,\mathrm{GeV}\,,
\end{equation}
where the superscript ``exp'' in the previous expressions remarks the fact that the mass and the decay width are obtained (experimentally) by fitting the standard Breit--Wigner distribution. The values in the case of a BSR scenario would be different from the true $M_Z$ and $\Gamma_Z$. Then, $M_Z^2$ is the value of $\mu^2$, and $(M_Z^{\text{exp}})^2$ is the value of $m^2$ at the peak of the distribution. According to Eq.~\eqref{eq:DCL_tp} one has
\begin{equation}
M_Z^2\,=\,(M_Z^{\text{exp}})^2\left(1+\epsilon\frac{(M_Z^{\text{exp}})^2}{\Lambda^2} \right)\,.
\end{equation}

If one assumes a maximum modification $\delta M_Z=M_Z-M_Z^{\text{exp}}$ of the $Z$ mass determination from LEP compatible with other observations, one can put a limit to the scale $\Lambda$
\begin{equation}\label{Lambda0}
\Lambda\,\geq\, \Lambda_0\,=\,M_Z^{\text{exp}}\left(\frac{M_Z^{\text{exp}}}{2\,\delta M_Z}\right)^{1/2} \,=\, 3,55 \,\left(\frac{30\,\text{MeV}}{\delta M_Z}\right)^{1/2} \,\text{TeV}\,.
\end{equation} 
For the determination of the bound on $\Lambda$, we have compared LEP data with the energy dependence of the first of the peaks of the modified Breit--Wigner distribution cross section. Due to the small contribution of the tail of the second peak, we have not taken into account its effect. 

We can see that an energy scale of the order of a few TeV could be compatible with LEP data bounds, which is of the order of magnitude found for the scale in the case of a canonical noncommutativity~\cite{Alboteanu:2006hh,Ohl:2010zf,YaserAyazi:2012ni}. As this energy can be reached in  a future VHE hadron collider, in the following subsection we will study how to implement this modification for such a case.

\subsection{Searches for BSR resonances in a VHE hadron collider}
\label{sec:limits_2}

There is no evidence of a modification of SR due to the corrections considered in our model in LEP observations of the $Z$ boson. This can be understood since the mass of the $Z$ boson is not comparable to the high energy scale, making the proposed effect of a double peak completely unobservable, since the energy of the resonance is not high enough (see Eq.~\eqref{Lambda0}).

Since the lower bound on $\Lambda$ from LEP is of a few TeV, a future electron-positron collider like ILC will not have enough energy to observe the two peaks we are proposing. So let us consider than some future VHE hadron (proton--proton, pp) collider will be able to reach such energy and then will observe the two peaks of a new resonance at $m^2={m_\pm^*}^2$ (we are considering the interesting case $\epsilon=-1$). Let us suppose that the resonance is due to the annihilation of a quark-antiquark of momenta $p$ and $\overline{p}$ respectively, and that it decays to two fermions of momenta $q$ and $\overline{q}$, although other particles will be produced due to the hadron scattering. From Eq.~\eqref{BSR exp}, the differential cross section with respect to $m^2=(q+\overline{q})^2$ for each peak is
\begin{equation} \label{eq:difcs}
\frac{d\sigma}{dm^2}\approx \mathcal{F}_\pm(s,{m_\pm^*}^2) \frac{K_\pm}{(m^2-{m_\pm^*}^2)^2+{m_\pm^*}^2{\Gamma_\pm^*}^2}\,,
\end{equation}
where the function $\mathcal{F}_\pm(s,{m_\pm^*}^2)$ can be obtained from the parton model as follows.

We start by the usual Mandelstam variable $s$ of the pp system for the ultra-relativistic case, 
\begin{equation}
s\,=\,(P+\overline{P})^2\,=\,2E\overline{E}(1-\cos\theta)\,=\, 4E\overline{E}\,,
\end{equation}
where $P$ and $\overline{P}$ are the momenta of the two protons in the initial state. One can write
\begin{equation}
P^\mu\,=\,\frac{\sqrt{s}}{2}(1,0,0,1)\,,\quad\quad\quad\quad \overline{P}^\mu\,=\,\frac{\sqrt{s}}{2}(1,0,0,-1)\,,
\end{equation}
for the momenta of the protons, and
 \begin{equation}
p^\mu\,=\,xP^\mu\,,\quad\quad\quad \overline{p}^\mu\,=\,\overline{x}\overline{P}^\mu\,,
\end{equation}
($0<x,\overline{x}<1$) for the momenta of the quark--antiquark pair producing the resonance.

The squared invariant mass of the quark--antiquark system in SR is, according to Eq.~\eqref{eq:s2},
\be
m^2\,=\,(p+\overline{p})^2\,=\,4E_p E_{\overline{p}}\,=\,4 x \overline{x} E \overline{E} \,=\, x \overline{x} s\,,
\label{eq:mxs}
\ee
where we have used the same symbol ($m^2$) at the initial and final states since it is a conserved quantity, i.e. $(p+\overline{p})^2=(q+\overline{q})^2$. In the case of a DCL the energy-momentum conservation law requires that  $(p\oplus\overline{p})^2=(q\oplus\overline{q})^2=\mu^2$, $\mu^2$ to be conserved but one can easily see from Eq.~\eqref{eq:DCL_tp} that the conservation of  $\mu^2$ implies the conservation of $m^2$.

Using the relation $m^2=x\overline{x}s$ from Eq.~\eqref{eq:mxs} we can write $\mathcal{F}_\pm(s,{m_\pm^*}^2)$  as 
\begin{equation}\label{eq:F}
\mathcal{F}_\pm(s,{m_\pm^*}^2)\,=\,\int_0^1\int_0^1 dx\, d\overline{x} \,f_q(x,{m_\pm^*}^2) \, f_{\overline{q}}(\overline{x},{m_\pm^*}^2) \, \delta\left(x\overline{x}-\frac{{m_\pm^*}^2}{s}\right)\,,
\end{equation}
where $f_q(x,{m_\pm^*}^2)$ is the parton distribution function. It is defined as the probability density to find a parton (quark) in a hadron (proton) with a fraction $x$ of its momentum when one probes the hadron at an energy scale $m^2 \sim {m_\pm^*}^2$, where ${m_\pm^*}^2$ is given by Eq.~\eqref{epsilon-}.

The $K_\pm$ factors in Eq.~\eqref{eq:difcs} take into account  the coupling dependence and all the details of the annihilation of the quark--antiquark pair. Using the previous expressions, one can estimate the expected number of events for different pp colliders, the mass values ($M_X$) of different resonances and the characteristic scale ($\Lambda$). Also, the observation of a double peak would lead us to extract the true mass and width of the resonance through Eqs.~\eqref{changes} and~\eqref{eq:width}.

\section{Cross section calculation in a QFT approach BSR}
\label{sec:cross-section}

In this section we are trying to justify the ansatz of Eq.~\eqref{eq:BW-BSR}. In order to do so, we will consider the process $e^-(k) e^+(\overline{k})\rightarrow Z \rightarrow \mu^-(p) \mu^+(\overline{p})$ and study the modification through a DCL in the DSR scenarios. We will use modified dynamical squared matrix elements in which the ingredient of a DCL is introduced through new Mandelstam variables which replace the SR invariants. It is unavoidable  to use some ad hoc prescriptions since we do not know how to introduce a DCL in the QFT framework. Anyway, it suggests that Eq.~\eqref{eq:BW-BSR} can be a good approximation, which is obvious since the main modification of the peak of the resonance is due to the variation of the propagator. Also, it helps us to show how to handle the problem of different \emph{channels}, since a generic DCL is not symmetric.

\subsection{Phase-space momentum integrals}
\label{integral}

Let us start with the modification of the two-particle phase-space integral in SR for the massless case:
\be\label{PS2}
PS_2 \,\,=\,\, \int \frac{d^4p}{(2\pi)^3}  \delta(p^2) \theta(p_0) \,\frac{d^4\overline{p}}{(2\pi)^3} \delta(\overline{p}^2) \theta(\overline{p}_0) \,(2\pi)^4 \delta^{(4)}[(k+\overline{k}) - (p+\overline{p})] \,. 
\ee

For simplicity, we will consider a DCL related to the Snyder algebra~\cite{Battisti:2010sr} (obtained  by geometrical arguments in Ch.~\ref{chapter_curved_momentum_space})
\begin{equation}\label{BSRDCL}
\left[l\oplus q\right]^{\mu}\,=\,l^{\mu}\sqrt{1+\frac{q^{2}}{\overline{\Lambda}^{2}}}+\frac{1}{\overline{\Lambda}^{2}\left(1+\sqrt{1+{l^{2}}/{\overline{\Lambda}^{2}}}\right)}l^{\mu}\left(l\cdot q\right)+q^{\mu}\approx l^{\mu}+q^{\mu}+\frac{1}{2\overline{\Lambda}^{2}}l^{\mu}\left(l \cdot q\right)\,,
\end{equation}
where we have used the fact that the particles involved in the process are relativistic particles. One gets the relation used in Sec.~\ref{sec:twin-peaks} with $\epsilon\,=\,+1$ from $(l\oplus q)^2$ of Eq.~\eqref{BSRDCL}. The negative sign of the parameter $\epsilon\,=\,-1$ can be also found in Eq.~\eqref{BSRDCL} if one considers the two possible signs in the commutator of space-time coordinates of the Snyder algebra: $\left[x^\mu, x^\nu \right]\,=\,\pm J^{\mu\nu}/\overline{\Lambda}^2$. This is understood as well from the geometrical perspective, considering the DCL related to de Sitter or anti-de Sitter momentum spaces. From now on, we will use Eq.~\eqref{BSRDCL}. Note that Eq.~\eqref{eq:DCL_tp} can be recovered with $\Lambda^2\,=\,2\overline{\Lambda}^2$. We can now justify the proposed model of the previous section from this covariant DCL. 

As the DRK we are considering has the same dispersion relation of SR, the only modification of the phase-space integral is due to the DCL which, not being symmetric, leads to four possible conservation laws (channels) in which the process\\*  \mbox{$e^-(k) e^+(\overline{k}) \to Z \to \mu^-(p) \mu^+(\overline{p})$} can be produced. Then, we have four phase-space integrals, one for each channel ($\alpha\,=\,1,2,3,4$)
\be\label{PS2bar}
\overline{PS}_2^{(\alpha)} \,=\, \int \frac{d^4p}{(2\pi)^3}  \delta(p^2) \theta(p_0) \,\frac{d^4\overline{p}}{(2\pi)^3} \delta(\overline{p}^2) \theta(\overline{p}_0) \,(2\pi)^4 \delta^{(4)}_\alpha(k, \overline{k}; p, {\overline p}) \,,
\ee
where
\be
\begin{split}
\delta^{(4)}_1(k, \overline{k}; p, {\overline p}) &\,=\, \delta^{(4)}[(k\oplus\overline{k}) - (p\oplus\overline{p})]\,, \qquad
\delta^{(4)}_2(k, \overline{k}; p, {\overline p}) \,=\, \delta^{(4)}[(\overline{k}\oplus k) - (p\oplus\overline{p})]\,, \\
\delta^{(4)}_3(k, \overline{k}; p, {\overline p}) &\,=\, \delta^{(4)}[(k\oplus\overline{k}) - (\overline{p}\oplus p)]\,, \qquad 
\delta^{(4)}_4(k, \overline{k}; p, {\overline p}) \,=\, \delta^{(4)}[(\overline{k}\oplus k) - (\overline{p}\oplus p)]\,.
\end{split}
\ee    
  
\subsection{Choice of the dynamical factor with a DCL}
\label{sec:dynamical}

As the scattering process is produced in a collider, one can assume that both particles of the initial state have the same modulus of the momentum (where we have neglected the masses since we are working in the ultra-relativistic limit) 
\be
k_\mu\,=\,\left(E_{0},\,\vec{k}\right)\,,\, \overline{k}_\mu\,=\,\left(E_{0},\,-\vec{k}\right)\,,\mbox{ with }E_{0}\,=\,|\vec{k}|\,.
\ee

We take as starting point the SR cross section at the lowest order, which is a product of four factors: the kinematic factor of the initial state, the phase-space integral of the two-particle, the propagator of the $Z$-boson and the dynamical factor $A$
\be
\sigma \,=\,\frac{1}{8 E_0^2} \, PS_2 \,\frac{1}{\left[(s - M_Z^2)^2 + \Gamma_Z^2 M_Z^2\right]} \,A\,.
\label{eq:sigma}
\ee

The SM dynamical factor for this process is~\cite{Atsue2015}
\be\label{A}
A \,=\, \frac{e^4}{2 \sin^4\theta_W \cos^4\theta_W} \, \left(\left[C_V^2+C_A^2\right]^2 \left[\left(\frac{t}{2}\right)^2 + \left(\frac{u}{2}\right)^2\right] - 4 C_V^2 C_A^2 \left[\left(\frac{t}{2}\right)^2 - \left(\frac{u}{2}\right)^2\right]\right)\,,
\ee
where  ($C_V$) and ($C_A$) are the corrections to the vector and axial weak charges respectively, ($\theta_W$) is the Weinberg angle and ($s$, $t$, $u$) the Mandelstam variables 
\begin{align}
s&\,=\,\left(k+\overline{k}\right)^{2}\,=\,\left(p+\overline{p}\right)^{2}\,\,, \\
t&\,=\,\left(k-p\right)^{2}\,=\,\left(\overline{p}-\overline{k}\right)^{2}\,, \\
u&\,=\,\left(k-\overline{p}\right)^{2}\,=\,\left(p-\overline{k}\right)^{2}\,.
\end{align}

As we do not have a BSR-QFT, we do not know how the SR cross section of  Eq.~\eqref{eq:sigma} is modified. But we can assume that the generalization should be compatible with Lorentz invariance and in the limit $\Lambda \to \infty$ we should recover the SR cross section $\sigma$ of Eq.~\eqref{eq:sigma}. So we can consider that, since the two-particle phase-space integral is modified as in Eq.~\eqref{PS2bar},  the generalization of the cross section will be\footnote{Note that this expression is corrected from the one used in~\cite{Albalate:2018kcf}, as the factor used there of $8 E_0^2$ is not relativistic invariant when a DCL is considered, while $2s$ is. However, the results barely change since the main contribution is due to the modification of the propagator.}   
\be\label{sigmaalpha}
\overline{\sigma}_\alpha \,=\, \frac{1}{2\,s} \, \overline{PS}_2^{(\alpha)} \,\frac{1}{\left[(\overline{s} - \overline{M}_Z^2)^2 + \overline{\Gamma}_Z^2 \overline{M}_Z^2\right]} \,\overline{A}_\alpha\,,
\ee
for each channel $\alpha$. In our simple choice of  the DCL, the squared total mass is the same for every channel
\be
(k\oplus\overline{k})^2 \,=\, (\overline{k}\oplus k)^2 \,=\, (p\oplus\overline{p})^2 \,=\, (\overline{p}\oplus p)^2 \doteq \overline{s}\,.
\ee

As we do not have a dynamic framework, we will consider two possible different assumptions for the dynamical factor $\overline{A}_\alpha$: 
\begin{enumerate}[leftmargin=*,labelsep=4.9mm]
\item One can consider that the dynamical factor does not depend on the DCL, and then $\overline{A}_\alpha=A$. However, since the DCL implies that $(k-p)\neq (\overline{p}-\overline{k})$, $(k-\overline{p})\neq (p-\overline{k})$, one has to consider
\begin{align}
t &\,=\,  \frac{1}{2} \left[(k-p)^2 + (\overline{p} - \overline{k})^2\right] \,=\, - k\cdot p - \overline{k}\cdot \overline{p}\,, \\
u &\,=\, \frac{1}{2} \left[(k-\overline{p})^2 + (p-\overline{k})^2\right] \,=\, - k\cdot \overline{p} - \overline{k}\cdot p\,.
\end{align} 
\item Also, it is possible to regard that the generalization of $A$ is carried out by the replacement of the usual Mandelstam variables $t$, $u$ by new invariants $\overline{t}$, $\overline{u}$. Since we cannot find how to  associate new invariants for each channel, we will consider that the dynamical factor is channel independent ($\overline{A}_\alpha=\overline{A}$), obtained from $A$ just replacing the usual Mandelstam variables \mbox{$t$, $u$ } by 
\begin{align}
\overline{t} &\,=\, \frac{1}{2} \left[(k\oplus\hat{p})^2 + (\overline{p}\oplus \hat{\overline{k}})^2\right] \,=\, - k\cdot p - \overline{k}\cdot \overline{p} + \frac{(k\cdot p)^2}{2\overline{\Lambda}^2} + \frac{(\overline{k}\cdot \overline{p})^2}{2\overline{\Lambda}^2} \,,\\
\overline{u} &\,=\, \frac{1}{2} \left[(k\oplus\hat{\overline{p}})^2 + (p\oplus \hat{\overline{k}})^2\right] \,=\, - k\cdot \overline{p} - \overline{k}\cdot p + \frac{(k\cdot \overline{p})^2}{2\overline{\Lambda}^2} + \frac{(\overline{k}\cdot p)^2}{2\overline{\Lambda}^2}\,,
\end{align}
(note that in the particular case we are considering, the squared of a composition of two momenta is symmetric, regardless of the asymmetry of the DCL), where we have used the \textit{antipode} $\hat{p}$ defined in Ch.~\ref{chapter_curved_momentum_space}. One can check that for our choice of the DCL  of Eq.~\eqref{BSRDCL}, the antipode is the usual of SR, i.e. just $-p$  
\[
\begin{split}
\left[p\oplus \hat{p}\right]^{\mu}\,=\,\left[p\oplus -p\right]^{\mu}&\,=\, p^\mu \left\lbrace \sqrt{1+\frac{p^{2}}{\overline{\Lambda}^{2}}}-\frac{p^2}{\overline{\Lambda}^{2}\left(1+\sqrt{1+{p^{2}}/{\overline{\Lambda}^{2}}}\right)}-1 \right\rbrace  \\
&\,=\, p^\mu\left\lbrace \frac{\overline{\Lambda}^2\left( \sqrt{1+p^2/\overline{\Lambda}^2}+1\right)+\overline{\Lambda}^2 p^2/\overline{\Lambda}^2 -p^2}{\overline{\Lambda}^{2}\left(1+\sqrt{1+p^2/\overline{\Lambda}^2}\right)}-1 \right\rbrace \,=\, 0\,. 
\end{split}
\]
\end{enumerate}

We will consider that the scattering can be produced in all the possible channels with the same probability, so in order to compute the whole cross section, we will take the average of all channels
\be\label{sigmabar}
\overline{\sigma} \,\doteq\, \frac{1}{4}\sum_\alpha \overline{\sigma}_\alpha\,,
\ee 
with $\overline{\sigma}_\alpha$ in Eq.~\eqref{sigmaalpha}. In fact, one has two assumptions from the two choices of the modified dynamical factor $\overline{A}$
\begin{align}
\overline{\sigma}^{(1)} &\,=\, \frac{1}{32 E_0^2} \,\frac{1}{\left[(\overline{s} - \overline{M}_Z^2)^2 + \overline{\Gamma}_Z^2 \overline{M}_Z^2\right]} \,\sum_\alpha \overline{PS}_2^{(\alpha)} \,A(t,u),  \\
\overline{\sigma}^{(2)} &\,=\, \frac{1}{32 E_0^2} \,\frac{1}{\left[(\overline{s} - \overline{M}_Z^2)^2 + \overline{\Gamma}_Z^2 \overline{M}_Z^2\right]} \,\sum_\alpha \overline{PS}_2^{(\alpha)} \,A(\overline{t},\overline{u}).
\end{align}

In Appendix~\ref{appendix_cross_sections} it is showed how to obtain these two cross sections, finding
\begingroup\makeatletter\def\f@size{9}\check@mathfonts
\def\maketag@@@#1{\hbox{\m@th\fontsize{10}{10}\selectfont \normalfont#1}}%
\begin{align}
 \overline{\sigma}^{(1)} &\,=\, \frac{e^4}{48 \pi \sin^4\theta_W \cos^4\theta_W\left(1+\frac{E_0^2}{\overline{\Lambda}^2}\right)} \,\frac{E_0^2}{\left[\left(4E_0^2(1+E_0^2/\overline{\Lambda}^2) - \overline{M}_Z^2\right)^2 + \overline{\Gamma}_Z^2 \overline{M}_Z^2\right]} \, \left((C_V^2+C_A^2)^2 \left[1 -\frac{E_0^2}{2\overline{\Lambda} ^2}\right] \right)\,,
  \label{eq:cross_section_final_1}\\
  \overline{\sigma}^{(2)} &\,=\,  \frac{e^4}{48 \pi \sin^4\theta_W \cos^4\theta_W\left(1+\frac{E_0^2}{\overline{\Lambda}^2}\right)} \,\frac{E_0^2}{\left[\left(4E_0^2(1+E_0^2/\overline{\Lambda}^2) - \overline{M}_Z^2\right)^2 + \overline{\Gamma}_Z^2 \overline{M}_Z^2\right]} \,\left((C_V^2+C_A^2)^2 \left[1 -\frac{2 E_0^2}{\overline{\Lambda} ^2}\right] \right)\,.
 \label{eq:cross_section_final_2}
\end{align}
\endgroup

\subsection{Constraints on \texorpdfstring{$\overline{\Lambda}$}{Lg}}

Now we are able to find the constraints on $\overline{\Lambda}$ due to the modified cross section, taking into account the Particle Data Group (PDG) data~\cite{PhysRevD.98.030001}. We require the cross section  $\overline{\sigma}$ to be compatible with the PDG data for a value of $\overline{M}_Z$ and $\overline{\Gamma}_Z$ in an interval $\pm \,30\,\text{MeV}$ around their central values given by the PDG\footnote{It can be seen that for bigger values of $\delta\overline{M}_Z$ and $\delta\overline{\Gamma}_Z$ there is not a significant variation in the constraint for $\overline{\Lambda}$.}. As the SM is really successful, we will use the SR cross section, with the PDG values of $M_Z$ and $\Gamma_Z$ at one or two standard deviations, as a good approximation to the experimental data. 

In Table~\ref{table:two} the obtained results are shown, where we have denoted by $\overline{\sigma}^{(j)}_{i}$ the cross section taking $i$ standard deviations in the data. Note that the constraints are independent of the sign in the DCL.
\begin{table}[H]
\caption{Bounds on the scale of new physics $\overline{\Lambda}$ from LEP 
data of the $Z$ boson.}
\centering
\begin{tabular}{ccccc}
\toprule
{\bf Constraints} & \boldmath{$\overline{\sigma}^{(1)}_{1}$} & \boldmath{$\overline{\sigma}^{(1)}_{2}$} & \boldmath{$\overline{\sigma}^{(2)}_{1}$} & \boldmath{$\overline{\sigma}^{(2)}_{2}$}  \\
\midrule
$\overline{\Lambda}${[}TeV{]} & 2.4 &1.8 & 2.7 & 1.9 \\
\bottomrule
\end{tabular}

\label{table:two}
\end{table}

As we can see from the previous table, although the computations developed along this section help us to understand better the BSR framework of QFT, the results show that the simple approximation used in  Eq.~\eqref{eq:DCL_tp} gives a good estimate.

Once we have studied in the last two chapters some of the phenomenological consequences of DRK in flat spacetime, we can try to see how such modification of SR can be generalized in the context of a curved spacetime. This is necessary to study the possible existence or absence of time delays, since the universe expansion is not negligible due to the long distances photons travel from where they are emitted to our telescopes. In the next chapter we will see a new proposal incorporating a curvature of spacetime in the discussion carried out in Ch.~\ref{chapter_curved_momentum_space}.

\chapter{Cotangent bundle geometry}
\label{ch:cotangent}
\ifpdf
    \graphicspath{{Chapter8/Figs/Raster/}{Chapter8/Figs/PDF/}{Chapter8/Figs/}}
\else
    \graphicspath{{Chapter8/Figs/Vector/}{Chapter8/Figs/}}
\fi

\epigraph{Only those who risk going too far can possibly find out how far one can go.}{T.S.Eliot}
 
When we have considered the DRK as a way to go beyond SR, we have not taken into account its possible effects on the space-time metric (we  studied how locality can be implemented when there is a DCL using some noncommutative coordinates, but we did not mention the space-time geometry). In fact, in Ch.~\ref{chapter_curved_momentum_space}  we have seen how a DRK can be understood through a curved momentum space with a flat spacetime. 

There are a lot of works in the literature studying the space-time consequences of a DDR in LIV scenarios~\cite{Kostelecky:2011qz,Barcelo:2001cp,Weinfurtner:2006wt,Hasse:2019zqi,Stavrinos:2016xyg}. Most of them have been developed by considering Finsler geometries, formulated by Finsler in 1918~\cite{zbMATH02613491} (these geometries are a generalization of Riemannian spaces where the space-time metric can depend also on vectors of the tangent space). However, in those works the introduction of a velocity dependent metric is considered out of the DSR context since there is no mention to a DLT and DCL, hence precluding the possibility to have a relativity principle.

In the DSR framework the starting point is also a DDR. However, there is a crucial difference between the LIV and DSR scenarios, since the latter implements a deformed Lorentz transformations (in the one-particle system) which makes the DDR invariant for different observers related by such transformation. The case of Finsler geometries in this context was considered for the cases of flat~\cite{Girelli:2006fw,Amelino-Camelia:2014rga} and curved spacetimes~\cite{Letizia:2016lew}, provoking a velocity dependence on the space-time metric. Besides Finsler geometries, which starts from a Lagrangian (in fact, Finsler geometries are particular realizations of Lagrange spaces~\cite{2012arXiv1203.4101M}), there is another possible approach to define a deformed metric from the Hamiltonian. This leads to Hamilton geometry~\cite{2012arXiv1203.4101M}, considered in~\cite{Barcaroli:2015xda}. In this kind of approach, the space-time metric depends on the phase-space coordinates (momentum and positions), instead of the tangent bundle coordinates (velocities and positions).  Both geometries are particular cases of geometries in the tangent and cotangent bundle respectively. 

Moreover, in~\cite{Rosati:2015pga} another way to consider possible phenomenology based on time delays in an expanding universe due to deformations of SR is explored. In that paper, it is studied both LIV and DSR scenarios, starting with a DDR and considering nontrivial translations (in a similar way in which such effect was studied in the first model of Ch.~\ref{chapter_time_delay}). In order to do so, they considered the expansion of the universe by gluing slices of de Sitter spacetimes, finding difficult to formulate such study in a direct way. 

As we have mentioned, the DDR and the one-particle DLT are the only ingredients in all previous works. But as we saw in Ch.~\ref{chapter_second_order}, there is a particular basis in  $\kappa$-Poincaré, the classical basis, in which the DDR and DLT are just the ones of SR, so, following the prescription used in these works, there would be no effect on the space-time metric. 

Another geometrical interpretation was considered in ~\cite{Freidel:2018apz}. In that work, considering a Born geometry of a double phase space leads to  a  modified  action of a quantum model that describes the propagation of a free relativistic particle, what they called a metaparticle. There, the DDR is obtained from the poles of the momentum integral representation of the metaparticle quantum propagator, instead of reading it from the constraint in the classical action and interpreting it as the squared distance in a curved momentum space, which is the approach we have discussed along this work.

Here we will study the case of  curved spacetime and momentum spaces by considering a geometry in the cotangent bundle, i.e. a geometrical structure for all the phase-space coordinates~\cite{Relancio:2020zok}. As we will see, this is mandatory in order to make compatible a de Sitter momentum space and a generic space-time geometry. With our prescription, we find a nontrivial (momentum dependent) metric for whatever form of the DDR (as long as there is a nontrivial geometry for the momentum space), being able to describe the  propagation of a free particle in the canonical variables. This differs from the perspective of~\cite{Rosati:2015pga}, where the considered metric for the spacetime is the one given by GR. However, as we will see, the existence or not of a time delay for photons in an expanding universe is still an open problem that deserves further research. 

We will start by making clear our proposal of constructing a metric in the cotangent bundle, in such a way that the resulting momentum curvature tensor corresponds to a maximally symmetric momentum space. As in the flat space-time case, we will see that one can also identify 10 transformations in momentum space for a fixed point of spacetime. After that, we will introduce the main ingredients of the cotangent geometry~\cite{2012arXiv1203.4101M} that we will use along this chapter. Then, we will show the connection between this formalism and the common approach followed in the literature that considers an action with a DDR.  Finally, we will study the phenomenological implications in two different space-time geometries, a Friedmann-Robertson-Walker universe and a Schwarzschild black hole.

\section{Metric in the cotangent bundle}
\label{sec:metric}

In this section, we will present a simple way to generalize the results obtained in Ch.~\ref{chapter_curved_momentum_space} taking into account a curvature of spacetime, characterized by a metric $g_{\mu\nu}^x(x)$, and we will explain, from our point of view, how to deal with a metric in the cotangent bundle depending on both momentum and space-time coordinates.  

\subsection{Curved momentum and space-time spaces}
We start with the action of a free particle in SR 
\begin{equation}
S\,=\,\int{\dot{x}^\mu k_\mu-\mathcal{N} \left(C(k)-m^2\right)}\,,
\label{eq:SR_action}
\end{equation}
where $C(k)=k_\alpha \eta^{\alpha\beta }k_\beta$. It is easy to check that the same equation of motions derived  in GR by solving the geodesic equation can be obtained just by replacing $\bar{k}_\alpha=\bar{e}^\nu_\alpha (x) k_\nu$\footnote{To avoid confusions, we will use the symbol $\bar{e}$ in order to denote the inverse of the tetrad.} in Eq.~\eqref{eq:SR_action}
\begin{equation}
S\,=\,\int{\dot{x}^\mu k_\mu-\mathcal{N} \left(C(\bar{k})-m^2\right)}\,,
\label{eq:GR_action}
\end{equation}
where $\bar{e}^\nu_\alpha(x)$ is the inverse of the tetrad of the space-time metric  $e^\nu_\alpha(x)$, so that 
\begin{equation}
g^x_{\mu\nu}(x)\,= \, e^\alpha_\mu (x) \eta_{\alpha\beta} e^\beta_\nu (x)\,,
\label{eq:metric-st}
\end{equation}
and then, the dispersion relation is 
\begin{equation}
C(\bar{k})\,=\,\bar{k}_\alpha \eta^{\alpha\beta }\bar{k}_\beta\,=\,k_\mu g_x^{\mu\nu}(x) k_\nu\,.
\label{eq:cass_GR}
\end{equation}

As we saw in Ch.~\ref{chapter_curved_momentum_space}, the dispersion relation can be interpreted as the distance in momentum space from the origin to a point $k$, so one can consider the following line element for momenta\footnote{We will use along the chapter the symbol $\varphi$ as the inverse of the momentum tetrad $\bar{\varphi}$ (remember the relation found in Ch.~\ref{chapter_locality} between the inverse of the momentum space tetrad and the functions allowing us to implement locality).}
\begin{equation}
d\sigma^2\,=\,dk_{ \alpha}g_k^{\alpha\beta}(k)dk_{ \beta}\,=\,dk_{ \alpha}\bar{\varphi}^\alpha_\gamma(k)\eta^{\gamma\delta}\bar{\varphi}^\beta_\delta(k)dk_{ \beta}\,,
\label{eq:line_m1}
\end{equation}
where $\bar{\varphi}^\alpha_\beta(k)$ is the inverse of the tetrad in momentum space $\varphi^\alpha_\beta(k)$. This can be easily extended to the curved space-time case introducing the variables $\bar{k}$ in the previous momentum line element, obtaining
\begin{equation}
d\sigma^2\,\coloneqq\,d \bar{k}_{\alpha}g_{\bar{k}}^{\alpha\beta}( \bar{k})d \bar{k}_{ \beta}\,=\,dk_{\mu}g^{\mu\nu}(x,k)dk_{\nu}\,,
\end{equation}
where in the second equality we have taken into account that the distance is computed along a fiber, i.e. the Casimir is viewed as the squared distance from the point $(x,0)$ to $(x,k)$ (we will see this in more detail in Sec.~\ref{sec:fb_properties}). The metric tensor $g^{\mu\nu}(x,k)$ in momentum space depending on space-time coordinates is constructed with the tetrad of spacetime and the original metric in momentum space, explicitly
\begin{equation}
g_{\mu\nu}(x,k)\,=\,\Phi^\alpha_\mu(x,k) \eta_{\alpha\beta}\Phi^\beta_\nu(x,k)\,,
\end{equation}
where 
\begin{equation}
\Phi^\alpha_\mu(x,k)\,=\,e^\lambda_\mu(x)\varphi^\alpha_\lambda(\bar{k})\,.
\end{equation}
We can check that, in the way we have constructed this metric, it is invariant under space-time diffeomorphisms. A canonical transformation in phase space  $(x, k) \to (x',k')$ of the form 
\begin{equation}
x^{\prime \mu}\,=\,f^\mu(x)\,,\qquad k'_\mu\,=\,\frac{\partial x^\nu}{\partial x^{\prime \mu}} k_\nu\,,
\label{eq:canonical_transformation}
\end{equation} 
makes the tetrad transforms as 
\begin{equation}
\Phi^{\prime \mu}_\rho(x',k')\,=\, \frac{\partial x^\nu}{\partial x^{\prime \rho}} \Phi^\mu_\nu(x,k)\,,
\label{eq:complete_tetrad_transformation}
\end{equation}
because
\begin{equation}
\frac{\partial x^\mu}{\partial x^{\prime \rho}}  e^\lambda_\mu(x)\varphi^\alpha_\lambda(\bar{k})\,=\,e^{\prime \kappa}_\rho(x')\varphi^{\prime \alpha}_\kappa(\bar{k}')
\label{eq:tetrad_transformation}
\end{equation}
holds, since the barred momentum variables are invariant under space-time diffeomorphisms 
\begin{equation}
\bar{k}'_ \mu\,=\,k'_\nu \bar{e}^{\prime \nu}_\mu(x')\,=\,k_\nu \bar{e}^\nu_\mu(x) \,=\,\bar{k}_ \mu\,,
\end{equation} 
due to the fact that the space-time tetrad transforms as 
\begin{equation}
\bar{e}^{\prime \nu}_\mu(x')\,=\,\frac{\partial x^{\prime \nu}}{\partial x^\rho}\bar{e}^\rho_\mu(x)\,,
\end{equation} 
and then, the tetrad of momentum space is invariant under this kind of transformations, as its argument does not change. Then, the metric is invariant under the same space-time diffeomorphisms of GR.

In the flat space-time case, we have seen that the momentum metric is invariant under momentum coordinate transformations. In the way we propose to construct the momentum metric when a curvature in both momentum and coordinate spaces is present, one loses this kind of invariance, i.e. a canonical transformation 
\be
k_\mu \,=\, h_\mu(k')\,,\quad x^\mu \,=\, x^{\prime \nu} j^\mu_\nu(k')\,,
\label{eq:primes}
\ee
 with
\be
j^\mu_\rho(k') \frac{\partial h_\nu(k')}{\partial k'_\rho} \,=\, \delta^\mu_\nu \,,
\ee
does not leave the metric invariant (as it happens in the GR case, where the metric is not invariant under these transformations). However, we have seen that this metric is invariant under space-time diffeomorphisms.

With this proposal, we are selecting somehow a particular choice of momentum variables against others, in contrast with the flat space-time formulation, where there was an independence on this choice. As it was shown in Ch.~\ref{chapter_intro}, there is a vast discussion about the possible existence of some ``physical'' momentum variables, the ones that nature ``prefers''. Within this framework it seems natural to think that, since the model is not invariant under the choice of momentum coordinates, there should be a preferred basis in which to formulate the physics. 

In the rest of this subsection we will see that, from the definition of the metric, one can easily generalize for a curved spacetime the momentum transformations obtained in Ch.~\ref{chapter_curved_momentum_space} for a maximally symmetric momentum space: as in the flat space-time case, there are still 10 momentum isometries for a fixed spacetime point $x$, 4  translations and 6 transformations leaving the point in phase space $(x,0)$ invariant, and we can also understand the dispersion relation as the distance from the point $(x,0)$ to $(x,k)$. The fact that with this procedure we have also 10 momentum isometries can be understood since, if one considers as the starting point a momentum space with a constant scalar of curvature, then the new metric in momentum space will have also a constant momentum scalar of curvature (see Appendix~\ref{appendix:cotangent}).

\subsubsection{Modified  translations}

In Ch.~\ref{chapter_curved_momentum_space} we have found the translations from Eq.~\eqref{T(a,k)}
\begin{equation}
\varphi^\mu_\nu(p\oplus q) \,=\, \frac{\partial (p\oplus q)_\nu}{\partial q_\rho} \, \varphi^\mu_\rho(q)\,,
\end{equation}
so we should find the new translations by replacing $p\rightarrow \bar{p}_\mu=\bar{e}_\mu^\nu(x) p_\nu$, $q\rightarrow \bar{q}_\mu=\bar{e}_\mu^\nu(x) q_\nu$ on it
\begin{equation}
\varphi^\mu_\nu(\bar{p} \oplus \bar{q}) \,=\,  \frac{\partial (\bar{p} \oplus \bar{q})_\nu}{\partial \bar{q}_\rho} \, \varphi_\rho^{\,\mu}( \bar{q})\,.
\label{eq:tetrad_composition2}
\end{equation}
This leads us to introduce a generalized composition law ($\bar{\oplus}$) for a curved spacetime such that 
\be
(\bar{p} \oplus \bar{q})_\mu \,=\, \bar{e}_\mu^\nu(x) (p \bar{\oplus} q)_\nu\,.
\label{eq:composition_cotangent}
\ee
Then, one has 

\begin{equation}
\begin{split}
e^\tau_\nu(x)\varphi^\mu_\tau(\bar{p} \oplus \bar{q}) \,=&\,e^\tau_\nu(x)\frac{\partial (\bar{p} \oplus \bar{q})_\tau}{\partial \bar{q}_\sigma}\varphi_\sigma^{\,\mu}(\bar{q})\,=\,e^\tau_\nu(x) \bar{e}^\lambda_\tau(x) \,\frac{\partial (p \bar{\oplus} q)_\lambda}{\partial \bar{q}_\sigma}\varphi_\sigma^{\,\mu}(\bar{q})\\
=&\,\frac{\partial (p \bar{\oplus} q)_\nu}{\partial\bar{q}_\sigma} \varphi_\sigma^{\,\mu}(\bar{q})\,=\, 
\frac{\partial (p \bar{\oplus} q)_\nu}{\partial q_\rho}\frac{\partial q_\rho}{\partial\bar{q}_\sigma} \varphi_\sigma^{\,\mu}(\bar{q})\,=\,
\frac{\partial (p \bar{\oplus} q)_\nu}{\partial q_\rho} e_\rho^\sigma(x) 
\varphi_\sigma^{\,\mu}(\bar{q})\,,
\end{split}
\end{equation}
i.e.
\begin{equation}
\Phi^\mu_\nu(x,(p \bar{\oplus} q)) \,=\,  \frac{\partial (p \bar{\oplus} q)_\nu}{\partial q_\rho} \, \Phi_\rho^{\,\mu}(x,q)\,.
\label{eq:tetrad_cotangent}
\end{equation}
We have obtained, for a fixed $x$, the momentum isometries of the metric leaving the form of the tetrad invariant in the same way we did in Ch.~\ref{chapter_curved_momentum_space}. 

As we saw in the same chapter, the translations defined in this way form by construction a group, so the composition law must be associative. We can now show that the barred composition law is also associative from the fact that  the composition law $\oplus$ is associative. If we define $\bar{r}=(\bar{k}\oplus \bar{q})$ and $\bar{l}=(\bar{p}\oplus \bar{k})$, then we have  $r=(k \bar{\oplus} q)$ and $l=(p \bar{\oplus} k)$. Hence
\begin{equation}
(\bar{p}\oplus \bar{r})_\mu\,=\,\bar{e}^\alpha_\mu(x)(p\bar{\oplus}r)_\alpha\,=\,\bar{e}^\alpha_\mu(x)(p\bar{\oplus}(k\bar{\oplus}q))_\alpha\,,
\end{equation}
and 
\begin{equation}
(\bar{l}\oplus \bar{q})_\mu\,=\,\bar{e}^\alpha_\mu(x)(l\bar{\oplus}q)_\alpha\,=\,\bar{e}^\alpha_\mu(x)((p\bar{\oplus}k)\bar{\oplus}q)_\alpha\,,
\end{equation}
but due to the associativity of $\oplus$
\begin{equation}
(\bar{p}\oplus \bar{r})_\mu\,=\,(\bar{l}\oplus \bar{q})_\mu\,,
\end{equation}
and then 
\begin{equation}
(p\bar{\oplus}(k\bar{\oplus}q))_\alpha\,=\,((p\bar{\oplus}k)\bar{\oplus}q)_\alpha\,,
\end{equation}
we conclude that $\bar{\oplus}$ is also associative. We have then shown that in the cotangent bundle with a constant scalar of curvature in momentum space, one can also define associative momentum translations. 

\subsubsection{Modified Lorentz transformations}
One can also replace $k$ by $\bar{k}_\mu=\bar{e}_\mu^\nu(x) k_\nu$ in  Eq.~\eqref{cal(J)} 
\be
\frac{\partial g^k_{\mu\nu}(k)}{\partial k_\rho} {\cal J}^{\beta\gamma}_\rho(k) \,=\,
\frac{\partial{\cal J}^{\beta\gamma}_\mu(k)}{\partial k_\rho} g^k_{\rho\nu}(k) +
\frac{\partial{\cal J}^{\beta\gamma}_\nu(k)}{\partial k_\rho} g^k_{\mu\rho}(k)\,,
\ee
obtaining 
\be
\frac{\partial g^{\bar{k}}_{\mu\nu}(\bar{k})}{\partial \bar{k}_\rho} {\cal J}^{\beta\gamma}_\rho(\bar{k}) \,=\,
\frac{\partial{\cal J}^{\beta\gamma}_\mu(\bar{k})}{\partial \bar{k}_\rho} g^{\bar{k}}_{\rho\nu}(\bar{k}) +
\frac{\partial{\cal J}^{\beta\gamma}_\nu(\bar{k})}{\partial \bar{k}_\rho} g^{\bar{k}}_{\mu\rho}(\bar{k})\,.
\ee
From here, we have
\begin{equation}
\frac{\partial g^{\bar{k}}_{\mu\nu}(\bar{k})}{\partial k_\sigma}e^\rho_\sigma(x) {\cal J}^{\alpha\beta}_\rho(\bar{k}) \,=\,
\frac{\partial{\cal J}^{\alpha\beta}_\mu(\bar{k})}{\partial k_\sigma}e^\rho_\sigma(x) g^{\bar{k}}_{\rho\nu}(\bar{k}) +
\frac{\partial{\cal J}^{\alpha\beta}_\nu(\bar{k})}{\partial k_\sigma}e^\rho_\sigma(x) g^{\bar{k}}_{\mu\rho}(\bar{k})\,.
\end{equation}
Multiplying by $e_\lambda^\mu(x)e_\tau^\nu(x)$ both sides of the previous equation one finds 
\begin{equation}
\frac{\partial g_{\lambda \tau}(x,k)}{\partial k_\rho} \bar{{\cal J}}^{\alpha\beta}_\rho(x,k) \,=\,
\frac{\partial\bar{{\cal J}}^{\alpha\beta}_\lambda(x,k)}{\partial k_\rho} g_{\rho\tau}(x,k) +
\frac{\partial\bar{{\cal J}}^{\alpha\beta}_\tau (x,k)}{\partial k_\rho}g_{\lambda\rho}(x,k)\,,
\end{equation}
where we have defined
\begin{equation}
 \bar{{\cal J}}^{\alpha\beta}_\mu(x,k) \,=\,e^\mu_\nu(x){\cal J}^{\alpha\beta}_\nu(\bar{k})\,.
\end{equation}
We see that $ \bar{{\cal J}}^{\alpha\beta}_\mu(x,k)$ are the new isometries of the momentum metric that leave the momentum origin invariant for a fixed point $x$.
\subsubsection{Modified dispersion relation}

With our prescription, the generalization to Eq.~\eqref{eq:casimir_J} 
\be
\frac{\partial C(k)}{\partial k_\mu} \,{\cal J}^{\alpha\beta}_\mu(k) \,=\, 0\, ,
\ee
in presence of a curved spacetime is 
 \begin{equation}
\frac{\partial C(\bar{k})}{\partial \bar{k}_\mu}{\cal J}^{\alpha\beta}_\mu(\bar{k})\,=\,0\,.
\end{equation}
The generalized infinitesimal Lorentz transformation in curved spacetime, defined by $\bar{{\cal J}}^{\alpha\beta}_\lambda(x,k)$, when acting on $C(\bar{k})$ is
\be
\begin{split}
\delta C(\bar{k}) \,=&\, \omega_{\alpha\beta} \frac{\partial C(\bar{k})}{\partial k_\lambda}\,\bar{{\cal J}}^{\alpha\beta}_\lambda(x,k) \,=\, \omega_{\alpha\beta} \frac{\partial C(\bar{k})}{\partial \bar{k}_\rho}\,\frac{\partial\bar{k}_\rho}{\partial k_\lambda}\,\bar{{\cal J}}^{\alpha\beta}_\lambda(x,k) \\
=&\,\omega_{\alpha\beta} \frac{\partial C(\bar{k})}{\partial \bar{k}_\rho}\,\bar{e}^\lambda_\rho(x)\,\bar{{\cal J}}^{\alpha\beta}_\lambda(x,k) \,=\, 
\omega_{\alpha\beta} \frac{\partial C(\bar{k})}{\partial \bar{k}_\rho}\,{\cal J}^{\alpha\beta}_\rho(\bar{k}) \,=\, 0.
\end{split}
\ee

At the beginning of this section we have proposed to take into account the space-time curvature in an action with a deformed  Casimir considering the substitution $k\rightarrow \bar{k}=\bar{e}k$ in the DDR, as this works for the transition from SR to GR. We have just seen that if $C(k)$ can be viewed as the distance from the origin to a point $k$ of the momentum metric $g^k_{\mu\nu} (k)$,  $C(\bar{k})$ is the distance from $(x,0)$ to $(x,k)$ of the momentum metric $g_{\mu\nu} (x,k)$, since the new DLT leave the new DDR invariant, so it can be considered as (a function of) the squared distance calculated with the new metric. This is in accordance with our initial assumption of taking $C(\bar{k})$  as the DDR in presence of a curved spacetime. We will study deeper the relationship between the action with $C(\bar{k})$ and this metric in Sec.~\ref{subsec_action_metric}.

\subsection{Main properties of the geometry in the cotangent bundle}
\label{sec:fb_properties}
We have proposed a way to generalize the momentum metric studied in Ch.~\ref{chapter_curved_momentum_space} including a nontrivial curvature in spacetime. This metric can be considered as a metric in the whole cotangent bundle (also for the particular case of flat spacetime) following the formalism presented in Ch.~4 of~\cite{2012arXiv1203.4101M}. In this subsection we summarize the main ingredients of this prescription that we will use along this chapter. 

We denote by $H^\rho_{\mu\nu}$ the space-time affine connection of the metric defined asking that the covariant derivative of the metric vanishes 
\begin{equation}
g_{\mu\nu;\rho}(x,k)\,=\,  \frac{\delta g_{\mu\nu}(x,k)}{\delta x^\rho}-g_{\sigma\nu}(x,k)H^\sigma_{\rho\mu}(x,k)-g_{\sigma\mu}(x,k)H^\sigma_{\rho\nu}(x,k)\,=\,0\,,
\label{eq:cov_der}
\end{equation} 
where we use a new derivative 
\begin{equation}
\frac{\delta}{\delta x^\mu}\, \doteq \,\frac{\partial}{\partial x^\mu}+N_{\rho\mu}(x,k)\frac{ \partial}{\partial k_\rho}\,,
\label{eq:delta_derivative}
\end{equation}   
and  $N_{\mu\nu}(x,k)$ are the coefficients of the \textit{nonlinear connection} $N$, which is known as horizontal distribution. The cotangent bundle manifold can be decomposed into vertical and  horizontal distributions, generated by  $\partial/\partial k_\mu$ and  $\delta/\delta x^\mu$ respectively, being the last one constructed from the horizontal distribution to be the 
supplementary to the vertical distribution (the fiber). In the GR case, the \textit{nonlinear connection} coefficients are given by
\begin{equation}
N_{\mu\nu}(x,k)\, = \, k_\rho H^\rho_{\mu\nu}(x)\,.
\label{eq:nonlinear_connection}
\end{equation}

As in GR, the relation between the metric and the affine connection
\begin{equation}
H^\rho_{\mu\nu}(x,k)\,=\,\frac{1}{2}g^{\rho\sigma}(x,k)\left(\frac{\delta g_{\sigma\nu}(x,k)}{\delta x^\mu} +\frac{\delta g_{\sigma\mu}(x,k)}{\delta  x^\nu} -\frac{\delta g_{\mu\nu}(x,k)}{\delta x^\sigma} \right)\,,
\label{eq:affine_connection_st}
\end{equation} 
is still satisfied. The \textit{d-curvature tensor} is defined as
\begin{equation}
R_{\mu\nu\rho}(x,k)\,=\,\frac{\delta N_{\nu\mu}(x,k)}{\delta x^\rho}-\frac{\delta N_{\rho\mu}(x,k)}{\delta x^\nu}\,.
\label{eq:dtensor}
\end{equation} 
This tensor represents the curvature of the phase space, measuring the integrability of spacetime as a subspace of the cotangent bundle and can be defined as the commutator between the horizontal vector fields
\begin{equation}
\left\{ \frac{\delta}{\delta x^\mu}\,,\frac{\delta}{\delta x^\nu}\right\}\,=\, R_{\mu\nu\rho}(x,k)\frac{\partial}{\partial k_\rho}\,.
\end{equation} 
Also, it is easy to see that
\begin{equation}
R_{\mu\nu\rho}(x,k)\,=\,k_\sigma R^{*\sigma}_{\mu\nu\rho}(x,k)\,,
\end{equation} 
where 
\begin{equation}
R^{*\sigma}_{\mu\nu\rho}(x,k)\,=\,\left(\frac{\delta H_{\mu\nu}^\sigma(x,k)}{\delta x^\rho} -\frac{\delta H_{\mu \rho}^\sigma(x,k)}{\delta x^\nu} +H^\sigma_{\lambda\rho}(x,k)H_{\mu\nu}^\lambda(x,k)-H^\sigma_{\lambda\nu}(x,k)H_{\mu\rho}^\lambda(x,k)\right)\,.
\end{equation} 
In the GR case,  $R_{\mu\nu\rho}(x,k)=k_\sigma R^{\sigma}_{\mu\nu\rho}(x)$, being $R^{\sigma}_{\mu\nu\rho}(x)$ the Riemann tensor. The horizontal bundle would be integrable if and only if $R_{\mu\nu\rho}=0$ (see Refs.~\cite{2012arXiv1203.4101M}-\cite{Barcaroli:2015xda} for more details).

The  momentum affine connection is defined as  
\begin{equation}
C_\rho^{\mu\nu}(x,k)\,=\,\frac{1}{2}g_{\rho\sigma}\left(\frac{\partial g^{\sigma\nu}(x,k)}{\partial k_ \mu}+\frac{\partial g^{\sigma\mu}(x,k)}{\partial k_ \nu}-\frac{\partial g^{\mu \nu}(x,k)}{\partial k_ \sigma}\right)\,,
\label{eq:affine_connection_p}
\end{equation} 
and then, the following covariant derivative can be defined
\begin{equation}
v_{\nu}^{\,;\mu}\,=\,  \frac{\partial v_\nu}{\partial k_\mu}-v_\rho C^{\rho\mu}_\nu(x,k)\,.
\label{eq:k_covariant_derivative}
\end{equation}
The space-time curvature tensor is 
\begin{equation}
R^{\sigma}_{\mu\nu\rho}(x,k)\,=\,R^{*\sigma}_{\mu\nu\rho}(x,k)+C^{\sigma\lambda}_\mu (x,k)R_{\lambda\nu\rho}(x,k)\,,
\label{eq:Riemann_st}
\end{equation} 
and the corresponding one in momentum space is 
\begin{equation}
S_{\sigma}^{\mu\nu\rho}(x,k)\,=\, \frac{\partial C^{\mu\nu}_\sigma(x,k)}{\partial k_\rho}-\frac{\partial C^{\mu\rho}_\sigma(x,k)}{\partial k_\nu}+C_\sigma^{\lambda\nu}(x,k)C^{\mu\rho}_\lambda(x,k)-C_\sigma^{\lambda\rho}(x,k)C^{\mu\nu}_\lambda(x,k)\,.
\label{eq:Riemann_p}
\end{equation} 

The line element in the cotangent bundle is defined as  
\begin{equation}
\mathcal{G}\,=\, g_{\mu\nu}(x,k) dx^\mu dx^\nu+g^{\mu\nu}(x,k) \delta k_\mu \delta k_\nu\,, 
\end{equation}
where 
\begin{equation}
\delta k_\mu \,=\, d k_\mu - N_{\nu\mu}(x,k)\,dx^\nu\,. 
\end{equation}
Then, a vertical path is defined as a curve in the cotangent bundle with a fixed space-time point and the momentum satisfying the geodesic equation characterized by the affine connection of momentum space,
\begin{equation}
x^\mu\left(\tau\right)\,=\,x^\mu_0\,,\qquad \frac{d^2k_\mu}{d\tau^2}+C_\mu^{\nu\sigma}(x,k)\frac{dk_\nu}{d\tau}\frac{dk_\sigma}{d\tau}\,=\,0\,,
\end{equation} 
while for an horizontal curve one has
\begin{equation}
\frac{d^2x^\mu}{d\tau^2}+H^\mu_{\nu\sigma}(x,k)\frac{dx^\nu}{d\tau}\frac{dx^\sigma}{d\tau}\,=\,0\,,\qquad \frac{\delta k_\lambda}{\delta \tau}\,=\,\frac{dk_\lambda}{d\tau}-N_{\sigma\lambda} (x,k)\frac{dx^\sigma}{d\tau}\,=\,0\,.
\label{eq:horizontal_geodesics}
\end{equation} 
These equations are a generalization of the ones appearing in GR, obtaining them in the limit where the momentum affine connection vanishes and there is no momentum dependence in the space-time affine connection.

\subsection{Modified Killing equation} 
Here we derive the deformed  Killing equation for a generic metric in the cotangent bundle. The variation of the space-time coordinates $x^\alpha$ along a vector field $\chi^\alpha$ is 
\begin{equation}
\left(x'\right)^\alpha\,=\,x^\alpha+\chi^\alpha \Delta\lambda\,,
\label{eq:x_variation}
\end{equation}
where $\Delta\lambda$ is the infinitesimal parameter characterizing the variation. This variation of $x^\alpha$ provokes a variation on $k_\alpha$ given by
\begin{equation}
\left(k'\right)_\alpha\,=\,k_\beta \frac{\partial x^\beta}{\partial x^{\prime \alpha}}\,=\,k_\alpha-\frac{\partial\chi^\beta}{\partial x^\alpha}k_\beta \Delta\lambda\,,
\end{equation}
since $k$ transforms as a covector. The variation for a generic vector field depending on the phase-space variables $X^\alpha\left(x,k\right)$ is 
\begin{equation}
\Delta X^\alpha\,=\,\frac{\partial X^\alpha}{\partial x^\beta} \Delta x^\beta+\frac{\partial X^\alpha}{\partial k_\beta} \Delta k_\beta\,=\,\frac{\partial X^\alpha}{\partial x^\beta}\chi^\beta \Delta\lambda-\frac{\partial X^\alpha}{\partial k_\beta}\frac{\partial\chi^\gamma}{\partial x^\beta}k_\gamma \,\Delta\lambda \,.
\label{eq:vector_variation}
\end{equation}
We obtain the Killing equation imposing the invariance of the line element along a vector field   $\chi^\alpha$
\begin{equation}
\Delta\left(ds^2\right)\,=\, \Delta (g_{\mu\nu}dx^\mu dx^\nu)\,=\, \Delta(g_{\mu\nu})dx^\mu dx^\nu+g_{\mu\nu} \Delta(dx^\mu) dx^\nu +g_{\mu\nu} \Delta(dx^\nu) dx^\mu\,=\,0\,.
\label{eq:line_variation}
\end{equation}
From Eq.~\eqref{eq:vector_variation} we obtain the variation of the metric tensor
\begin{equation}
\Delta(g_{\mu\nu})\,=\,\frac{\partial g_ {\mu\nu}}{\partial x^\alpha} \chi^\alpha \Delta\lambda -\frac{\partial g_ {\mu\nu}}{\partial k_\alpha}\frac{\partial \chi^\gamma}{\partial x^\alpha}k_\gamma \,\Delta\lambda\,,
\end{equation}
while from  Eq.~\eqref{eq:x_variation}
\begin{equation}
\Delta(dx^\alpha)\,=\,d(\Delta x^\alpha)\,=\,d(\chi^\alpha \Delta\lambda)\,=\,\frac{\partial\chi^\alpha}{\partial x^\beta}dx^\beta\Delta\lambda\,,
\end{equation}
and then, Eq.~\eqref{eq:line_variation} becomes 
\begin{equation}
\begin{split}
\Delta\left(ds^2\right)\,=\,\left(\frac{\partial g_ {\mu\nu}}{\partial x^\alpha} \chi^\alpha  -\frac{\partial g_ {\mu\nu}}{\partial k_\alpha}\frac{\partial \chi^\gamma}{\partial x^\alpha}k_\gamma \right) dx^\mu dx^\nu \Delta\lambda \\+ g_{\mu\nu}\left(\frac{\partial\chi^\mu}{\partial x^\beta}dx^\beta dx^\nu+\frac{\partial\chi^\nu}{\partial x^\beta}dx^\beta dx^\mu\right)\Delta\lambda\,,
\end{split}
\end{equation}
so the Killing equation is
\begin{equation}
\frac{\partial g_ {\mu\nu}}{\partial x^\alpha} \chi^\alpha  -\frac{\partial g_ {\mu\nu}}{\partial k_\alpha}\frac{\partial \chi^\gamma}{\partial x^\alpha}k_\gamma + g_{\alpha\nu}\frac{\partial\chi^\alpha}{\partial x^\mu}+ g_{\alpha\mu}\frac{\partial\chi^\alpha}{\partial x^\nu}\,=\,0\,,
\label{eq:killing}
\end{equation}
which is the same result obtained in Ref.~\cite{Barcaroli:2015xda}. 

\subsection{Relationship between metric and action formalisms}
\label{subsec_action_metric}
In this subsection we will start by seeing the relationship between the metric and the distance from the origin to a point $k$. We will study this relation for the momentum metric, but it can be done for GR for space-time coordinates instead of momentum variables. After that, we will prove that there is a direct relation between the free action of a particle with a DDR and the line element of a momentum dependent metric for spacetime. 

In~\cite{Bhattacharya2012RelationshipBG} it is showed that the following relation holds for the distance of a Riemannian manifold
 \begin{equation}
\frac{\partial D(0,k)}{\partial k_\mu}\,=\,\frac{k_\nu g_k^{\mu\nu}(k)}{\sqrt{k_\rho g_k^{\rho\sigma}(k) k_\sigma}}\,,
\end{equation}
where $D(0,k)$ is the distance from a fixed point $0$ to $k$. This leads to 
 \begin{equation}
\frac{\partial D(0,k)}{\partial k_ \mu}g^k_{\mu\nu}(k) \frac{\partial D(0,k)}{\partial k_ \nu}\,=\,1\,.
\end{equation}
Moreover, this property is also checked in Ch.~3 of~\cite{Petersen2006} for the Minkowski space (inside the light cone and extended on the light cone by continuity) and so, by the Whitney embedding theorem~\cite{Burns1985}, valid for any pseudo-Riemannian manifold of dimension $n$, since they can be embedded in a Minkowski space of at most dimension $2n+1$. Through this property, we can establish a direct relationship between the metric and the Casimir defined as the square of the distance
 \begin{equation}
\frac{\partial C(k)}{\partial k_ \mu}g^k_{\mu\nu}(k) \frac{\partial C(k)}{\partial k_ \nu}\,=\,4 C(k)\,.
\label{eq:casimir_definition}
\end{equation}

On the other hand, from the action with a generic DDR
\begin{equation}
S\,=\,\int{\left(\dot{x}^\mu k_\mu-\mathcal{N} \left(C(k)-m^2\right)\right)d\tau}\,,
\label{eq:DSR_action}
\end{equation}
 one can find 
\begin{equation}
\dot{x}^\mu\,=\,\mathcal{N}\frac{\partial C(k)}{\partial k_\mu}\,,
\label{eq:velocity_action}
\end{equation}
being $\mathcal{N}=1/2m$ or  $1$ when the geodesic is timelike or null respectively.

As we have seen in the previous subsection, the momentum metric can be considered as a metric for the whole cotangent bundle, so one can take the following line element in spacetime for an horizontal curve 
\begin{equation}
ds^2\,=\,  dx^\mu g^k_{\mu\nu}(k) dx^\nu\,.
\end{equation}
For the timelike case, one can choose the parameter of the curve to be the natural parameter $s$, and then 
\begin{equation}
1\,=\, \dot{x}^\mu g^k_{\mu\nu}(k) \dot{x}^\nu\,.
\end{equation}
Substituting Eq.~\eqref{eq:velocity_action} in the previous equation one finds
\begin{equation}
\left. \frac{1}{4 m^2} \frac{\partial C(k)}{\partial k_\mu} g^k_{\mu\nu}(k) \frac{\partial C(k)}{\partial k_\nu}\right\rvert_{C(k)=m^2}\,=\, \frac{1}{4 m^2} 4 m^2\,=\,1 \,,
\end{equation}
where Eq.~\eqref{eq:casimir_definition} have been used. For a null geodesic one has 
\begin{equation}
0\,=\, \dot{x}^\mu g^k_{\mu\nu}(k) \dot{x}^\nu\,,
\end{equation}
and therefore, using Eq.~\eqref{eq:velocity_action} one finds
\begin{equation}
\left. \frac{\partial C(k)}{\partial k_\mu} g^k_{\mu\nu}(k) \frac{\partial C(k)}{\partial k_\nu}\right\rvert_{C(k)=0}\,=\,0 \,,
\end{equation}
where again Eq.~\eqref{eq:casimir_definition} was used in the last step. One can see that considering an action with a DDR and the line element of spacetime with a momentum dependent metric whose squared distance is the DDR gives the same results\footnote{One arrives also to the same equations if a function of the squared distance is considered as the Casimir (for timelike geodesics one would have to redefine the mass with the same function).}. 

One can also arrive to the same relation between these two formalisms for the generalization proposed of the cotangent bundle metric. The relation Eq.~\eqref{eq:casimir_definition} is generalized to
 \begin{equation}
\frac{\partial C(\bar{k})}{\partial \bar{k}_ \mu}g^{\bar{k}}_{\mu\nu}(\bar{k}) \frac{\partial C(\bar{k})}{\partial \bar{k}_ \nu}\,=\,4 C(\bar{k})\,=\,\frac{\partial C(\bar{k})}{\partial k_ \mu}g_{\mu\nu}(x,k) \frac{\partial C(\bar{k})}{\partial k_\nu}\,.
\label{eq:casimir_definition_cst}
\end{equation}
From the action 
\begin{equation}
S\,=\,\int{\dot{x}^\mu k_\mu-\mathcal{N} \left(C(\bar{k})-m^2\right)}
\label{eq:DGR_action}
\end{equation}
with the same Casimir function but with argument the barred momenta, one can find 
\begin{equation}
\dot{x}^\mu\,=\,\mathcal{N}\frac{\partial C(\bar{k})}{\partial k_\mu}\,,
\label{eq:velocity_action_curved}
\end{equation}
where again $\mathcal{N}=1/2m$ or $1$ when the geodesic is timelike or null respectively. We see that the same relation found for the flat space-time case holds also for curved spacetime. 

\subsection{Velocity in physical coordinates}
\label{sec:velocity_curved_physical}
From the previous subsection, we know that the photon trajectory is given by
\begin{equation}
ds^2\,=\, dx^\mu g^k_{\mu\nu}(k) dx^\nu \,=\,  dx^\mu \varphi_\mu^\alpha(k) \eta_{\alpha \beta}\varphi_\nu^\beta(k) dx^\nu\,=\,0 \,.
\end{equation}

Then, since  $\dot{k}=0$ along the trajectory, we have
\begin{equation}
 d\tilde{x}^\alpha \eta_{\alpha \beta} d\tilde{x}^\beta\,=\,0 \,,
\end{equation}
with
\begin{equation}
\tilde{x}^\alpha\,=\,x^\mu \varphi^\alpha_\mu (k)\,,
\end{equation}
which are the physical coordinates found in Ch.~\ref{chapter_locality}.  Now we can understand the result showed in Ch.~\ref{chapter_time_delay} of absence of a momentum dependence on times of flight for massless particles.

\section{Friedmann-Robertson-Walker metric}
\label{sec:rw}

Once we have defined our proposal of considering a nontrivial geometry for momentum and space-time coordinates, we can study its phenomenological implications. We will start by looking for possible effects on the Friedmann-Robertson-Walker universe. First of all, we will compute the velocity and the time dependence of momenta for photons both from the action of Eq.~\eqref{eq:GR_action} and through the line element of spacetime, checking that the same results are obtained. After that, we will study some phenomenological results in the Friedmann-Robertson-Walker universe. 

In order to construct the metric in the cotangent bundle, we will use the tetrad for momentum space considered in Ch.~\ref{chapter_curved_momentum_space}
\begin{equation}
\varphi^0_0(k)\,=\,1\,,\qquad \varphi^0_i(k)\,=\,\varphi^i_0(k)\,=\,0\,,\qquad \varphi^i_j(k)\,=\,\delta^i_j e^{-k_0/\Lambda}\,,
\label{eq:RW_tetrad_p}
\end{equation}
and the space-time tetrad
\begin{equation}
e^0_0(x)\,=\,1\,,\qquad e^0_i(x)\,=\,e^i_0(x)\,=\,0\,,\qquad e^i_j(x)\,=\,\delta^i_j R(x^0)\,,
\label{eq:RW_tetrad_st}
\end{equation}
where $R(x^0)$ is the scale factor. With these ingredients, we can construct the cotangent bundle metric 
\begin{equation}
g_{00}(x,k)\,=\,1\,,\qquad g_{0i}(x,k)\,=\,0\,, \qquad  g_{ij}(x,k)\,=\,\eta_{ij}\, R^2(x^0) e^{-2k_0/\Lambda}\,.
\label{eq:RW_metric}
\end{equation}

One can easily check from Eq.~\eqref{eq:Riemann_p} that the scalar of curvature in momentum space is constant, $S=12/\Lambda^2$, and that the  momentum curvature tensor corresponds to a maximally symmetric space
\begin{equation}
S_{\rho\sigma\mu\nu}\,\propto \, g_{\rho\mu}g_{\sigma\nu}-g_{\rho\nu}g_{\sigma\mu}\,,
\end{equation}
which is obvious from the result of Appendix~\ref{appendix:cotangent}.

\subsection{Velocities for photons}
We will compute the velocity of photons from the action
\begin{equation}
S\,=\,\int{\left(\dot{x}^\mu k_\mu-\mathcal{N} C(\bar{k})\right) }d\tau
\label{eq:action1}
\end{equation} 
with the deformed Casimir of the bicrossproduct basis depending of $x$ and $k$
\begin{equation}
C(\bar{k})\,=\,\Lambda^2\left(e^{\bar{k}_0/\Lambda}+e^{-\bar{k}_0/\Lambda}-2\right)- \vec{\bar{k}}^2e^{\bar{k}_0/\Lambda}\,=\,\Lambda^2\left(e^{k_0/\Lambda}+e^{-k_0/\Lambda}-2\right)-\frac{ \vec{k}^2e^{k_0/\Lambda}}{R^2(x^0)} \,.
\label{eq:Casimir_RW}
\end{equation} 
Setting $\dot{x}^0=1$, i.e. using time as the proper time, we can obtain the value of $\mathcal{N}$ as a function of position and momenta and then, we can obtain the velocity for massless particles (in 1+1 dimensions) as 
\begin{equation}
v\,=\,\dot{x}^1\,=\,-\frac{4 \Lambda ^3 k_1 e^{2 k_0/\Lambda} \left(e^{k_0/\Lambda}-1\right) R(x^0)^2}
{\left(k_1^2 e^{2 k_0/\Lambda}-\Lambda ^2 e^{2k_0/\Lambda} R(x^0)^2+\Lambda ^2
   R(x^0)^2\right)^2}\,.
\label{eq:velocity_RW_1}
\end{equation} 
When one uses the Casimir in order to obtain $k_1$ as a function of $k_0$, one finds
\begin{equation}
k_1\,= -\,\Lambda \, e^{-k_0/\Lambda}
   \left(e^{k_0/\Lambda }-1\right) R(x^0)\,,
\label{eq:RW_k}
\end{equation} 
and then, by substitution of Eq.~\eqref{eq:RW_k} in Eq.~\eqref{eq:velocity_RW_1}, one can see that the velocity is 
\begin{equation}
v\,=\, \frac{e^{k_0/\Lambda}}{R(x^0)}\,,
\label{eq:velocity_RW_casimir}
\end{equation} 
so we see an energy dependent velocity in these coordinates. When $\Lambda$ goes to infinity one gets $v=1/R(x^0)$, which is the GR result. 

This result can also be derived directly from the line element of the metric  
\begin{equation}
0\,=\, (dx^0)^2-R(x^0)e^{-2 k_0/\Lambda}(dx^1)^2\,,
\end{equation}
which agrees with the discussion of the previous subsection: the same velocity is obtained from the action and from the line element of the metric.  

Whether or not Eq.~\eqref{eq:velocity_RW_casimir} implies a time delay would require to consider the propagation in a generalization for a curved spacetime of the ``physical'' spacetime studied in Ch.~\ref{chapter_locality}.

\subsection{Momenta for photons}
We can obtain the momentum as a function of time looking for the extrema of the action~\eqref{eq:action1}
\begin{equation}
\dot{k}_0\,=\, - \frac{\Lambda\left(e^{k_0/\Lambda}-1\right)R'(x^0)}{R(x^0)}\,, \qquad  \dot{k}_1\,=\,0\,.
\label{eq:momenta_RW}
\end{equation}
Solving the differential equation, one obtains the energy as a function of time
\begin{equation}
k_0\,=\, -\Lambda \log\left(1+\frac{e^{-E/\Lambda}-1}{R(x^0)}\right) \,,
\label{eq:energy_RW}
\end{equation} 
where the constant of integration is the conserved energy along the geodesic, since when taking the limit $\Lambda$ going to infinity one has $E=k_0\, R(x^0)$, which is the barred energy.

\subsection{Redshift}
From the line element for photons we see that
\begin{equation}
0\,=\,(dx^0)^2-R^2(x^0)e^{-2k_0/\Lambda}d\vec{x}^2\,,
\end{equation} 
and then 
\begin{equation}
\int^{t_0}_{t_1}{\frac{dx^0\, e^{k_0/\Lambda}}{R(x^0)}}\,=\,\int^x_0 dx\,=\,x\,.
\label{eq:step}
\end{equation} 
We can now write Eq.~\eqref{eq:step} as a function of $x^0$ using Eq.~\eqref{eq:energy_RW}, obtaining  the quotient in frequencies (see Ch.~14 of Ref.~\cite{Weinberg:1972kfs})
\begin{equation}
\frac{\nu_0}{\nu_1}\,=\,\frac{\delta t_1}{\delta t_0}\,=\,\frac{R(t_1)\left(1+(e^{-E/\Lambda}-1)/R(t_1)\right)}{R(t_0)\left(1+(e^{-E/\Lambda}-1)/R(t_0)\right)}\,=\,\frac{R(t_1)+e^{-E/\Lambda}-1}{R(t_0)+e^{-E/\Lambda}-1}\,,
\end{equation} 
and then, the redshift is 
\begin{equation}
z\,=\,\frac{R(t_0)+e^{-E/\Lambda}-1}{R(t_1)+e^{-E/\Lambda}-1}-1\,.
\label{eq:redshift}
\end{equation} 
Taking the limit $\Lambda\rightarrow \infty$ in the redshift one recovers the usual expression of GR for a Friedmann-Robertson-Walker space~\cite{Weinberg:1972kfs}. This equation reveals an energy dependence of the redshift and then, two particles with different energies suffer different redshifts. To illustrate it,  let us suppose two particles emitted from a distant source with energies zero and $E$ such that $E\ll\Lambda$. The redshift for both particles at the detection point will be different for each one. Taking only the first term in the series expansion in $\Lambda$ we find
\begin{equation}
\frac{1+z(0)}{1+z(E)}\,=\, 1+\frac{E}{\Lambda}\left(\frac{1}{R(t_0)}-\frac{1}{R(t_1)}\right)\,.
\end{equation}
i.e. for the more energetic particle there is more redshift,
\begin{equation}
1+z(E)\,=\,(1+z(0))\left( 1-\frac{E}{\Lambda }\left(\frac{1}{R(t_0)}-\frac{1}{R(t_1)}\right)\right)\,,
\end{equation}
since the last factor is always greater than unity since, as the universe is expanding, $R(t_1)<R(t_0)$.

\subsection{Luminosity distance}
Following the procedure showed in Ch.~14 of Ref.~\cite{Weinberg:1972kfs}, we will obtain the luminosity distance for this metric. We start by considering a circular telescope mirror of radius $b$, with its center placed at the origin and its normal along the line of sight $x$ to the light source. The light rays that reach the limits of the mirror edge form a cone that, for a system of locally inertial coordinates at the source, and have a half-angle $|\epsilon|$ given by 
\begin{equation}
b\,\approx\, R(t_0) e^{-k_0/\Lambda} x |\epsilon|\,,
\end{equation}
where $b$ is expressed as a proper distance. The solid angle that encompass this cone is 
\begin{equation}
\pi |\epsilon|^2\,=\, \frac{\pi b^2}{R^2(t_0) e^{-2 k_0/\Lambda} x^2}\,,
\end{equation}
and then, the fraction of the photons that are emitted isotropically that arrives to the mirror is the ratio of this solid angle to $4\pi$, i.e. 
\begin{equation}
\frac{|\epsilon|^2}{4}\,=\, \frac{ A}{4 \pi R^2(t_0) e^{-2k_0/\Lambda}x^2}\,,
\label{eq:fract}
\end{equation}
where $A$ is the proper area of the mirror
\begin{equation}
A\,=\,\pi b^2\,.
\end{equation}
But if a photon is emitted with an energy $h \nu_1$, it will be red-shifted to an energy 
\begin{equation}
h \nu_1 \frac{R(t_1)+e^{-E/\Lambda}-1}{R(t_0)+e^{-E/\Lambda}-1}\,,
\end{equation}
and there will be a difference in the time of arrival for photons emitted at time intervals $\delta t_1$ given by
\begin{equation}
\delta t_1\frac{R(t_1)+e^{-E/\Lambda}-1}{R(t_0)+e^{-E/\Lambda}-1}\,,
\end{equation}
where $t_1$ is the time when  the photon is emitted from the source and $t_0$ is the time of arrival at the mirror. Then, the fraction of the total power from the source which is received by the mirror $P$, is given by the absolute luminosity $L$, times a factor 
\begin{equation}
\left(\frac{R(t_1)+e^{-E/\Lambda}-1}{R(t_0)+e^{-E/\Lambda}-1}\right)^2\,,
\end{equation}
multiplied by the fraction Eq.~\eqref{eq:fract}:
\begin{equation}
P\,=\,L\,A \frac{\left(R(t_1)+e^{-E/\Lambda}-1\right)^2}{4 \pi R^2(t_0)\left(R(t_0)+e^{-E/\Lambda}-1\right)^2 e^{-2 k_0/\Lambda}x^2}\,.
\end{equation}
The apparent luminosity $l$ is defined as the power per unit mirror area, so using Eq.~\eqref{eq:energy_RW} we obtain
\begin{equation}
l\,\equiv\,\frac{P}{A}\,=\, L \frac{\left(R(t_1)+e^{-E/\Lambda}-1\right)^2}{4 \pi \left(R(t_0)+e^{-E/\Lambda}-1\right)^4 x^2}\,.
\label{eq:apk_luminosity}
\end{equation}
The apparent luminosity of a source at rest placed at distance $d$, for an Euclidean space,  is given by $L/4\pi d^2$, so in general one may define the luminosity distance $d_L$ of a light source as 
\begin{equation}
d_L\,=\,\left(\frac{L}{4\pi l}\right)^{1/2}\,,
\end{equation}
and then Eq.~\eqref{eq:apk_luminosity} can be written
\begin{equation} 
d_L\,=\, \frac{\left(R(t_0)+e^{-E/\Lambda}-1\right)^2\,x}{R(t_1)+e^{-E/\Lambda}-1}\,.
\end{equation}
We can find from the previous equation that  
\begin{equation} 
d_L\,=\,\left( \frac{R(t_0)+e^{-E/\Lambda}-1}{R(t_1)+e^{-E/\Lambda}-1}\right)^2 d\,,
\end{equation}
where 
\begin{equation} 
d\,=\,\left(R(t_1)+e^{-E/\Lambda}-1\right)x\,,
\end{equation}
is the proper distance between the source and us. From here, we can write the luminosity distance from the redshift expression 
 \begin{equation} 
d_L\,=\,\left(1+z\right)^2 d\,,
\end{equation}
finding the same expression of GR. 
We can calculate as we did for the redshift, the difference on the luminosity distance for photons with different energies, obtaining 
\begin{equation}
\frac{d_L (0)}{d_L (E)}\,=\,\left(\frac{1+z(0)}{1+z(E)}\right)^2\,,
\end{equation}
so the luminosity distance will be greater for higher energies. This is an interesting result that perhaps could be tested in cosmographic analyses.

\subsection{Congruence of geodesics}
We study in this subsection the congruence of null geodesics for the metric of the cotangent bundle from the definition~\cite{Poisson:2009pwt}
\begin{equation}
\theta \,=\,\frac{1}{\delta S}\frac{d }{d \lambda}\delta S\,,
\end{equation}
where $\delta S$ is the infinitesimal change of area. For the metric~\eqref{eq:RW_metric} one obtains
\begin{equation}
\theta\,=\,2\frac{e^{k_0/\Lambda}R^\prime(t)}{R^2(t)}\,,
\label{eq:theta_RW}
\end{equation}
Making a series expansion in $1/\Lambda$ we get
\begin{equation}
\frac{\theta(0)}{\theta(E)} \,=\, 1-\frac{E}{R(t) \Lambda }\,.
\end{equation}
The expansion of the congruence is energy dependent, in such a way that is greater for larger energies. 

Note that in Ch.~\ref{chapter_curved_momentum_space} we have mentioned that there are two possible choices of the sign of $\Lambda$ for the de Sitter metric~\eqref{bicross-metric} making that, for the other sign, all the previous results change: the speed, redshift, luminosity distance and congruence of geodesics of high energy photons would be smaller than the low energy ones.

\section{Schwarzschild metric}
\label{sec:sch}

In this section, we study the Schwarzschild black hole with a curvature in momentum space. We will use the tetrad corresponding to Lemaître coordinates~\cite{Landau:1982dva}
\begin{equation}
e^t_t\,=\,1\,,\qquad e^x_x\,=\, \sqrt{\frac{2M}{r}}\,,\qquad e^\theta_\theta(x)\,=\, r\,,\qquad e^\phi_\phi(x)\,=\, r \sin{\theta}\,,
\label{eq:Sch_tetrad}
\end{equation}
where 
\begin{equation}
r\,=\,\left(\frac{3}{2}\left(x-t\right)\right)^{(2/3)}\left(2M\right)^{(1/3)}\,.
\end{equation}
Using the same momentum tetrad of Sec.~\ref{sec:rw}, one obtains the metric in the cotangent bundle
\begin{equation}
\begin{split}
g_{tt}(x,k)\,&=\,1\,, \qquad  g_{xx}(x,k)\,=\,-\frac{2M}{r}e^{-2 k_0/\Lambda}\,, \\
g_{\theta\theta}(x,k)\,&=\, -r^2e^{-2 k_0/\Lambda}\,, \qquad  g_{\phi\phi}(x,k)\,=\,- r^2 \sin^2{\theta}e^{-2 k_0/\Lambda}\,.
\label{eq:Sch_metric}
\end{split}
\end{equation}
As for the Friedmann-Robertson-Walker case, one can check that the momentum scalar of curvature is constant, $S=12/\Lambda^2$, and that the momentum curvature tensor corresponds to a maximally symmetric momentum space. 

The purpose of this subsection is to study the event horizon for Schwarzschild black hole when there is a curvature in momentum space. In order to do so, we first compute the conserved energy along geodesics. After that, we will represent graphically the null geodesics, obtaining the same event horizon for every particle, independently of their energy. Besides this fact, we will find an energy dependent surface gravity, pointing to a possible dependence on the energy of the Hawking radiation. 

\subsection{Energy from Killing equation} 
Using Eq.~\eqref{eq:killing} for this metric one obtains 
\begin{equation}
\chi^0\,=\,1\,,\qquad \chi^1\,=\,1\,,
\end{equation}
which gives the same Killing vector obtained in GR~\footnote{This can be easily understood from Eq.~\eqref{eq:killing}. If in GR a constant Killing vector exists for a given space-time geometry, then the same vector will be a Killing one for the deformed  cotangent metric.}.

The same result can be obtained from the action Eq.~\eqref{eq:action1} with the Casimir  
\begin{equation}
C(\bar{k})\,=\,\Lambda^2\left(e^{\bar{k}_0/\Lambda}+e^{-\bar{k}_0/\Lambda}-2\right)- \vec{\bar{k}}^2e^{\bar{k}_0/\Lambda}\,=\,\Lambda^2\left(e^{k_0/\Lambda}+e^{-k_0/\Lambda}-2\right)- \vec{k}^2e^{k_0/\Lambda}\frac{r}{2M} \,.
\label{eq:Casimir_Sch}
\end{equation} 
Choosing $\tau=x^0$, $\mathcal{N}$ of Eq.~\eqref{eq:action1} can be expressed as a function of the phase-space variables, and then one can check that the derivatives of the momenta satisfy (in 1+1 dimensions)
\begin{equation}
\dot{k}_0+\dot{k}_1\,=\,0\,.
\end{equation} 
From the Casimir one can obtain the relation between the spatial and zero momentum component for massless particles
\begin{equation}
k_1\,=\,\sqrt{\frac{2M}{r}}\Lambda \left(1-e^{-k_0/\Lambda}\right)\,,
\end{equation} 
so the conserved energy is 
\begin{equation}
E\,=\,k_0+k_1\,=\,k_0+\sqrt{\frac{2M}{r}}\Lambda \left(1-e^{-k_0/\Lambda}\right)\,.
\label{eq:energy_Sch}
\end{equation} 

\subsection{Event horizon} 
We can obtain the event horizon from the representation of the null ingoing and outgoing geodesics. In GR the horizon in the Lemaître coordinates is in $x-t=4M/3$ and the singularity is at  $x=t$~\cite{Landau:1982dva}.  From the line element of the metric Eq.~\eqref{eq:Sch_metric}, one can solve the differential equation 
\begin{equation}
ds^2\,=\,0\,\implies\,\frac{dx}{dt}\,=\,\pm \left(\frac{3(x-t)}{4M}\right)^{(1/3)}e^{k_0/\Lambda}\,,
\label{eq:eq_geo}
\end{equation}
where $+$ stands for outgoing and $-$ for ingoing geodesics. Solving numerically this differential equation expressing $k_0$ as a function of the conserved energy (inverting Eq.~\eqref{eq:energy_Sch}), we can plot the geodesics  for different energies, observing that there is no modification in the horizon: all the particles see the same horizon independently of their energy\footnote{Different trajectories appear in the following plots because we are using different initial conditions for different energies.}. Despite of the momentum dependence of the metric, we find that there is a common horizon for all particles, even if the velocity is energy dependent. For $M=1$ we plot the ingoing geodesics in Fig.~\ref{fig:ingoing}. 
\begin{figure}[H]
  \includegraphics[width=\linewidth]{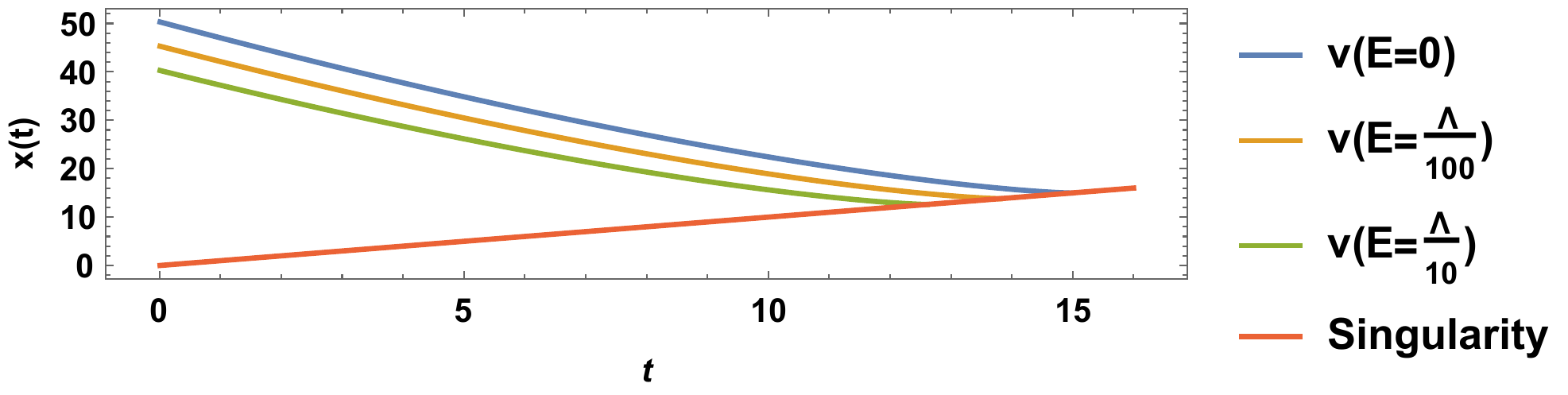}
  \caption{Particles with three different velocities coming from outside the horizon, crossing it and finally arriving to the singularity.}
  \label{fig:ingoing}
\end{figure}
Null particles emitted outside the horizon but near to it will escape in a finite time, see Fig.~\ref{fig:outgoing}.
\begin{figure}[H]
  \includegraphics[width=\linewidth]{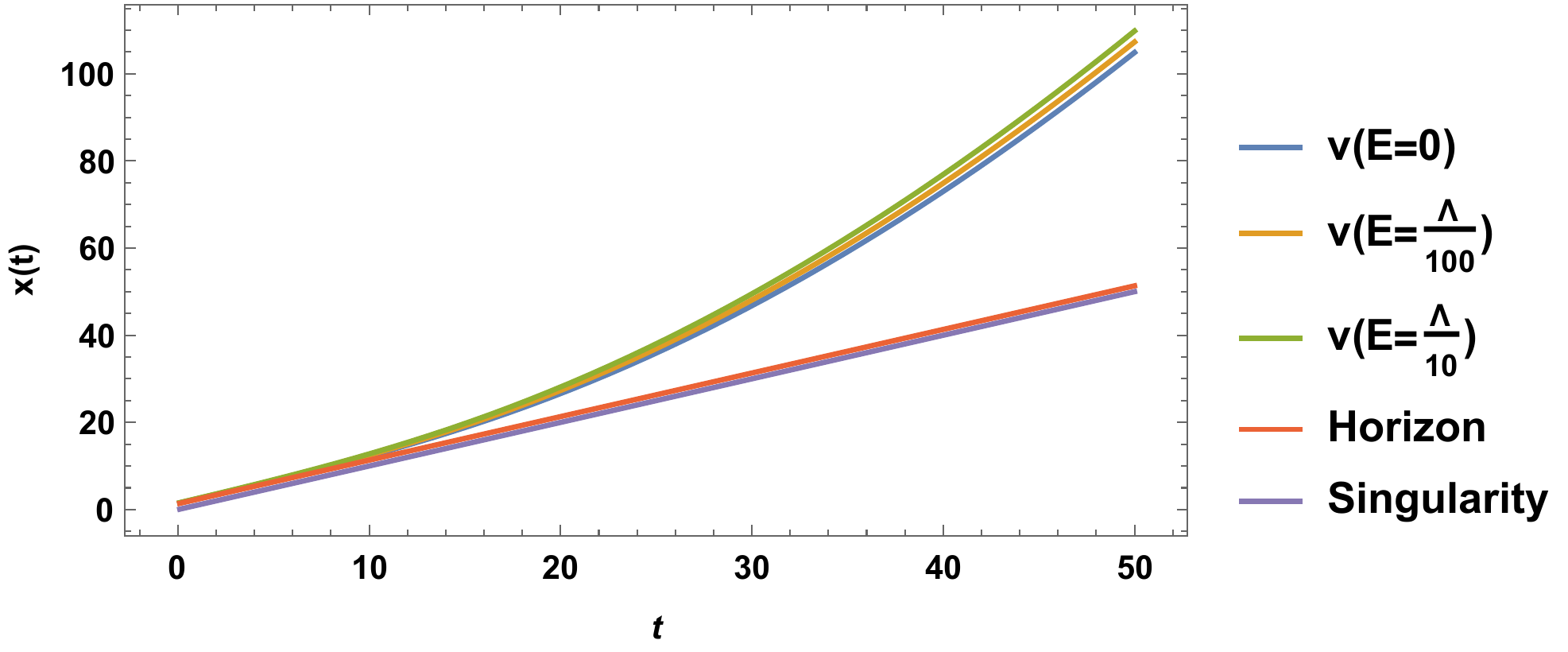}
  \caption{Outgoing null geodesics from outside the horizon.}
  \label{fig:outgoing}
\end{figure}
One can also represent the geodesics starting inside the horizon and falling to the singularity, as in Fig.~\ref{fig:outgoing_2}.
\begin{figure}[H]
  \includegraphics[width=\linewidth]{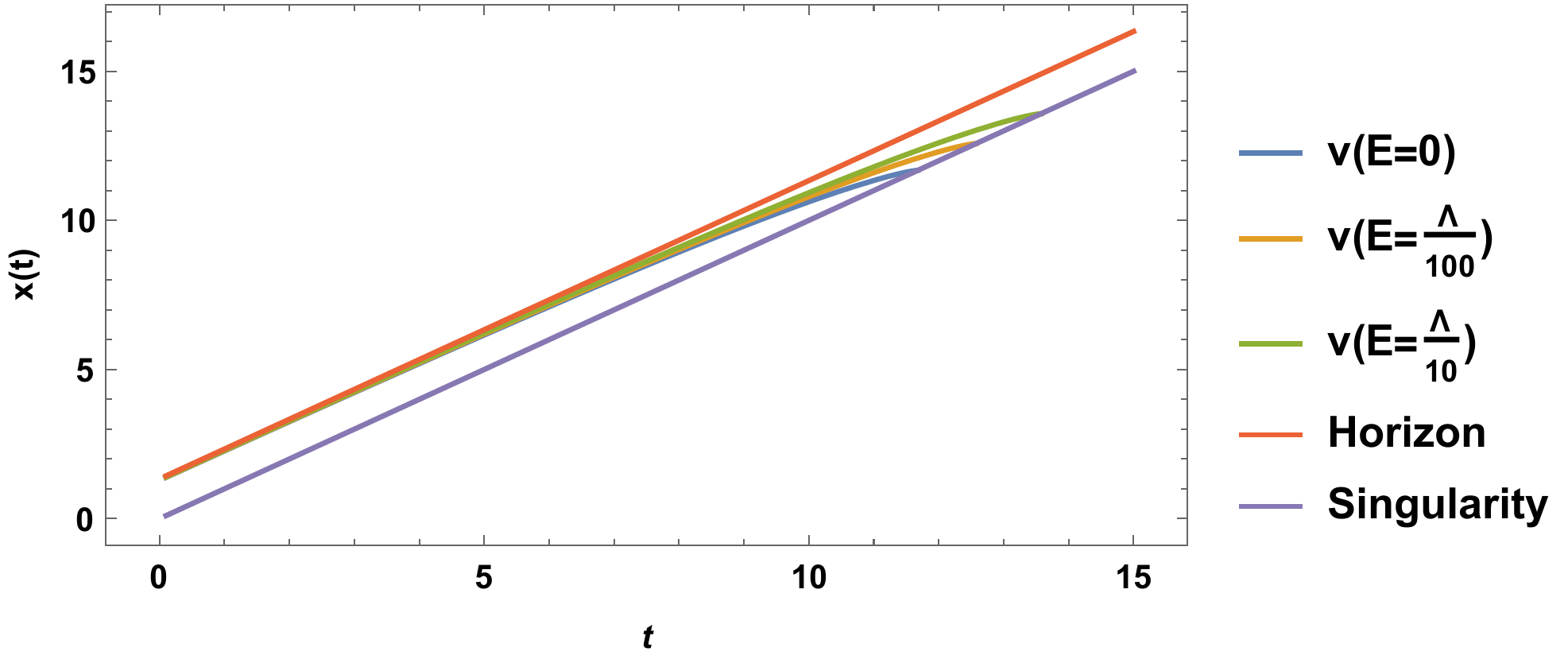}
  \caption{Null geodesics from inside the horizon falling at the singularity.}
  \label{fig:outgoing_2}
\end{figure}
In Refs.~\cite{Dubovsky:2006vk,Barausse:2011pu,Blas:2011ni,Bhattacharyya:2015gwa} it is shown that in a LIV scenario, particles with different energies see different horizons. It can be proved that due to this effect,  there is a violation of the second law of the black hole thermodynamics, leading to the possibility of construction of a perpetuum mobile (see however~\cite{Benkel:2018abt} for a possible resolution of this problem). With our prescription, we have found that there is a unique horizon for all particles, which is in agreement with the fact that in DSR theories there is a relativity principle, in contrast with the LIV scenarios.

\subsection{Surface gravity}
As it was pointed in~\cite{Cropp:2013zxi}, there are different procedures to obtain the surface gravity. One of them is related with the peeling off properties of null geodesics near the horizon
\begin{equation}
\frac{d |x_1(t)-x_2(t)| }{d t}\,\approx \,  \kappa_{\text{peeling}}(t) |x_1(t)-x_2(t)|\,,
\label{eq:surface_gravity}
\end{equation}
where $x_1(t)$ and $x_2(t)$ represent two null geodesics on the same side of the horizon and the absence of factors is due to $\kappa_{\text{peeling}}=\kappa_{\text{inaffinity}}$ in the GR limit.  From Eq.~\eqref{eq:eq_geo}, we obtain for two null geodesics with the same energy $k_0$:
\begin{equation}
\frac{d |x_1(t)-x_2(t)| }{d t}\,\approx \,\frac{e^{k_0/\Lambda}}{4M} |x_1(t)-x_2(t)|\,,
\end{equation}
and then, 
\begin{equation}
 \kappa_{\text{peeling}}\,=\,\frac{e^{k_0/\Lambda}}{4M}\,,
\end{equation}
with a dependence on the energy. This points to the possibility that the Hawking temperature~\cite{Poisson:2009pwt}
\begin{equation}
T\,=\,\frac{\kappa}{2 \pi}\,,
\end{equation}
could generally depend on the energy of the outgoing particles.
\chapter{Conclusions}

\ifpdf
    \graphicspath{{Chapter9/Figs/Raster/}{Chapter9/Figs/PDF/}{Chapter9/Figs/}}
\else
    \graphicspath{{Chapter9/Figs/Vector/}{Chapter9/Figs/}}
\fi
\epigraph{Cry in the dojo, laugh on the battlefield.}{Japanese proverb}

In this thesis we have studied a possible deviation of special relativity baptized as Doubly Special Relativity (DSR), in which a deformed  relativistic kinematics, parametrized by a high energy scale, appears. This theory is proposed, not as a fundamental quantum gravity theory, but as a low energy limit of it, trying to provide some phenomenological observations that could point towards the correct approach to such a theory.

Firstly, we have extended to second order in an expansion in powers of the inverse of a high energy scale, a previous study  at first order of a deformed kinematics compatible with the relativity principle (DRK). We have shown that the results can be obtained from a simple trick: a change of basis (modifying the Casimir and the Lorentz transformation in the one-particle system) and a change momentum variables in the two-particle system (changing the composition law and the Lorentz transformation in the two-particle system). The same method can be easily generalized to higher orders providing a way to obtain in a systematic way a DRK. But in doing so, one finds an enormous arbitrariness to go beyond special relativity,  so that an additional requirement, physical or mathematical, may be needed.   

In order to look for some additional ingredient to reduce this arbitrariness, we have considered two different perspectives. From a geometrical point of view, we have arrived to the conclusion that only a maximally symmetric momentum space could lead to a deformed  relativistic kinematics when one identifies the composition law and the Lorentz transformations as the isometries of the metric. Since one wants 4 translations and 6 Lorentz generators, the momentum space must have 10 isometries, leaving only place for a maximally symmetric space. We have found that the most common examples of deformed kinematics appearing in the literature ($\kappa$-Poincaré, Snyder and hybrid models) can be reproduced and understood from this geometrical perspective by choosing properly the algebra of the generators of translations.  

The previous study does not show the possible implications that a deformed kinematics provoke on spacetime. In previous works it is shown that, from an action in which the conservation of momentum is imposed through a deformed composition law, there is a non-locality of interactions for any observer not placed at the interaction point, making this effect larger when the observer sees the interaction farther and farther away. However, it is possible to find new noncommutative coordinates (we call them ``physical'') in which interactions are local. We have found different ways to impose locality and, in one of them, associativity in the composition law is required in order to have local interactions. This seems to select, among the previous kinematics obtained from a geometrical perspective, $\kappa$-Poincaré kinematics. Also, we have found a relation between the different perspectives, observing that the tetrad characterizing the momentum space curvature can be used to define the functions of momentum that determine the physical coordinates.

Once the spacetime consequences of a deformed kinematics are better understood, we have explored two of its phenomenological consequences. In DSR, the only current phenomenological observation due to a deformed kinematics is a possible time delay for massless particles with different energies. We have proposed three different models in the DSR context, showing that time delay is not necessarily a possible effect in this framework, depending on the model and on the kinematics (and basis) used. This removes the strong constraints on the high energy scale that parametrizes the kinematics of DSR, which could be many orders of magnitude below the Planck scale. 

Considering this possibility, we have analyzed a process in QFT when a deformed kinematics is present. Despite the lack of a dynamical theory, the computations showed here can shed some light on how the usual QFT should be modified. On the one hand, we have shown that in the presence of a covariant composition law, instead of one peak associated to a resonance, another peak could appear, correlated to the former, allowing us to determine not only the mass of the particle, but also the high energy scale. We have baptized this effect as \emph{twin peaks}. If this scale is  sufficiently small, this effect could be observed in a future high energy particle accelerator. 

Besides, we have considered a simple process, an electron-positron pair going to $Z$ boson and decaying in a muon-antimuon pair. In order to do so, we have introduced  simple prescriptions to take into account the effects of a (covariant) deformed composition law. We have shown that an scale of the order of some TeV could be compatible with the experimental data of such process.  

Finally, we have investigated how a curvature of spacetime could modify the kinematics. This is a crucial ingredient in the study of time delays, since the expansion of the universe must be taken into account for photons coming from astrophysical sources. In order to do so, we have studied, from a geometrical point of view, how to consider simultaneously a curvature in spacetime and in momentum space. This can be done in the so-called cotangent bundle geometry, taking into account a nontrivial geometry for all the phase space.  In this framework, we have analyzed the phenomenological consequences of a maximally symmetric momentum space combined with an expanding universe (Friedmann-Robertson-Walker) and a black hole (Schwarzschild) geometries for spacetime. In the first case, we have computed the velocity for massless particles, the redshift, the luminosity distance and the congruence of geodesics. Whether or not there is a time delay in this proposal is still an open question which deserves further study.  For the black hole space-time geometry, we have seen that there is a common event horizon for all particles, in contrast with the result of a Lorentz violating scenario, where particles with different energies would see different horizons. This is in agreement with the relativity principle imposed in DSR. However, the surface gravity is energy dependent, what seems to indicate that the Hawking temperature could also show such behavior.

\chapter*{Conclusiones}

En esta tesis hemos estudiado una posible desviación de la relatividad especial en la que aparece una cinemática relativista deformada parametrizada por una escala de alta energía, bautizada como Relatividad Doblemente Especial (DSR). Esta teoría se propone, no como una teoría fundamental de gravedad cuántica, sino como un límite a bajas energías de ella, intentando proporcionar observaciones experimentales que podrían señalarnos cuál es el enfoque correcto de tal teoría.  

Primero hemos extendido, a segundo orden en una expansión en potencias del inverso de una escala de alta energía, un estudio previo de una cinemática modificada compatible con el principio de la relatividad (DRK) a primer orden. Hemos mostrado que los resultados pueden obtenerse a partir de un simple truco: un cambio de variables (modificando el Casimir y las transformaciones de Lorentz en el sistema de una partícula) y un cambio de variables momento en el sistema de dos partículas (cambiando la ley de composición y las transformaciones de Lorentz en el sistema de dos partículas). El mismo método puede generalizarse fácilmente a órdenes superiores proporcionando una forma de conseguir de forma sistemática una DRK. Sin embargo, encontramos una enorme arbitrariedad para ir más allá de relatividad especial, de modo que puede ser necesario introducir un requerimiento adicional, físico o matemático.  

Para buscar algún ingrediente que reduzca esta arbitrariedad, hemos considerado dos perspectivas distintas. Desde un punto de vista geométrico, hemos llegado a la conclusión de que sólo un espacio de momentos maximalmente simétrico podría conducirnos a una DRK cuando uno identifica la ley de composición y las transformaciones de Lorentz como isometrías de la métrica. Como uno quiere 4 traslaciones y 6 generadores Lorentz, el espacio de momentos debe tener 10 isometrías, dejando solo sitio para un espacio de momentos maximalmente simétrico. Hemos encontrado que los ejemplos más comunes de cinemáticas deformadas que aparecen en la literatura ($\kappa$-Poincaré, Snyder y modelos híbridos) pueden reproducirse y entenderse a partir de esta perspectiva geométrica eligiendo apropiadamente el álgebra de los generadores de traslaciones.  

En el estudio previo no se ha mostrado las posibles implicaciones que una cinemática modificada provoca en el espacio-tiempo. En trabajos anteriores se ha mostrado que, a partir de una acción en la que la ley de conservación de momento está impuesta a través de una ley de composición, hay una no localidad de las interacciones para cualquier observador que no se encuentra en el punto de la interacción, haciéndose este efecto mayor cuando el observador ve la interacción más y más lejos. Sin embargo, es posible encontrar unas nuevas coordenadas no conmutativas de espacio-tiempo (que llamamos físicas) en las que las interacciones son locales. Hemos encontrado formas diferentes de imponer localidad y, en una de ellas, se requiere la asociatividad de la ley de composición para tener interacciones locales. Esto parece seleccionar de entre las cinemáticas obtenidas previamente desde una perspectiva geométrica el modelo de $\kappa$-Poincaré. Además, hemos encontrado una relación entre estos dos estudios, observado que la tétrada que caracteriza la curvatura del espacio de momentos puede usarse para definir las funciones de momento que determinan las coordenadas físicas. 

Una vez que se han entendido mejor las consecuencias sobre el espacio-tiempo debido a una cinemática deformada, hemos desarrollado dos estudios fenomenológicos. En DSR, la única observación fenomenológica actual debido a una cinemática deformada es un posible retraso en el tiempo de vuelo para partículas sin masa con diferentes energías. Hemos propuesto tres modelos distintos en el contexto de DSR, mostrando que el retraso de tiempo no es necesariamente un efecto posible en este marco, dependiendo del modelo y de la cinemática (y de la base) utilizados. Esto elimina las fuertes restricciones en la escala de alta energía que parametriza las cinemáticas de DSR, que podría estar muchos órdenes de magnitud por debajo de la escala de Planck.  

Considerando esta posibilidad, hemos analizado un proceso en QFT en presencia de una cinemática deformada. A pesar de la falta de una teoría dinámica, los cálculos mostrados aquí pueden arrojar algo de luz en cómo la QFT usual debería modificarse. Por otro lado, se ha mostrado que en presencia de una ley de composición covariante, en vez de tener un pico asociado a una resonancia, podría aparecer otro pico correlacionado con el primero, permitiéndonos determinar no solo la masa de la partícula, sino también la escala de alta energía. Hemos bautizado este efecto como \emph{twin peaks}. Si esta escala es lo suficientemente pequeña, este efecto podría ser observado en un futuro acelerador de partículas de altas energías.

Además, hemos estudiado un proceso simple, un par electrón-positrón yendo a un bosón $Z$ y desintegrándose en un par muón-antimuón. Para hacerlo, hemos implementado unas nuevas reglas de Feynman teniendo en cuenta una ley de composición covariante. Hemos mostrado que, en ambos casos, una escala del orden de algunos TeV podría ser compatible con los datos experimentales obtenidos de aceleradores de partículas para este proceso. 

Finalmente, hemos investigado cómo una curvatura del espacio-tiempo podría modificar la cinemática. Esto es un ingrediente crucial en el estudio de retraso de tiempos de vuelo, ya que la expansión del universo debe tenerse en cuenta para fotones que provienen de fuentes astrofísicas. Para hacerlo, hemos estudiado, desde un punto de vista geométrico, cómo considerar simultáneamente una curvatura en el espacio-tiempo y en el espacio de momentos. Esto puede hacerse en la conocida como geometría en el fibrado cotangente, teniendo en cuenta una geometría no trivial para todo el espacio de fases. En este marco, hemos analizado las consecuencias fenomenológicas de un espacio de momentos maximalmente simétrico combinado con la geometría para el espacio-tiempo de un universo en expansión (Friedmann-Robertson-Walker) y de un agujero negro (Schwarzschild). En el primer caso, hemos calculado la velocidad para partículas sin masa, el corrimiento al rojo, la distancia lumínica y la congruencia de geodésicas. Si podría o no haber un retraso en el tiempo de vuelo en esta propuesta es todavía una pregunta abierta, que merece un trabajo futuro. Para el agujero negro, hemos visto que hay un horizonte de sucesos común para todas las partículas, lo que difiere del resultado para el caso de violación de invariancia Lorentz, donde las partículas con distintas energías verían distintos horizontes. Esto está de acuerdo con el principio de la relatividad impuesto en DSR. Sin embargo, la gravedad superficial depende de la energía, lo que parece indicar que la temperatura de Hawking podría mostrar también el mismo comportamiento.


\begin{spacing}{0.9}


\bibliographystyle{apsrev4-1}

\begin{thebibliography}{150}%
\makeatletter
\providecommand \@ifxundefined [1]{%
 \@ifx{#1\undefined}
}%
\providecommand \@ifnum [1]{%
 \ifnum #1\expandafter \@firstoftwo
 \else \expandafter \@secondoftwo
 \fi
}%
\providecommand \@ifx [1]{%
 \ifx #1\expandafter \@firstoftwo
 \else \expandafter \@secondoftwo
 \fi
}%
\providecommand \natexlab [1]{#1}%
\providecommand \enquote  [1]{``#1''}%
\providecommand \bibnamefont  [1]{#1}%
\providecommand \bibfnamefont [1]{#1}%
\providecommand \citenamefont [1]{#1}%
\providecommand \href@noop [0]{\@secondoftwo}%
\providecommand \href [0]{\begingroup \@sanitize@url \@href}%
\providecommand \@href[1]{\@@startlink{#1}\@@href}%
\providecommand \@@href[1]{\endgroup#1\@@endlink}%
\providecommand \@sanitize@url [0]{\catcode `\\12\catcode `\$12\catcode
  `\&12\catcode `\#12\catcode `\^12\catcode `\_12\catcode `\%12\relax}%
\providecommand \@@startlink[1]{}%
\providecommand \@@endlink[0]{}%
\providecommand \url  [0]{\begingroup\@sanitize@url \@url }%
\providecommand \@url [1]{\endgroup\@href {#1}{\urlprefix }}%
\providecommand \urlprefix  [0]{URL }%
\providecommand \Eprint [0]{\href }%
\providecommand \doibase [0]{http://dx.doi.org/}%
\providecommand \selectlanguage [0]{\@gobble}%
\providecommand \bibinfo  [0]{\@secondoftwo}%
\providecommand \bibfield  [0]{\@secondoftwo}%
\providecommand \translation [1]{[#1]}%
\providecommand \BibitemOpen [0]{}%
\providecommand \bibitemStop [0]{}%
\providecommand \bibitemNoStop [0]{.\EOS\space}%
\providecommand \EOS [0]{\spacefactor3000\relax}%
\providecommand \BibitemShut  [1]{\csname bibitem#1\endcsname}%
\let\auto@bib@innerbib\@empty
\bibitem [{\citenamefont {Feynman}(1996)}]{Feynman:1996kb}%
  \BibitemOpen
  \bibfield  {author} {\bibinfo {author} {\bibfnamefont {R.~P.}\ \bibnamefont
  {Feynman}},\ }\href@noop {} {\emph {\bibinfo {title} {{Feynman lectures on
  gravitation}}}},\ edited by\ \bibinfo {editor} {\bibfnamefont {F.~B.}\
  \bibnamefont {Morinigo}}, \bibinfo {editor} {\bibfnamefont {W.~G.}\
  \bibnamefont {Wagner}}, \ and\ \bibinfo {editor} {\bibfnamefont
  {B.}~\bibnamefont {Hatfield}}\ (\bibinfo {year} {1996})\BibitemShut {NoStop}%
\bibitem [{\citenamefont {Hawking}(1976)}]{Hawking:1976ra}%
  \BibitemOpen
  \bibfield  {author} {\bibinfo {author} {\bibfnamefont {S.~W.}\ \bibnamefont
  {Hawking}},\ }\href {\doibase 10.1103/PhysRevD.14.2460} {\bibfield  {journal}
  {\bibinfo  {journal} {Phys. Rev.}\ }\textbf {\bibinfo {volume} {D14}},\
  \bibinfo {pages} {2460} (\bibinfo {year} {1976})}\BibitemShut {NoStop}%
\bibitem [{\citenamefont {Hawking}(1975)}]{Hawking:1974sw}%
  \BibitemOpen
  \bibfield  {author} {\bibinfo {author} {\bibfnamefont {S.~W.}\ \bibnamefont
  {Hawking}},\ }\bibfield  {booktitle} {\emph {\bibinfo {booktitle} {{Euclidean
  quantum gravity}}},\ }\href {\doibase 10.1007/BF02345020, 10.1007/BF01608497}
  {\bibfield  {journal} {\bibinfo  {journal} {Commun. Math. Phys.}\ }\textbf
  {\bibinfo {volume} {43}},\ \bibinfo {pages} {199} (\bibinfo {year} {1975})},\
  \bibinfo {note} {[,167(1975)]}\BibitemShut {NoStop}%
\bibitem [{\citenamefont {Almheiri}\ \emph {et~al.}(2013)\citenamefont
  {Almheiri}, \citenamefont {Marolf}, \citenamefont {Polchinski},\ and\
  \citenamefont {Sully}}]{Almheiri:2012rt}%
  \BibitemOpen
  \bibfield  {author} {\bibinfo {author} {\bibfnamefont {A.}~\bibnamefont
  {Almheiri}}, \bibinfo {author} {\bibfnamefont {D.}~\bibnamefont {Marolf}},
  \bibinfo {author} {\bibfnamefont {J.}~\bibnamefont {Polchinski}}, \ and\
  \bibinfo {author} {\bibfnamefont {J.}~\bibnamefont {Sully}},\ }\href
  {\doibase 10.1007/JHEP02(2013)062} {\bibfield  {journal} {\bibinfo  {journal}
  {JHEP}\ }\textbf {\bibinfo {volume} {02}},\ \bibinfo {pages} {062} (\bibinfo
  {year} {2013})},\ \Eprint {http://arxiv.org/abs/1207.3123} {arXiv:1207.3123
  [hep-th]} \BibitemShut {NoStop}%
\bibitem [{\citenamefont {Einstein}(1905)}]{Einstein1905}%
  \BibitemOpen
  \bibfield  {author} {\bibinfo {author} {\bibfnamefont {A.}~\bibnamefont
  {Einstein}},\ }\href {\doibase 10.1002/andp.19053221004} {\bibfield
  {journal} {\bibinfo  {journal} {Annalen der Physik}\ }\textbf {\bibinfo
  {volume} {322}},\ \bibinfo {pages} {891} (\bibinfo {year} {1905})},\ \Eprint
  {http://arxiv.org/abs/https://onlinelibrary.wiley.com/doi/pdf/10.1002/andp.19053221004}
  {https://onlinelibrary.wiley.com/doi/pdf/10.1002/andp.19053221004}
  \BibitemShut {NoStop}%
\bibitem [{\citenamefont {Amelino-Camelia}\ \emph
  {et~al.}(2011{\natexlab{a}})\citenamefont {Amelino-Camelia}, \citenamefont
  {Freidel}, \citenamefont {Kowalski-Glikman},\ and\ \citenamefont
  {Smolin}}]{AmelinoCamelia:2011bm}%
  \BibitemOpen
  \bibfield  {author} {\bibinfo {author} {\bibfnamefont {G.}~\bibnamefont
  {Amelino-Camelia}}, \bibinfo {author} {\bibfnamefont {L.}~\bibnamefont
  {Freidel}}, \bibinfo {author} {\bibfnamefont {J.}~\bibnamefont
  {Kowalski-Glikman}}, \ and\ \bibinfo {author} {\bibfnamefont
  {L.}~\bibnamefont {Smolin}},\ }\href {\doibase 10.1103/PhysRevD.84.084010}
  {\bibfield  {journal} {\bibinfo  {journal} {Phys. Rev.}\ }\textbf {\bibinfo
  {volume} {D84}},\ \bibinfo {pages} {084010} (\bibinfo {year}
  {2011}{\natexlab{a}})},\ \Eprint {http://arxiv.org/abs/1101.0931}
  {arXiv:1101.0931 [hep-th]} \BibitemShut {NoStop}%
\bibitem [{\citenamefont {Amelino-Camelia}\ \emph
  {et~al.}(2011{\natexlab{b}})\citenamefont {Amelino-Camelia}, \citenamefont
  {Freidel}, \citenamefont {Kowalski-Glikman},\ and\ \citenamefont
  {Smolin}}]{AmelinoCamelia:2011pe}%
  \BibitemOpen
  \bibfield  {author} {\bibinfo {author} {\bibfnamefont {G.}~\bibnamefont
  {Amelino-Camelia}}, \bibinfo {author} {\bibfnamefont {L.}~\bibnamefont
  {Freidel}}, \bibinfo {author} {\bibfnamefont {J.}~\bibnamefont
  {Kowalski-Glikman}}, \ and\ \bibinfo {author} {\bibfnamefont
  {L.}~\bibnamefont {Smolin}},\ }\href {\doibase 10.1142/S0218271811020743,
  10.1007/s10714-011-1212-8} {\bibfield  {journal} {\bibinfo  {journal}
  {Gen.Rel.Grav.}\ }\textbf {\bibinfo {volume} {43}},\ \bibinfo {pages} {2547}
  (\bibinfo {year} {2011}{\natexlab{b}})},\ \Eprint
  {http://arxiv.org/abs/1106.0313} {arXiv:1106.0313 [hep-th]} \BibitemShut
  {NoStop}%
\bibitem [{\citenamefont {Mukhi}(2011)}]{Mukhi:2011zz}%
  \BibitemOpen
  \bibfield  {author} {\bibinfo {author} {\bibfnamefont {S.}~\bibnamefont
  {Mukhi}},\ }\href {\doibase 10.1088/0264-9381/28/15/153001} {\bibfield
  {journal} {\bibinfo  {journal} {Class. Quant. Grav.}\ }\textbf {\bibinfo
  {volume} {28}},\ \bibinfo {pages} {153001} (\bibinfo {year} {2011})},\
  \Eprint {http://arxiv.org/abs/1110.2569} {arXiv:1110.2569 [physics.pop-ph]}
  \BibitemShut {NoStop}%
\bibitem [{\citenamefont {Aharony}(2000)}]{Aharony:1999ks}%
  \BibitemOpen
  \bibfield  {author} {\bibinfo {author} {\bibfnamefont {O.}~\bibnamefont
  {Aharony}},\ }\bibfield  {booktitle} {\emph {\bibinfo {booktitle} {{Strings
  '99. Proceedings, Conference, Potsdam, Germany, July 19-24, 1999}}},\ }\href
  {\doibase 10.1088/0264-9381/17/5/302} {\bibfield  {journal} {\bibinfo
  {journal} {Class. Quant. Grav.}\ }\textbf {\bibinfo {volume} {17}},\ \bibinfo
  {pages} {929} (\bibinfo {year} {2000})},\ \Eprint
  {http://arxiv.org/abs/hep-th/9911147} {arXiv:hep-th/9911147 [hep-th]}
  \BibitemShut {NoStop}%
\bibitem [{\citenamefont {Dienes}(1997)}]{Dienes:1996du}%
  \BibitemOpen
  \bibfield  {author} {\bibinfo {author} {\bibfnamefont {K.~R.}\ \bibnamefont
  {Dienes}},\ }\bibfield  {booktitle} {\emph {\bibinfo {booktitle} {{Institute
  for Theoretical Physics Conference on Unification: From the Weak Scale to the
  Planck Scale Santa Barbara, California, October 23-27, 1995}}},\ }\href
  {\doibase 10.1016/S0370-1573(97)00009-4} {\bibfield  {journal} {\bibinfo
  {journal} {Phys. Rept.}\ }\textbf {\bibinfo {volume} {287}},\ \bibinfo
  {pages} {447} (\bibinfo {year} {1997})},\ \Eprint
  {http://arxiv.org/abs/hep-th/9602045} {arXiv:hep-th/9602045 [hep-th]}
  \BibitemShut {NoStop}%
\bibitem [{\citenamefont {Sahlmann}(2010)}]{Sahlmann:2010zf}%
  \BibitemOpen
  \bibfield  {author} {\bibinfo {author} {\bibfnamefont {H.}~\bibnamefont
  {Sahlmann}},\ }in\ \href
  {https://inspirehep.net/record/843661/files/arXiv:1001.4188.pdf} {\emph
  {\bibinfo {booktitle} {{Proceedings, Foundations of Space and Time:
  Reflections on Quantum Gravity: Cape Town, South Africa}}}}\ (\bibinfo {year}
  {2010})\ pp.\ \bibinfo {pages} {185--210},\ \Eprint
  {http://arxiv.org/abs/1001.4188} {arXiv:1001.4188 [gr-qc]} \BibitemShut
  {NoStop}%
\bibitem [{\citenamefont {Dupuis}\ \emph {et~al.}(2012)\citenamefont {Dupuis},
  \citenamefont {Ryan},\ and\ \citenamefont {Speziale}}]{Dupuis:2012yw}%
  \BibitemOpen
  \bibfield  {author} {\bibinfo {author} {\bibfnamefont {M.}~\bibnamefont
  {Dupuis}}, \bibinfo {author} {\bibfnamefont {J.~P.}\ \bibnamefont {Ryan}}, \
  and\ \bibinfo {author} {\bibfnamefont {S.}~\bibnamefont {Speziale}},\ }\href
  {\doibase 10.3842/SIGMA.2012.052} {\bibfield  {journal} {\bibinfo  {journal}
  {SIGMA}\ }\textbf {\bibinfo {volume} {8}},\ \bibinfo {pages} {052} (\bibinfo
  {year} {2012})},\ \Eprint {http://arxiv.org/abs/1204.5394} {arXiv:1204.5394
  [gr-qc]} \BibitemShut {NoStop}%
\bibitem [{\citenamefont {Van~Nieuwenhuizen}(1981)}]{VanNieuwenhuizen:1981ae}%
  \BibitemOpen
  \bibfield  {author} {\bibinfo {author} {\bibfnamefont {P.}~\bibnamefont
  {Van~Nieuwenhuizen}},\ }\href {\doibase 10.1016/0370-1573(81)90157-5}
  {\bibfield  {journal} {\bibinfo  {journal} {Phys. Rept.}\ }\textbf {\bibinfo
  {volume} {68}},\ \bibinfo {pages} {189} (\bibinfo {year} {1981})}\BibitemShut
  {NoStop}%
\bibitem [{\citenamefont {Taylor}(1984)}]{Taylor:1983su}%
  \BibitemOpen
  \bibfield  {author} {\bibinfo {author} {\bibfnamefont {J.~G.}\ \bibnamefont
  {Taylor}},\ }\href {\doibase 10.1016/0146-6410(84)90002-4} {\bibfield
  {journal} {\bibinfo  {journal} {Prog. Part. Nucl. Phys.}\ }\textbf {\bibinfo
  {volume} {12}},\ \bibinfo {pages} {1} (\bibinfo {year} {1984})}\BibitemShut
  {NoStop}%
\bibitem [{\citenamefont {Wallden}(2010)}]{Wallden:2010sh}%
  \BibitemOpen
  \bibfield  {author} {\bibinfo {author} {\bibfnamefont {P.}~\bibnamefont
  {Wallden}},\ }\bibfield  {booktitle} {\emph {\bibinfo {booktitle} {{Classical
  and quantum gravity. Proceedings, 1st Mediterranean Conference, MCCQG 2009,
  Kolymbari, Crete, Greece, September 14-18, 2009}}},\ }\href {\doibase
  10.1088/1742-6596/222/1/012053} {\bibfield  {journal} {\bibinfo  {journal}
  {J. Phys. Conf. Ser.}\ }\textbf {\bibinfo {volume} {222}},\ \bibinfo {pages}
  {012053} (\bibinfo {year} {2010})},\ \Eprint {http://arxiv.org/abs/1001.4041}
  {arXiv:1001.4041 [gr-qc]} \BibitemShut {NoStop}%
\bibitem [{\citenamefont {Wallden}(2013)}]{Wallden:2013kka}%
  \BibitemOpen
  \bibfield  {author} {\bibinfo {author} {\bibfnamefont {P.}~\bibnamefont
  {Wallden}},\ }\bibfield  {booktitle} {\emph {\bibinfo {booktitle}
  {{Proceedings, 15th Conference on Recent Developments in Gravity (NEB 15):
  Chania, Crete, Greece, June 20-23, 2012}}},\ }\href {\doibase
  10.1088/1742-6596/453/1/012023} {\bibfield  {journal} {\bibinfo  {journal}
  {J. Phys. Conf. Ser.}\ }\textbf {\bibinfo {volume} {453}},\ \bibinfo {pages}
  {012023} (\bibinfo {year} {2013})}\BibitemShut {NoStop}%
\bibitem [{\citenamefont {Henson}(2009)}]{Henson:2006kf}%
  \BibitemOpen
  \bibfield  {author} {\bibinfo {author} {\bibfnamefont {J.}~\bibnamefont
  {Henson}},\ }in\ \href@noop {} {\emph {\bibinfo {booktitle} {Approaches to
  Quantum Gravity: Toward a New Understanding of Space, Time and Matter}}},\
  \bibinfo {editor} {edited by\ \bibinfo {editor} {\bibfnamefont
  {D.}~\bibnamefont {Oriti}}}\ (\bibinfo  {publisher} {Cambridge University
  Press},\ \bibinfo {year} {2009})\ pp.\ \bibinfo {pages} {393--413},\ \Eprint
  {http://arxiv.org/abs/gr-qc/0601121} {arXiv:gr-qc/0601121 [gr-qc]}
  \BibitemShut {NoStop}%
\bibitem [{\citenamefont {Ehlers}\ and\ \citenamefont
  {Lammerzahl}(2006)}]{LectNotes702}%
  \BibitemOpen
  \bibinfo {editor} {\bibfnamefont {J.}~\bibnamefont {Ehlers}}\ and\ \bibinfo
  {editor} {\bibfnamefont {C.}~\bibnamefont {Lammerzahl}},\ eds.,\ \href
  {\doibase 10.1007/3-540-34523-X} {\emph {\bibinfo {title} {{Special
  Relativity: Will It Survive the Next 100 Years?}}}},\ Vol.\ \bibinfo {volume}
  {702}\ (\bibinfo  {publisher} {Lect. Notes Phys., Springer},\ \bibinfo {year}
  {2006})\BibitemShut {NoStop}%
\bibitem [{\citenamefont {Gross}\ and\ \citenamefont
  {Mende}(1988)}]{Gross:1987ar}%
  \BibitemOpen
  \bibfield  {author} {\bibinfo {author} {\bibfnamefont {D.~J.}\ \bibnamefont
  {Gross}}\ and\ \bibinfo {author} {\bibfnamefont {P.~F.}\ \bibnamefont
  {Mende}},\ }\href {\doibase 10.1016/0550-3213(88)90390-2} {\bibfield
  {journal} {\bibinfo  {journal} {Nucl. Phys.}\ }\textbf {\bibinfo {volume}
  {B303}},\ \bibinfo {pages} {407} (\bibinfo {year} {1988})}\BibitemShut
  {NoStop}%
\bibitem [{\citenamefont {Amati}\ \emph {et~al.}(1989)\citenamefont {Amati},
  \citenamefont {Ciafaloni},\ and\ \citenamefont {Veneziano}}]{Amati:1988tn}%
  \BibitemOpen
  \bibfield  {author} {\bibinfo {author} {\bibfnamefont {D.}~\bibnamefont
  {Amati}}, \bibinfo {author} {\bibfnamefont {M.}~\bibnamefont {Ciafaloni}}, \
  and\ \bibinfo {author} {\bibfnamefont {G.}~\bibnamefont {Veneziano}},\ }\href
  {\doibase 10.1016/0370-2693(89)91366-X} {\bibfield  {journal} {\bibinfo
  {journal} {Phys. Lett.}\ }\textbf {\bibinfo {volume} {B216}},\ \bibinfo
  {pages} {41} (\bibinfo {year} {1989})}\BibitemShut {NoStop}%
\bibitem [{\citenamefont {Garay}(1995)}]{Garay1995}%
  \BibitemOpen
  \bibfield  {author} {\bibinfo {author} {\bibfnamefont {L.~J.}\ \bibnamefont
  {Garay}},\ }\href {\doibase 10.1142/S0217751X95000085} {\bibfield  {journal}
  {\bibinfo  {journal} {Int. J. Mod. Phys.}\ }\textbf {\bibinfo {volume}
  {A10}},\ \bibinfo {pages} {145} (\bibinfo {year} {1995})},\ \Eprint
  {http://arxiv.org/abs/gr-qc/9403008} {arXiv:gr-qc/9403008 [gr-qc]}
  \BibitemShut {NoStop}%
\bibitem [{\citenamefont {Schiller}(1996)}]{Schiller:1996fw}%
  \BibitemOpen
  \bibfield  {author} {\bibinfo {author} {\bibfnamefont {C.}~\bibnamefont
  {Schiller}},\ }\href@noop {} {\  (\bibinfo {year} {1996})},\ \Eprint
  {http://arxiv.org/abs/gr-qc/9610066} {arXiv:gr-qc/9610066 [gr-qc]}
  \BibitemShut {NoStop}%
\bibitem [{\citenamefont {Wheeler}(1955)}]{Wheeler:1955zz}%
  \BibitemOpen
  \bibfield  {author} {\bibinfo {author} {\bibfnamefont {J.~A.}\ \bibnamefont
  {Wheeler}},\ }\href {\doibase 10.1103/PhysRev.97.511} {\bibfield  {journal}
  {\bibinfo  {journal} {Phys. Rev.}\ }\textbf {\bibinfo {volume} {97}},\
  \bibinfo {pages} {511} (\bibinfo {year} {1955})}\BibitemShut {NoStop}%
\bibitem [{\citenamefont {Ng}(2011)}]{Ng:2011rn}%
  \BibitemOpen
  \bibfield  {author} {\bibinfo {author} {\bibfnamefont {Y.~J.}\ \bibnamefont
  {Ng}},\ }in\ \href
  {http://inspirehep.net/record/890213/files/arXiv:1102.4109.pdf} {\emph
  {\bibinfo {booktitle} {{Time and Matter: Proceedings, 3rd International
  Conference, TAM2010, Budva, Montenegro, 4-8 October, 2010}}}}\ (\bibinfo
  {year} {2011})\ pp.\ \bibinfo {pages} {103--122},\ \Eprint
  {http://arxiv.org/abs/1102.4109} {arXiv:1102.4109 [gr-qc]} \BibitemShut
  {NoStop}%
\bibitem [{\citenamefont {Hossenfelder}(2013)}]{Hossenfelder:2012jw}%
  \BibitemOpen
  \bibfield  {author} {\bibinfo {author} {\bibfnamefont {S.}~\bibnamefont
  {Hossenfelder}},\ }\href {\doibase 10.12942/lrr-2013-2} {\bibfield  {journal}
  {\bibinfo  {journal} {Living Rev.Rel.}\ }\textbf {\bibinfo {volume} {16}},\
  \bibinfo {pages} {2} (\bibinfo {year} {2013})},\ \Eprint
  {http://arxiv.org/abs/1203.6191} {arXiv:1203.6191 [gr-qc]} \BibitemShut
  {NoStop}%
\bibitem [{\citenamefont {Kato}(1990)}]{Kato:1990bd}%
  \BibitemOpen
  \bibfield  {author} {\bibinfo {author} {\bibfnamefont {M.}~\bibnamefont
  {Kato}},\ }\href {\doibase 10.1016/0370-2693(90)90162-Y} {\bibfield
  {journal} {\bibinfo  {journal} {Phys. Lett.}\ }\textbf {\bibinfo {volume}
  {B245}},\ \bibinfo {pages} {43} (\bibinfo {year} {1990})}\BibitemShut
  {NoStop}%
\bibitem [{\citenamefont {Susskind}(1993)}]{Susskind:1993ki}%
  \BibitemOpen
  \bibfield  {author} {\bibinfo {author} {\bibfnamefont {L.}~\bibnamefont
  {Susskind}},\ }\href {\doibase 10.1103/PhysRevLett.71.2367} {\bibfield
  {journal} {\bibinfo  {journal} {Phys. Rev. Lett.}\ }\textbf {\bibinfo
  {volume} {71}},\ \bibinfo {pages} {2367} (\bibinfo {year} {1993})},\ \Eprint
  {http://arxiv.org/abs/hep-th/9307168} {arXiv:hep-th/9307168 [hep-th]}
  \BibitemShut {NoStop}%
\bibitem [{\citenamefont {Snyder}(1947)}]{Snyder:1946qz}%
  \BibitemOpen
  \bibfield  {author} {\bibinfo {author} {\bibfnamefont {H.~S.}\ \bibnamefont
  {Snyder}},\ }\href {\doibase 10.1103/PhysRev.71.38} {\bibfield  {journal}
  {\bibinfo  {journal} {Phys. Rev.}\ }\textbf {\bibinfo {volume} {71}},\
  \bibinfo {pages} {38} (\bibinfo {year} {1947})}\BibitemShut {NoStop}%
\bibitem [{\citenamefont {Szabo}(2003)}]{Szabo:2001kg}%
  \BibitemOpen
  \bibfield  {author} {\bibinfo {author} {\bibfnamefont {R.~J.}\ \bibnamefont
  {Szabo}},\ }\bibfield  {booktitle} {\emph {\bibinfo {booktitle} {{Frontiers
  of Mathematical Physics: Summer Workshop on Particles, Fields and Strings
  Burnaby, Canada, July 16-27, 2001}}},\ }\href {\doibase
  10.1016/S0370-1573(03)00059-0} {\bibfield  {journal} {\bibinfo  {journal}
  {Phys. Rept.}\ }\textbf {\bibinfo {volume} {378}},\ \bibinfo {pages} {207}
  (\bibinfo {year} {2003})},\ \Eprint {http://arxiv.org/abs/hep-th/0109162}
  {arXiv:hep-th/0109162 [hep-th]} \BibitemShut {NoStop}%
\bibitem [{\citenamefont {Douglas}\ and\ \citenamefont
  {Nekrasov}(2001)}]{Douglas:2001ba}%
  \BibitemOpen
  \bibfield  {author} {\bibinfo {author} {\bibfnamefont {M.~R.}\ \bibnamefont
  {Douglas}}\ and\ \bibinfo {author} {\bibfnamefont {N.~A.}\ \bibnamefont
  {Nekrasov}},\ }\href {\doibase 10.1103/RevModPhys.73.977} {\bibfield
  {journal} {\bibinfo  {journal} {Rev. Mod. Phys.}\ }\textbf {\bibinfo {volume}
  {73}},\ \bibinfo {pages} {977} (\bibinfo {year} {2001})},\ \Eprint
  {http://arxiv.org/abs/hep-th/0106048} {arXiv:hep-th/0106048 [hep-th]}
  \BibitemShut {NoStop}%
\bibitem [{\citenamefont {Smoli{\'{n}}ski}(1994)}]{Smolinski1994}%
  \BibitemOpen
  \bibfield  {author} {\bibinfo {author} {\bibfnamefont {K.~A.}\ \bibnamefont
  {Smoli{\'{n}}ski}},\ }\href {\doibase 10.1007/BF01690462} {\bibfield
  {journal} {\bibinfo  {journal} {Czechoslovak Journal of Physics}\ }\textbf
  {\bibinfo {volume} {44}},\ \bibinfo {pages} {1101} (\bibinfo {year}
  {1994})}\BibitemShut {NoStop}%
\bibitem [{\citenamefont {Kallosh}\ \emph {et~al.}(1995)\citenamefont
  {Kallosh}, \citenamefont {Linde}, \citenamefont {Linde},\ and\ \citenamefont
  {Susskind}}]{Kallosh:1995hi}%
  \BibitemOpen
  \bibfield  {author} {\bibinfo {author} {\bibfnamefont {R.}~\bibnamefont
  {Kallosh}}, \bibinfo {author} {\bibfnamefont {A.~D.}\ \bibnamefont {Linde}},
  \bibinfo {author} {\bibfnamefont {D.~A.}\ \bibnamefont {Linde}}, \ and\
  \bibinfo {author} {\bibfnamefont {L.}~\bibnamefont {Susskind}},\ }\href
  {\doibase 10.1103/PhysRevD.52.912} {\bibfield  {journal} {\bibinfo  {journal}
  {Phys. Rev.}\ }\textbf {\bibinfo {volume} {D52}},\ \bibinfo {pages} {912}
  (\bibinfo {year} {1995})},\ \Eprint {http://arxiv.org/abs/hep-th/9502069}
  {arXiv:hep-th/9502069 [hep-th]} \BibitemShut {NoStop}%
\bibitem [{\citenamefont {Kostelecky}\ and\ \citenamefont
  {Russell}(2011)}]{Kostelecky:2008ts}%
  \BibitemOpen
  \bibfield  {author} {\bibinfo {author} {\bibfnamefont {V.~A.}\ \bibnamefont
  {Kostelecky}}\ and\ \bibinfo {author} {\bibfnamefont {N.}~\bibnamefont
  {Russell}},\ }\href {\doibase 10.1103/RevModPhys.83.11} {\bibfield  {journal}
  {\bibinfo  {journal} {Rev. Mod. Phys.}\ }\textbf {\bibinfo {volume} {83}},\
  \bibinfo {pages} {11} (\bibinfo {year} {2011})},\ \Eprint
  {http://arxiv.org/abs/0801.0287} {arXiv:0801.0287 [hep-ph]} \BibitemShut
  {NoStop}%
\bibitem [{\citenamefont {Long}\ and\ \citenamefont
  {Kostelecky}(2015)}]{Long:2014swa}%
  \BibitemOpen
  \bibfield  {author} {\bibinfo {author} {\bibfnamefont {J.~C.}\ \bibnamefont
  {Long}}\ and\ \bibinfo {author} {\bibfnamefont {V.~A.}\ \bibnamefont
  {Kostelecky}},\ }\href {\doibase 10.1103/PhysRevD.91.092003} {\bibfield
  {journal} {\bibinfo  {journal} {Phys. Rev.}\ }\textbf {\bibinfo {volume}
  {D91}},\ \bibinfo {pages} {092003} (\bibinfo {year} {2015})},\ \Eprint
  {http://arxiv.org/abs/1412.8362} {arXiv:1412.8362 [hep-ex]} \BibitemShut
  {NoStop}%
\bibitem [{\citenamefont {Kostelecky}\ \emph
  {et~al.}(2016{\natexlab{a}})\citenamefont {Kostelecky}, \citenamefont
  {Lunghi},\ and\ \citenamefont {Vieira}}]{Kostelecky:2016pyx}%
  \BibitemOpen
  \bibfield  {author} {\bibinfo {author} {\bibfnamefont {A.}~\bibnamefont
  {Kostelecky}}, \bibinfo {author} {\bibfnamefont {E.}~\bibnamefont {Lunghi}},
  \ and\ \bibinfo {author} {\bibfnamefont {A.~R.}\ \bibnamefont {Vieira}},\
  }\href@noop {} {\  (\bibinfo {year} {2016}{\natexlab{a}})},\ \Eprint
  {http://arxiv.org/abs/1610.08755} {arXiv:1610.08755 [hep-ph]} \BibitemShut
  {NoStop}%
\bibitem [{\citenamefont {Kostelecky}\ \emph
  {et~al.}(2016{\natexlab{b}})\citenamefont {Kostelecky}, \citenamefont
  {Melissinos},\ and\ \citenamefont {Mewes}}]{Kostelecky:2016kkn}%
  \BibitemOpen
  \bibfield  {author} {\bibinfo {author} {\bibfnamefont {V.~A.}\ \bibnamefont
  {Kostelecky}}, \bibinfo {author} {\bibfnamefont {A.~C.}\ \bibnamefont
  {Melissinos}}, \ and\ \bibinfo {author} {\bibfnamefont {M.}~\bibnamefont
  {Mewes}},\ }\href {\doibase 10.1016/j.physletb.2016.08.001} {\bibfield
  {journal} {\bibinfo  {journal} {Phys. Lett.}\ }\textbf {\bibinfo {volume}
  {B761}},\ \bibinfo {pages} {1} (\bibinfo {year} {2016}{\natexlab{b}})},\
  \Eprint {http://arxiv.org/abs/1608.02592} {arXiv:1608.02592 [gr-qc]}
  \BibitemShut {NoStop}%
\bibitem [{\citenamefont {Amelino-Camelia}(2013)}]{AmelinoCamelia:2008qg}%
  \BibitemOpen
  \bibfield  {author} {\bibinfo {author} {\bibfnamefont {G.}~\bibnamefont
  {Amelino-Camelia}},\ }\href {\doibase 10.12942/lrr-2013-5} {\bibfield
  {journal} {\bibinfo  {journal} {Living Rev.Rel.}\ }\textbf {\bibinfo {volume}
  {16}},\ \bibinfo {pages} {5} (\bibinfo {year} {2013})},\ \Eprint
  {http://arxiv.org/abs/0806.0339} {arXiv:0806.0339 [gr-qc]} \BibitemShut
  {NoStop}%
\bibitem [{\citenamefont {Colladay}\ and\ \citenamefont
  {Kostelecky}(1998)}]{Colladay:1998fq}%
  \BibitemOpen
  \bibfield  {author} {\bibinfo {author} {\bibfnamefont {D.}~\bibnamefont
  {Colladay}}\ and\ \bibinfo {author} {\bibfnamefont {V.~A.}\ \bibnamefont
  {Kostelecky}},\ }\href {\doibase 10.1103/PhysRevD.58.116002} {\bibfield
  {journal} {\bibinfo  {journal} {Phys. Rev.}\ }\textbf {\bibinfo {volume}
  {D58}},\ \bibinfo {pages} {116002} (\bibinfo {year} {1998})},\ \Eprint
  {http://arxiv.org/abs/hep-ph/9809521} {arXiv:hep-ph/9809521 [hep-ph]}
  \BibitemShut {NoStop}%
\bibitem [{\citenamefont {Dirac}(1951)}]{Dirac:1951:TA}%
  \BibitemOpen
  \bibfield  {author} {\bibinfo {author} {\bibfnamefont {P.~A.~M.}\
  \bibnamefont {Dirac}},\ }\href {\doibase https://doi.org/10.1038/168906a0} {\
  \textbf {\bibinfo {volume} {168}},\ \bibinfo {pages} {906} (\bibinfo {year}
  {1951})}\BibitemShut {NoStop}%
\bibitem [{\citenamefont {Bjorken}(1963)}]{Bjorken:1963vg}%
  \BibitemOpen
  \bibfield  {author} {\bibinfo {author} {\bibfnamefont {J.~D.}\ \bibnamefont
  {Bjorken}},\ }\href {\doibase 10.1016/0003-4916(63)90069-1} {\bibfield
  {journal} {\bibinfo  {journal} {Annals Phys.}\ }\textbf {\bibinfo {volume}
  {24}},\ \bibinfo {pages} {174} (\bibinfo {year} {1963})}\BibitemShut
  {NoStop}%
\bibitem [{\citenamefont {Phillips}(1966)}]{Phillips:1966zzc}%
  \BibitemOpen
  \bibfield  {author} {\bibinfo {author} {\bibfnamefont {P.~R.}\ \bibnamefont
  {Phillips}},\ }\href {\doibase 10.1103/PhysRev.146.966} {\bibfield  {journal}
  {\bibinfo  {journal} {Phys. Rev.}\ }\textbf {\bibinfo {volume} {146}},\
  \bibinfo {pages} {966} (\bibinfo {year} {1966})}\BibitemShut {NoStop}%
\bibitem [{\citenamefont {Pavlopoulos}(1967)}]{Pavlopoulos:1967dm}%
  \BibitemOpen
  \bibfield  {author} {\bibinfo {author} {\bibfnamefont {T.~G.}\ \bibnamefont
  {Pavlopoulos}},\ }\href {\doibase 10.1103/PhysRev.159.1106} {\bibfield
  {journal} {\bibinfo  {journal} {Phys. Rev.}\ }\textbf {\bibinfo {volume}
  {159}},\ \bibinfo {pages} {1106} (\bibinfo {year} {1967})}\BibitemShut
  {NoStop}%
\bibitem [{\citenamefont {Redei}(1967)}]{Redei:1967zz}%
  \BibitemOpen
  \bibfield  {author} {\bibinfo {author} {\bibfnamefont {L.~B.}\ \bibnamefont
  {Redei}},\ }\href {\doibase 10.1103/PhysRev.162.1299} {\bibfield  {journal}
  {\bibinfo  {journal} {Phys. Rev.}\ }\textbf {\bibinfo {volume} {162}},\
  \bibinfo {pages} {1299} (\bibinfo {year} {1967})}\BibitemShut {NoStop}%
\bibitem [{\citenamefont {Nielsen}\ and\ \citenamefont
  {Ninomiya}(1978)}]{Nielsen:1978is}%
  \BibitemOpen
  \bibfield  {author} {\bibinfo {author} {\bibfnamefont {H.~B.}\ \bibnamefont
  {Nielsen}}\ and\ \bibinfo {author} {\bibfnamefont {M.}~\bibnamefont
  {Ninomiya}},\ }\href {\doibase 10.1016/0550-3213(78)90341-3} {\bibfield
  {journal} {\bibinfo  {journal} {Nucl. Phys.}\ }\textbf {\bibinfo {volume}
  {B141}},\ \bibinfo {pages} {153} (\bibinfo {year} {1978})}\BibitemShut
  {NoStop}%
\bibitem [{\citenamefont {Ellis}\ \emph {et~al.}(1980)\citenamefont {Ellis},
  \citenamefont {Gaillard}, \citenamefont {Nanopoulos},\ and\ \citenamefont
  {Rudaz}}]{Ellis:1980jm}%
  \BibitemOpen
  \bibfield  {author} {\bibinfo {author} {\bibfnamefont {J.~R.}\ \bibnamefont
  {Ellis}}, \bibinfo {author} {\bibfnamefont {M.~K.}\ \bibnamefont {Gaillard}},
  \bibinfo {author} {\bibfnamefont {D.~V.}\ \bibnamefont {Nanopoulos}}, \ and\
  \bibinfo {author} {\bibfnamefont {S.}~\bibnamefont {Rudaz}},\ }\href
  {\doibase 10.1016/0550-3213(80)90064-4} {\bibfield  {journal} {\bibinfo
  {journal} {Nucl. Phys.}\ }\textbf {\bibinfo {volume} {B176}},\ \bibinfo
  {pages} {61} (\bibinfo {year} {1980})}\BibitemShut {NoStop}%
\bibitem [{\citenamefont {Zee}(1982)}]{Zee:1981sy}%
  \BibitemOpen
  \bibfield  {author} {\bibinfo {author} {\bibfnamefont {A.}~\bibnamefont
  {Zee}},\ }\href {\doibase 10.1103/PhysRevD.25.1864} {\bibfield  {journal}
  {\bibinfo  {journal} {Phys. Rev.}\ }\textbf {\bibinfo {volume} {D25}},\
  \bibinfo {pages} {1864} (\bibinfo {year} {1982})}\BibitemShut {NoStop}%
\bibitem [{\citenamefont {Nielsen}\ and\ \citenamefont
  {Picek}(1982)}]{Nielsen:1982kx}%
  \BibitemOpen
  \bibfield  {author} {\bibinfo {author} {\bibfnamefont {H.~B.}\ \bibnamefont
  {Nielsen}}\ and\ \bibinfo {author} {\bibfnamefont {I.}~\bibnamefont
  {Picek}},\ }\href {\doibase 10.1016/0370-2693(82)90133-2} {\bibfield
  {journal} {\bibinfo  {journal} {Phys. Lett.}\ }\textbf {\bibinfo {volume}
  {114B}},\ \bibinfo {pages} {141} (\bibinfo {year} {1982})}\BibitemShut
  {NoStop}%
\bibitem [{\citenamefont {Chadha}\ and\ \citenamefont
  {Nielsen}(1983)}]{Chadha:1982qq}%
  \BibitemOpen
  \bibfield  {author} {\bibinfo {author} {\bibfnamefont {S.}~\bibnamefont
  {Chadha}}\ and\ \bibinfo {author} {\bibfnamefont {H.~B.}\ \bibnamefont
  {Nielsen}},\ }\href {\doibase 10.1016/0550-3213(83)90081-0} {\bibfield
  {journal} {\bibinfo  {journal} {Nucl. Phys.}\ }\textbf {\bibinfo {volume}
  {B217}},\ \bibinfo {pages} {125} (\bibinfo {year} {1983})}\BibitemShut
  {NoStop}%
\bibitem [{\citenamefont {Nielsen}\ and\ \citenamefont
  {Picek}(1983)}]{Nielsen:1982sz}%
  \BibitemOpen
  \bibfield  {author} {\bibinfo {author} {\bibfnamefont {H.~B.}\ \bibnamefont
  {Nielsen}}\ and\ \bibinfo {author} {\bibfnamefont {I.}~\bibnamefont
  {Picek}},\ }\href {\doibase 10.1016/0550-3213(83)90409-1,
  10.1016/0550-3213(84)90408-5} {\bibfield  {journal} {\bibinfo  {journal}
  {Nucl. Phys.}\ }\textbf {\bibinfo {volume} {B211}},\ \bibinfo {pages} {269}
  (\bibinfo {year} {1983})},\ \bibinfo {note} {[Addendum: Nucl.
  Phys.B242,542(1984)]}\BibitemShut {NoStop}%
\bibitem [{\citenamefont {Mattingly}(2005)}]{Mattingly:2005re}%
  \BibitemOpen
  \bibfield  {author} {\bibinfo {author} {\bibfnamefont {D.}~\bibnamefont
  {Mattingly}},\ }\href@noop {} {\bibfield  {journal} {\bibinfo  {journal}
  {Living Rev.Rel.}\ }\textbf {\bibinfo {volume} {8}},\ \bibinfo {pages} {5}
  (\bibinfo {year} {2005})},\ \Eprint {http://arxiv.org/abs/gr-qc/0502097}
  {arXiv:gr-qc/0502097 [gr-qc]} \BibitemShut {NoStop}%
\bibitem [{\citenamefont {Liberati}(2013)}]{Liberati2013}%
  \BibitemOpen
  \bibfield  {author} {\bibinfo {author} {\bibfnamefont {S.}~\bibnamefont
  {Liberati}},\ }\href {\doibase 10.1088/0264-9381/30/13/133001} {\bibfield
  {journal} {\bibinfo  {journal} {Class.Quant.Grav.}\ }\textbf {\bibinfo
  {volume} {30}},\ \bibinfo {pages} {133001} (\bibinfo {year} {2013})},\
  \Eprint {http://arxiv.org/abs/1304.5795} {arXiv:1304.5795 [gr-qc]}
  \BibitemShut {NoStop}%
\bibitem [{\citenamefont {Kostelecky}\ and\ \citenamefont
  {Lane}(1999)}]{Kostelecky:1999mr}%
  \BibitemOpen
  \bibfield  {author} {\bibinfo {author} {\bibfnamefont {V.~A.}\ \bibnamefont
  {Kostelecky}}\ and\ \bibinfo {author} {\bibfnamefont {C.~D.}\ \bibnamefont
  {Lane}},\ }\href {\doibase 10.1103/PhysRevD.60.116010} {\bibfield  {journal}
  {\bibinfo  {journal} {Phys. Rev.}\ }\textbf {\bibinfo {volume} {D60}},\
  \bibinfo {pages} {116010} (\bibinfo {year} {1999})},\ \Eprint
  {http://arxiv.org/abs/hep-ph/9908504} {arXiv:hep-ph/9908504 [hep-ph]}
  \BibitemShut {NoStop}%
\bibitem [{\citenamefont {Amelino-Camelia}\ \emph {et~al.}(1998)\citenamefont
  {Amelino-Camelia}, \citenamefont {Ellis}, \citenamefont {Mavromatos},
  \citenamefont {Nanopoulos},\ and\ \citenamefont
  {Sarkar}}]{Amelino-Camelia1998}%
  \BibitemOpen
  \bibfield  {author} {\bibinfo {author} {\bibfnamefont {G.}~\bibnamefont
  {Amelino-Camelia}}, \bibinfo {author} {\bibfnamefont {J.~R.}\ \bibnamefont
  {Ellis}}, \bibinfo {author} {\bibfnamefont {N.}~\bibnamefont {Mavromatos}},
  \bibinfo {author} {\bibfnamefont {D.~V.}\ \bibnamefont {Nanopoulos}}, \ and\
  \bibinfo {author} {\bibfnamefont {S.}~\bibnamefont {Sarkar}},\ }\href
  {\doibase 10.1038/31647} {\bibfield  {journal} {\bibinfo  {journal} {Nature}\
  }\textbf {\bibinfo {volume} {393}},\ \bibinfo {pages} {763} (\bibinfo {year}
  {1998})},\ \Eprint {http://arxiv.org/abs/astro-ph/9712103}
  {arXiv:astro-ph/9712103 [astro-ph]} \BibitemShut {NoStop}%
\bibitem [{\citenamefont {Maccione}\ \emph {et~al.}(2008)\citenamefont
  {Maccione}, \citenamefont {Liberati}, \citenamefont {Celotti}, \citenamefont
  {Kirk},\ and\ \citenamefont {Ubertini}}]{Maccione:2008tq}%
  \BibitemOpen
  \bibfield  {author} {\bibinfo {author} {\bibfnamefont {L.}~\bibnamefont
  {Maccione}}, \bibinfo {author} {\bibfnamefont {S.}~\bibnamefont {Liberati}},
  \bibinfo {author} {\bibfnamefont {A.}~\bibnamefont {Celotti}}, \bibinfo
  {author} {\bibfnamefont {J.~G.}\ \bibnamefont {Kirk}}, \ and\ \bibinfo
  {author} {\bibfnamefont {P.}~\bibnamefont {Ubertini}},\ }\href {\doibase
  10.1103/PhysRevD.78.103003} {\bibfield  {journal} {\bibinfo  {journal} {Phys.
  Rev.}\ }\textbf {\bibinfo {volume} {D78}},\ \bibinfo {pages} {103003}
  (\bibinfo {year} {2008})},\ \Eprint {http://arxiv.org/abs/0809.0220}
  {arXiv:0809.0220 [astro-ph]} \BibitemShut {NoStop}%
\bibitem [{\citenamefont {Jacobson}\ \emph {et~al.}(2003)\citenamefont
  {Jacobson}, \citenamefont {Liberati},\ and\ \citenamefont
  {Mattingly}}]{Jacobson:2002hd}%
  \BibitemOpen
  \bibfield  {author} {\bibinfo {author} {\bibfnamefont {T.}~\bibnamefont
  {Jacobson}}, \bibinfo {author} {\bibfnamefont {S.}~\bibnamefont {Liberati}},
  \ and\ \bibinfo {author} {\bibfnamefont {D.}~\bibnamefont {Mattingly}},\
  }\href {\doibase 10.1103/PhysRevD.67.124011} {\bibfield  {journal} {\bibinfo
  {journal} {Phys.Rev.}\ }\textbf {\bibinfo {volume} {D67}},\ \bibinfo {pages}
  {124011} (\bibinfo {year} {2003})},\ \Eprint
  {http://arxiv.org/abs/hep-ph/0209264} {arXiv:hep-ph/0209264 [hep-ph]}
  \BibitemShut {NoStop}%
\bibitem [{\citenamefont {Greisen}(1966)}]{Greisen:1966jv}%
  \BibitemOpen
  \bibfield  {author} {\bibinfo {author} {\bibfnamefont {K.}~\bibnamefont
  {Greisen}},\ }\href {\doibase 10.1103/PhysRevLett.16.748} {\bibfield
  {journal} {\bibinfo  {journal} {Phys. Rev. Lett.}\ }\textbf {\bibinfo
  {volume} {16}},\ \bibinfo {pages} {748} (\bibinfo {year} {1966})}\BibitemShut
  {NoStop}%
\bibitem [{\citenamefont {Zatsepin}\ and\ \citenamefont
  {Kuzmin}(1966)}]{Zatsepin:1966jv}%
  \BibitemOpen
  \bibfield  {author} {\bibinfo {author} {\bibfnamefont {G.~T.}\ \bibnamefont
  {Zatsepin}}\ and\ \bibinfo {author} {\bibfnamefont {V.~A.}\ \bibnamefont
  {Kuzmin}},\ }\href@noop {} {\bibfield  {journal} {\bibinfo  {journal} {JETP
  Lett.}\ }\textbf {\bibinfo {volume} {4}},\ \bibinfo {pages} {78} (\bibinfo
  {year} {1966})},\ \bibinfo {note} {[Pisma Zh. Eksp. Teor.
  Fiz.4,114(1966)]}\BibitemShut {NoStop}%
\bibitem [{\citenamefont {Roth}(2007)}]{Roth:2007in}%
  \BibitemOpen
  \bibfield  {author} {\bibinfo {author} {\bibfnamefont {M.}~\bibnamefont
  {Roth}} (\bibinfo {collaboration} {Pierre Auger}),\ }in\ \href
  {http://lss.fnal.gov/cgi-bin/find_paper.pl?conf-07-372} {\emph {\bibinfo
  {booktitle} {{Proceedings, 30th International Cosmic Ray Conference (ICRC
  2007): Merida, Yucatan, Mexico, July 3-11, 2007}}}},\ Vol.~\bibinfo {volume}
  {4}\ (\bibinfo {year} {2007})\ pp.\ \bibinfo {pages} {327--330},\ \Eprint
  {http://arxiv.org/abs/0706.2096} {arXiv:0706.2096 [astro-ph]} \BibitemShut
  {NoStop}%
\bibitem [{\citenamefont {Thomson}(2006)}]{Thomson:2006mm}%
  \BibitemOpen
  \bibfield  {author} {\bibinfo {author} {\bibfnamefont {G.~B.}\ \bibnamefont
  {Thomson}} (\bibinfo {collaboration} {HiRes}),\ }in\ \href@noop {} {\emph
  {\bibinfo {booktitle} {{14th International Seminar on High Energy Physics:
  Quarks 2006 St. Petersburg, Russia, May 19-25, 2006}}}}\ (\bibinfo {year}
  {2006})\ \Eprint {http://arxiv.org/abs/astro-ph/0609403}
  {arXiv:astro-ph/0609403 [astro-ph]} \BibitemShut {NoStop}%
\bibitem [{\citenamefont {Takeda}\ \emph {et~al.}(1998)\citenamefont {Takeda}
  \emph {et~al.}}]{Takeda:1998ps}%
  \BibitemOpen
  \bibfield  {author} {\bibinfo {author} {\bibfnamefont {M.}~\bibnamefont
  {Takeda}} \emph {et~al.},\ }\href {\doibase 10.1103/PhysRevLett.81.1163}
  {\bibfield  {journal} {\bibinfo  {journal} {Phys. Rev. Lett.}\ }\textbf
  {\bibinfo {volume} {81}},\ \bibinfo {pages} {1163} (\bibinfo {year}
  {1998})},\ \Eprint {http://arxiv.org/abs/astro-ph/9807193}
  {arXiv:astro-ph/9807193 [astro-ph]} \BibitemShut {NoStop}%
\bibitem [{\citenamefont {Amelino-Camelia}(2001)}]{Amelino-Camelia2001}%
  \BibitemOpen
  \bibfield  {author} {\bibinfo {author} {\bibfnamefont {G.}~\bibnamefont
  {Amelino-Camelia}},\ }\href@noop {} {\bibfield  {journal} {\bibinfo
  {journal} {Physics Letters B}\ }\textbf {\bibinfo {volume} {510}},\ \bibinfo
  {pages} {255} (\bibinfo {year} {2001})}\BibitemShut {NoStop}%
\bibitem [{\citenamefont
  {Amelino-Camelia}(2002{\natexlab{a}})}]{Amelino-Camelia2002a}%
  \BibitemOpen
  \bibfield  {author} {\bibinfo {author} {\bibfnamefont {G.}~\bibnamefont
  {Amelino-Camelia}},\ }\href {\doibase 10.1142/S0218271802001330} {\bibfield
  {journal} {\bibinfo  {journal} {Int.J.Mod.Phys.}\ }\textbf {\bibinfo {volume}
  {D11}},\ \bibinfo {pages} {35} (\bibinfo {year} {2002}{\natexlab{a}})},\
  \Eprint {http://arxiv.org/abs/gr-qc/0012051} {arXiv:gr-qc/0012051 [gr-qc]}
  \BibitemShut {NoStop}%
\bibitem [{\citenamefont {Borowiec}\ and\ \citenamefont
  {Pachol}(2010)}]{Borowiec2010}%
  \BibitemOpen
  \bibfield  {author} {\bibinfo {author} {\bibfnamefont {A.}~\bibnamefont
  {Borowiec}}\ and\ \bibinfo {author} {\bibfnamefont {A.}~\bibnamefont
  {Pachol}},\ }\href {\doibase 10.1088/1751-8113/43/4/045203} {\bibfield
  {journal} {\bibinfo  {journal} {J. Phys.}\ }\textbf {\bibinfo {volume}
  {A43}},\ \bibinfo {pages} {045203} (\bibinfo {year} {2010})},\ \Eprint
  {http://arxiv.org/abs/0903.5251} {arXiv:0903.5251 [hep-th]} \BibitemShut
  {NoStop}%
\bibitem [{\citenamefont
  {Amelino-Camelia}(2002{\natexlab{b}})}]{Amelino-Camelia2002}%
  \BibitemOpen
  \bibfield  {author} {\bibinfo {author} {\bibfnamefont {G.}~\bibnamefont
  {Amelino-Camelia}},\ }\href {\doibase 10.1142/S021827180200302X} {\bibfield
  {journal} {\bibinfo  {journal} {Int.J.Mod.Phys.}\ }\textbf {\bibinfo {volume}
  {D11}},\ \bibinfo {pages} {1643} (\bibinfo {year} {2002}{\natexlab{b}})},\
  \Eprint {http://arxiv.org/abs/gr-qc/0210063} {arXiv:gr-qc/0210063 [gr-qc]}
  \BibitemShut {NoStop}%
\bibitem [{\citenamefont {Amelino-Camelia}(2010)}]{AmelinoCamelia:2010pd}%
  \BibitemOpen
  \bibfield  {author} {\bibinfo {author} {\bibfnamefont {G.}~\bibnamefont
  {Amelino-Camelia}},\ }\href {\doibase 10.3390/sym2010230} {\bibfield
  {journal} {\bibinfo  {journal} {Symmetry}\ }\textbf {\bibinfo {volume} {2}},\
  \bibinfo {pages} {230} (\bibinfo {year} {2010})},\ \Eprint
  {http://arxiv.org/abs/1003.3942} {arXiv:1003.3942 [gr-qc]} \BibitemShut
  {NoStop}%
\bibitem [{\citenamefont {Majid}(1995)}]{Majid:1995qg}%
  \BibitemOpen
  \bibfield  {author} {\bibinfo {author} {\bibfnamefont {S.}~\bibnamefont
  {Majid}},\ }\href@noop {} {\emph {\bibinfo {title} {Foundations of Quantum
  Group Theory}}}\ (\bibinfo  {publisher} {Cambridge University Press},\
  \bibinfo {year} {1995})\BibitemShut {NoStop}%
\bibitem [{\citenamefont {Lukierski}\ \emph {et~al.}(1992)\citenamefont
  {Lukierski}, \citenamefont {Nowicki},\ and\ \citenamefont
  {Ruegg}}]{Lukierski:1992dt}%
  \BibitemOpen
  \bibfield  {author} {\bibinfo {author} {\bibfnamefont {J.}~\bibnamefont
  {Lukierski}}, \bibinfo {author} {\bibfnamefont {A.}~\bibnamefont {Nowicki}},
  \ and\ \bibinfo {author} {\bibfnamefont {H.}~\bibnamefont {Ruegg}},\ }\href
  {\doibase 10.1016/0370-2693(92)90894-A} {\bibfield  {journal} {\bibinfo
  {journal} {Phys. Lett.}\ }\textbf {\bibinfo {volume} {B293}},\ \bibinfo
  {pages} {344} (\bibinfo {year} {1992})}\BibitemShut {NoStop}%
\bibitem [{\citenamefont {Lukierski}\ \emph {et~al.}(1993)\citenamefont
  {Lukierski}, \citenamefont {Ruegg},\ and\ \citenamefont
  {Ruhl}}]{Lukierski:1993df}%
  \BibitemOpen
  \bibfield  {author} {\bibinfo {author} {\bibfnamefont {J.}~\bibnamefont
  {Lukierski}}, \bibinfo {author} {\bibfnamefont {H.}~\bibnamefont {Ruegg}}, \
  and\ \bibinfo {author} {\bibfnamefont {W.}~\bibnamefont {Ruhl}},\ }\href
  {\doibase 10.1016/0370-2693(93)90004-2} {\bibfield  {journal} {\bibinfo
  {journal} {Phys. Lett.}\ }\textbf {\bibinfo {volume} {B313}},\ \bibinfo
  {pages} {357} (\bibinfo {year} {1993})}\BibitemShut {NoStop}%
\bibitem [{\citenamefont {Majid}\ and\ \citenamefont
  {Ruegg}(1994)}]{Majid1994}%
  \BibitemOpen
  \bibfield  {author} {\bibinfo {author} {\bibfnamefont {S.}~\bibnamefont
  {Majid}}\ and\ \bibinfo {author} {\bibfnamefont {H.}~\bibnamefont {Ruegg}},\
  }\href {\doibase 10.1016/0370-2693(94)90699-8} {\bibfield  {journal}
  {\bibinfo  {journal} {Phys. Lett.}\ }\textbf {\bibinfo {volume} {B334}},\
  \bibinfo {pages} {348} (\bibinfo {year} {1994})},\ \Eprint
  {http://arxiv.org/abs/hep-th/9405107} {arXiv:hep-th/9405107 [hep-th]}
  \BibitemShut {NoStop}%
\bibitem [{\citenamefont {Lukierski}\ \emph {et~al.}(1995)\citenamefont
  {Lukierski}, \citenamefont {Ruegg},\ and\ \citenamefont
  {Zakrzewski}}]{Lukierski1995}%
  \BibitemOpen
  \bibfield  {author} {\bibinfo {author} {\bibfnamefont {J.}~\bibnamefont
  {Lukierski}}, \bibinfo {author} {\bibfnamefont {H.}~\bibnamefont {Ruegg}}, \
  and\ \bibinfo {author} {\bibfnamefont {W.~J.}\ \bibnamefont {Zakrzewski}},\
  }\href {\doibase 10.1006/aphy.1995.1092} {\bibfield  {journal} {\bibinfo
  {journal} {Annals Phys.}\ }\textbf {\bibinfo {volume} {243}},\ \bibinfo
  {pages} {90} (\bibinfo {year} {1995})},\ \Eprint
  {http://arxiv.org/abs/hep-th/9312153} {arXiv:hep-th/9312153 [hep-th]}
  \BibitemShut {NoStop}%
\bibitem [{\citenamefont {Kowalski-Glikman}\ and\ \citenamefont
  {Nowak}(2002)}]{KowalskiGlikman:2002we}%
  \BibitemOpen
  \bibfield  {author} {\bibinfo {author} {\bibfnamefont {J.}~\bibnamefont
  {Kowalski-Glikman}}\ and\ \bibinfo {author} {\bibfnamefont {S.}~\bibnamefont
  {Nowak}},\ }\href {\doibase 10.1016/S0370-2693(02)02063-4} {\bibfield
  {journal} {\bibinfo  {journal} {Phys. Lett.}\ }\textbf {\bibinfo {volume}
  {B539}},\ \bibinfo {pages} {126} (\bibinfo {year} {2002})},\ \Eprint
  {http://arxiv.org/abs/hep-th/0203040} {arXiv:hep-th/0203040 [hep-th]}
  \BibitemShut {NoStop}%
\bibitem [{\citenamefont {Lukierski}\ \emph {et~al.}(1991)\citenamefont
  {Lukierski}, \citenamefont {Ruegg}, \citenamefont {Nowicki},\ and\
  \citenamefont {Tolstoi}}]{Lukierski:1991pn}%
  \BibitemOpen
  \bibfield  {author} {\bibinfo {author} {\bibfnamefont {J.}~\bibnamefont
  {Lukierski}}, \bibinfo {author} {\bibfnamefont {H.}~\bibnamefont {Ruegg}},
  \bibinfo {author} {\bibfnamefont {A.}~\bibnamefont {Nowicki}}, \ and\
  \bibinfo {author} {\bibfnamefont {V.~N.}\ \bibnamefont {Tolstoi}},\ }\href
  {\doibase 10.1016/0370-2693(91)90358-W} {\bibfield  {journal} {\bibinfo
  {journal} {Phys. Lett.}\ }\textbf {\bibinfo {volume} {B264}},\ \bibinfo
  {pages} {331} (\bibinfo {year} {1991})}\BibitemShut {NoStop}%
\bibitem [{\citenamefont {Carmona}\ \emph {et~al.}(2010)\citenamefont
  {Carmona}, \citenamefont {Cort\'es},\ and\ \citenamefont
  {Maz\'on}}]{Carmona:2010ze}%
  \BibitemOpen
  \bibfield  {author} {\bibinfo {author} {\bibfnamefont {J.~M.}\ \bibnamefont
  {Carmona}}, \bibinfo {author} {\bibfnamefont {J.~L.}\ \bibnamefont
  {Cort\'es}}, \ and\ \bibinfo {author} {\bibfnamefont {D.}~\bibnamefont
  {Maz\'on}},\ }\href {\doibase 10.1103/PhysRevD.82.085012} {\bibfield
  {journal} {\bibinfo  {journal} {Phys.Rev.}\ }\textbf {\bibinfo {volume}
  {D82}},\ \bibinfo {pages} {085012} (\bibinfo {year} {2010})},\ \Eprint
  {http://arxiv.org/abs/1007.3190} {arXiv:1007.3190 [gr-qc]} \BibitemShut
  {NoStop}%
\bibitem [{\citenamefont {Carmona}\ \emph {et~al.}(2015)\citenamefont
  {Carmona}, \citenamefont {Cort\'es},\ and\ \citenamefont
  {Romeo}}]{Carmona:2014aba}%
  \BibitemOpen
  \bibfield  {author} {\bibinfo {author} {\bibfnamefont {J.~M.}\ \bibnamefont
  {Carmona}}, \bibinfo {author} {\bibfnamefont {J.~L.}\ \bibnamefont
  {Cort\'es}}, \ and\ \bibinfo {author} {\bibfnamefont {B.}~\bibnamefont
  {Romeo}},\ }\href {\doibase 10.1103/PhysRevD.91.085036} {\bibfield  {journal}
  {\bibinfo  {journal} {Phys.Rev.}\ }\textbf {\bibinfo {volume} {D91}},\
  \bibinfo {pages} {085036} (\bibinfo {year} {2015})},\ \Eprint
  {http://arxiv.org/abs/1412.6449} {arXiv:1412.6449 [hep-ph]} \BibitemShut
  {NoStop}%
\bibitem [{\citenamefont {Vasileiou}\ \emph {et~al.}(2015)\citenamefont
  {Vasileiou}, \citenamefont {Granot}, \citenamefont {Piran},\ and\
  \citenamefont {Amelino-Camelia}}]{Vasileiou:2015wja}%
  \BibitemOpen
  \bibfield  {author} {\bibinfo {author} {\bibfnamefont {V.}~\bibnamefont
  {Vasileiou}}, \bibinfo {author} {\bibfnamefont {J.}~\bibnamefont {Granot}},
  \bibinfo {author} {\bibfnamefont {T.}~\bibnamefont {Piran}}, \ and\ \bibinfo
  {author} {\bibfnamefont {G.}~\bibnamefont {Amelino-Camelia}},\ }\href
  {\doibase 10.1038/nphys3270} {\bibfield  {journal} {\bibinfo  {journal}
  {Nature Phys.}\ }\textbf {\bibinfo {volume} {11}},\ \bibinfo {pages} {344}
  (\bibinfo {year} {2015})}\BibitemShut {NoStop}%
\bibitem [{\citenamefont {Abdalla}\ \emph {et~al.}(2019)\citenamefont {Abdalla}
  \emph {et~al.}}]{Abdalla:2019krx}%
  \BibitemOpen
  \bibfield  {author} {\bibinfo {author} {\bibfnamefont {H.}~\bibnamefont
  {Abdalla}} \emph {et~al.} (\bibinfo {collaboration} {H.E.S.S.}),\ }\href
  {\doibase 10.3847/1538-4357/aaf1c4} {\bibfield  {journal} {\bibinfo
  {journal} {Astrophys. J.}\ }\textbf {\bibinfo {volume} {870}},\ \bibinfo
  {pages} {93} (\bibinfo {year} {2019})},\ \Eprint
  {http://arxiv.org/abs/1901.05209} {arXiv:1901.05209 [astro-ph.HE]}
  \BibitemShut {NoStop}%
\bibitem [{\citenamefont {Ellis}\ \emph {et~al.}(2019)\citenamefont {Ellis},
  \citenamefont {Konoplich}, \citenamefont {Mavromatos}, \citenamefont
  {Nguyen}, \citenamefont {Sakharov},\ and\ \citenamefont
  {Sarkisyan-Grinbaum}}]{Ellis:2018lca}%
  \BibitemOpen
  \bibfield  {author} {\bibinfo {author} {\bibfnamefont {J.}~\bibnamefont
  {Ellis}}, \bibinfo {author} {\bibfnamefont {R.}~\bibnamefont {Konoplich}},
  \bibinfo {author} {\bibfnamefont {N.~E.}\ \bibnamefont {Mavromatos}},
  \bibinfo {author} {\bibfnamefont {L.}~\bibnamefont {Nguyen}}, \bibinfo
  {author} {\bibfnamefont {A.~S.}\ \bibnamefont {Sakharov}}, \ and\ \bibinfo
  {author} {\bibfnamefont {E.~K.}\ \bibnamefont {Sarkisyan-Grinbaum}},\ }\href
  {\doibase 10.1103/PhysRevD.99.083009} {\bibfield  {journal} {\bibinfo
  {journal} {Phys. Rev.}\ }\textbf {\bibinfo {volume} {D99}},\ \bibinfo {pages}
  {083009} (\bibinfo {year} {2019})},\ \Eprint
  {http://arxiv.org/abs/1807.00189} {arXiv:1807.00189 [astro-ph.HE]}
  \BibitemShut {NoStop}%
\bibitem [{\citenamefont {Carmona}\ \emph
  {et~al.}(2018{\natexlab{a}})\citenamefont {Carmona}, \citenamefont {Cortes},\
  and\ \citenamefont {Relancio}}]{Carmona:2017oit}%
  \BibitemOpen
  \bibfield  {author} {\bibinfo {author} {\bibfnamefont {J.}~\bibnamefont
  {Carmona}}, \bibinfo {author} {\bibfnamefont {J.}~\bibnamefont {Cortes}}, \
  and\ \bibinfo {author} {\bibfnamefont {J.}~\bibnamefont {Relancio}},\ }\href
  {\doibase 10.1088/1361-6382/aa9ef8} {\bibfield  {journal} {\bibinfo
  {journal} {Class.\ Quant.\ Grav.}\ }\textbf {\bibinfo {volume} {35}},\
  \bibinfo {pages} {025014} (\bibinfo {year} {2018}{\natexlab{a}})},\ \Eprint
  {http://arxiv.org/abs/1702.03669} {arXiv:1702.03669 [hep-th]} \BibitemShut
  {NoStop}%
\bibitem [{\citenamefont {Carmona}\ \emph
  {et~al.}(2018{\natexlab{b}})\citenamefont {Carmona}, \citenamefont
  {Cortés},\ and\ \citenamefont {Relancio}}]{Carmona:2018xwm}%
  \BibitemOpen
  \bibfield  {author} {\bibinfo {author} {\bibfnamefont {J.~M.}\ \bibnamefont
  {Carmona}}, \bibinfo {author} {\bibfnamefont {J.~L.}\ \bibnamefont
  {Cortés}}, \ and\ \bibinfo {author} {\bibfnamefont {J.~J.}\ \bibnamefont
  {Relancio}},\ }\href {\doibase 10.3390/sym10070231} {\bibfield  {journal}
  {\bibinfo  {journal} {Symmetry}\ }\textbf {\bibinfo {volume} {10}},\ \bibinfo
  {pages} {231} (\bibinfo {year} {2018}{\natexlab{b}})},\ \Eprint
  {http://arxiv.org/abs/1806.01725} {arXiv:1806.01725 [hep-th]} \BibitemShut
  {NoStop}%
\bibitem [{\citenamefont {Carmona}\ \emph
  {et~al.}(2019{\natexlab{a}})\citenamefont {Carmona}, \citenamefont {Cortes},\
  and\ \citenamefont {Relancio}}]{Carmona:2019oph}%
  \BibitemOpen
  \bibfield  {author} {\bibinfo {author} {\bibfnamefont {J.~M.}\ \bibnamefont
  {Carmona}}, \bibinfo {author} {\bibfnamefont {J.~L.}\ \bibnamefont {Cortes}},
  \ and\ \bibinfo {author} {\bibfnamefont {J.~J.}\ \bibnamefont {Relancio}},\
  }\href {\doibase 10.3390/sym11111401} {\bibfield  {journal} {\bibinfo
  {journal} {Symmetry}\ }\textbf {\bibinfo {volume} {11}},\ \bibinfo {pages}
  {1401} (\bibinfo {year} {2019}{\natexlab{a}})},\ \Eprint
  {http://arxiv.org/abs/1911.12700} {arXiv:1911.12700 [hep-th]} \BibitemShut
  {NoStop}%
\bibitem [{\citenamefont {Amelino-Camelia}\ \emph
  {et~al.}(2011{\natexlab{c}})\citenamefont {Amelino-Camelia}, \citenamefont
  {Loret},\ and\ \citenamefont {Rosati}}]{AmelinoCamelia:2011cv}%
  \BibitemOpen
  \bibfield  {author} {\bibinfo {author} {\bibfnamefont {G.}~\bibnamefont
  {Amelino-Camelia}}, \bibinfo {author} {\bibfnamefont {N.}~\bibnamefont
  {Loret}}, \ and\ \bibinfo {author} {\bibfnamefont {G.}~\bibnamefont
  {Rosati}},\ }\href {\doibase 10.1016/j.physletb.2011.04.054} {\bibfield
  {journal} {\bibinfo  {journal} {Phys. Lett.}\ }\textbf {\bibinfo {volume}
  {B700}},\ \bibinfo {pages} {150} (\bibinfo {year} {2011}{\natexlab{c}})},\
  \Eprint {http://arxiv.org/abs/1102.4637} {arXiv:1102.4637 [hep-th]}
  \BibitemShut {NoStop}%
\bibitem [{\citenamefont {Freidel}\ and\ \citenamefont
  {Smolin}(2011)}]{Freidel:2011}%
  \BibitemOpen
  \bibfield  {author} {\bibinfo {author} {\bibfnamefont {L.}~\bibnamefont
  {Freidel}}\ and\ \bibinfo {author} {\bibfnamefont {L.}~\bibnamefont
  {Smolin}},\ }\href@noop {} {\  (\bibinfo {year} {2011})},\ \Eprint
  {http://arxiv.org/abs/1103.5626} {arXiv:1103.5626 [hep-th]} \BibitemShut
  {NoStop}%
\bibitem [{\citenamefont {Born}(1938)}]{Born:1938}%
  \BibitemOpen
  \bibfield  {author} {\bibinfo {author} {\bibfnamefont {M.}~\bibnamefont
  {Born}},\ }\href {\doibase 10.1098/rspa.1938.0060} {\bibfield  {journal}
  {\bibinfo  {journal} {Proceedings of the Royal Society of London. Series A.
  Mathematical and Physical Sciences}\ }\textbf {\bibinfo {volume} {165}},\
  \bibinfo {pages} {291} (\bibinfo {year} {1938})}\BibitemShut {NoStop}%
\bibitem [{\citenamefont {Amelino-Camelia}\ \emph {et~al.}(2012)\citenamefont
  {Amelino-Camelia}, \citenamefont {Arzano}, \citenamefont {Kowalski-Glikman},
  \citenamefont {Rosati},\ and\ \citenamefont
  {Trevisan}}]{AmelinoCamelia:2011nt}%
  \BibitemOpen
  \bibfield  {author} {\bibinfo {author} {\bibfnamefont {G.}~\bibnamefont
  {Amelino-Camelia}}, \bibinfo {author} {\bibfnamefont {M.}~\bibnamefont
  {Arzano}}, \bibinfo {author} {\bibfnamefont {J.}~\bibnamefont
  {Kowalski-Glikman}}, \bibinfo {author} {\bibfnamefont {G.}~\bibnamefont
  {Rosati}}, \ and\ \bibinfo {author} {\bibfnamefont {G.}~\bibnamefont
  {Trevisan}},\ }\href {\doibase 10.1088/0264-9381/29/7/075007} {\bibfield
  {journal} {\bibinfo  {journal} {Class. Quant. Grav.}\ }\textbf {\bibinfo
  {volume} {29}},\ \bibinfo {pages} {075007} (\bibinfo {year} {2012})},\
  \Eprint {http://arxiv.org/abs/1107.1724} {arXiv:1107.1724 [hep-th]}
  \BibitemShut {NoStop}%
\bibitem [{\citenamefont {Lobo}\ and\ \citenamefont
  {Palmisano}(2016)}]{Lobo:2016blj}%
  \BibitemOpen
  \bibfield  {author} {\bibinfo {author} {\bibfnamefont {I.~P.}\ \bibnamefont
  {Lobo}}\ and\ \bibinfo {author} {\bibfnamefont {G.}~\bibnamefont
  {Palmisano}},\ }\bibfield  {booktitle} {\emph {\bibinfo {booktitle}
  {{Proceedings, 9th Alexander Friedmann International Seminar on Gravitation
  and Cosmology and 3rd Satellite Symposium on the Casimir Effect: St.
  Petersburg, Russia, June 21-27, 2015}}},\ }\href {\doibase
  10.1142/S2010194516601265} {\bibfield  {journal} {\bibinfo  {journal} {Int.
  J. Mod. Phys. Conf. Ser.}\ }\textbf {\bibinfo {volume} {41}},\ \bibinfo
  {pages} {1660126} (\bibinfo {year} {2016})},\ \Eprint
  {http://arxiv.org/abs/1612.00326} {arXiv:1612.00326 [hep-th]} \BibitemShut
  {NoStop}%
\bibitem [{\citenamefont {Jafari}\ and\ \citenamefont
  {Shariati}(2006)}]{Jafari:2006rr}%
  \BibitemOpen
  \bibfield  {author} {\bibinfo {author} {\bibfnamefont {N.}~\bibnamefont
  {Jafari}}\ and\ \bibinfo {author} {\bibfnamefont {A.}~\bibnamefont
  {Shariati}},\ }\bibfield  {booktitle} {\emph {\bibinfo {booktitle} {{A
  Century of relativity physics, Proceedings, 28th Spanish Relativity Meeting,
  ERE 2005, Oviedo, Spain, September 6-10, 2005}}},\ }\href {\doibase
  10.1063/1.2218214} {\bibfield  {journal} {\bibinfo  {journal} {AIP Conf.
  Proc.}\ }\textbf {\bibinfo {volume} {841}},\ \bibinfo {pages} {462} (\bibinfo
  {year} {2006})},\ \Eprint {http://arxiv.org/abs/gr-qc/0602075}
  {arXiv:gr-qc/0602075 [gr-qc]} \BibitemShut {NoStop}%
\bibitem [{\citenamefont {Albalate}\ \emph {et~al.}(2018)\citenamefont
  {Albalate}, \citenamefont {Carmona}, \citenamefont {Cortés},\ and\
  \citenamefont {Relancio}}]{Albalate:2018kcf}%
  \BibitemOpen
  \bibfield  {author} {\bibinfo {author} {\bibfnamefont {G.}~\bibnamefont
  {Albalate}}, \bibinfo {author} {\bibfnamefont {J.}~\bibnamefont {Carmona}},
  \bibinfo {author} {\bibfnamefont {J.~L.}\ \bibnamefont {Cortés}}, \ and\
  \bibinfo {author} {\bibfnamefont {J.}~\bibnamefont {Relancio}},\ }\href
  {\doibase 10.3390/sym10100432} {\bibfield  {journal} {\bibinfo  {journal}
  {Symmetry}\ }\textbf {\bibinfo {volume} {10}},\ \bibinfo {pages} {432}
  (\bibinfo {year} {2018})},\ \Eprint {http://arxiv.org/abs/1809.08167}
  {arXiv:1809.08167 [hep-ph]} \BibitemShut {NoStop}%
\bibitem [{\citenamefont {Amelino-Camelia}(2012)}]{AmelinoCamelia:2011yi}%
  \BibitemOpen
  \bibfield  {author} {\bibinfo {author} {\bibfnamefont {G.}~\bibnamefont
  {Amelino-Camelia}},\ }\href {\doibase 10.1103/PhysRevD.85.084034} {\bibfield
  {journal} {\bibinfo  {journal} {Phys.Rev.}\ }\textbf {\bibinfo {volume}
  {D85}},\ \bibinfo {pages} {084034} (\bibinfo {year} {2012})},\ \Eprint
  {http://arxiv.org/abs/1110.5081} {arXiv:1110.5081 [hep-th]} \BibitemShut
  {NoStop}%
\bibitem [{\citenamefont {Carmona}\ \emph {et~al.}(2012)\citenamefont
  {Carmona}, \citenamefont {Cortes},\ and\ \citenamefont
  {Mercati}}]{Carmona:2012un}%
  \BibitemOpen
  \bibfield  {author} {\bibinfo {author} {\bibfnamefont {J.}~\bibnamefont
  {Carmona}}, \bibinfo {author} {\bibfnamefont {J.}~\bibnamefont {Cortes}}, \
  and\ \bibinfo {author} {\bibfnamefont {F.}~\bibnamefont {Mercati}},\ }\href
  {\doibase 10.1103/PhysRevD.86.084032} {\bibfield  {journal} {\bibinfo
  {journal} {Phys.\ Rev.\ D}\ }\textbf {\bibinfo {volume} {86}},\ \bibinfo
  {pages} {084032} (\bibinfo {year} {2012})},\ \Eprint
  {http://arxiv.org/abs/1206.5961} {arXiv:1206.5961 [hep-th]} \BibitemShut
  {NoStop}%
\bibitem [{\citenamefont {Carmona}\ \emph {et~al.}(2016)\citenamefont
  {Carmona}, \citenamefont {Cortes},\ and\ \citenamefont
  {Relancio}}]{Carmona:2016obd}%
  \BibitemOpen
  \bibfield  {author} {\bibinfo {author} {\bibfnamefont {J.}~\bibnamefont
  {Carmona}}, \bibinfo {author} {\bibfnamefont {J.}~\bibnamefont {Cortes}}, \
  and\ \bibinfo {author} {\bibfnamefont {J.}~\bibnamefont {Relancio}},\ }\href
  {\doibase 10.1103/PhysRevD.94.084008} {\bibfield  {journal} {\bibinfo
  {journal} {Phys.\ Rev.\ D}\ }\textbf {\bibinfo {volume} {94}},\ \bibinfo
  {pages} {084008} (\bibinfo {year} {2016})},\ \Eprint
  {http://arxiv.org/abs/1609.01347} {arXiv:1609.01347 [hep-th]} \BibitemShut
  {NoStop}%
\bibitem [{\citenamefont {Battisti}\ and\ \citenamefont
  {Meljanac}(2010)}]{Battisti:2010sr}%
  \BibitemOpen
  \bibfield  {author} {\bibinfo {author} {\bibfnamefont {M.~V.}\ \bibnamefont
  {Battisti}}\ and\ \bibinfo {author} {\bibfnamefont {S.}~\bibnamefont
  {Meljanac}},\ }\href {\doibase 10.1103/PhysRevD.82.024028} {\bibfield
  {journal} {\bibinfo  {journal} {Phys. Rev.}\ }\textbf {\bibinfo {volume}
  {D82}},\ \bibinfo {pages} {024028} (\bibinfo {year} {2010})},\ \Eprint
  {http://arxiv.org/abs/1003.2108} {arXiv:1003.2108 [hep-th]} \BibitemShut
  {NoStop}%
\bibitem [{\citenamefont {Stecker}\ \emph {et~al.}(2015)\citenamefont
  {Stecker}, \citenamefont {Scully}, \citenamefont {Liberati},\ and\
  \citenamefont {Mattingly}}]{Stecker:2014oxa}%
  \BibitemOpen
  \bibfield  {author} {\bibinfo {author} {\bibfnamefont {F.~W.}\ \bibnamefont
  {Stecker}}, \bibinfo {author} {\bibfnamefont {S.~T.}\ \bibnamefont {Scully}},
  \bibinfo {author} {\bibfnamefont {S.}~\bibnamefont {Liberati}}, \ and\
  \bibinfo {author} {\bibfnamefont {D.}~\bibnamefont {Mattingly}},\ }\href
  {\doibase 10.1103/PhysRevD.91.045009} {\bibfield  {journal} {\bibinfo
  {journal} {Phys. Rev.}\ }\textbf {\bibinfo {volume} {D91}},\ \bibinfo {pages}
  {045009} (\bibinfo {year} {2015})},\ \Eprint {http://arxiv.org/abs/1411.5889}
  {arXiv:1411.5889 [hep-ph]} \BibitemShut {NoStop}%
\bibitem [{\citenamefont {Carmona}\ \emph
  {et~al.}(2019{\natexlab{b}})\citenamefont {Carmona}, \citenamefont {Cortes},
  \citenamefont {Relancio},\ and\ \citenamefont {Reyes}}]{Carmona:2019xxp}%
  \BibitemOpen
  \bibfield  {author} {\bibinfo {author} {\bibfnamefont {J.~M.}\ \bibnamefont
  {Carmona}}, \bibinfo {author} {\bibfnamefont {J.~L.}\ \bibnamefont {Cortes}},
  \bibinfo {author} {\bibfnamefont {J.~J.}\ \bibnamefont {Relancio}}, \ and\
  \bibinfo {author} {\bibfnamefont {M.~A.}\ \bibnamefont {Reyes}},\ }\href
  {\doibase 10.3390/sym11111419} {\bibfield  {journal} {\bibinfo  {journal}
  {Symmetry}\ }\textbf {\bibinfo {volume} {11}},\ \bibinfo {pages} {1419}
  (\bibinfo {year} {2019}{\natexlab{b}})},\ \Eprint
  {http://arxiv.org/abs/1911.12710} {arXiv:1911.12710 [hep-ph]} \BibitemShut
  {NoStop}%
\bibitem [{\citenamefont {Collins}\ \emph {et~al.}(2004)\citenamefont
  {Collins}, \citenamefont {Perez}, \citenamefont {Sudarsky}, \citenamefont
  {Urrutia},\ and\ \citenamefont {Vucetich}}]{Collins2004}%
  \BibitemOpen
  \bibfield  {author} {\bibinfo {author} {\bibfnamefont {J.}~\bibnamefont
  {Collins}}, \bibinfo {author} {\bibfnamefont {A.}~\bibnamefont {Perez}},
  \bibinfo {author} {\bibfnamefont {D.}~\bibnamefont {Sudarsky}}, \bibinfo
  {author} {\bibfnamefont {L.}~\bibnamefont {Urrutia}}, \ and\ \bibinfo
  {author} {\bibfnamefont {H.}~\bibnamefont {Vucetich}},\ }\href {\doibase
  10.1103/PhysRevLett.93.191301} {\bibfield  {journal} {\bibinfo  {journal}
  {Phys. Rev. Lett.}\ }\textbf {\bibinfo {volume} {93}},\ \bibinfo {pages}
  {191301} (\bibinfo {year} {2004})},\ \Eprint
  {http://arxiv.org/abs/gr-qc/0403053} {arXiv:gr-qc/0403053 [gr-qc]}
  \BibitemShut {NoStop}%
\bibitem [{\citenamefont {Bolokhov}\ \emph {et~al.}(2005)\citenamefont
  {Bolokhov}, \citenamefont {Groot~Nibbelink},\ and\ \citenamefont
  {Pospelov}}]{Bolokhov2005}%
  \BibitemOpen
  \bibfield  {author} {\bibinfo {author} {\bibfnamefont {P.~A.}\ \bibnamefont
  {Bolokhov}}, \bibinfo {author} {\bibfnamefont {S.}~\bibnamefont
  {Groot~Nibbelink}}, \ and\ \bibinfo {author} {\bibfnamefont {M.}~\bibnamefont
  {Pospelov}},\ }\href {\doibase 10.1103/PhysRevD.72.015013} {\bibfield
  {journal} {\bibinfo  {journal} {Phys. Rev.}\ }\textbf {\bibinfo {volume}
  {D72}},\ \bibinfo {pages} {015013} (\bibinfo {year} {2005})},\ \Eprint
  {http://arxiv.org/abs/hep-ph/0505029} {arXiv:hep-ph/0505029 [hep-ph]}
  \BibitemShut {NoStop}%
\bibitem [{\citenamefont {Kislat}\ and\ \citenamefont
  {Krawczynski}(2015)}]{Kislat2015}%
  \BibitemOpen
  \bibfield  {author} {\bibinfo {author} {\bibfnamefont {F.}~\bibnamefont
  {Kislat}}\ and\ \bibinfo {author} {\bibfnamefont {H.}~\bibnamefont
  {Krawczynski}},\ }\href {\doibase 10.1103/PhysRevD.92.045016} {\bibfield
  {journal} {\bibinfo  {journal} {Phys. Rev.}\ }\textbf {\bibinfo {volume}
  {D92}},\ \bibinfo {pages} {045016} (\bibinfo {year} {2015})},\ \Eprint
  {http://arxiv.org/abs/1505.02669} {arXiv:1505.02669 [astro-ph.HE]}
  \BibitemShut {NoStop}%
\bibitem [{\citenamefont {Judes}\ and\ \citenamefont
  {Visser}(2003)}]{Judes:2002bw}%
  \BibitemOpen
  \bibfield  {author} {\bibinfo {author} {\bibfnamefont {S.}~\bibnamefont
  {Judes}}\ and\ \bibinfo {author} {\bibfnamefont {M.}~\bibnamefont {Visser}},\
  }\href {\doibase 10.1103/PhysRevD.68.045001} {\bibfield  {journal} {\bibinfo
  {journal} {Phys.Rev.}\ }\textbf {\bibinfo {volume} {D68}},\ \bibinfo {pages}
  {045001} (\bibinfo {year} {2003})},\ \Eprint
  {http://arxiv.org/abs/gr-qc/0205067} {arXiv:gr-qc/0205067 [gr-qc]}
  \BibitemShut {NoStop}%
\bibitem [{\citenamefont {Ivetic}\ \emph {et~al.}(2016)\citenamefont {Ivetic},
  \citenamefont {Mignemi},\ and\ \citenamefont {Samsarov}}]{Ivetic:2016qtz}%
  \BibitemOpen
  \bibfield  {author} {\bibinfo {author} {\bibfnamefont {B.}~\bibnamefont
  {Ivetic}}, \bibinfo {author} {\bibfnamefont {S.}~\bibnamefont {Mignemi}}, \
  and\ \bibinfo {author} {\bibfnamefont {A.}~\bibnamefont {Samsarov}},\
  }\href@noop {} {\  (\bibinfo {year} {2016})},\ \Eprint
  {http://arxiv.org/abs/1606.04692} {arXiv:1606.04692 [hep-th]} \BibitemShut
  {NoStop}%
\bibitem [{\citenamefont {Majid}(2000)}]{Majid:1999tc}%
  \BibitemOpen
  \bibfield  {author} {\bibinfo {author} {\bibfnamefont {S.}~\bibnamefont
  {Majid}},\ }\bibfield  {booktitle} {\emph {\bibinfo {booktitle} {{Towards
  quantum gravity. Proceedings, 35th International Winter School on theoretical
  physics, Polanica, Poland, February 2-11, 1999}}},\ }\href@noop {} {\bibfield
   {journal} {\bibinfo  {journal} {Lect. Notes Phys.}\ }\textbf {\bibinfo
  {volume} {541}},\ \bibinfo {pages} {227} (\bibinfo {year} {2000})},\ \Eprint
  {http://arxiv.org/abs/hep-th/0006166} {arXiv:hep-th/0006166 [hep-th]}
  \BibitemShut {NoStop}%
\bibitem [{\citenamefont {Kowalski-Glikman}(2002)}]{KowalskiGlikman:2002ft}%
  \BibitemOpen
  \bibfield  {author} {\bibinfo {author} {\bibfnamefont {J.}~\bibnamefont
  {Kowalski-Glikman}},\ }\href {\doibase 10.1016/S0370-2693(02)02762-4}
  {\bibfield  {journal} {\bibinfo  {journal} {Phys. Lett.}\ }\textbf {\bibinfo
  {volume} {B547}},\ \bibinfo {pages} {291} (\bibinfo {year} {2002})},\ \Eprint
  {http://arxiv.org/abs/hep-th/0207279} {arXiv:hep-th/0207279 [hep-th]}
  \BibitemShut {NoStop}%
\bibitem [{\citenamefont {Amelino-Camelia}\ \emph {et~al.}(2016)\citenamefont
  {Amelino-Camelia}, \citenamefont {Gubitosi},\ and\ \citenamefont
  {Palmisano}}]{Amelino-Camelia:2013sba}%
  \BibitemOpen
  \bibfield  {author} {\bibinfo {author} {\bibfnamefont {G.}~\bibnamefont
  {Amelino-Camelia}}, \bibinfo {author} {\bibfnamefont {G.}~\bibnamefont
  {Gubitosi}}, \ and\ \bibinfo {author} {\bibfnamefont {G.}~\bibnamefont
  {Palmisano}},\ }\href {\doibase 10.1142/S0218271816500279} {\bibfield
  {journal} {\bibinfo  {journal} {Int. J. Mod. Phys.}\ }\textbf {\bibinfo
  {volume} {D25}},\ \bibinfo {pages} {1650027} (\bibinfo {year} {2016})},\
  \Eprint {http://arxiv.org/abs/1307.7988} {arXiv:1307.7988 [gr-qc]}
  \BibitemShut {NoStop}%
\bibitem [{\citenamefont {Carmona}\ \emph
  {et~al.}(2019{\natexlab{c}})\citenamefont {Carmona}, \citenamefont
  {Cortés},\ and\ \citenamefont {Relancio}}]{Carmona:2019fwf}%
  \BibitemOpen
  \bibfield  {author} {\bibinfo {author} {\bibfnamefont {J.~M.}\ \bibnamefont
  {Carmona}}, \bibinfo {author} {\bibfnamefont {J.~L.}\ \bibnamefont
  {Cortés}}, \ and\ \bibinfo {author} {\bibfnamefont {J.~J.}\ \bibnamefont
  {Relancio}},\ }\href {\doibase 10.1103/PhysRevD.100.104031} {\bibfield
  {journal} {\bibinfo  {journal} {Phys. Rev.}\ }\textbf {\bibinfo {volume}
  {D100}},\ \bibinfo {pages} {104031} (\bibinfo {year} {2019}{\natexlab{c}})},\
  \Eprint {http://arxiv.org/abs/1907.12298} {arXiv:1907.12298 [hep-th]}
  \BibitemShut {NoStop}%
\bibitem [{\citenamefont {Meljanac}\ \emph {et~al.}(2009)\citenamefont
  {Meljanac}, \citenamefont {Meljanac}, \citenamefont {Samsarov},\ and\
  \citenamefont {Stojic}}]{Meljanac:2009ej}%
  \BibitemOpen
  \bibfield  {author} {\bibinfo {author} {\bibfnamefont {S.}~\bibnamefont
  {Meljanac}}, \bibinfo {author} {\bibfnamefont {D.}~\bibnamefont {Meljanac}},
  \bibinfo {author} {\bibfnamefont {A.}~\bibnamefont {Samsarov}}, \ and\
  \bibinfo {author} {\bibfnamefont {M.}~\bibnamefont {Stojic}},\ }\href@noop {}
  {\  (\bibinfo {year} {2009})},\ \Eprint {http://arxiv.org/abs/0909.1706}
  {arXiv:0909.1706 [math-ph]} \BibitemShut {NoStop}%
\bibitem [{\citenamefont {Chern}\ \emph {et~al.}(1999)\citenamefont {Chern},
  \citenamefont {Chen},\ and\ \citenamefont {Lam}}]{Chern:1999jn}%
  \BibitemOpen
  \bibfield  {author} {\bibinfo {author} {\bibfnamefont {S.~S.}\ \bibnamefont
  {Chern}}, \bibinfo {author} {\bibfnamefont {W.~H.}\ \bibnamefont {Chen}}, \
  and\ \bibinfo {author} {\bibfnamefont {K.~S.}\ \bibnamefont {Lam}},\
  }\href@noop {} {\emph {\bibinfo {title} {{Lectures on differential
  geometry}}}}\ (\bibinfo  {publisher} {World Scientific},\ \bibinfo {year}
  {1999})\ \bibinfo {note} {, see Eqs. (1.30) and (1.31) of Chapter
  6}\BibitemShut {NoStop}%
\bibitem [{\citenamefont {Gubitosi}\ and\ \citenamefont
  {Mercati}(2013)}]{Gubitosi:2013rna}%
  \BibitemOpen
  \bibfield  {author} {\bibinfo {author} {\bibfnamefont {G.}~\bibnamefont
  {Gubitosi}}\ and\ \bibinfo {author} {\bibfnamefont {F.}~\bibnamefont
  {Mercati}},\ }\href {\doibase 10.1088/0264-9381/30/14/145002} {\bibfield
  {journal} {\bibinfo  {journal} {Class. Quant. Grav.}\ }\textbf {\bibinfo
  {volume} {30}},\ \bibinfo {pages} {145002} (\bibinfo {year} {2013})},\
  \Eprint {http://arxiv.org/abs/1106.5710} {arXiv:1106.5710 [gr-qc]}
  \BibitemShut {NoStop}%
\bibitem [{\citenamefont {Banburski}\ and\ \citenamefont
  {Freidel}(2014)}]{Banburski:2013jfa}%
  \BibitemOpen
  \bibfield  {author} {\bibinfo {author} {\bibfnamefont {A.}~\bibnamefont
  {Banburski}}\ and\ \bibinfo {author} {\bibfnamefont {L.}~\bibnamefont
  {Freidel}},\ }\href {\doibase 10.1103/PhysRevD.90.076010} {\bibfield
  {journal} {\bibinfo  {journal} {Phys. Rev.}\ }\textbf {\bibinfo {volume}
  {D90}},\ \bibinfo {pages} {076010} (\bibinfo {year} {2014})},\ \Eprint
  {http://arxiv.org/abs/1308.0300} {arXiv:1308.0300 [gr-qc]} \BibitemShut
  {NoStop}%
\bibitem [{\citenamefont {Kosinski}\ and\ \citenamefont
  {Maslanka}(2012)}]{Kosinski_paring}%
  \BibitemOpen
  \bibfield  {author} {\bibinfo {author} {\bibfnamefont {P.}~\bibnamefont
  {Kosinski}}\ and\ \bibinfo {author} {\bibfnamefont {P.}~\bibnamefont
  {Maslanka}},\ }\enquote {\bibinfo {title} {The kappa-weyl group and its
  algebra},}\ in\ \href {\doibase 10.1142/9789812830425_0003} {\emph {\bibinfo
  {booktitle} {From field theory to quantum groups}}}\ (\bibinfo  {publisher}
  {World Scientific},\ \bibinfo {year} {2012})\ pp.\ \bibinfo {pages}
  {41--51}\BibitemShut {NoStop}%
\bibitem [{\citenamefont {Carmona}\ \emph
  {et~al.}(2018{\natexlab{c}})\citenamefont {Carmona}, \citenamefont {Cortes},\
  and\ \citenamefont {Relancio}}]{Carmona:2017cry}%
  \BibitemOpen
  \bibfield  {author} {\bibinfo {author} {\bibfnamefont {J.~M.}\ \bibnamefont
  {Carmona}}, \bibinfo {author} {\bibfnamefont {J.~L.}\ \bibnamefont {Cortes}},
  \ and\ \bibinfo {author} {\bibfnamefont {J.~J.}\ \bibnamefont {Relancio}},\
  }\href {\doibase 10.1103/PhysRevD.97.064025} {\bibfield  {journal} {\bibinfo
  {journal} {Phys. Rev.}\ }\textbf {\bibinfo {volume} {D97}},\ \bibinfo {pages}
  {064025} (\bibinfo {year} {2018}{\natexlab{c}})},\ \Eprint
  {http://arxiv.org/abs/1711.08403} {arXiv:1711.08403 [hep-th]} \BibitemShut
  {NoStop}%
\bibitem [{\citenamefont {Meljanac}\ \emph {et~al.}(2017)\citenamefont
  {Meljanac}, \citenamefont {Meljanac}, \citenamefont {Mercati},\ and\
  \citenamefont {PikutiÄ‡}}]{Meljanac:2016jwk}%
  \BibitemOpen
  \bibfield  {author} {\bibinfo {author} {\bibfnamefont {S.}~\bibnamefont
  {Meljanac}}, \bibinfo {author} {\bibfnamefont {D.}~\bibnamefont {Meljanac}},
  \bibinfo {author} {\bibfnamefont {F.}~\bibnamefont {Mercati}}, \ and\
  \bibinfo {author} {\bibfnamefont {D.}~\bibnamefont {PikutiÄ‡}},\ }\href
  {\doibase 10.1016/j.physletb.2017.01.006} {\bibfield  {journal} {\bibinfo
  {journal} {Phys. Lett.}\ }\textbf {\bibinfo {volume} {B766}},\ \bibinfo
  {pages} {181} (\bibinfo {year} {2017})},\ \Eprint
  {http://arxiv.org/abs/1610.06716} {arXiv:1610.06716 [hep-th]} \BibitemShut
  {NoStop}%
\bibitem [{\citenamefont {Loret}\ \emph {et~al.}(2017)\citenamefont {Loret},
  \citenamefont {Meljanac}, \citenamefont {Mercati},\ and\ \citenamefont
  {Pikuti\'{c}}}]{Loret:2016jrg}%
  \BibitemOpen
  \bibfield  {author} {\bibinfo {author} {\bibfnamefont {N.}~\bibnamefont
  {Loret}}, \bibinfo {author} {\bibfnamefont {S.}~\bibnamefont {Meljanac}},
  \bibinfo {author} {\bibfnamefont {F.}~\bibnamefont {Mercati}}, \ and\
  \bibinfo {author} {\bibfnamefont {D.}~\bibnamefont {Pikuti\'{c}}},\ }\href
  {\doibase 10.1142/S0218271817501231} {\bibfield  {journal} {\bibinfo
  {journal} {Int. J. Mod. Phys.}\ }\textbf {\bibinfo {volume} {D26}},\ \bibinfo
  {pages} {1750123} (\bibinfo {year} {2017})},\ \Eprint
  {http://arxiv.org/abs/1610.08310} {arXiv:1610.08310 [hep-th]} \BibitemShut
  {NoStop}%
\bibitem [{\citenamefont {Kowalski-Glikman}\ and\ \citenamefont
  {Nowak}(2003)}]{KowalskiGlikman:2002jr}%
  \BibitemOpen
  \bibfield  {author} {\bibinfo {author} {\bibfnamefont {J.}~\bibnamefont
  {Kowalski-Glikman}}\ and\ \bibinfo {author} {\bibfnamefont {S.}~\bibnamefont
  {Nowak}},\ }\href {\doibase 10.1142/S0218271803003050} {\bibfield  {journal}
  {\bibinfo  {journal} {Int. J. Mod. Phys.}\ }\textbf {\bibinfo {volume}
  {D12}},\ \bibinfo {pages} {299} (\bibinfo {year} {2003})},\ \Eprint
  {http://arxiv.org/abs/hep-th/0204245} {arXiv:hep-th/0204245 [hep-th]}
  \BibitemShut {NoStop}%
\bibitem [{\citenamefont {Carmona}\ \emph {et~al.}(2020)\citenamefont
  {Carmona}, \citenamefont {Cortés},\ and\ \citenamefont
  {Relancio}}]{Carmona:2019vsh}%
  \BibitemOpen
  \bibfield  {author} {\bibinfo {author} {\bibfnamefont {J.~M.}\ \bibnamefont
  {Carmona}}, \bibinfo {author} {\bibfnamefont {J.~L.}\ \bibnamefont
  {Cortés}}, \ and\ \bibinfo {author} {\bibfnamefont {J.~J.}\ \bibnamefont
  {Relancio}},\ }\href {\doibase 10.1103/PhysRevD.101.044057} {\bibfield
  {journal} {\bibinfo  {journal} {Phys. Rev.}\ }\textbf {\bibinfo {volume}
  {D101}},\ \bibinfo {pages} {044057} (\bibinfo {year} {2020})},\ \Eprint
  {http://arxiv.org/abs/1912.12885} {arXiv:1912.12885 [hep-th]} \BibitemShut
  {NoStop}%
\bibitem [{\citenamefont {Loret}(2014)}]{Loret:2014uia}%
  \BibitemOpen
  \bibfield  {author} {\bibinfo {author} {\bibfnamefont {N.}~\bibnamefont
  {Loret}},\ }\href {\doibase 10.1103/PhysRevD.90.124013} {\bibfield  {journal}
  {\bibinfo  {journal} {Phys. Rev.}\ }\textbf {\bibinfo {volume} {D90}},\
  \bibinfo {pages} {124013} (\bibinfo {year} {2014})},\ \Eprint
  {http://arxiv.org/abs/1404.5093} {arXiv:1404.5093 [hep-th]} \BibitemShut
  {NoStop}%
\bibitem [{\citenamefont {Mignemi}\ and\ \citenamefont
  {Samsarov}(2017)}]{Mignemi:2016ilu}%
  \BibitemOpen
  \bibfield  {author} {\bibinfo {author} {\bibfnamefont {S.}~\bibnamefont
  {Mignemi}}\ and\ \bibinfo {author} {\bibfnamefont {A.}~\bibnamefont
  {Samsarov}},\ }\href {\doibase 10.1016/j.physleta.2017.03.033} {\bibfield
  {journal} {\bibinfo  {journal} {Phys. Lett.}\ }\textbf {\bibinfo {volume}
  {A381}},\ \bibinfo {pages} {1655} (\bibinfo {year} {2017})},\ \Eprint
  {http://arxiv.org/abs/1610.09692} {arXiv:1610.09692 [hep-th]} \BibitemShut
  {NoStop}%
\bibitem [{\citenamefont {Maggiore}(1993)}]{Maggiore:1993kv}%
  \BibitemOpen
  \bibfield  {author} {\bibinfo {author} {\bibfnamefont {M.}~\bibnamefont
  {Maggiore}},\ }\href {\doibase 10.1016/0370-2693(93)90785-G} {\bibfield
  {journal} {\bibinfo  {journal} {Phys. Lett.}\ }\textbf {\bibinfo {volume}
  {B319}},\ \bibinfo {pages} {83} (\bibinfo {year} {1993})},\ \Eprint
  {http://arxiv.org/abs/hep-th/9309034} {arXiv:hep-th/9309034 [hep-th]}
  \BibitemShut {NoStop}%
\bibitem [{\citenamefont {Hewett}\ \emph {et~al.}(2001)\citenamefont {Hewett},
  \citenamefont {Petriello},\ and\ \citenamefont {Rizzo}}]{Hewett:2000zp}%
  \BibitemOpen
  \bibfield  {author} {\bibinfo {author} {\bibfnamefont {J.~L.}\ \bibnamefont
  {Hewett}}, \bibinfo {author} {\bibfnamefont {F.~J.}\ \bibnamefont
  {Petriello}}, \ and\ \bibinfo {author} {\bibfnamefont {T.~G.}\ \bibnamefont
  {Rizzo}},\ }\href {\doibase 10.1103/PhysRevD.64.075012} {\bibfield  {journal}
  {\bibinfo  {journal} {Phys. Rev.}\ }\textbf {\bibinfo {volume} {D64}},\
  \bibinfo {pages} {075012} (\bibinfo {year} {2001})},\ \Eprint
  {http://arxiv.org/abs/hep-ph/0010354} {arXiv:hep-ph/0010354 [hep-ph]}
  \BibitemShut {NoStop}%
\bibitem [{\citenamefont {Hewett}\ \emph {et~al.}(2002)\citenamefont {Hewett},
  \citenamefont {Petriello},\ and\ \citenamefont {Rizzo}}]{Hewett:2001im}%
  \BibitemOpen
  \bibfield  {author} {\bibinfo {author} {\bibfnamefont {J.~L.}\ \bibnamefont
  {Hewett}}, \bibinfo {author} {\bibfnamefont {F.~J.}\ \bibnamefont
  {Petriello}}, \ and\ \bibinfo {author} {\bibfnamefont {T.~G.}\ \bibnamefont
  {Rizzo}},\ }\href {\doibase 10.1103/PhysRevD.66.036001} {\bibfield  {journal}
  {\bibinfo  {journal} {Phys. Rev.}\ }\textbf {\bibinfo {volume} {D66}},\
  \bibinfo {pages} {036001} (\bibinfo {year} {2002})},\ \Eprint
  {http://arxiv.org/abs/hep-ph/0112003} {arXiv:hep-ph/0112003 [hep-ph]}
  \BibitemShut {NoStop}%
\bibitem [{\citenamefont {Mathews}(2001)}]{Mathews:2000we}%
  \BibitemOpen
  \bibfield  {author} {\bibinfo {author} {\bibfnamefont {P.}~\bibnamefont
  {Mathews}},\ }\href {\doibase 10.1103/PhysRevD.63.075007} {\bibfield
  {journal} {\bibinfo  {journal} {Phys. Rev.}\ }\textbf {\bibinfo {volume}
  {D63}},\ \bibinfo {pages} {075007} (\bibinfo {year} {2001})},\ \Eprint
  {http://arxiv.org/abs/hep-ph/0011332} {arXiv:hep-ph/0011332 [hep-ph]}
  \BibitemShut {NoStop}%
\bibitem [{\citenamefont {Alboteanu}\ \emph {et~al.}(2006)\citenamefont
  {Alboteanu}, \citenamefont {Ohl},\ and\ \citenamefont
  {Ruckl}}]{Alboteanu:2006hh}%
  \BibitemOpen
  \bibfield  {author} {\bibinfo {author} {\bibfnamefont {A.}~\bibnamefont
  {Alboteanu}}, \bibinfo {author} {\bibfnamefont {T.}~\bibnamefont {Ohl}}, \
  and\ \bibinfo {author} {\bibfnamefont {R.}~\bibnamefont {Ruckl}},\ }\href
  {\doibase 10.1103/PhysRevD.74.096004} {\bibfield  {journal} {\bibinfo
  {journal} {Phys. Rev.}\ }\textbf {\bibinfo {volume} {D74}},\ \bibinfo {pages}
  {096004} (\bibinfo {year} {2006})},\ \Eprint
  {http://arxiv.org/abs/hep-ph/0608155} {arXiv:hep-ph/0608155 [hep-ph]}
  \BibitemShut {NoStop}%
\bibitem [{\citenamefont {Ohl}\ and\ \citenamefont
  {Speckner}(2010)}]{Ohl:2010zf}%
  \BibitemOpen
  \bibfield  {author} {\bibinfo {author} {\bibfnamefont {T.}~\bibnamefont
  {Ohl}}\ and\ \bibinfo {author} {\bibfnamefont {C.}~\bibnamefont {Speckner}},\
  }\href {\doibase 10.1103/PhysRevD.82.116011} {\bibfield  {journal} {\bibinfo
  {journal} {Phys. Rev.}\ }\textbf {\bibinfo {volume} {D82}},\ \bibinfo {pages}
  {116011} (\bibinfo {year} {2010})},\ \Eprint {http://arxiv.org/abs/1008.4710}
  {arXiv:1008.4710 [hep-ph]} \BibitemShut {NoStop}%
\bibitem [{\citenamefont {Yaser~Ayazi}\ \emph {et~al.}(2012)\citenamefont
  {Yaser~Ayazi}, \citenamefont {Esmaeili},\ and\ \citenamefont
  {Mohammadi-Najafabadi}}]{YaserAyazi:2012ni}%
  \BibitemOpen
  \bibfield  {author} {\bibinfo {author} {\bibfnamefont {S.}~\bibnamefont
  {Yaser~Ayazi}}, \bibinfo {author} {\bibfnamefont {S.}~\bibnamefont
  {Esmaeili}}, \ and\ \bibinfo {author} {\bibfnamefont {M.}~\bibnamefont
  {Mohammadi-Najafabadi}},\ }\href {\doibase 10.1016/j.physletb.2012.04.063}
  {\bibfield  {journal} {\bibinfo  {journal} {Phys. Lett.}\ }\textbf {\bibinfo
  {volume} {B712}},\ \bibinfo {pages} {93} (\bibinfo {year} {2012})},\ \Eprint
  {http://arxiv.org/abs/1202.2505} {arXiv:1202.2505 [hep-ph]} \BibitemShut
  {NoStop}%
\bibitem [{\citenamefont {Hinchliffe}\ \emph {et~al.}(2004)\citenamefont
  {Hinchliffe}, \citenamefont {Kersting},\ and\ \citenamefont
  {Ma}}]{Hinchliffe:2002km}%
  \BibitemOpen
  \bibfield  {author} {\bibinfo {author} {\bibfnamefont {I.}~\bibnamefont
  {Hinchliffe}}, \bibinfo {author} {\bibfnamefont {N.}~\bibnamefont
  {Kersting}}, \ and\ \bibinfo {author} {\bibfnamefont {Y.~L.}\ \bibnamefont
  {Ma}},\ }\href {\doibase 10.1142/S0217751X04017094} {\bibfield  {journal}
  {\bibinfo  {journal} {Int. J. Mod. Phys.}\ }\textbf {\bibinfo {volume}
  {A19}},\ \bibinfo {pages} {179} (\bibinfo {year} {2004})},\ \Eprint
  {http://arxiv.org/abs/hep-ph/0205040} {arXiv:hep-ph/0205040 [hep-ph]}
  \BibitemShut {NoStop}%
\bibitem [{\citenamefont {Tanabashi}\ \emph {et~al.}(2018)\citenamefont
  {Tanabashi}, \citenamefont {Hagiwara}, \citenamefont {Hikasa}, \citenamefont
  {Nakamura}, \citenamefont {Sumino}, \citenamefont {Takahashi}, \citenamefont
  {Tanaka}, \citenamefont {Agashe}, \citenamefont {Aielli}, \citenamefont
  {Amsler}, \citenamefont {Antonelli}, \citenamefont {Asner}, \citenamefont
  {Baer}, \citenamefont {Banerjee}, \citenamefont {Barnett}, \citenamefont
  {Basaglia}, \citenamefont {Bauer}, \citenamefont {Beatty}, \citenamefont
  {Belousov}, \citenamefont {Beringer}, \citenamefont {Bethke}, \citenamefont
  {Bettini}, \citenamefont {Bichsel}, \citenamefont {Biebel}, \citenamefont
  {Black}, \citenamefont {Blucher}, \citenamefont {Buchmuller}, \citenamefont
  {Burkert}, \citenamefont {Bychkov}, \citenamefont {Cahn}, \citenamefont
  {Carena}, \citenamefont {Ceccucci}, \citenamefont {Cerri}, \citenamefont
  {Chakraborty}, \citenamefont {Chen}, \citenamefont {Chivukula}, \citenamefont
  {Cowan}, \citenamefont {Dahl}, \citenamefont {D'Ambrosio}, \citenamefont
  {Damour}, \citenamefont {de~Florian}, \citenamefont {de~Gouv\^ea},
  \citenamefont {DeGrand}, \citenamefont {de~Jong}, \citenamefont {Dissertori},
  \citenamefont {Dobrescu}, \citenamefont {D'Onofrio}, \citenamefont {Doser},
  \citenamefont {Drees}, \citenamefont {Dreiner}, \citenamefont {Dwyer},
  \citenamefont {Eerola}, \citenamefont {Eidelman}, \citenamefont {Ellis},
  \citenamefont {Erler}, \citenamefont {Ezhela}, \citenamefont {Fetscher},
  \citenamefont {Fields}, \citenamefont {Firestone}, \citenamefont {Foster},
  \citenamefont {Freitas}, \citenamefont {Gallagher}, \citenamefont {Garren},
  \citenamefont {Gerber}, \citenamefont {Gerbier}, \citenamefont {Gershon},
  \citenamefont {Gershtein}, \citenamefont {Gherghetta}, \citenamefont
  {Godizov}, \citenamefont {Goodman}, \citenamefont {Grab}, \citenamefont
  {Gritsan}, \citenamefont {Grojean}, \citenamefont {Groom}, \citenamefont
  {Gr\"unewald}, \citenamefont {Gurtu}, \citenamefont {Gutsche}, \citenamefont
  {Haber}, \citenamefont {Hanhart}, \citenamefont {Hashimoto}, \citenamefont
  {Hayato}, \citenamefont {Hayes}, \citenamefont {Hebecker}, \citenamefont
  {Heinemeyer}, \citenamefont {Heltsley}, \citenamefont {Hern\'andez-Rey},
  \citenamefont {Hisano}, \citenamefont {H\"ocker}, \citenamefont {Holder},
  \citenamefont {Holtkamp}, \citenamefont {Hyodo}, \citenamefont {Irwin},
  \citenamefont {Johnson}, \citenamefont {Kado}, \citenamefont {Karliner},
  \citenamefont {Katz}, \citenamefont {Klein}, \citenamefont {Klempt},
  \citenamefont {Kowalewski}, \citenamefont {Krauss}, \citenamefont {Kreps},
  \citenamefont {Krusche}, \citenamefont {Kuyanov}, \citenamefont {Kwon},
  \citenamefont {Lahav}, \citenamefont {Laiho}, \citenamefont {Lesgourgues},
  \citenamefont {Liddle}, \citenamefont {Ligeti}, \citenamefont {Lin},
  \citenamefont {Lippmann}, \citenamefont {Liss}, \citenamefont {Littenberg},
  \citenamefont {Lugovsky}, \citenamefont {Lugovsky}, \citenamefont {Lusiani},
  \citenamefont {Makida}, \citenamefont {Maltoni}, \citenamefont {Mannel},
  \citenamefont {Manohar}, \citenamefont {Marciano}, \citenamefont {Martin},
  \citenamefont {Masoni}, \citenamefont {Matthews}, \citenamefont
  {Mei\ss{}ner}, \citenamefont {Milstead}, \citenamefont {Mitchell},
  \citenamefont {M\"onig}, \citenamefont {Molaro}, \citenamefont {Moortgat},
  \citenamefont {Moskovic}, \citenamefont {Murayama}, \citenamefont {Narain},
  \citenamefont {Nason}, \citenamefont {Navas}, \citenamefont {Neubert},
  \citenamefont {Nevski}, \citenamefont {Nir}, \citenamefont {Olive},
  \citenamefont {Pagan~Griso}, \citenamefont {Parsons}, \citenamefont
  {Patrignani}, \citenamefont {Peacock}, \citenamefont {Pennington},
  \citenamefont {Petcov}, \citenamefont {Petrov}, \citenamefont {Pianori},
  \citenamefont {Piepke}, \citenamefont {Pomarol}, \citenamefont {Quadt},
  \citenamefont {Rademacker}, \citenamefont {Raffelt}, \citenamefont
  {Ratcliff}, \citenamefont {Richardson}, \citenamefont {Ringwald},
  \citenamefont {Roesler}, \citenamefont {Rolli}, \citenamefont {Romaniouk},
  \citenamefont {Rosenberg}, \citenamefont {Rosner}, \citenamefont {Rybka},
  \citenamefont {Ryutin}, \citenamefont {Sachrajda}, \citenamefont {Sakai},
  \citenamefont {Salam}, \citenamefont {Sarkar}, \citenamefont {Sauli},
  \citenamefont {Schneider}, \citenamefont {Scholberg}, \citenamefont
  {Schwartz}, \citenamefont {Scott}, \citenamefont {Sharma}, \citenamefont
  {Sharpe}, \citenamefont {Shutt}, \citenamefont {Silari}, \citenamefont
  {Sj\"ostrand}, \citenamefont {Skands}, \citenamefont {Skwarnicki},
  \citenamefont {Smith}, \citenamefont {Smoot}, \citenamefont {Spanier},
  \citenamefont {Spieler}, \citenamefont {Spiering}, \citenamefont {Stahl},
  \citenamefont {Stone}, \citenamefont {Sumiyoshi}, \citenamefont {Syphers},
  \citenamefont {Terashi}, \citenamefont {Terning}, \citenamefont {Thoma},
  \citenamefont {Thorne}, \citenamefont {Tiator}, \citenamefont {Titov},
  \citenamefont {Tkachenko}, \citenamefont {T\"ornqvist}, \citenamefont
  {Tovey}, \citenamefont {Valencia}, \citenamefont {Van~de Water},
  \citenamefont {Varelas}, \citenamefont {Venanzoni}, \citenamefont {Verde},
  \citenamefont {Vincter}, \citenamefont {Vogel}, \citenamefont {Vogt},
  \citenamefont {Wakely}, \citenamefont {Walkowiak}, \citenamefont {Walter},
  \citenamefont {Wands}, \citenamefont {Ward}, \citenamefont {Wascko},
  \citenamefont {Weiglein}, \citenamefont {Weinberg}, \citenamefont {Weinberg},
  \citenamefont {White}, \citenamefont {Wiencke}, \citenamefont {Willocq},
  \citenamefont {Wohl}, \citenamefont {Womersley}, \citenamefont {Woody},
  \citenamefont {Workman}, \citenamefont {Yao}, \citenamefont {Zeller},
  \citenamefont {Zenin}, \citenamefont {Zhu}, \citenamefont {Zhu},
  \citenamefont {Zimmermann}, \citenamefont {Zyla}, \citenamefont {Anderson},
  \citenamefont {Fuller}, \citenamefont {Lugovsky},\ and\ \citenamefont
  {Schaffner}}]{PhysRevD.98.030001}%
  \BibitemOpen
  \bibfield  {author} {\bibinfo {author} {\bibfnamefont {M.}~\bibnamefont
  {Tanabashi}}, \bibinfo {author} {\bibfnamefont {K.}~\bibnamefont {Hagiwara}},
  \bibinfo {author} {\bibfnamefont {K.}~\bibnamefont {Hikasa}}, \bibinfo
  {author} {\bibfnamefont {K.}~\bibnamefont {Nakamura}}, \bibinfo {author}
  {\bibfnamefont {Y.}~\bibnamefont {Sumino}}, \bibinfo {author} {\bibfnamefont
  {F.}~\bibnamefont {Takahashi}}, \bibinfo {author} {\bibfnamefont
  {J.}~\bibnamefont {Tanaka}}, \bibinfo {author} {\bibfnamefont
  {K.}~\bibnamefont {Agashe}}, \bibinfo {author} {\bibfnamefont
  {G.}~\bibnamefont {Aielli}}, \bibinfo {author} {\bibfnamefont
  {C.}~\bibnamefont {Amsler}}, \bibinfo {author} {\bibfnamefont
  {M.}~\bibnamefont {Antonelli}}, \bibinfo {author} {\bibfnamefont {D.~M.}\
  \bibnamefont {Asner}}, \bibinfo {author} {\bibfnamefont {H.}~\bibnamefont
  {Baer}}, \bibinfo {author} {\bibfnamefont {S.}~\bibnamefont {Banerjee}},
  \bibinfo {author} {\bibfnamefont {R.~M.}\ \bibnamefont {Barnett}}, \bibinfo
  {author} {\bibfnamefont {T.}~\bibnamefont {Basaglia}}, \bibinfo {author}
  {\bibfnamefont {C.~W.}\ \bibnamefont {Bauer}}, \bibinfo {author}
  {\bibfnamefont {J.~J.}\ \bibnamefont {Beatty}}, \bibinfo {author}
  {\bibfnamefont {V.~I.}\ \bibnamefont {Belousov}}, \bibinfo {author}
  {\bibfnamefont {J.}~\bibnamefont {Beringer}}, \bibinfo {author}
  {\bibfnamefont {S.}~\bibnamefont {Bethke}}, \bibinfo {author} {\bibfnamefont
  {A.}~\bibnamefont {Bettini}}, \bibinfo {author} {\bibfnamefont
  {H.}~\bibnamefont {Bichsel}}, \bibinfo {author} {\bibfnamefont
  {O.}~\bibnamefont {Biebel}}, \bibinfo {author} {\bibfnamefont {K.~M.}\
  \bibnamefont {Black}}, \bibinfo {author} {\bibfnamefont {E.}~\bibnamefont
  {Blucher}}, \bibinfo {author} {\bibfnamefont {O.}~\bibnamefont {Buchmuller}},
  \bibinfo {author} {\bibfnamefont {V.}~\bibnamefont {Burkert}}, \bibinfo
  {author} {\bibfnamefont {M.~A.}\ \bibnamefont {Bychkov}}, \bibinfo {author}
  {\bibfnamefont {R.~N.}\ \bibnamefont {Cahn}}, \bibinfo {author}
  {\bibfnamefont {M.}~\bibnamefont {Carena}}, \bibinfo {author} {\bibfnamefont
  {A.}~\bibnamefont {Ceccucci}}, \bibinfo {author} {\bibfnamefont
  {A.}~\bibnamefont {Cerri}}, \bibinfo {author} {\bibfnamefont
  {D.}~\bibnamefont {Chakraborty}}, \bibinfo {author} {\bibfnamefont {M.-C.}\
  \bibnamefont {Chen}}, \bibinfo {author} {\bibfnamefont {R.~S.}\ \bibnamefont
  {Chivukula}}, \bibinfo {author} {\bibfnamefont {G.}~\bibnamefont {Cowan}},
  \bibinfo {author} {\bibfnamefont {O.}~\bibnamefont {Dahl}}, \bibinfo {author}
  {\bibfnamefont {G.}~\bibnamefont {D'Ambrosio}}, \bibinfo {author}
  {\bibfnamefont {T.}~\bibnamefont {Damour}}, \bibinfo {author} {\bibfnamefont
  {D.}~\bibnamefont {de~Florian}}, \bibinfo {author} {\bibfnamefont
  {A.}~\bibnamefont {de~Gouv\^ea}}, \bibinfo {author} {\bibfnamefont
  {T.}~\bibnamefont {DeGrand}}, \bibinfo {author} {\bibfnamefont
  {P.}~\bibnamefont {de~Jong}}, \bibinfo {author} {\bibfnamefont
  {G.}~\bibnamefont {Dissertori}}, \bibinfo {author} {\bibfnamefont {B.~A.}\
  \bibnamefont {Dobrescu}}, \bibinfo {author} {\bibfnamefont {M.}~\bibnamefont
  {D'Onofrio}}, \bibinfo {author} {\bibfnamefont {M.}~\bibnamefont {Doser}},
  \bibinfo {author} {\bibfnamefont {M.}~\bibnamefont {Drees}}, \bibinfo
  {author} {\bibfnamefont {H.~K.}\ \bibnamefont {Dreiner}}, \bibinfo {author}
  {\bibfnamefont {D.~A.}\ \bibnamefont {Dwyer}}, \bibinfo {author}
  {\bibfnamefont {P.}~\bibnamefont {Eerola}}, \bibinfo {author} {\bibfnamefont
  {S.}~\bibnamefont {Eidelman}}, \bibinfo {author} {\bibfnamefont
  {J.}~\bibnamefont {Ellis}}, \bibinfo {author} {\bibfnamefont
  {J.}~\bibnamefont {Erler}}, \bibinfo {author} {\bibfnamefont {V.~V.}\
  \bibnamefont {Ezhela}}, \bibinfo {author} {\bibfnamefont {W.}~\bibnamefont
  {Fetscher}}, \bibinfo {author} {\bibfnamefont {B.~D.}\ \bibnamefont
  {Fields}}, \bibinfo {author} {\bibfnamefont {R.}~\bibnamefont {Firestone}},
  \bibinfo {author} {\bibfnamefont {B.}~\bibnamefont {Foster}}, \bibinfo
  {author} {\bibfnamefont {A.}~\bibnamefont {Freitas}}, \bibinfo {author}
  {\bibfnamefont {H.}~\bibnamefont {Gallagher}}, \bibinfo {author}
  {\bibfnamefont {L.}~\bibnamefont {Garren}}, \bibinfo {author} {\bibfnamefont
  {H.-J.}\ \bibnamefont {Gerber}}, \bibinfo {author} {\bibfnamefont
  {G.}~\bibnamefont {Gerbier}}, \bibinfo {author} {\bibfnamefont
  {T.}~\bibnamefont {Gershon}}, \bibinfo {author} {\bibfnamefont
  {Y.}~\bibnamefont {Gershtein}}, \bibinfo {author} {\bibfnamefont
  {T.}~\bibnamefont {Gherghetta}}, \bibinfo {author} {\bibfnamefont {A.~A.}\
  \bibnamefont {Godizov}}, \bibinfo {author} {\bibfnamefont {M.}~\bibnamefont
  {Goodman}}, \bibinfo {author} {\bibfnamefont {C.}~\bibnamefont {Grab}},
  \bibinfo {author} {\bibfnamefont {A.~V.}\ \bibnamefont {Gritsan}}, \bibinfo
  {author} {\bibfnamefont {C.}~\bibnamefont {Grojean}}, \bibinfo {author}
  {\bibfnamefont {D.~E.}\ \bibnamefont {Groom}}, \bibinfo {author}
  {\bibfnamefont {M.}~\bibnamefont {Gr\"unewald}}, \bibinfo {author}
  {\bibfnamefont {A.}~\bibnamefont {Gurtu}}, \bibinfo {author} {\bibfnamefont
  {T.}~\bibnamefont {Gutsche}}, \bibinfo {author} {\bibfnamefont {H.~E.}\
  \bibnamefont {Haber}}, \bibinfo {author} {\bibfnamefont {C.}~\bibnamefont
  {Hanhart}}, \bibinfo {author} {\bibfnamefont {S.}~\bibnamefont {Hashimoto}},
  \bibinfo {author} {\bibfnamefont {Y.}~\bibnamefont {Hayato}}, \bibinfo
  {author} {\bibfnamefont {K.~G.}\ \bibnamefont {Hayes}}, \bibinfo {author}
  {\bibfnamefont {A.}~\bibnamefont {Hebecker}}, \bibinfo {author}
  {\bibfnamefont {S.}~\bibnamefont {Heinemeyer}}, \bibinfo {author}
  {\bibfnamefont {B.}~\bibnamefont {Heltsley}}, \bibinfo {author}
  {\bibfnamefont {J.~J.}\ \bibnamefont {Hern\'andez-Rey}}, \bibinfo {author}
  {\bibfnamefont {J.}~\bibnamefont {Hisano}}, \bibinfo {author} {\bibfnamefont
  {A.}~\bibnamefont {H\"ocker}}, \bibinfo {author} {\bibfnamefont
  {J.}~\bibnamefont {Holder}}, \bibinfo {author} {\bibfnamefont
  {A.}~\bibnamefont {Holtkamp}}, \bibinfo {author} {\bibfnamefont
  {T.}~\bibnamefont {Hyodo}}, \bibinfo {author} {\bibfnamefont {K.~D.}\
  \bibnamefont {Irwin}}, \bibinfo {author} {\bibfnamefont {K.~F.}\ \bibnamefont
  {Johnson}}, \bibinfo {author} {\bibfnamefont {M.}~\bibnamefont {Kado}},
  \bibinfo {author} {\bibfnamefont {M.}~\bibnamefont {Karliner}}, \bibinfo
  {author} {\bibfnamefont {U.~F.}\ \bibnamefont {Katz}}, \bibinfo {author}
  {\bibfnamefont {S.~R.}\ \bibnamefont {Klein}}, \bibinfo {author}
  {\bibfnamefont {E.}~\bibnamefont {Klempt}}, \bibinfo {author} {\bibfnamefont
  {R.~V.}\ \bibnamefont {Kowalewski}}, \bibinfo {author} {\bibfnamefont
  {F.}~\bibnamefont {Krauss}}, \bibinfo {author} {\bibfnamefont
  {M.}~\bibnamefont {Kreps}}, \bibinfo {author} {\bibfnamefont
  {B.}~\bibnamefont {Krusche}}, \bibinfo {author} {\bibfnamefont {Y.~V.}\
  \bibnamefont {Kuyanov}}, \bibinfo {author} {\bibfnamefont {Y.}~\bibnamefont
  {Kwon}}, \bibinfo {author} {\bibfnamefont {O.}~\bibnamefont {Lahav}},
  \bibinfo {author} {\bibfnamefont {J.}~\bibnamefont {Laiho}}, \bibinfo
  {author} {\bibfnamefont {J.}~\bibnamefont {Lesgourgues}}, \bibinfo {author}
  {\bibfnamefont {A.}~\bibnamefont {Liddle}}, \bibinfo {author} {\bibfnamefont
  {Z.}~\bibnamefont {Ligeti}}, \bibinfo {author} {\bibfnamefont {C.-J.}\
  \bibnamefont {Lin}}, \bibinfo {author} {\bibfnamefont {C.}~\bibnamefont
  {Lippmann}}, \bibinfo {author} {\bibfnamefont {T.~M.}\ \bibnamefont {Liss}},
  \bibinfo {author} {\bibfnamefont {L.}~\bibnamefont {Littenberg}}, \bibinfo
  {author} {\bibfnamefont {K.~S.}\ \bibnamefont {Lugovsky}}, \bibinfo {author}
  {\bibfnamefont {S.~B.}\ \bibnamefont {Lugovsky}}, \bibinfo {author}
  {\bibfnamefont {A.}~\bibnamefont {Lusiani}}, \bibinfo {author} {\bibfnamefont
  {Y.}~\bibnamefont {Makida}}, \bibinfo {author} {\bibfnamefont
  {F.}~\bibnamefont {Maltoni}}, \bibinfo {author} {\bibfnamefont
  {T.}~\bibnamefont {Mannel}}, \bibinfo {author} {\bibfnamefont {A.~V.}\
  \bibnamefont {Manohar}}, \bibinfo {author} {\bibfnamefont {W.~J.}\
  \bibnamefont {Marciano}}, \bibinfo {author} {\bibfnamefont {A.~D.}\
  \bibnamefont {Martin}}, \bibinfo {author} {\bibfnamefont {A.}~\bibnamefont
  {Masoni}}, \bibinfo {author} {\bibfnamefont {J.}~\bibnamefont {Matthews}},
  \bibinfo {author} {\bibfnamefont {U.-G.}\ \bibnamefont {Mei\ss{}ner}},
  \bibinfo {author} {\bibfnamefont {D.}~\bibnamefont {Milstead}}, \bibinfo
  {author} {\bibfnamefont {R.~E.}\ \bibnamefont {Mitchell}}, \bibinfo {author}
  {\bibfnamefont {K.}~\bibnamefont {M\"onig}}, \bibinfo {author} {\bibfnamefont
  {P.}~\bibnamefont {Molaro}}, \bibinfo {author} {\bibfnamefont
  {F.}~\bibnamefont {Moortgat}}, \bibinfo {author} {\bibfnamefont
  {M.}~\bibnamefont {Moskovic}}, \bibinfo {author} {\bibfnamefont
  {H.}~\bibnamefont {Murayama}}, \bibinfo {author} {\bibfnamefont
  {M.}~\bibnamefont {Narain}}, \bibinfo {author} {\bibfnamefont
  {P.}~\bibnamefont {Nason}}, \bibinfo {author} {\bibfnamefont
  {S.}~\bibnamefont {Navas}}, \bibinfo {author} {\bibfnamefont
  {M.}~\bibnamefont {Neubert}}, \bibinfo {author} {\bibfnamefont
  {P.}~\bibnamefont {Nevski}}, \bibinfo {author} {\bibfnamefont
  {Y.}~\bibnamefont {Nir}}, \bibinfo {author} {\bibfnamefont {K.~A.}\
  \bibnamefont {Olive}}, \bibinfo {author} {\bibfnamefont {S.}~\bibnamefont
  {Pagan~Griso}}, \bibinfo {author} {\bibfnamefont {J.}~\bibnamefont
  {Parsons}}, \bibinfo {author} {\bibfnamefont {C.}~\bibnamefont {Patrignani}},
  \bibinfo {author} {\bibfnamefont {J.~A.}\ \bibnamefont {Peacock}}, \bibinfo
  {author} {\bibfnamefont {M.}~\bibnamefont {Pennington}}, \bibinfo {author}
  {\bibfnamefont {S.~T.}\ \bibnamefont {Petcov}}, \bibinfo {author}
  {\bibfnamefont {V.~A.}\ \bibnamefont {Petrov}}, \bibinfo {author}
  {\bibfnamefont {E.}~\bibnamefont {Pianori}}, \bibinfo {author} {\bibfnamefont
  {A.}~\bibnamefont {Piepke}}, \bibinfo {author} {\bibfnamefont
  {A.}~\bibnamefont {Pomarol}}, \bibinfo {author} {\bibfnamefont
  {A.}~\bibnamefont {Quadt}}, \bibinfo {author} {\bibfnamefont
  {J.}~\bibnamefont {Rademacker}}, \bibinfo {author} {\bibfnamefont
  {G.}~\bibnamefont {Raffelt}}, \bibinfo {author} {\bibfnamefont {B.~N.}\
  \bibnamefont {Ratcliff}}, \bibinfo {author} {\bibfnamefont {P.}~\bibnamefont
  {Richardson}}, \bibinfo {author} {\bibfnamefont {A.}~\bibnamefont
  {Ringwald}}, \bibinfo {author} {\bibfnamefont {S.}~\bibnamefont {Roesler}},
  \bibinfo {author} {\bibfnamefont {S.}~\bibnamefont {Rolli}}, \bibinfo
  {author} {\bibfnamefont {A.}~\bibnamefont {Romaniouk}}, \bibinfo {author}
  {\bibfnamefont {L.~J.}\ \bibnamefont {Rosenberg}}, \bibinfo {author}
  {\bibfnamefont {J.~L.}\ \bibnamefont {Rosner}}, \bibinfo {author}
  {\bibfnamefont {G.}~\bibnamefont {Rybka}}, \bibinfo {author} {\bibfnamefont
  {R.~A.}\ \bibnamefont {Ryutin}}, \bibinfo {author} {\bibfnamefont {C.~T.}\
  \bibnamefont {Sachrajda}}, \bibinfo {author} {\bibfnamefont {Y.}~\bibnamefont
  {Sakai}}, \bibinfo {author} {\bibfnamefont {G.~P.}\ \bibnamefont {Salam}},
  \bibinfo {author} {\bibfnamefont {S.}~\bibnamefont {Sarkar}}, \bibinfo
  {author} {\bibfnamefont {F.}~\bibnamefont {Sauli}}, \bibinfo {author}
  {\bibfnamefont {O.}~\bibnamefont {Schneider}}, \bibinfo {author}
  {\bibfnamefont {K.}~\bibnamefont {Scholberg}}, \bibinfo {author}
  {\bibfnamefont {A.~J.}\ \bibnamefont {Schwartz}}, \bibinfo {author}
  {\bibfnamefont {D.}~\bibnamefont {Scott}}, \bibinfo {author} {\bibfnamefont
  {V.}~\bibnamefont {Sharma}}, \bibinfo {author} {\bibfnamefont {S.~R.}\
  \bibnamefont {Sharpe}}, \bibinfo {author} {\bibfnamefont {T.}~\bibnamefont
  {Shutt}}, \bibinfo {author} {\bibfnamefont {M.}~\bibnamefont {Silari}},
  \bibinfo {author} {\bibfnamefont {T.}~\bibnamefont {Sj\"ostrand}}, \bibinfo
  {author} {\bibfnamefont {P.}~\bibnamefont {Skands}}, \bibinfo {author}
  {\bibfnamefont {T.}~\bibnamefont {Skwarnicki}}, \bibinfo {author}
  {\bibfnamefont {J.~G.}\ \bibnamefont {Smith}}, \bibinfo {author}
  {\bibfnamefont {G.~F.}\ \bibnamefont {Smoot}}, \bibinfo {author}
  {\bibfnamefont {S.}~\bibnamefont {Spanier}}, \bibinfo {author} {\bibfnamefont
  {H.}~\bibnamefont {Spieler}}, \bibinfo {author} {\bibfnamefont
  {C.}~\bibnamefont {Spiering}}, \bibinfo {author} {\bibfnamefont
  {A.}~\bibnamefont {Stahl}}, \bibinfo {author} {\bibfnamefont {S.~L.}\
  \bibnamefont {Stone}}, \bibinfo {author} {\bibfnamefont {T.}~\bibnamefont
  {Sumiyoshi}}, \bibinfo {author} {\bibfnamefont {M.~J.}\ \bibnamefont
  {Syphers}}, \bibinfo {author} {\bibfnamefont {K.}~\bibnamefont {Terashi}},
  \bibinfo {author} {\bibfnamefont {J.}~\bibnamefont {Terning}}, \bibinfo
  {author} {\bibfnamefont {U.}~\bibnamefont {Thoma}}, \bibinfo {author}
  {\bibfnamefont {R.~S.}\ \bibnamefont {Thorne}}, \bibinfo {author}
  {\bibfnamefont {L.}~\bibnamefont {Tiator}}, \bibinfo {author} {\bibfnamefont
  {M.}~\bibnamefont {Titov}}, \bibinfo {author} {\bibfnamefont {N.~P.}\
  \bibnamefont {Tkachenko}}, \bibinfo {author} {\bibfnamefont {N.~A.}\
  \bibnamefont {T\"ornqvist}}, \bibinfo {author} {\bibfnamefont {D.~R.}\
  \bibnamefont {Tovey}}, \bibinfo {author} {\bibfnamefont {G.}~\bibnamefont
  {Valencia}}, \bibinfo {author} {\bibfnamefont {R.}~\bibnamefont {Van~de
  Water}}, \bibinfo {author} {\bibfnamefont {N.}~\bibnamefont {Varelas}},
  \bibinfo {author} {\bibfnamefont {G.}~\bibnamefont {Venanzoni}}, \bibinfo
  {author} {\bibfnamefont {L.}~\bibnamefont {Verde}}, \bibinfo {author}
  {\bibfnamefont {M.~G.}\ \bibnamefont {Vincter}}, \bibinfo {author}
  {\bibfnamefont {P.}~\bibnamefont {Vogel}}, \bibinfo {author} {\bibfnamefont
  {A.}~\bibnamefont {Vogt}}, \bibinfo {author} {\bibfnamefont {S.~P.}\
  \bibnamefont {Wakely}}, \bibinfo {author} {\bibfnamefont {W.}~\bibnamefont
  {Walkowiak}}, \bibinfo {author} {\bibfnamefont {C.~W.}\ \bibnamefont
  {Walter}}, \bibinfo {author} {\bibfnamefont {D.}~\bibnamefont {Wands}},
  \bibinfo {author} {\bibfnamefont {D.~R.}\ \bibnamefont {Ward}}, \bibinfo
  {author} {\bibfnamefont {M.~O.}\ \bibnamefont {Wascko}}, \bibinfo {author}
  {\bibfnamefont {G.}~\bibnamefont {Weiglein}}, \bibinfo {author}
  {\bibfnamefont {D.~H.}\ \bibnamefont {Weinberg}}, \bibinfo {author}
  {\bibfnamefont {E.~J.}\ \bibnamefont {Weinberg}}, \bibinfo {author}
  {\bibfnamefont {M.}~\bibnamefont {White}}, \bibinfo {author} {\bibfnamefont
  {L.~R.}\ \bibnamefont {Wiencke}}, \bibinfo {author} {\bibfnamefont
  {S.}~\bibnamefont {Willocq}}, \bibinfo {author} {\bibfnamefont {C.~G.}\
  \bibnamefont {Wohl}}, \bibinfo {author} {\bibfnamefont {J.}~\bibnamefont
  {Womersley}}, \bibinfo {author} {\bibfnamefont {C.~L.}\ \bibnamefont
  {Woody}}, \bibinfo {author} {\bibfnamefont {R.~L.}\ \bibnamefont {Workman}},
  \bibinfo {author} {\bibfnamefont {W.-M.}\ \bibnamefont {Yao}}, \bibinfo
  {author} {\bibfnamefont {G.~P.}\ \bibnamefont {Zeller}}, \bibinfo {author}
  {\bibfnamefont {O.~V.}\ \bibnamefont {Zenin}}, \bibinfo {author}
  {\bibfnamefont {R.-Y.}\ \bibnamefont {Zhu}}, \bibinfo {author} {\bibfnamefont
  {S.-L.}\ \bibnamefont {Zhu}}, \bibinfo {author} {\bibfnamefont
  {F.}~\bibnamefont {Zimmermann}}, \bibinfo {author} {\bibfnamefont {P.~A.}\
  \bibnamefont {Zyla}}, \bibinfo {author} {\bibfnamefont {J.}~\bibnamefont
  {Anderson}}, \bibinfo {author} {\bibfnamefont {L.}~\bibnamefont {Fuller}},
  \bibinfo {author} {\bibfnamefont {V.~S.}\ \bibnamefont {Lugovsky}}, \ and\
  \bibinfo {author} {\bibfnamefont {P.}~\bibnamefont {Schaffner}} (\bibinfo
  {collaboration} {Particle Data Group}),\ }\href {\doibase
  10.1103/PhysRevD.98.030001} {\bibfield  {journal} {\bibinfo  {journal} {Phys.
  Rev. D}\ }\textbf {\bibinfo {volume} {98}},\ \bibinfo {pages} {030001}
  (\bibinfo {year} {2018})}\BibitemShut {NoStop}%
\bibitem [{\citenamefont {Atsue}\ and\ \citenamefont
  {Oyewande}(2015)}]{Atsue2015}%
  \BibitemOpen
  \bibfield  {author} {\bibinfo {author} {\bibfnamefont {T.}~\bibnamefont
  {Atsue}}\ and\ \bibinfo {author} {\bibfnamefont {E.}~\bibnamefont
  {Oyewande}},\ }\href@noop {} {\bibfield  {journal} {\bibinfo  {journal}
  {International Journal of High Energy Physics}\ }\textbf {\bibinfo {volume}
  {2}},\ \bibinfo {pages} {56} (\bibinfo {year} {2015})}\BibitemShut {NoStop}%
\bibitem [{\citenamefont {Kostelecky}(2011)}]{Kostelecky:2011qz}%
  \BibitemOpen
  \bibfield  {author} {\bibinfo {author} {\bibfnamefont {A.}~\bibnamefont
  {Kostelecky}},\ }\href {\doibase 10.1016/j.physletb.2011.05.041} {\bibfield
  {journal} {\bibinfo  {journal} {Phys. Lett.}\ }\textbf {\bibinfo {volume}
  {B701}},\ \bibinfo {pages} {137} (\bibinfo {year} {2011})},\ \Eprint
  {http://arxiv.org/abs/1104.5488} {arXiv:1104.5488 [hep-th]} \BibitemShut
  {NoStop}%
\bibitem [{\citenamefont {Barcelo}\ \emph {et~al.}(2002)\citenamefont
  {Barcelo}, \citenamefont {Liberati},\ and\ \citenamefont
  {Visser}}]{Barcelo:2001cp}%
  \BibitemOpen
  \bibfield  {author} {\bibinfo {author} {\bibfnamefont {C.}~\bibnamefont
  {Barcelo}}, \bibinfo {author} {\bibfnamefont {S.}~\bibnamefont {Liberati}}, \
  and\ \bibinfo {author} {\bibfnamefont {M.}~\bibnamefont {Visser}},\ }\href
  {\doibase 10.1088/0264-9381/19/11/314} {\bibfield  {journal} {\bibinfo
  {journal} {Class. Quant. Grav.}\ }\textbf {\bibinfo {volume} {19}},\ \bibinfo
  {pages} {2961} (\bibinfo {year} {2002})},\ \Eprint
  {http://arxiv.org/abs/gr-qc/0111059} {arXiv:gr-qc/0111059 [gr-qc]}
  \BibitemShut {NoStop}%
\bibitem [{\citenamefont {Weinfurtner}\ \emph {et~al.}(2007)\citenamefont
  {Weinfurtner}, \citenamefont {Liberati},\ and\ \citenamefont
  {Visser}}]{Weinfurtner:2006wt}%
  \BibitemOpen
  \bibfield  {author} {\bibinfo {author} {\bibfnamefont {S.}~\bibnamefont
  {Weinfurtner}}, \bibinfo {author} {\bibfnamefont {S.}~\bibnamefont
  {Liberati}}, \ and\ \bibinfo {author} {\bibfnamefont {M.}~\bibnamefont
  {Visser}},\ }\bibfield  {booktitle} {\emph {\bibinfo {booktitle}
  {{Proceedings, International Workshop on Quantum Simulations via Analogues:
  Dresden, Germany, July 25-28, 2005}}},\ }\href {\doibase
  10.1007/3-540-70859-6_6} {\bibfield  {journal} {\bibinfo  {journal} {Lect.
  Notes Phys.}\ }\textbf {\bibinfo {volume} {718}},\ \bibinfo {pages} {115}
  (\bibinfo {year} {2007})},\ \Eprint {http://arxiv.org/abs/gr-qc/0605121}
  {arXiv:gr-qc/0605121 [gr-qc]} \BibitemShut {NoStop}%
\bibitem [{\citenamefont {Hasse}\ and\ \citenamefont
  {Perlick}(2019)}]{Hasse:2019zqi}%
  \BibitemOpen
  \bibfield  {author} {\bibinfo {author} {\bibfnamefont {W.}~\bibnamefont
  {Hasse}}\ and\ \bibinfo {author} {\bibfnamefont {V.}~\bibnamefont
  {Perlick}},\ }\href@noop {} {\  (\bibinfo {year} {2019})},\ \Eprint
  {http://arxiv.org/abs/1904.08521} {arXiv:1904.08521 [gr-qc]} \BibitemShut
  {NoStop}%
\bibitem [{\citenamefont {Stavrinos}\ and\ \citenamefont
  {Alexiou}(2017)}]{Stavrinos:2016xyg}%
  \BibitemOpen
  \bibfield  {author} {\bibinfo {author} {\bibfnamefont {P.~C.}\ \bibnamefont
  {Stavrinos}}\ and\ \bibinfo {author} {\bibfnamefont {M.}~\bibnamefont
  {Alexiou}},\ }\href {\doibase 10.1142/S0219887818500391} {\bibfield
  {journal} {\bibinfo  {journal} {Int. J. Geom. Meth. Mod. Phys.}\ }\textbf
  {\bibinfo {volume} {15}},\ \bibinfo {pages} {1850039} (\bibinfo {year}
  {2017})},\ \Eprint {http://arxiv.org/abs/1612.04554} {arXiv:1612.04554
  [gr-qc]} \BibitemShut {NoStop}%
\bibitem [{\citenamefont {{Finsler}}(1918)}]{zbMATH02613491}%
  \BibitemOpen
  \bibfield  {author} {\bibinfo {author} {\bibfnamefont {P.}~\bibnamefont
  {{Finsler}}},\ }\href@noop {} {\enquote {\bibinfo {title} {{\"Uber Kurven und
  Fl\"achen in allgemeinen R\"aumen.}}}\ }\bibinfo {howpublished}
  {{G\"ottingen, Z\"urich: O. F\"ussli, 120 S. \(8^{\circ}\) (1918).}}
  (\bibinfo {year} {1918})\BibitemShut {NoStop}%
\bibitem [{\citenamefont {Girelli}\ \emph {et~al.}(2007)\citenamefont
  {Girelli}, \citenamefont {Liberati},\ and\ \citenamefont
  {Sindoni}}]{Girelli:2006fw}%
  \BibitemOpen
  \bibfield  {author} {\bibinfo {author} {\bibfnamefont {F.}~\bibnamefont
  {Girelli}}, \bibinfo {author} {\bibfnamefont {S.}~\bibnamefont {Liberati}}, \
  and\ \bibinfo {author} {\bibfnamefont {L.}~\bibnamefont {Sindoni}},\ }\href
  {\doibase 10.1103/PhysRevD.75.064015} {\bibfield  {journal} {\bibinfo
  {journal} {Phys. Rev.}\ }\textbf {\bibinfo {volume} {D75}},\ \bibinfo {pages}
  {064015} (\bibinfo {year} {2007})},\ \Eprint
  {http://arxiv.org/abs/gr-qc/0611024} {arXiv:gr-qc/0611024 [gr-qc]}
  \BibitemShut {NoStop}%
\bibitem [{\citenamefont {Amelino-Camelia}\ \emph {et~al.}(2014)\citenamefont
  {Amelino-Camelia}, \citenamefont {Barcaroli}, \citenamefont {Gubitosi},
  \citenamefont {Liberati},\ and\ \citenamefont
  {Loret}}]{Amelino-Camelia:2014rga}%
  \BibitemOpen
  \bibfield  {author} {\bibinfo {author} {\bibfnamefont {G.}~\bibnamefont
  {Amelino-Camelia}}, \bibinfo {author} {\bibfnamefont {L.}~\bibnamefont
  {Barcaroli}}, \bibinfo {author} {\bibfnamefont {G.}~\bibnamefont {Gubitosi}},
  \bibinfo {author} {\bibfnamefont {S.}~\bibnamefont {Liberati}}, \ and\
  \bibinfo {author} {\bibfnamefont {N.}~\bibnamefont {Loret}},\ }\href
  {\doibase 10.1103/PhysRevD.90.125030} {\bibfield  {journal} {\bibinfo
  {journal} {Phys. Rev.}\ }\textbf {\bibinfo {volume} {D90}},\ \bibinfo {pages}
  {125030} (\bibinfo {year} {2014})},\ \Eprint {http://arxiv.org/abs/1407.8143}
  {arXiv:1407.8143 [gr-qc]} \BibitemShut {NoStop}%
\bibitem [{\citenamefont {Letizia}\ and\ \citenamefont
  {Liberati}(2017)}]{Letizia:2016lew}%
  \BibitemOpen
  \bibfield  {author} {\bibinfo {author} {\bibfnamefont {M.}~\bibnamefont
  {Letizia}}\ and\ \bibinfo {author} {\bibfnamefont {S.}~\bibnamefont
  {Liberati}},\ }\href {\doibase 10.1103/PhysRevD.95.046007} {\bibfield
  {journal} {\bibinfo  {journal} {Phys. Rev.}\ }\textbf {\bibinfo {volume}
  {D95}},\ \bibinfo {pages} {046007} (\bibinfo {year} {2017})},\ \Eprint
  {http://arxiv.org/abs/1612.03065} {arXiv:1612.03065 [gr-qc]} \BibitemShut
  {NoStop}%
\bibitem [{\citenamefont {{Miron}}(2012)}]{2012arXiv1203.4101M}%
  \BibitemOpen
  \bibfield  {author} {\bibinfo {author} {\bibfnamefont {R.}~\bibnamefont
  {{Miron}}},\ }\href@noop {} {\bibfield  {journal} {\bibinfo  {journal} {arXiv
  e-prints}\ ,\ \bibinfo {eid} {arXiv:1203.4101}} (\bibinfo {year} {2012})},\
  \Eprint {http://arxiv.org/abs/1203.4101} {arXiv:1203.4101 [math.DG]}
  \BibitemShut {NoStop}%
\bibitem [{\citenamefont {Barcaroli}\ \emph {et~al.}(2015)\citenamefont
  {Barcaroli}, \citenamefont {Brunkhorst}, \citenamefont {Gubitosi},
  \citenamefont {Loret},\ and\ \citenamefont {Pfeifer}}]{Barcaroli:2015xda}%
  \BibitemOpen
  \bibfield  {author} {\bibinfo {author} {\bibfnamefont {L.}~\bibnamefont
  {Barcaroli}}, \bibinfo {author} {\bibfnamefont {L.~K.}\ \bibnamefont
  {Brunkhorst}}, \bibinfo {author} {\bibfnamefont {G.}~\bibnamefont
  {Gubitosi}}, \bibinfo {author} {\bibfnamefont {N.}~\bibnamefont {Loret}}, \
  and\ \bibinfo {author} {\bibfnamefont {C.}~\bibnamefont {Pfeifer}},\ }\href
  {\doibase 10.1103/PhysRevD.92.084053} {\bibfield  {journal} {\bibinfo
  {journal} {Phys. Rev.}\ }\textbf {\bibinfo {volume} {D92}},\ \bibinfo {pages}
  {084053} (\bibinfo {year} {2015})},\ \Eprint
  {http://arxiv.org/abs/1507.00922} {arXiv:1507.00922 [gr-qc]} \BibitemShut
  {NoStop}%
\bibitem [{\citenamefont {Rosati}\ \emph {et~al.}(2015)\citenamefont {Rosati},
  \citenamefont {Amelino-Camelia}, \citenamefont {Marciano},\ and\
  \citenamefont {Matassa}}]{Rosati:2015pga}%
  \BibitemOpen
  \bibfield  {author} {\bibinfo {author} {\bibfnamefont {G.}~\bibnamefont
  {Rosati}}, \bibinfo {author} {\bibfnamefont {G.}~\bibnamefont
  {Amelino-Camelia}}, \bibinfo {author} {\bibfnamefont {A.}~\bibnamefont
  {Marciano}}, \ and\ \bibinfo {author} {\bibfnamefont {M.}~\bibnamefont
  {Matassa}},\ }\href {\doibase 10.1103/PhysRevD.92.124042} {\bibfield
  {journal} {\bibinfo  {journal} {Phys. Rev.}\ }\textbf {\bibinfo {volume}
  {D92}},\ \bibinfo {pages} {124042} (\bibinfo {year} {2015})},\ \Eprint
  {http://arxiv.org/abs/1507.02056} {arXiv:1507.02056 [hep-th]} \BibitemShut
  {NoStop}%
\bibitem [{\citenamefont {Freidel}\ \emph {et~al.}(2019)\citenamefont
  {Freidel}, \citenamefont {Kowalski-Glikman}, \citenamefont {Leigh},\ and\
  \citenamefont {Minic}}]{Freidel:2018apz}%
  \BibitemOpen
  \bibfield  {author} {\bibinfo {author} {\bibfnamefont {L.}~\bibnamefont
  {Freidel}}, \bibinfo {author} {\bibfnamefont {J.}~\bibnamefont
  {Kowalski-Glikman}}, \bibinfo {author} {\bibfnamefont {R.~G.}\ \bibnamefont
  {Leigh}}, \ and\ \bibinfo {author} {\bibfnamefont {D.}~\bibnamefont
  {Minic}},\ }\href {\doibase 10.1103/PhysRevD.99.066011} {\bibfield  {journal}
  {\bibinfo  {journal} {Phys. Rev.}\ }\textbf {\bibinfo {volume} {D99}},\
  \bibinfo {pages} {066011} (\bibinfo {year} {2019})},\ \Eprint
  {http://arxiv.org/abs/1812.10821} {arXiv:1812.10821 [hep-th]} \BibitemShut
  {NoStop}%
\bibitem [{\citenamefont {Relancio}\ and\ \citenamefont
  {Liberati}(2020)}]{Relancio:2020zok}%
  \BibitemOpen
  \bibfield  {author} {\bibinfo {author} {\bibfnamefont {J.~J.}\ \bibnamefont
  {Relancio}}\ and\ \bibinfo {author} {\bibfnamefont {S.}~\bibnamefont
  {Liberati}},\ }\href {\doibase 10.1103/PhysRevD.101.064062} {\bibfield
  {journal} {\bibinfo  {journal} {Phys. Rev.}\ }\textbf {\bibinfo {volume}
  {D101}},\ \bibinfo {pages} {064062} (\bibinfo {year} {2020})},\ \Eprint
  {http://arxiv.org/abs/2002.10833} {arXiv:2002.10833 [gr-qc]} \BibitemShut
  {NoStop}%
\bibitem [{\citenamefont {Bhattacharya}\ \emph {et~al.}(2012)\citenamefont
  {Bhattacharya}, \citenamefont {Ghrist},\ and\ \citenamefont
  {Kumar}}]{Bhattacharya2012RelationshipBG}%
  \BibitemOpen
  \bibfield  {author} {\bibinfo {author} {\bibfnamefont {S.}~\bibnamefont
  {Bhattacharya}}, \bibinfo {author} {\bibfnamefont {R.}~\bibnamefont
  {Ghrist}}, \ and\ \bibinfo {author} {\bibfnamefont {V.}~\bibnamefont
  {Kumar}},\ }in\ \href {https://doi.org/10.1007/978-3-642-36279-8} {\emph
  {\bibinfo {booktitle} {Algorithmic Foundations of Robotics X}}}\ (\bibinfo
  {year} {2012})\BibitemShut {NoStop}%
\bibitem [{\citenamefont {Petersen}(2006)}]{Petersen2006}%
  \BibitemOpen
  \bibfield  {author} {\bibinfo {author} {\bibfnamefont {P.}~\bibnamefont
  {Petersen}},\ }\href {https://books.google.es/books?id=9cekXdo52hEC} {\emph
  {\bibinfo {title} {Riemannian Geometry}}},\ Graduate Texts in Mathematics\
  (\bibinfo  {publisher} {Springer New York},\ \bibinfo {year}
  {2006})\BibitemShut {NoStop}%
\bibitem [{\citenamefont {Burns}\ \emph {et~al.}(1985)\citenamefont {Burns},
  \citenamefont {Dubrovin}, \citenamefont {Fomenko},\ and\ \citenamefont
  {Novikov}}]{Burns1985}%
  \BibitemOpen
  \bibfield  {author} {\bibinfo {author} {\bibfnamefont {R.}~\bibnamefont
  {Burns}}, \bibinfo {author} {\bibfnamefont {B.}~\bibnamefont {Dubrovin}},
  \bibinfo {author} {\bibfnamefont {A.}~\bibnamefont {Fomenko}}, \ and\
  \bibinfo {author} {\bibfnamefont {S.}~\bibnamefont {Novikov}},\ }\href
  {https://books.google.es/books?id=tlzc7xXYKd8C} {\emph {\bibinfo {title}
  {Modern Geometryâ€” Methods and Applications: Part II: The Geometry and
  Topology of Manifolds}}},\ Graduate Texts in Mathematics\ (\bibinfo
  {publisher} {Springer New York},\ \bibinfo {year} {1985})\BibitemShut
  {NoStop}%
\bibitem [{\citenamefont {Weinberg}(1972)}]{Weinberg:1972kfs}%
  \BibitemOpen
  \bibfield  {author} {\bibinfo {author} {\bibfnamefont {S.}~\bibnamefont
  {Weinberg}},\ }\href
  {http://www-spires.fnal.gov/spires/find/books/www?cl=QC6.W431} {\emph
  {\bibinfo {title} {{Gravitation and Cosmology}}}}\ (\bibinfo  {publisher}
  {John Wiley and Sons},\ \bibinfo {address} {New York},\ \bibinfo {year}
  {1972})\BibitemShut {NoStop}%
\bibitem [{\citenamefont {Poisson}(2009)}]{Poisson:2009pwt}%
  \BibitemOpen
  \bibfield  {author} {\bibinfo {author} {\bibfnamefont {E.}~\bibnamefont
  {Poisson}},\ }\href {\doibase 10.1017/CBO9780511606601} {\emph {\bibinfo
  {title} {{A Relativist's Toolkit: The Mathematics of Black-Hole
  Mechanics}}}}\ (\bibinfo  {publisher} {Cambridge University Press},\ \bibinfo
  {year} {2009})\BibitemShut {NoStop}%
\bibitem [{\citenamefont {Landau}\ and\ \citenamefont
  {Lifschits}(1975)}]{Landau:1982dva}%
  \BibitemOpen
  \bibfield  {author} {\bibinfo {author} {\bibfnamefont {L.~D.}\ \bibnamefont
  {Landau}}\ and\ \bibinfo {author} {\bibfnamefont {E.~M.}\ \bibnamefont
  {Lifschits}},\ }\href@noop {} {\emph {\bibinfo {title} {{The Classical Theory
  of Fields}}}},\ \bibinfo {series} {Course of Theoretical Physics}, Vol.\
  \bibinfo {volume} {Volume 2}\ (\bibinfo  {publisher} {Pergamon Press},\
  \bibinfo {address} {Oxford},\ \bibinfo {year} {1975})\BibitemShut {NoStop}%
\bibitem [{\citenamefont {Dubovsky}\ and\ \citenamefont
  {Sibiryakov}(2006)}]{Dubovsky:2006vk}%
  \BibitemOpen
  \bibfield  {author} {\bibinfo {author} {\bibfnamefont {S.~L.}\ \bibnamefont
  {Dubovsky}}\ and\ \bibinfo {author} {\bibfnamefont {S.~M.}\ \bibnamefont
  {Sibiryakov}},\ }\href {\doibase 10.1016/j.physletb.2006.05.074} {\bibfield
  {journal} {\bibinfo  {journal} {Phys. Lett.}\ }\textbf {\bibinfo {volume}
  {B638}},\ \bibinfo {pages} {509} (\bibinfo {year} {2006})},\ \Eprint
  {http://arxiv.org/abs/hep-th/0603158} {arXiv:hep-th/0603158 [hep-th]}
  \BibitemShut {NoStop}%
\bibitem [{\citenamefont {Barausse}\ \emph {et~al.}(2011)\citenamefont
  {Barausse}, \citenamefont {Jacobson},\ and\ \citenamefont
  {Sotiriou}}]{Barausse:2011pu}%
  \BibitemOpen
  \bibfield  {author} {\bibinfo {author} {\bibfnamefont {E.}~\bibnamefont
  {Barausse}}, \bibinfo {author} {\bibfnamefont {T.}~\bibnamefont {Jacobson}},
  \ and\ \bibinfo {author} {\bibfnamefont {T.~P.}\ \bibnamefont {Sotiriou}},\
  }\href {\doibase 10.1103/PhysRevD.83.124043} {\bibfield  {journal} {\bibinfo
  {journal} {Phys. Rev.}\ }\textbf {\bibinfo {volume} {D83}},\ \bibinfo {pages}
  {124043} (\bibinfo {year} {2011})},\ \Eprint {http://arxiv.org/abs/1104.2889}
  {arXiv:1104.2889 [gr-qc]} \BibitemShut {NoStop}%
\bibitem [{\citenamefont {Blas}\ and\ \citenamefont
  {Sibiryakov}(2011)}]{Blas:2011ni}%
  \BibitemOpen
  \bibfield  {author} {\bibinfo {author} {\bibfnamefont {D.}~\bibnamefont
  {Blas}}\ and\ \bibinfo {author} {\bibfnamefont {S.}~\bibnamefont
  {Sibiryakov}},\ }\href {\doibase 10.1103/PhysRevD.84.124043} {\bibfield
  {journal} {\bibinfo  {journal} {Phys. Rev.}\ }\textbf {\bibinfo {volume}
  {D84}},\ \bibinfo {pages} {124043} (\bibinfo {year} {2011})},\ \Eprint
  {http://arxiv.org/abs/1110.2195} {arXiv:1110.2195 [hep-th]} \BibitemShut
  {NoStop}%
\bibitem [{\citenamefont {Bhattacharyya}\ \emph {et~al.}(2016)\citenamefont
  {Bhattacharyya}, \citenamefont {Colombo},\ and\ \citenamefont
  {Sotiriou}}]{Bhattacharyya:2015gwa}%
  \BibitemOpen
  \bibfield  {author} {\bibinfo {author} {\bibfnamefont {J.}~\bibnamefont
  {Bhattacharyya}}, \bibinfo {author} {\bibfnamefont {M.}~\bibnamefont
  {Colombo}}, \ and\ \bibinfo {author} {\bibfnamefont {T.~P.}\ \bibnamefont
  {Sotiriou}},\ }\href {\doibase 10.1088/0264-9381/33/23/235003} {\bibfield
  {journal} {\bibinfo  {journal} {Class. Quant. Grav.}\ }\textbf {\bibinfo
  {volume} {33}},\ \bibinfo {pages} {235003} (\bibinfo {year} {2016})},\
  \Eprint {http://arxiv.org/abs/1509.01558} {arXiv:1509.01558 [gr-qc]}
  \BibitemShut {NoStop}%
\bibitem [{\citenamefont {Benkel}\ \emph {et~al.}(2018)\citenamefont {Benkel},
  \citenamefont {Bhattacharyya}, \citenamefont {Louko}, \citenamefont
  {Mattingly},\ and\ \citenamefont {Sotiriou}}]{Benkel:2018abt}%
  \BibitemOpen
  \bibfield  {author} {\bibinfo {author} {\bibfnamefont {R.}~\bibnamefont
  {Benkel}}, \bibinfo {author} {\bibfnamefont {J.}~\bibnamefont
  {Bhattacharyya}}, \bibinfo {author} {\bibfnamefont {J.}~\bibnamefont
  {Louko}}, \bibinfo {author} {\bibfnamefont {D.}~\bibnamefont {Mattingly}}, \
  and\ \bibinfo {author} {\bibfnamefont {T.~P.}\ \bibnamefont {Sotiriou}},\
  }\href {\doibase 10.1103/PhysRevD.98.024034} {\bibfield  {journal} {\bibinfo
  {journal} {Phys. Rev.}\ }\textbf {\bibinfo {volume} {D98}},\ \bibinfo {pages}
  {024034} (\bibinfo {year} {2018})},\ \Eprint
  {http://arxiv.org/abs/1803.01624} {arXiv:1803.01624 [gr-qc]} \BibitemShut
  {NoStop}%
\bibitem [{\citenamefont {Cropp}\ \emph {et~al.}(2013)\citenamefont {Cropp},
  \citenamefont {Liberati},\ and\ \citenamefont {Visser}}]{Cropp:2013zxi}%
  \BibitemOpen
  \bibfield  {author} {\bibinfo {author} {\bibfnamefont {B.}~\bibnamefont
  {Cropp}}, \bibinfo {author} {\bibfnamefont {S.}~\bibnamefont {Liberati}}, \
  and\ \bibinfo {author} {\bibfnamefont {M.}~\bibnamefont {Visser}},\ }\href
  {\doibase 10.1088/0264-9381/30/12/125001} {\bibfield  {journal} {\bibinfo
  {journal} {Class. Quant. Grav.}\ }\textbf {\bibinfo {volume} {30}},\ \bibinfo
  {pages} {125001} (\bibinfo {year} {2013})},\ \Eprint
  {http://arxiv.org/abs/1302.2383} {arXiv:1302.2383 [gr-qc]} \BibitemShut
  {NoStop}%
\end{thebibliography}%
\cleardoublepage
%



\end{spacing}


\begin{appendices} 

\chapter{Change of variables at first and second order} 
\label{appendix_second_order_a}
In this Appendix, we will obtain the DCL and DLT from a generic change of variables up to second order. We first start by taking into account the terms proportional to $(1/\Lambda)^2$ coming from the first order change of variables of Eq.~\eqref{p,q} to $p^2$, $q^2$:
\be
\begin{split}
P^2 & \,=\, p^2 + \frac{v_1^L v_1^L}{\Lambda^2} \left[q^2 (n\cdot p)^2 - 2 (p\cdot q)(n\cdot p)(n\cdot q) + (p\cdot q)^2n^2\right] \\ & + \frac{v_2^L v_2^L}{\Lambda^2} \left[p^2 q^2 n^2 + 2 (p\cdot q)(n\cdot p) (n\cdot q) - p^2 (n\cdot q)^2 - q^2 (n\cdot p)^2 - (p\cdot q)^2n^2\right]\,, \\
Q^2 &\,= \,q^2 + \frac{v_1^R v_1^R}{\Lambda^2} \left[p^2 (n\cdot q)^2 - 2 (p\cdot q)(n\cdot p)(n\cdot q) + (p\cdot q)^2n^2\right] \\ & + \frac{v_2^R v_2^R}{\Lambda^2} \left[p^2 q^2 n^2 + 2 (p\cdot q) (n\cdot p) (n\cdot q) - p^2 (n\cdot q)^2 - q^2 (n\cdot p)^2 - (p\cdot q)^2n^2\right]\,.
\end{split}
\ee 
The following change of variables is compatible with $p^2=P^2$ up to second order
{\small
\be
\begin{split}
P_\mu \,&=\,  p_\mu + \frac{v_1^L}{\Lambda} \left[q_\mu (n\cdot p) - n_\mu (p\cdot q)\right] + \frac{v_2^L}{\Lambda} \epsilon_{\mu\nu\rho\sigma} p^\nu q^\rho n^\sigma - \frac{v_1^L v_1^L}{2 \Lambda^2} \left[n_\mu q^2 (n\cdot p) -\right. \\
& \left.  2 n_\mu (p\cdot q)(n\cdot q) +q_\mu (p\cdot q) n^2 \right] 
 - \frac{v_2^L v_2^L}{2 \Lambda^2} \left[p_\mu q^2 n^2 + 2 n_\mu (p\cdot q) (n\cdot q) - p_\mu (n\cdot q)^2 - n_\mu q^2 (n\cdot p)\right.\\ 
& \left.  - q_\mu (p\cdot q) n^2\right]+ \frac{v_3^L}{\Lambda^2} \left[p_\mu (n\cdot p) - n_\mu p^2\right] (n\cdot q) + \frac{v_4^L}{\Lambda^2} \left[q_\mu (n\cdot p) - n_\mu (p\cdot q)\right] (n\cdot p) +\\ 
&
\frac{v_5^L}{\Lambda^2} \left[q_\mu (n\cdot p) - n_\mu (p\cdot q)\right] (n\cdot q)  +
\frac{v_6^L}{\Lambda^2} (n\cdot p) \epsilon_{\mu\nu\rho\sigma} p^\nu q^\rho n^\sigma + 
\frac{v_7^L}{\Lambda^2} (n\cdot q) \epsilon_{\mu\nu\rho\sigma} p^\nu q^\rho n^\sigma \,,
\end{split}
\label{P->p}
\ee}
while for the variable $Q$ we obtain
{\small
\be
\begin{split}
Q_\mu \,&=\,  q_\mu + \frac{v_1^R}{\Lambda} \left[p_\mu (n\cdot q) - n_\mu (p\cdot q)\right] + \frac{v_2^R}{\Lambda} \epsilon_{\mu\nu\rho\sigma} q^\nu p^\rho n^\sigma - \frac{v_1^R v_1^R}{2 \Lambda^2} \left[n_\mu p^2 (n\cdot q)  \right. \\
& \left.  - 2 n_\mu (p\cdot q)(n\cdot p) + p_\mu (p\cdot q) n^2 \right] 
 - \frac{v_2^R v_2^R}{2 \Lambda^2} \left[q_\mu p^2 n^2 + 2 n_\mu (p\cdot q) (n\cdot p) - q_\mu (n\cdot p)^2 -\right. \\ 
& \left.  n_\mu p^2 (n\cdot q)- p_\mu (p\cdot q) n^2\right] + \frac{v_3^R}{\Lambda^2} \left[q_\mu (n\cdot q) - n_\mu q^2\right] (n\cdot p) + \frac{v_4^R}{\Lambda^2} \left[p_\mu (n\cdot q) - n_\mu (p\cdot q)\right] (n\cdot q) 
 \\ &+\frac{v_5^R}{\Lambda^2} \left[p_\mu (n\cdot q) - n_\mu (p\cdot q)\right] (n\cdot p) +
\frac{v_6^R}{\Lambda^2} (n\cdot q) \epsilon_{\mu\nu\rho\sigma} q^\nu p^\rho n^\sigma + 
\frac{v_7^R}{\Lambda^2} (n\cdot p) \epsilon_{\mu\nu\rho\sigma} q^\nu p^\rho n^\sigma \,.
\end{split}
\label{Q->q}
\ee}

We see we have a total of 14 parameters $(v_1^L,\ldots,v_7^L;v_1^R,\ldots,v_7^R)$ that form a generic change of variables up to second order. In order to obtain the DCL in the new variables $(p, q)$, we apply it to the composition law of the variables $(P, Q)$. As these variables transform linearly, the composition law  must be composed of  covariant terms under linear Lorentz transformations:
\be
\left[P\bigoplus Q\right]_\mu \,= \,P_\mu + Q_\mu + \frac{c_1}{\Lambda^2} P_\mu Q^2 + \frac{c_2}{\Lambda^2} Q_\mu P^2 + \frac{c_3}{\Lambda^2} P_\mu (P\cdot Q) + \frac{c_4}{\Lambda^2} Q_\mu (P\cdot Q) \,.
\label{ccl2a}
\ee
Then, applying~\eqref{P->p}-\eqref{Q->q} to Eq.~\eqref{ccl2a} one obtains the DCL
\be
\begin{split}
\left[p\oplus q\right]_\mu &\,= \,p_\mu + q_\mu + \frac{v_1^L}{\Lambda} \left[q_\mu (n\cdot p) - n_\mu (p\cdot q)\right] + \frac{v_1^R}{\Lambda} \left[p_\mu (n\cdot q) - n_\mu (p\cdot q)\right] + \\\ 
&\frac{(v_2^L-v_2^R)}{\Lambda} \epsilon_{\mu\nu\rho\sigma} p^\nu q^\rho n^\sigma  + \frac{c_1}{\Lambda^2} p_\mu q^2 + \frac{c_2}{\Lambda^2} q_\mu p^2 + \frac{c_3}{\Lambda^2} p_\mu (p\cdot q) + \frac{c_4}{\Lambda^2} q_\mu (p\cdot q) \\ 
& - \frac{v_1^L v_1^L}{2 \Lambda^2} \left[n_\mu q^2 (n\cdot p) - 2 n_\mu (p\cdot q)(n\cdot q) + q_\mu (p\cdot q) n^2\right]   - \frac{v_1^R v_1^R}{2 \Lambda^2} \left[n_\mu p^2 (n\cdot q) -\right. \\ 
& \left.  2 n_\mu (p\cdot q)(n\cdot p) +p_\mu (p\cdot q) n^2\right]  - \frac{v_2^L v_2^L}{2 \Lambda^2} \left[p_\mu q^2 n^2 + 2 n_\mu (p\cdot q) (n\cdot q) - p_\mu (n\cdot q)^2\right. \\ 
&\left. - n_\mu q^2 (n\cdot p) - q_\mu (p\cdot q) n^2 \right]  - \frac{v_2^R v_2^R}{2 \Lambda^2} \left[q_\mu p^2 n^2 + 2 n_\mu (p\cdot q) (n\cdot p) - q_\mu (n\cdot p)^2 \right.\\ 
& \left. - n_\mu p^2 (n\cdot q) - p_\mu (p\cdot q) n^2\right] + \frac{v_3^L}{\Lambda^2} \left[p_\mu (n\cdot p) - n_\mu p^2\right] (n\cdot q) +  \\ &  \frac{v_3^R}{\Lambda^2} \left[q_\mu (n\cdot q) - n_\mu q^2\right] (n\cdot p)+ \frac{v_4^L}{\Lambda^2} \left[q_\mu (n\cdot p) - n_\mu (p\cdot q)\right] (n\cdot p) + \\ 
& \frac{v_4^R}{\Lambda^2} \left[p_\mu (n\cdot q) - n_\mu (p\cdot q)\right] (n\cdot q) + \frac{v_5^L}{\Lambda^2} \left[q_\mu (n\cdot p) - n_\mu (p\cdot q)\right] (n\cdot q) + \\ 
&  \frac{v_5^R}{\Lambda^2} \left[p_\mu (n\cdot q) - n_\mu (p\cdot q)\right] (n\cdot p)+ \frac{(v_6^L - v_7^R)}{\Lambda^2} (n\cdot p) \epsilon_{\mu\nu\rho\sigma} p^\nu q^\rho n^\sigma \\ 
&+ \frac{(v_7^L - v_6^R)}{\Lambda^2} (n\cdot q) \epsilon_{\mu\nu\rho\sigma} p^\nu q^\rho n^\sigma \,.
\end{split}
\label{cl2}
\ee
In order to consider the rotational invariant case, we take $n_\mu=(1, 0, 0, 0)$ in Eq.~\eqref{cl2}, obtaining Eq.~\ref{generalCL}.

In order to obtain the DLT in the two-particle system, $(p,q) \to (p',q')$, one can follow the same procedure used to obtain Eqs.~\eqref{p'1}-\eqref{q'1} in Sec.~\ref{sec:covariant}; after some algebra, one obtains $p'$
{\small
\be
\begin{split}
p'_\mu &\,=\, \tilde{p}_\mu + \omega^{\alpha\beta} n_\beta \left[\frac{v_1^L}{\Lambda} p_\alpha q_\mu - \frac{v_1^L}{\Lambda} \eta_{\alpha\mu} (p\cdot q) - \frac{v_2^L}{\Lambda} \epsilon_{\alpha\mu\nu\rho} p^\nu q^\rho\right] \\
 & + \omega^{\alpha\beta} n_\beta \left[-\frac{v_1^L v_1^R}{\Lambda^2} \left(q_\alpha p_\mu - \eta_{\alpha\mu} (p\cdot q)\right) (n\cdot p) - \frac{v_1^L v_2^R}{\Lambda^2} \epsilon_{\alpha\mu\nu\rho} p^\nu q^\rho (n\cdot p) - \frac{v_1^L v_1^L}{\Lambda^2} p_\alpha q_\mu (n\cdot q)  \right. \\
 &  \left. + \frac{v_1^L v_2^L}{\Lambda^2} \epsilon_{\alpha\nu\rho\sigma} q_\mu p^\nu q^\rho n^\sigma + \frac{v_1^L v_1^L}{\Lambda^2} \left(p_\alpha q^2 - q_\alpha (p\cdot q)\right) n_\mu + \frac{v_1^L v_1^R}{\Lambda^2} \left(q_\alpha p^2 - p_\alpha (p\cdot q)\right) n_\mu \right. \\
 &  \left. - \frac{v_2^L v_1^L}{\Lambda^2} \epsilon_ {\alpha\mu\nu\rho} q^\nu n^\rho (p\cdot q) + \frac{v_2^L v_1^R}{\Lambda^2} \epsilon_ {\alpha\mu\nu\rho} p^\nu n^\rho (p\cdot q) + \frac{v_2^L v_2^L}{\Lambda^2} \left[q_\alpha \left(p_\mu (n\cdot q) - q_\mu (n\cdot p)\right) - \right.\right. \\ 
&  \left. \left.
 \eta_{\alpha\mu} \left((n\cdot q)(p\cdot q) - (n\cdot p) q^2\right)\right]  - \frac{v_2^L v_2^R}{\Lambda^2} \left[p_\alpha \left(q_\mu (n\cdot p) - p_\mu (n\cdot q)\right) - \eta_{\alpha\mu} \left((n\cdot p)(p\cdot q)+ \right. \right. \right. \\ 
&  \left.\left.\left.- (n\cdot q) p^2\right)\right] \frac{v_2^L v_2^L}{\Lambda^2} q_\alpha p_\mu (n\cdot q)  - \frac{(v_1^L v_1^L - v_2^L v_2^L)}{2 \Lambda^2} \left(p_\alpha n_\mu + \eta_{\alpha\mu} (n\cdot p)\right) q^2 + \right. \\ 
&  \left. \frac{(v_1^L v_1^L - v_2^L v_2^L - v_5^L)}{\Lambda^2} \left(q_\alpha n_\mu + \eta_{\alpha\mu} (n\cdot q)\right) (p\cdot q) + \frac{v_3^L}{\Lambda^2} \left(q_\alpha (n\cdot p) + p_\alpha (n\cdot q)\right) p_\mu - \right. \\ 
&  \left. \frac{v_3^L}{\Lambda^2} \left(q_\alpha n_\mu + \eta_{\alpha\mu} (n\cdot q)\right) p^2 + \frac{v_4^L}{\Lambda^2} 2 p_\alpha q_\mu (n\cdot p) - \frac{v_4^L}{\Lambda^2} \left(p_\alpha n_\mu + \eta_{\alpha\mu} (n\cdot p)\right) (p\cdot q) \right. \\ 
&  \left. + \frac{v_5^L}{\Lambda^2} \left(q_\alpha (n\cdot p) + p_\alpha (n\cdot q)\right) q_\mu + \frac{v_6^L}{\Lambda^2} \left[p_\alpha \epsilon_{\mu\nu\rho\sigma} p^\nu q^\rho n^\sigma - \epsilon_{\alpha\mu\nu\rho} p^\nu q^\rho (n\cdot p)\right]  \right. \\
 &  \left.+ \frac{v_7^L}{\Lambda^2} \left[q_\alpha \epsilon_{\mu\nu\rho\sigma} p^\nu q^\rho n^\sigma - \epsilon_{\alpha\mu\nu\rho} p^\nu q^\rho (n\cdot q)\right]\right] \,,
\end{split}
\label{p'2}
\ee}\normalsize 
and for the second momentum variable ($q'$), one gets the same expression but interchanging $p\leftrightarrow q$ and $v^i_L \leftrightarrow v^i_R$
{\small
\be
\begin{split}
q'_\mu &\,= \,\tilde{q}_\mu + \omega^{\alpha\beta} n_\beta \left[\frac{v_1^R}{\Lambda} q_\alpha p_\mu - \frac{v_1^R}{\Lambda} \eta_{\alpha\mu} (p\cdot q) - \frac{v_2^R}{\Lambda} \epsilon_{\alpha\mu\nu\rho} q^\nu p^\rho\right] \\ 
& + \omega^{\alpha\beta} n_\beta \left[-\frac{v_1^R v_1^L}{\Lambda^2} \left(p_\alpha q_\mu - \eta_{\alpha\mu} (p\cdot q)\right) (n\cdot q) - \frac{v_1^R v_2^L}{\Lambda^2} \epsilon_{\alpha\mu\nu\rho} q^\nu p^\rho (n\cdot q) - \frac{v_1^R v_1^R}{\Lambda^2} q_\alpha p_\mu (n\cdot p) \right. \\ 
&  \left. + \frac{v_1^R v_2^R}{\Lambda^2} \epsilon_{\alpha\nu\rho\sigma} p_\mu q^\nu p^\rho n^\sigma + \frac{v_1^R v_1^R}{\Lambda^2} \left(q_\alpha p^2 - p_\alpha (p\cdot q)\right) n_\mu + \frac{v_1^R v_1^L}{\Lambda^2} \left(p_\alpha q^2 - q_\alpha (p\cdot q)\right) n_\mu \right. \\
&  \left. - \frac{v_2^R v_1^R}{\Lambda^2} \epsilon_ {\alpha\mu\nu\rho} p^\nu n^\rho (p\cdot q) + \frac{v_2^R v_1^L}{\Lambda^2} \epsilon_ {\alpha\mu\nu\rho} q^\nu n^\rho (p\cdot q) + \frac{v_2^R v_2^R}{\Lambda^2} \left[p_\alpha \left(q_\mu (n\cdot p) - p_\mu (n\cdot q)\right)  \right. \right. \\ 
&   \left. \left. - \eta_{\alpha\mu} \left((n\cdot p)(p\cdot q) - (n\cdot q) p^2\right)\right]  - \frac{v_2^Rv_2^L}{\Lambda^2} \left[q_\alpha \left(p_\mu (n\cdot q) - q_\mu (n\cdot p)\right) - \eta_{\alpha\mu} \left((n\cdot q)(p\cdot q)\right.\right.\right. \\ 
&  \left.\left.\left. - (n\cdot p) q^2\right)\right]  + \frac{v_2^R v_2^R}{\Lambda^2} p_\alpha q_\mu (n\cdot p)  - \frac{(v_1^R v_1^R - v_2^R v_2^R)}{2 \Lambda^2} \left(q_\alpha n_\mu + \eta_{\alpha\mu} (n\cdot q)\right) p^2  \right. \\
 &  \left. + \frac{(v_1^R v_1^R- v_2^R v_2^R - v_5^R)}{\Lambda^2} \left(p_\alpha n_\mu + \eta_{\alpha\mu} (n\cdot p)\right) (p\cdot q)+ \frac{v_3^R}{\Lambda^2} \left(p_\alpha (n\cdot q) + q_\alpha (n\cdot p)\right) q_\mu  \right. \\ 
&  \left. - \frac{v_3^R}{\Lambda^2} \left(p_\alpha n_\mu + \eta_{\alpha\mu} (n\cdot p)\right) q^2+ \frac{v_4^R}{\Lambda^2} 2 q_\alpha p_\mu (n\cdot q) - \frac{v_4^R}{\Lambda^2} \left(q_\alpha n_\mu + \eta_{\alpha\mu} (n\cdot q)\right) (p\cdot q)   \right. \\ 
&  \left. + \frac{v_5^R}{\Lambda^2} \left(p_\alpha (n\cdot q) + q_\alpha (n\cdot p)\right) p_\mu+  \frac{v_6^R}{\Lambda^2} \left[q_\alpha \epsilon_{\mu\nu\rho\sigma} q^\nu p^\rho n^\sigma - \epsilon_{\alpha\mu\nu\rho} q^\nu p^\rho (n\cdot q)\right]  \right. \\ 
&\left.+ \frac{v_7^R}{\Lambda^2} \left[p_\alpha \epsilon_{\mu\nu\rho\sigma} q^\nu p^\rho n^\sigma - \epsilon_{\alpha\mu\nu\rho} q^\nu p^\rho (n\cdot p)\right]\right] \,.
\end{split}
\label{q'2}
\ee}
\normalsize

As one could expect, the coefficients of the DLT are determined by the 14 parameters $(v_i^L;v_i^R),\, i=1,\ldots 7$, appearing in the change of variables. 

If we take again $n_\mu=(1, 0, 0, 0)$ in Eqs.~\eqref{p'2} and~\eqref{q'2} we find
\be
\begin{split}
p_{0}^{\prime}&\,=\,p_{0}+\vec{p}\cdot \vec{\xi}-\frac{v_{1}^{L}}{\Lambda}q_{0}\left(\vec{p}\cdot \vec{\xi}\right)+\frac{v_{2}^{L}}{\Lambda}\left(\vec{p}\wedge\vec{q}\right)\cdot \vec{\xi}+\frac{v_{1}^{L}v_{1}^{L}-v_{2}^{L}v_{2}^{L}-2v_{5}^{L}}{2\Lambda^{2}}q_{0}^{2}\left(\vec{p}\cdot \vec{\xi}\right) \\
&+\frac{v_{1}^{L}v_{1}^{L}+v_{2}^{L}v_{2}^{L}}{2\Lambda^{2}}\vec{q}^{2}\left(\vec{p}\cdot \vec{\xi}\right) +\frac{v_{1}^{L}v_{1}^{R}-v_{3}^{L}}{\Lambda^{2}}\vec{p}^{2}\left(\vec{q}\cdot \vec{\xi}\right) +
\frac{v_{1}^{L}v_{1}^{R}-v_{3}^{L}-v_{4}^{L}}{\Lambda^{2}}p_{0}q_{0}\left(\vec{p}\cdot \vec{\xi}\right) \\
&  -\frac{v_{1}^{L}v_{1}^{R}+v_{4}^{L}}{\Lambda^{2}}\left(\vec{p}\cdot \vec{q}\right)\left(\vec{p}\cdot \vec{\xi}\right)-\frac{v_{2}^{L}v_{2}^{L}+v_{5}^{L}}{\Lambda^{2}}\left(\vec{p}\cdot\vec{q}\right)\left(\vec{q}\cdot\vec{\xi}\right)+\frac{v_{1}^{L}v_{2}^{R}+v_{6}^{L}}{\Lambda^{2}}p_{0}\left(\vec{p}\wedge\vec{q}\right)\vec{\xi}\\
&+\frac{-v_{1}^{L}v_{2}^{L}+v_{7}^{L}}{\Lambda^{2}}q_{0}\left(\vec{p}\wedge\vec{q}\right)\vec{\xi} \, ,
\end{split}
\label{generaltr1}
\ee
\be
\begin{split}
p_{i}^{\prime}&\,=\,p_{i}+p_{0}\xi_{i}-\frac{v_{1}^{L}}{\Lambda}\left[q_{i}\left(\vec{p}\cdot \vec{\xi}\right)+\left(p\cdot q\right)\xi_{i}\right]-\frac{v_{2}^{L}}{\Lambda^{2}}\left(q_{0}\epsilon_{ijk}p_{j}\xi_{k}-p_{0}\epsilon_{ijk}q_{j}\xi_{k}\right)+ \\
&\frac{v_{1}^{L}v_{1}^{R}-v_{3}^{L}-v_{4}^{L}}{\Lambda^{2}}p_{0}^{2}q_{0}\xi_{i}+
\frac{v_{1}^{L}v_{1}^{L}-v_{2}^{L}v_{2}^{L}-2v_{5}^{L}}{2\Lambda^{2}}p_{0}q_{0}^{2}\xi_{i}+\frac{-v_{1}^{L}v_{1}^{R}-v_{2}^{L}v_{2}^{R}+v_{4}^{L}}{\Lambda^{2}}\left(\vec{p}\cdot \vec{q}\right)p_{0}\xi_{i}+ \\
&\frac{-v_{1}^{L}v_{1}^{L}+2v_{2}^{L}v_{2}^{L}+v_{5}^{L}}{\Lambda^{2}}\left(\vec{p}\cdot \vec{q}\right)q_{0}\xi_{i}+\frac{v_{2}^{L}v_{2}^{R}+v_{3}^{L}}{\Lambda^{2}}\vec{p}^{2}q_{0}\xi_{i} +
\frac{v_{1}^{L}v_{1}^{L}-3v_{2}^{L}v_{2}^{L}}{2\Lambda^{2}}p_{0}\vec{q}^{2}\xi_{i}+
\\ &\frac{v_{2}^{L}v_{2}^{R}-2v_{4}^{L}}{\Lambda^{2}}p_{0}q_{i}\left(\vec{p}\cdot \vec{\xi}\right)-\frac{v_{2}^{L}v_{2}^{R}+v_{3}^{L}}{\Lambda^{2}}p_{i}q_{0}\left(\vec{p}\cdot \vec{\xi}\right)+\frac{v_{2}^{L}v_{2}^{L}-v_{5}^{L}}{\Lambda^{2}}p_{0}q_{i}\left(\vec{q}\cdot \vec{\xi}\right)-\\
&\frac{2v_{2}^{L}v_{2}^{L}}{\Lambda^{2}}p_{i}q_{0}\left(\vec{q}\cdot \vec{\xi}\right) + \frac{v_{1}^{L}v_{1}^{R}-v_{3}^{L}}{\Lambda^{2}}p_{0}p_{i}\left(\vec{q}\cdot \vec{\xi}\right)+\frac{v_{1}^{L}v_{1}^{L}-v_{5}^{L}}{\Lambda^{2}}q_{0}q_{i}\left(\vec{p}\cdot \vec{\xi}\right)-\frac{v_{1}^{L}v_{2}^{L}}{\Lambda^{2}}q_{i}\left(\vec{p}\wedge\vec{q}\right)\vec{\xi}+ \\
&\frac{v_{1}^{L}v_{2}^{R}+v_{6}^{L}}{\Lambda^{2}}p_{0}^{2}\epsilon_{ijk}q_{j}\xi_{k}-\frac{v_{6}^{L}}{\Lambda^{2}}\left(\vec{p}\cdot\vec{\xi}\right)\epsilon_{ijk}p_{j}q_{k}+\frac{v_{7}^{L}}{\Lambda^{2}}p_{0}q_{0}\epsilon_{ijk}q_{j}\xi_{k}-\frac{v_{7}^{L}}{\Lambda^{2}}\left(\vec{q}\cdot \vec{\xi}\right)\epsilon_{ijk}p_{j}q_{k}-\\
&\frac{v_{7}^{L}}{\Lambda^{2}}q_{0}^{2}\epsilon_{ijk}p_{j}\xi_{k}-\frac{v_{1}^{L}v_{2}^{R}+v_{6}^{L}}{\Lambda^{2}}p_{0}q_{0}\epsilon_{ijk}p_{j}\xi_{k}+\frac{v_{1}^{L}v_{2}^{L}}{\Lambda^{2}}\left(p\cdot q\right)\epsilon_{ijk}q_{j}\xi_{k} -\frac{v_{1}^{R}v_{2}^{L}}{\Lambda^{2}}\left(p\cdot q\right)\epsilon_{ijk}p_{j}\xi_{k} \,,
\end{split}
\ee
\be
\begin{split}
q_{0}^{\prime}&\,=\,q_{0}+\vec{q}\cdot \vec{\xi}-\frac{v_{1}^{R}}{\Lambda}p_{0}\left(\vec{q}\cdot \vec{\xi}\right)+\frac{v_{2}^{R}}{\Lambda}\left(\vec{q}\wedge\vec{p}\right)\cdot \vec{\xi}+\frac{v_{1}^{R}v_{1}^{R}-v_{2}^{R}v_{2}^{R}-2v_{5}^{R}}{2\Lambda^{2}}p_{0}^{2}\left(\vec{q}\cdot \vec{\xi}\right)+ \\
& \frac{v_{1}^{R}v_{1}^{R}+v_{2}^{R}v_{2}^{R}}{2\Lambda^{2}}\vec{p}^{2}\left(\vec{q}\cdot \vec{\xi}\right)+\frac{v_{1}^{L}v_{1}^{R}-v_{3}^{R}}{\Lambda^{2}}\vec{q}^{2}\left(\vec{p}\cdot \vec{\xi}\right) +
\frac{v_{1}^{L}v_{1}^{R}-v_{3}^{R}-v_{4}^{R}}{\Lambda^{2}}q_{0}p_{0}\left(\vec{q}\cdot \vec{\xi}\right) - \\
&
\frac{v_{1}^{L}v_{1}^{R}+v_{4}^{R}}{\Lambda^{2}}\left(\vec{p}\cdot \vec{q}\right)\left(\vec{q}\cdot \vec{\xi}\right) -
\frac{v_{2}^{R}v_{2}^{R}+v_{5}^{R}}{\Lambda^{2}}\left(\vec{p}\cdot\vec{q}\right)\left(\vec{p}\cdot\vec{\xi}\right)-\frac{v_{1}^{R}v_{2}^{L}+v_{6}^{R}}{\Lambda^{2}}q_{0}\left(\vec{p}\wedge\vec{q}\right)\vec{\xi}\\
&+\frac{v_{1}^{R}v_{2}^{R}-v_{7}^{R}}{\Lambda^{2}}p_{0}\left(\vec{p}\wedge\vec{q}\right)\vec{\xi} \, ,
\end{split}
\ee
\be
\begin{split}
q_{i}^{\prime}&\,=\,q_{i}+q_{0}\xi_{i}-\frac{v_{1}^{R}}{\Lambda}\left[p_{i}\left(\vec{q}\cdot \vec{\xi}\right)+\left(p\cdot q\right)\xi_{i}\right]-\frac{v_{2}^{R}}{\Lambda^{2}}\left(p_{0}\epsilon_{ijk}q_{j}\xi_{k}-q_{0}\epsilon_{ijk}p_{j}\xi_{k}\right)+\\
&\frac{v_{1}^{L}v_{1}^{R}-v_{3}^{R}-v_{4}^{R}}{\Lambda^{2}}q_{0}^{2}p_{0}\xi_{i} +
\frac{v_{1}^{R}v_{1}^{R}-v_{2}^{R}v_{2}^{R}-2v_{5}^{R}}{2\Lambda^{2}}q_{0}p_{0}^{2}\xi_{i}+\frac{-v_{1}^{L}v_{1}^{R}-v_{2}^{R}v_{2}^{L}+v_{4}^{R}}{\Lambda^{2}}\left(\vec{p}\cdot \vec{q}\right)q_{0}\xi_{i}+ \\
&\frac{-v_{1}^{R}v_{1}^{R}+2v_{2}^{R}v_{2}^{R}+v_{5}^{R}}{\Lambda^{2}}\left(\vec{p}\cdot \vec{q}\right)p_{0}\xi_{i}+\frac{v_{2}^{L}v_{2}^{R}+v_{3}^{R}}{\Lambda^{2}}\vec{q}^{2}p_{0}\xi_{i} +
\frac{v_{1}^{R}v_{1}^{R}-3v_{2}^{R}v_{2}^{R}}{2\Lambda^{2}}q_{0}\vec{p}^{2}\xi_{i}+
\\ &\frac{v_{2}^{L}v_{2}^{R}-2v_{4}^{R}}{\Lambda^{2}}q_{0}p_{i}\left(\vec{q}\cdot \vec{\xi}\right)-\frac{v_{2}^{L}v_{2}^{R}+v_{3}^{R}}{\Lambda^{2}}q_{i}p_{0}\left(\vec{q}\cdot \vec{\xi}\right)+\frac{v_{2}^{R}v_{2}^{R}-v_{5}^{R}}{\Lambda^{2}}q_{0}p_{i}\left(\vec{p}\cdot \vec{\xi}\right)- \\
&
 \frac{2v_{2}^{R}v_{2}^{R}}{\Lambda^{2}}q_{i}p_{0}\left(\vec{p}\cdot \vec{\xi}\right)+\frac{v_{1}^{L}v_{1}^{R}-v_{3}^{R}}{\Lambda^{2}}q_{0}q_{i}\left(\vec{p}\cdot \vec{\xi}\right)+\frac{v_{1}^{R}v_{1}^{R}-v_{5}^{R}}{\Lambda^{2}}p_{0}p_{i}\left(\vec{q}\cdot \vec{\xi}\right)+\frac{v_{1}^{R}v_{2}^{R}}{\Lambda^{2}}p_{i}\left(\vec{p}\wedge\vec{q}\right)\vec{\xi}+ \\
&\frac{v_{1}^{R}v_{2}^{L}+v_{6}^{R}}{\Lambda^{2}}q_{0}^{2}\epsilon_{ijk}p_{j}\xi_{k}+\frac{v_{6}^{R}}{\Lambda^{2}}\left(\vec{q}\cdot\vec{\xi}\right)\epsilon_{ijk}p_{j}q_{k}+\frac{v_{7}^{R}}{\Lambda^{2}}p_{0}q_{0}\epsilon_{ijk}p_{j}\xi_{k}+
\frac{v_{7}^{R}}{\Lambda^{2}}\left(\vec{p}\cdot \vec{\xi}\right)\epsilon_{ijk}p_{j}q_{k}- \\
&\frac{v_{7}^{R}}{\Lambda^{2}}p_{0}^{2}\epsilon_{ijk}q_{j}\xi_{k}-\frac{v_{1}^{R}v_{2}^{L}+v_{6}^{R}}{\Lambda^{2}}p_{0}q_{0}\epsilon_{ijk}q_{j}\xi_{k}+\frac{v_{1}^{R}v_{2}^{R}}{\Lambda^{2}}\left(p\cdot q\right)\epsilon_{ijk}p_{j}\xi_{k}-\frac{v_{1}^{L}v_{2}^{R}}{\Lambda^{2}}\left(p\cdot q\right)\epsilon_{ijk}q_{j}\xi_{k} \,.
\end{split}
\label{generaltr4}
\ee
These are the DLT generalizing Eq.~\eqref{DLT-1st} up to order $(1/\Lambda)^2$ making $p^2$ and $q^2$ invariant, and their coefficients depend on the 14 parameters that characterize a generic change of variables in the two-particle system.

\chapter{Momentum space geometry}

\section{Translations in de Sitter}
\label{appendix_translations}
We are going to obtain the translations of de Sitter space Sec.~\ref{subsection_kappa_desitter} when the tetrad is the one proposed in Eq.~\eqref{bicross-tetrad}. The condition~\eqref{T(a,k)} can be written in terms of the composition law as 
\be
e_\mu^\alpha(p \oplus q) \,=\, \frac{\partial (p \oplus q)_\mu}{\partial q_\nu} \,e_\nu^\alpha(q)\,.
\ee
Then, the system of equations we need to solve is  
\be
\begin{split}
\text{For}\quad \mu\,&=\,0,\quad \nu\,=\,0 \qquad 1\,=\,\frac{\partial (p\oplus q)_0}{\partial q_0} \,.\\
\text{For}\quad \mu\,&=\,i,\quad \nu\,=\,0 \qquad 0\,=\,\frac{\partial (p\oplus q)_0}{\partial q_j}\,\delta^i_j \,e^{\pm q_0/\Lambda} \,.\\
\text{For}\quad \mu\,&=\,0,\quad \nu\,=\,i \qquad 0\,=\,\frac{\partial (p\oplus q)_i}{\partial q_0}\,.\\
\text{For}\quad \mu\,&=\,i,\quad \nu\,=\,j \qquad \delta^i_j\, e^{\pm (p \oplus q)_0/\Lambda}\,=\,\frac{\partial (p\oplus q)_j}{\partial q_k} \delta^i_k\, e^{\pm q_0/\Lambda}\,.\\
\end{split}
\ee
The first equation implies that 
\be
(p\oplus q)_0\,=\,p_0\, f \left(\frac{p_0}{\Lambda},\frac{\vec{p}^2}{\Lambda^2},\frac{\vec{q}^2}{\Lambda^2},\frac{\vec{p}\cdot \vec{q}}{\Lambda^2}\right)+q_0\,,
\ee
but the second one requires the component zero of the composition law to be independent of the spatial components of the momentum $q$, so  by condition~\eqref{eq:cl0}, $f=1$ and then
 \be
(p\oplus q)_0\,=\,p_0+q_0\,.
\ee
The fourth equation can be now written as 
\be
 \delta^i_j\, e^{\pm (p_0+q_0)/\Lambda}\,=\,\frac{\partial (p\oplus q)_j}{\partial q_k} \delta^i_k\, e^{\pm q_0/\Lambda}\,,
\ee
so 
\be
(p\oplus q)_i\,=\,p_i\, g\left(\frac{p_0}{\Lambda},\frac{q_0}{\Lambda},\frac{\vec{p}^2}{\Lambda^2}\right)+q_i e^{\pm p_0/\Lambda}\,,
\ee
but by virtue of the third equation, the  condition~\eqref{eq:cl0} gives that $g=1$, so the composition law finally is
\be
(p\oplus q)_0\,=\,p_0+q_0\,,\qquad (p\oplus q)_i\,=\,p_i+q_i e^{-p_0/\Lambda}\,.
\label{DCL-bcb}
\ee

\section{Algebra of isometry generators in de Sitter and anti-de Sitter spaces} 
\label{appendix_algebra}

In this appendix, we want to study the algebra of isometries of the maximally symmetric spaces of de Sitter and anti-de Sitter. We can start in de Sitter space with the algebra of generators of translations $(T^\alpha_S, J^{\beta\gamma})$ which is Lorentz covariant
\be
\begin{split}
&\{T^\alpha_S, T^\beta_S\} \,=\, \frac{J^{\alpha\beta}}{\Lambda^2}\,, \quad\quad \{T^\alpha_S, J^{\beta\gamma}\} \,=\, \eta^{\alpha\beta} T^\gamma_S - \eta^{\alpha\gamma} T^\beta_S\,,\\
 &\lbrace J^{\alpha\beta},J^{\gamma\delta}\rbrace\,=\, \eta^{\beta\gamma}J^{\alpha\delta} - \eta^{\alpha\gamma}J^{\beta\delta} - \eta^{\beta\delta}J^{\alpha\gamma} + \eta^{\alpha\delta}J^{\beta\gamma}.
\label{cov_generators}
\end{split}
\ee
In order to have an associative composition law, we saw in Ch.~\ref{chapter_curved_momentum_space} that the translation generators must form a four-dimensional subalgebra, so we can consider the following change of basis of the spatial translations
\be
T^0_\kappa \,=\, T^0_S, \quad\quad\quad  T^i_\kappa \,=\, T^i_S \pm \frac{J^{0i}}{\Lambda},
\label{eq:change_generators}
\ee
and then, the new algebra of the translations is
\be
\lbrace T_\kappa^0, T_\kappa^i\rbrace\,=\, \mp \frac{T_\kappa^i}{\Lambda}\,,
\ee
which is the algebra from which one can deduce the $\kappa$-Poincar\'e kinematics, as we saw in Sec.~\ref{sec:examples}. 

However, one can not obtain a closed subalgebra of the generators of translations for the anti-de Sitter algebra. The difference resides in the minus sign appearing in the algebra of anti-de Sitter space isometries (which corresponds to substitute $\Lambda^2$ by $-\Lambda^2$ in (\ref{cov_generators})). If one tries to make a similar change on the basis to Eq.~\eqref{eq:change_generators} of these generators, one sees that one can not find a closed subalgebra. Then we can conclude that there is no way to obtain a DRK with an associative composition law in anti-de Sitter momentum space.

From a generic change of translation generators (always maintaining isotropy) 
\be
T^0_H \,=\, T^0_S, \quad\quad\quad T^i_H \,=\, T^i_S + \alpha \frac{J^{0i}}{\Lambda},
\ee
where $\alpha$ is an arbitrary parameter, one can find the DCL corresponding to the hybrids models which have a Lorentz covariant term as in Snyder kinematics and a non-covariant one as in $\kappa$-Poincar\'e kinematics.

\chapter{Locality and noncommutative spacetime}
\label{appendix:locality}
\section{Different representations of a noncommutative spacetime}
\label{ncst-rep}

One can make a canonical transformation in phase space $(x, k) \to (x', k')$ 
\be
k_\mu \,=\, f_\mu(k')\,,\quad x^\mu \,=\, x^{\prime\nu} g^\mu_\nu(k')\,,
\ee
for any set of momentum dependent functions $f_\mu$, with
\be
g^\mu_\rho(k') \frac{\partial f_\nu(k')}{\partial k'_\rho} \,=\, \delta^\mu_\nu \,.
\ee
One can write the noncommutative space-time coordinates as a function of the new canonical phase-space coordinates
\be
\tilde{x}^\mu \doteq x^\nu \varphi^\mu_\nu(k) \,=\, x^{\prime\rho} g^\nu_\rho(k') \varphi^\mu_\nu(f(k')) \,.
\ee
Introducing
\be
\varphi^{\prime\mu}_\rho(k') \doteq g^\nu_\rho(k') \varphi^\mu_\nu(f(k')) \,=\, \frac{\partial k'_\rho}{\partial k_\nu} \varphi^\mu_\nu(k)\,,
\ee
the noncommutative spacetime is 
\be
\tilde{x}^\mu \,=\, x^{\prime\rho} \varphi^{\prime \mu}_\rho(k') \,.
\ee
We see then that different canonical coordinates related by the choice of momentum variables lead to  different representations of the same noncommutative spacetime with different functions $\varphi^\mu_\nu(k)$.

\section{SR spacetime from the locality condition for a commutative DCL}
\label{append-commut}

In subsection~\ref{sec:firstattempt} we saw that the first way to try to implement locality requires the condition of Eq.~\eqref{eq:limitsym}, which is not valid for a generic DCL (in particular, it is valid for a commutative DCL).  We show here that the spacetime defined by Eq.~\eqref{eq:firstspt} is indeed the SR spacetime. We start by checking that the new space-time coordinates $\tilde{x}^\mu$ are in fact commutative coordinates. If one derives with respect to $p_\sigma$ the first equality of Eq.~\eqref{loc1}, one gets
\be
\frac{\partial \varphi^\mu_\nu(p\oplus q)}{\partial p_\sigma} \,=\, \frac{\partial}{\partial q_\rho} \left(\frac{\partial [p\oplus q]_\nu}{\partial p_\sigma}\right) \,\varphi^\mu_\rho(q) \,.
\label{eq:pre1}
\ee
Using that the left hand side of the previous equation is in fact (applying the chain rule)
\be
\frac{\partial \varphi^\mu_\nu(p\oplus q)}{\partial p_\sigma} \,=\, \frac{\partial \varphi^\mu_\nu(p\oplus q)}{\partial [p\oplus q]_\rho} \,\frac{\partial [p\oplus q]_\rho}{\partial p_\sigma}\,,
\label{eq:pre2}
\ee
and taking the limit $p\to 0$ of the right hand sides of Eqs.~\eqref{eq:pre1} and Eq.~\eqref{eq:pre2}, one gets, using also Eq.~\eqref{eq:limit2},
\be
 \frac{\partial \varphi^\mu_\nu(q)}{\partial q_\rho} \,\varphi^\sigma_\rho(q) \,=\,  \frac{\partial \varphi^\sigma_\nu(q)}{\partial q_\rho} \,\varphi^\mu_\rho(q) \,.
\label{eq:pre3}
\ee
Comparing Eq.~\eqref{eq:pre3} with Eq.~\eqref{eq:commNCspt}, we see that $\{\tilde{x}^\mu, \tilde{x}^\sigma\}=0.$

As the space-time coordinates $\tilde{x}$ commute,  one can define new momentum variables $\tilde{p}_\mu=g_\mu(p)$ such that $(\tilde{x},\tilde{p})$ are canonical conjugate variables, as in the case of $(x,p)$, that is: a canonical transformation that relates both phase-space coordinates exists. Indeed, one can prove that the composition of the new momentum variables are in fact the sum
\be
[\tilde{p}\,\tilde{\oplus} \,\tilde{q}]_\mu \,\doteq\, g_\mu(p\oplus q) \,=\, \tilde{p}_\mu + \tilde{q}_\mu\,,
\label{lcl}
\ee
(where the DCL $\tilde{\oplus}$ has been defined as in Eq.~\eqref{eq:DCLdef}), so that $(\tilde{x},\tilde{p})$ is in fact the phase space of SR.

In order to do that, we will firstly check that Eq.~\eqref{loc1} is invariant under different choices of momentum variables. If we consider new momentum variables $\tilde{p}_\mu = g_\mu(p)$, we can calculate the derivative of the DCL with respect to $\tilde{p}_\rho$
\begin{equation}
\frac{\partial[\tilde{p}\,\tilde{\oplus}\, \tilde{q}]_\nu}{\partial \tilde{p}_\rho} \,=\, \frac{\partial g_\nu (p\oplus q)}{\partial g_\rho (p)}\,=\,\frac{\partial g_\nu (p\oplus q)}{\partial [p\oplus q]_\lambda}\,\frac{\partial [p\oplus q]_\lambda}{\partial p_\sigma }\,\frac{\partial p_\sigma}{\partial g_\rho (p)}\,,
\end{equation}
and also 
\begin{equation}
\tilde{\varphi}^\mu _\rho (\tilde{p}) \,= \, \lim\limits_{g(l)\rightarrow 0}\,\frac{\partial g_\rho (l\oplus p)}{\partial g_\mu (l)}\,=\,\lim\limits_{l\rightarrow 0}\,\frac{\partial g_\rho (l\oplus p)}{\partial [l\oplus p]_\xi}\,\frac{\partial [l\oplus p]_\xi }{\partial l_\eta}\,\frac{\partial l_\eta}{\partial g_\mu (l)}\,=\,\,\frac{\partial g_\rho (p)}{\partial p_\xi}\,\varphi _\xi ^\mu (p)\,,
\label{phi_transformation}
\end{equation}
where we used  Eq.~\eqref{eq:limit2} and also the condition in the change of variables $\tilde{p}=0\implies p=0$. Taking both results and using Eq.~\eqref{loc1} one finds 
\begin{equation}
\frac{\partial[\tilde{p}\,\tilde{\oplus}\, \tilde{q}]_\nu}{\partial \tilde{p}_\rho} \,\tilde{\varphi}^\mu _\rho (\tilde{p}) \,= \, \frac{\partial g_\nu (p\oplus q)}{\partial [p\oplus q]_\lambda} \,\frac{\partial[p\oplus q]_\lambda}{\partial p_\sigma} \,\varphi^\mu _\sigma (p) \,.
\end{equation}
One can do the same for the third term of Eq.~\eqref{loc1} obtaining
\begin{equation}
\frac{\partial[\tilde{p}\,\tilde{\oplus}\, \tilde{q}]_\nu}{\partial \tilde{q}_\rho} \,\tilde{\varphi}^\mu _\rho (\tilde{q}) \,= \, \frac{\partial g_\nu (p\oplus q)}{\partial [p\oplus q]_\lambda} \,\frac{\partial[p\oplus q]_\lambda}{\partial q_\sigma} \,\varphi^\mu _\sigma (q) \,.
\end{equation}
And finally, the first term of Eq.~\eqref{loc1}, following Eq.~\eqref{phi_transformation}, transforms as 
\begin{equation}
\tilde{\varphi}^\mu _\nu (\tilde{p}\,\tilde{\oplus}\, \tilde{q}) \,= \,\frac{\partial g_\nu (p\oplus q)}{\partial [p\oplus q]_\lambda}\,\varphi _\lambda ^\mu (p\oplus q)\,.
\end{equation}

To summarize, we have seen that if Eq.~\eqref{loc1} is satisfied for the variables ($p$, $q$), also will be by ($\tilde{p}$, $\tilde{q}$) obtained by $\tilde{p}_\mu = g_\mu(p)$, $\tilde{q}_\mu = g_\mu(q)$. This means that one can always choose this change of basis in such a way that $\tilde{\varphi}^\mu_\nu(\tilde{k}) = \delta^\mu_\nu$, leading to
\be
\frac{\partial[\tilde{p}\,\tilde{\oplus}\, \tilde{q}]_\nu}{\partial \tilde{p}_\mu} \,=\,  \frac{\partial[\tilde{p}\,\tilde{\oplus}\, \tilde{q}]_\nu}{\partial \tilde{q}_\mu} \,=\, \delta^\mu_\nu\,,
\ee
and then $[\tilde{p}\,\tilde{\oplus}\, \tilde{q}]_\nu = \tilde{p}_\nu + \tilde{q}_\nu$, as it was stated in Sect.~\ref{sec:firstattempt}.

\section{DCL of \texorpdfstring{$\kappa$}{k}-Poincaré in the bicrossproduct basis through locality}
\label{append-bicross}

As we did in \ref{appendix_translations}, we will compute the composition law obtained from  the locality condition when  $\varphi_{L\,\nu}^{\:\:\mu}(p, q)\,=\, \varphi^\mu_\nu(p)$ taking $\varphi^\mu_\nu(p)$ of Eq.~\eqref{eq:phibicross}, which corresponds to  $\kappa$-Poincaré in the bicrossproduct basis. From Eq.~\eqref{varphi-oplus}, we have the following system of equations to solve
\be
\begin{split}
\text{For}\quad \mu\,&=\,0,\quad \nu\,=\,0 \qquad 1\,=\,\frac{\partial (p\oplus q)_0}{\partial p_0}-\frac{\partial (p\oplus q)_0}{\partial p_i}\,\frac{p_i}{\Lambda} \,.\\
\text{For}\quad \mu\,&=\,i,\quad \nu\,=\,0 \qquad 0\,=\,\frac{\partial (p\oplus q)_0}{\partial p_j}\,\delta^i_j \,.\\
\text{For}\quad \mu\,&=\,0,\quad \nu\,=\,i \qquad -\frac{(p \oplus q)_i}{\Lambda}\,=\,\frac{\partial (p\oplus q)_i}{\partial p_0}-\frac{\partial (p\oplus q)_i}{\partial p_j}\,\frac{p_j}{\Lambda}\,.\\
\text{For}\quad \mu\,&=\,i,\quad \nu\,=\,j \qquad \delta^i_j\,=\,\frac{\partial (p\oplus q)_j}{\partial p_i}\,.\\
\end{split}
\ee 
We follow a similar strategy to find the composition law. The first two equations give the result 
\be
(p\oplus q)_0\,=\,p_0+q_0\,.
\ee
The fourth equation requires the composition law to be linear in the spatial components of $p$, so the composition can be written as
\be (p\oplus q)_i\,=\,p_i+q_i \cdot f\left(\frac{p_0}{\Lambda},\frac{q_0}{\Lambda},\frac{\vec{q}^2}{\Lambda^2}\right)\,.
\ee
Introducing this into the third equation, we obtain a differential equation to solve
\be
f\,=\,e^{-p_0/\Lambda}\,g\left(\frac{q_0}{\Lambda},\frac{\vec{q}^2}{\Lambda^2}\right)\,.
\ee
Taking into account the condition~\eqref{eq:cl0}, we conclude that $g\,=\,1$ and then the composition reads
\be
(p\oplus q)_0\,=\,p_0+q_0\,,\qquad (p\oplus q)_i\,=\,p_i+q_i e^{-p_0/\Lambda}\,.
\label{DCL-bcb-locality}
\ee

\section{Lorentz transformation in the bicrossproduct basis from locality}
\label{append-2pLT}

In order to obtain the Lorentz transformations in the one-particle system, we will impose that the Lorentz generators, together with the noncommutative space-time coordinates, should close a 10 dimensional algebra. The modified Poisson brackets are the ones involving the boost generators $J^{0i}$:
\be
\lbrace \tilde{x}^0, J^{0i} \rbrace \,=\, \tilde{x}^i + \frac{1}{\Lambda} J^{0i} \,,\qquad \lbrace \tilde{x}^j, J^{0i}\rbrace \,=\, \delta^i_j \tilde{x}^0 + \frac{1}{\Lambda} J^{ji}\,.
\label{eq:boost_pb}
\ee
Using the notation
\be
\tilde{x}^\mu \,=\, x^\nu \varphi^\mu_\nu(k) \,,\quad J^{0i} \,=\, x^\mu  {\cal J}^{0i}_\mu(k) \,, \quad J^{ij} \,=\, x^\mu  {\cal J}^{ij}_\mu (k) = x^j k_i - x^i k_j\,,
\ee
Eq.~\eqref{eq:boost_pb} leads to the system of equations
\be
\frac{\partial\varphi^0_\mu}{\partial k_\nu}  {\cal J}^{0i}_\nu - \frac{\partial {\cal J}^{0i}_\mu}{\partial k_\nu} \varphi^0_\nu \,=\, \varphi^i_\mu + \frac{1}{\Lambda}  {\cal J}^{0i}_\mu \,,\,\,\,
\frac{\partial\varphi^j_\mu}{\partial k_\nu}  {\cal J}^{0i}_\nu - \frac{\partial {\cal J}^{0i}_\mu}{\partial k_\nu} \varphi^j_\nu \,=\, \delta^i_j \varphi^0_\mu + \frac{1}{\Lambda} \left(\delta^i_\mu k_j - \delta^j_\mu k_i\right)\,.
\label{J0i-xtilde}
\ee
Inserting the expression of $\varphi^\mu_\nu(k)$ of Eq.~\eqref{eq:phibicross} which corresponds to  $\kappa$-Poincaré in the bicrossproduct basis we can obtain  $ {\cal J}^{0i}_\mu$ from  (\ref{J0i-xtilde}). One finally obtains
\be
 {\cal J}^{0i}_0 \,=\, - k_i \,,\qquad  {\cal J}^{0i}_j \,=\, \delta^i_j \,\frac{\Lambda}{2} \left[e^{-2 k_0/\Lambda} - 1 - \frac{\vec{k}^2}{\Lambda^2}\right] + \,\frac{k_i k_j}{\Lambda}\,.
\label{psi(0i)}
\ee
Now we can write the Poisson brackets of $k_\mu$ and $J^{0i}$
\be
\lbrace k_0, J^{0i}\rbrace \,=\, - k_i \,,\qquad \lbrace k_j, J^{0i}\rbrace  \,=\, \delta^i_j \,\frac{\Lambda}{2} \left[e^{-2 k_0/\Lambda} - 1 - \frac{\vec{k}^2}{\Lambda^2}\right] + \,\frac{k_i k_j}{\Lambda} \,.
\ee

\section{Lorentz transformation in the one-particle system of the local DCL1 kinematics}
\label{LT-one-particle}

We consider that the Lorentz generators $J^{\alpha\beta}$ are given by imposing that the space-time coordinates $\tilde{x}^\alpha$ with them form a ten-dimensional Lie algebra. From the Lorentz algebra  generated by $J^{\alpha\beta}$ and the space-time coordinates algebra $\tilde{x}^\alpha$ (when there is no mixing of phase-space coordinates in $\tilde{y}^\alpha$) ,
\be
\{\tilde{x}^i, \tilde{x}^0\} \,=\, - (\epsilon/\Lambda) \,\tilde{x}^i\,, \quad\quad\quad \{\tilde{x}^i, \tilde{x}^j\} \,=\, 0\,,
\ee
one can determine the rest of the Poisson brackets through Jacobi identities:
\be
\begin{split}
    \{\tilde{x}^0, J^{0j}\} \,&=\, \tilde{x}^j - (\epsilon/\Lambda) J^{0j}\,, \quad\quad \{\tilde{x}^i, J^{0j}\} \,=\, \delta^{ij} \tilde{x}^0 - (\epsilon/\Lambda) J^{ij}\,,\\  \{\tilde{x}^0, J^{jk}\} \,&=\, 0\,, \quad\quad \{\tilde{x}^i, J^{jk}\} \,=\, \delta^{ik} \tilde{x}^j - \delta^{ij} \tilde{x}^k\,.
\end{split}
\label{xtilde-J}
\ee
Using
\be
\{\tilde{x}^\alpha, J^{\beta\gamma}\} \,=\, \{x^\nu \varphi^\alpha_\nu(k), x^\rho {\cal J}^{\beta\gamma}_\rho(k)\} \,=\, x^\mu \left(\frac{\partial\varphi^\alpha_\mu(k)}{\partial k_\rho} {\cal J}^{\beta\gamma}_\rho(k) - \frac{\partial J^{\beta\gamma}_\mu(k)}{\partial k_\nu} \varphi^\alpha_\nu(k)\right)
\ee
in the algebra (\ref{xtilde-J}), one obtains 
\begin{align}
\left(\frac{\partial\varphi^0_\mu(k)}{\partial k_\rho} {\cal J}^{0j}_\rho(k) - \frac{\partial {\cal J}^{0j}_\mu(k)}{\partial k_\nu} \varphi^0_\nu(k)\right) \,=& \varphi^j_\mu(k) \blue{-} (\epsilon/\Lambda) {\cal J}^{0j}_\mu(k)\,, \nonumber \\
 \left(\frac{\partial\varphi^i_\mu(k)}{\partial k_\rho} {\cal J}^{0j}_\rho(k) - \frac{\partial {\cal J}^{0j}_\mu(k)}{\partial k_\nu} \varphi^i_\nu(k)\right) \,=& \delta^{ij} \varphi^0_\mu(k) \blue{-} (\epsilon/\Lambda) {\cal J}^{ij}_\mu(k)\,, \nonumber \\
\left(\frac{\partial\varphi^0_\mu(k)}{\partial k_\rho} {\cal J}^{jk}_\rho(k) - \frac{\partial {\cal J}^{jk}_\mu(k)}{\partial k_\nu} \varphi^0_\nu(k)\right) \,=& 0\,, \nonumber \\
\left(\frac{\partial\varphi^i_\mu(k)}{\partial k_\rho} {\cal J}^{jk}_\rho(k) - \frac{\partial {\cal J}^{jk}_\mu(k)}{\partial k_\nu} \varphi^i_\nu(k)\right) \,=& \delta^{ik} \varphi^j_\mu(k) - \delta^{ij} \varphi^k_\mu(k)\,.
\end{align}

From the functions determining the generalized space-time coordinates for the one-particle system of the local DCL1 kinematics according to Eq.~\eqref{magicformula},
\begin{align}
& \varphi^0_0(k) \,=\, \lim_{l\to 0} \frac{\partial(l\oplus k)_0}{\partial l_0} \,=\, 1 + \epsilon k_0/\Lambda\,, &\quad\quad  &\varphi^j_0(k) \,=\, \lim_{l\to 0} \frac{\partial(l\oplus k)_0}{\partial l_j} \,=\, 0 \,,\nonumber \\
& \varphi^0_i(k) \,=\,   \lim_{l\to 0} \frac{\partial(l\oplus k)_i}{\partial k_0} \,=\, \epsilon k_i/\Lambda\,, &\quad\quad &\varphi^j_i(k) \,=\, \lim_{l\to 0} \frac{\partial(l\oplus k)_i}{\partial k_j} \,=\, \delta_i^j\,,
\end{align}
we find the following system of equations for ${\cal J}^{\alpha\beta}_\mu(k)$:
\begin{align}
  & \frac{\partial{\cal J}^{0j}_0}{\partial k_0} \,=\, (\epsilon/\Lambda) \,\left[2 {\cal J}^{0j}_0 - k_0 \frac{\partial{\cal J}^{0j}_0}{\partial k_0} - k_k \frac{\partial{\cal J}^{0j}_0}{\partial k_k}\right]\,, \nonumber \\
 & \frac{\partial{\cal J}^{0j}_0}{\partial k_i} \,=\, - \delta^{ij} (1 + \epsilon k_0/\Lambda) + (\epsilon/\Lambda) {\cal J}^{ij}_0\,, \nonumber \\
  & \frac{\partial{\cal J}^{0j}_l}{\partial k_0} \,=\, - \delta^j_l + (\epsilon/\Lambda) \,\left[2 {\cal J}^{0j}_l - k_0 \frac{\partial{\cal J}^{0j}_l}{\partial k_0} - k_k \frac{\partial{\cal J}^{0j}_l}{\partial k_k}\right]\,, \nonumber \\
 & \frac{\partial{\cal J}^{0j}_l}{\partial k_i} \,=\, - \delta^{ij} \epsilon k_l/\Lambda + (\epsilon/\Lambda) {\cal J}^{ij}_l\,, \nonumber \\
  & \frac{\partial{\cal J}^{jk}_0}{\partial k_0} \,=\, (\epsilon/\Lambda) \,\left[{\cal J}^{jk}_0 - k_0 \frac{\partial{\cal J}^{jk}_0}{\partial k_0} - k_m \frac{\partial{\cal J}^{jk}_0}{\partial k_m}\right]\,, \quad\quad 
  \frac{\partial{\cal J}^{jk}_0}{\partial k_i} \,=\, 0\,, \nonumber \\
& \frac{\partial{\cal J}^{jk}_l}{\partial k_0} \,=\, (\epsilon/\Lambda) \,\left[{\cal J}^{jk}_l - k_0 \frac{\partial{\cal J}^{jk}_l}{\partial k_0} - k_m \frac{\partial{\cal J}^{jk}_l}{\partial k_m}\right]\,, \quad\quad 
  \frac{\partial{\cal J}^{jk}_l}{\partial k_i} \,=\, \delta^{ij} \delta^k_l - \delta^{ik} \delta^j_l\,.
  \end{align}

Imposing the condition that in the limit $(k_0^2/\Lambda^2)\to 0$, $(\vec{k}^2/\Lambda^2)\to 0$, we should recover linear Lorentz transformation
\be
   {\cal J}^{0j}_0 \to - k_j\,, \quad\quad {\cal J}^{0j}_k \to - \delta^j_k \,k_0\,, \quad\quad {\cal J}^{jk}_0 \to 0\,, \quad\quad {\cal J}^{jk}_l \to (\delta^k_l k_j - \delta^j_l k_k)\,,
   \ee
we find a unique solution:
\begin{align}
  & {\cal J}^{ij}_0(k) \,=\, 0\,, &\quad&{\cal J}^{ij}_k(k) \,=\, \delta^j_k \, k_i - \delta^i_k \, k_j\,, \nonumber \\
  & {\cal J}^{0j}_0(k) \,=\, - k_j (1+\epsilon k_0/\Lambda)\,, &\quad&{\cal J}^{0j}_k(k) \,=\, \delta^j_k \left[-k_0 - \epsilon k_0^2/2\Lambda\right] + (\epsilon/\Lambda) \left[\vec{k}^2/2 - k_j k_k\right]\,.
\end{align} 
\chapter{Resonances and cross sections with a DRK}

\section{BSR Extension of the Breit--Wigner Distribution} 
\label{appendix:B-W}

We start from Eq.~\eqref{eq:BW-BSR},
\begin{equation}
f_{\text{BSR}}(m^2)\,=\,\frac {K}{\left[\mu^{2}(m^2)-M_X^{2}\right]^{2}+M_X^{2}\Gamma_X ^{2}}.
\label{eqapp:BW-BSR}
\end{equation}
In order to simplify future expressions, we introduce the dimensionless variables
\begin{equation}\label{dimensionless}
\tau:=\frac{m^2}{M_X^2}\quad \quad\quad \gamma:=\frac{\Gamma_X^2}{M_X^2}\quad\quad \quad \lambda:=\frac{\Lambda^2}{M_X^2},
\end{equation}
so we can write Eq.~\eqref{eqapp:BW-BSR} as
\begin{equation}
\label{simplification}
f_{\text{BSR}}(m^2)=\frac{K}{M_X^4F(\tau)}\,,
\end{equation}
where
\begin{equation}
F(\tau):=\left[\tau\left(1+\epsilon\frac{\tau}{\lambda}\right)-1\right]^2+\gamma\,,
\end{equation}
for the BSR we are considering Eq.~\eqref{eq:DCL_tp}.

For a resonance, we need $\gamma\ll 1$, so in order to have a peak we need that the following equation holds 
\begin{equation}
\tau\left(1+\epsilon\frac{\tau}{\lambda}\right)-1=0\,,
\end{equation}
with solutions
\begin{equation}
\tau^*=-\frac{\lambda}{2\epsilon}\left[1\pm\left(1+4\frac{\epsilon}{\lambda}\right)^{1/2}\right].
\end{equation}

We can make a Taylor expansion at $\tau^*$ in order to study the distribution close to the peaks. Evaluating the derivatives of $F(\tau)$ up to second order
\begin{equation}
\left.\frac{dF}{d\tau}\right\vert_{\tau=\tau^*}=2\left[\tau^*\left(1+\epsilon\frac{\tau^*}{\lambda}\right)-1 \right]\left(1+2\frac{\epsilon}{\lambda}\tau^*\right)=0\,,
\end{equation}
\begin{equation}
\left.\frac{d^2F}{d\tau^2}\right\vert_{\tau=\tau^*}=2\left(1+2\frac{\epsilon}{\lambda}\tau^* \right)^2=2\left(1+4\frac{\epsilon}{\lambda}\right)\,,
\end{equation}
one obtains
\begin{equation}
F(\tau)\approx \gamma+\left(1+4\frac{\epsilon}{\lambda}\right)(\tau-\tau^*)^2\,,
\end{equation}
and substituting in Eq.~\eqref{simplification}, and using Eq.~\eqref{dimensionless}, one finds
\begin{equation}
f_{\text{BSR}}(m^2)\approx \frac{1}{M^4_X\left(1+4\epsilon M_X^2/\Lambda^2 \right)}\cdot \frac{K}{((m^2-{m^*}^2))^2+M_X^2\Gamma_X^2 \left(1+4\epsilon M_X^2/\Lambda^2 \right)^{-1}}\,,
\label{BSR exp}
\end{equation}
where
\begin{equation}\label{m_*}
{m^*}^2:=M_X^2\tau^*=\frac{\Lambda^2}{2\epsilon}\left[-1\pm\left(1+4\epsilon\frac{M_X^2}{\Lambda^2}\right)^{1/2} \right]\,.
\end{equation}

One can see that the maximum value of the distribution~\eqref{BSR exp} is reached at $m^2={m^*}^2$, and one must study separately if the value of $\epsilon$ of Eq.~\eqref{eq:DCL_tp} is positive or negative.

For $\epsilon=+1$, one obtains a unique solution for the pole 
\begin{equation}\label{epsilon+}
{m^*}^2=\frac{\Lambda^2}{2}\left[\left(1+4\frac{M_X^2}{\Lambda^2}\right)^{1/2}-1 \right].
\end{equation}

For this case, one can see that the shape of the distribution of Eq.~\eqref{BSR exp} is the same as in SR, but the difference resides in that the position of the peak $(m^2={m^*}^2)$ is not the squared mass of the resonance. One can easily see from Eq.~\eqref{epsilon+} that one recovers the peak in the SR case when $M_X\ll \Lambda$,
\begin{equation}
{m^*}^2\approx\frac{\Lambda^2}{2}\left[1+2\frac{M_X^2}{\Lambda^2}-1 \right]=M_X^2.
\end{equation}

The width of the peak can be computed from Eq.~\eqref{BSR exp} (to be compared with the Breit--Wigner distribution~\eqref{eq:BW}):
\begin{equation}
{\Gamma^*}^2=\frac{M_X^2\Gamma_X^2}{{m^*}^2\left(1+4{M_X^2}/{\Lambda^2}\right)}=\Gamma_X^2\frac{2{M_X^2}/{\Lambda^2}}{\left(1+4{M_X^2}/{\Lambda^2}\right)\left[\left(1+4{M_X^2}/{\Lambda^2}\right)^{1/2}-1 \right]},
\label{eq:width+}
\end{equation}
which also leads to the SR decay width of the resonance when $M_X^2\ll \Lambda^2$.

The $\epsilon=-1$ case is much more interesting. If
\begin{equation}
1-4\frac{M_X^2}{\Lambda^2}>0,
\end{equation}
that is, when $M_X<\Lambda/2$, Eq.~\eqref{m_*} gives two solutions for  ${m^*}^2>0$,
\begin{equation}\label{epsilon-}
{m^*_{\pm}}^2=\frac{\Lambda^2}{2}\left[1\pm\left(1-4\frac{M_X^2}{\Lambda^2}\right)^{1/2} \right].
\end{equation} 

Then, one finds two peaks (at $m^2={m^*_{\pm}}^2$) in the squared mass distribution in contrast to the SR case, where there is only one. One can read from the position of these two peaks $\Lambda^2$ and $M_X^2$ using Eq.~\eqref{epsilon-}:
\begin{equation}\label{changes}
\Lambda^2=({m^*_+}^2+{m^*_-}^2)\,, \quad \quad\quad M_X^2=\frac{{m^*_+}^2{m^*_-}^2}{({m^*_+}^2+{m^*_-}^2)}\,.
\end{equation}

As in Eq.~\eqref{eq:width+}, the widths of the two peaks are
\begin{equation}\label{gamma1}
{\Gamma_\pm^*}^2=\frac{M_X^2\Gamma_X^2}{{m_\pm^*}^2\left(1-4{M_X^2}/{\Lambda^2}\right)}.
\end{equation}
From $M_X^2$ and $\Lambda^2$ of Eq.~\eqref{changes}, we get 
\be
1-4\frac{M_X^2}{\Lambda^2}=\frac{({m_+^*}^2-{m_-^*}^2)^2}{({m_+^*}^2+{m_-^*}^2)^2}.
\ee
Substituting in Eq.~\eqref{gamma1} one finds
\begin{equation}\label{gamma}
{\Gamma_\pm^*}^2=\Gamma_X^2\frac{({m_+^*}^2+{m_-^*}^2){m_\mp^*}^2}{({m_+^*}^2-{m_-^*}^2)^2}.
\end{equation}

From Eq.~\eqref{gamma}, one can find the decay width of the resonance $X$:
\begin{equation} \label{eq:width}
\Gamma_X^2={\Gamma_+^*}^2\frac{({m_+^*}^2-{m_-^*}^2)^2}{{m_-^*}^2({m_+^*}^2+{m_-^*}^2)}={\Gamma_-^*}^2\frac{({m_+^*}^2-{m_-^*}^2)^2}{{m_+^*}^2({m_+^*}^2+{m_-^*}^2)}=({\Gamma_+^*}^2+{\Gamma_-^*}^2)\left[\frac{{m_+^*}^2-{m_-^*}^2}{{m_+^*}^2+{m_-^*}^2}\right]^2.
\end{equation}
When $M_X^2\ll\Lambda^2$, the expressions for the poles (Eq.~\eqref{epsilon-}) and the widths (Eq.~\eqref{gamma1}) are
\begin{equation}
{m^*_{\pm}}^2\approx\frac{\Lambda^2}{2}\left[1\pm \left(1-2\frac{M_X^2}{\Lambda^2}\right)\right],
\end{equation}
\begin{equation}
{\Gamma_\pm^*}^2\approx\Gamma_X^2\frac{2{M_X^2}/{\Lambda^2}}{\left[1\pm\left(1-2{M_X^2}/{\Lambda^2}\right) \right]},
\end{equation}
so in this limit
\begin{equation}
\begin{array}{ll}
{m_+^*}^2\approx \Lambda^2 \,,& {\Gamma_+^*}^2\approx \Gamma_X^2 {M_X^2}/{\Lambda^2}\,, \\
{m_-^*}^2 \approx M_X^2 \,,& {\Gamma_-^*}^2\approx \Gamma_X^2\,,
\end{array}
\end{equation}
and one can see that for one of the peaks ($-$) one finds the result of SR, while the other peak ($+$) is shifted by a factor $\Lambda/M_X$, and its width reduced by a factor $M_X/\Lambda$ with respect to the SR peak. 

We can note that for $M_X>\Lambda/2$ there is not any peak, since the square root of Eq.~\eqref{epsilon-} becomes negative. This would lead to an ``invisible'' resonance. In the limit $M_X\rightarrow \Lambda/2$ one can see that the two poles coincide with their width tending to infinite.

We have not considered the dependence on $m^2$ of the $K$ factor taking into account how the resonance is produced and the decay width of the two particles because we assume that the analysis is carried out near the peaks, where $K\approx K({m^*}^2)$, and then we can neglect the variation of $K$ with respect to $m^2$.

\section{Cross sections with a DCL}
\label{appendix_cross_sections}

In this part of the appendix we show how to obtain the cross section of the process $e^-(k) e^+(\overline{k}) \rightarrow Z \rightarrow \mu^-(p) \mu^+(\overline{p})$ in the BSR case. Firstly, we need to compute the two-particle phase-space integral 
\be
\widehat{F}^{(\alpha)}(E_0) \,\doteq\, \overline{PS}_2^{(\alpha)} F(k, \overline{k}, p, \overline{p})
\ee
for different Lorentz invariant functions $F$ of the four momenta $k$, $\overline{k}$, $p$, $\overline{p}$. First, we are going to use the Dirac delta function $\delta_\alpha^{(4)}(k, \overline{k}; p, \overline{p})$ that takes into account the conservation law for each channel $\alpha$ and that will let us express $\overline{p}$ as a function $\overline{p}^{(\alpha)}(k, \overline{k}, p)$ of the other remaining three momenta $k$, $\overline{k}$, $p$. Therefore, we have
\be
\widehat{F}^{(\alpha)}(E_0) \,=\, \frac{1}{(2\pi)^2} \int d^4p \,\delta(p^2) \theta(p_0) \,\delta\left(\overline{p}^{(\alpha)\,2}(k, \overline{k}, p)\right) \theta\left(\overline{p}_0^{(\alpha)}(k, \overline{k}, p)\right) F_\alpha(k, \overline{k}, p)\,,
\ee
where
\be
F_\alpha(k, \overline{k}, p) \,=\, F(k, \overline{k}, p, \overline{p}^{(\alpha)}(k, \overline{k}, p))\,.
\ee

Then, integrating over $p_0$ and $|\vec{p}|$ with the remaining two Dirac delta functions we find
\be
\widehat{F}^{(\alpha)}(E_0) \,=\, \frac{1}{8\pi^2} \int d\Omega_{\hat{p}} \frac{E^{(\alpha)}(k, \overline{k}, \hat{p})}{|\frac{\partial\overline{p}^{(\alpha)\,2}}{\partial p_0}|_{p_0=E^{(\alpha)}(k, \overline{k}, \hat{p})}} \,F_\alpha(k, \overline{k}, p)|_{|\vec{p}|=p_0=E^{(\alpha)}(k, \overline{k}\,, \hat{p})}\,,
\ee
where $E^{(\alpha)}(k, \overline{k}, \hat{p})$ is the positive value of $p_0$ such that $\overline{p}^{(\alpha)\,2}=0$. Due to the rotational invariance and the choice $\vec{\overline{k}}=-\vec{k}$ one can show that $E^{(\alpha)}$ is a function of the energy $E_0$ of the particles in the initial state and the angle $\theta$ between the directions of $\vec{k}$ and $\vec{p}$. So we have
\be
\widehat{F}^{(\alpha)}(E_0) \,=\, \frac{1}{4\pi} \int d\cos\theta \frac{E^{(\alpha)}(E_0, \cos\theta)}{|\frac{\partial\overline{p}^{(\alpha)\,2}}{\partial p_0}|_{p_0=E^{(\alpha)}(E_0, \cos\theta)}} F_\alpha(k, \overline{k}, p)|_{|\vec{p}|=p_0=E^{(\alpha)}(E_0, \cos\theta)}\, .
\label{eq:Falpha}
\ee
 In order to obtain the expression of  $E^{(\alpha)}(E_0, \cos\theta)$ and $|\frac{\partial\overline{p}^{(\alpha)\,2}}{\partial p_0}|_{p_0=E^{(\alpha)}(E_0, \cos\theta)}$, we need to use the explicit form of the conservation law for each channel.  

For the first channel, we have $k\oplus\overline{k}=p\oplus\overline{p}$, and then 
\be
k_\mu + \overline{k}_\mu + \frac{k\cdot \overline{k}}{2\overline{\Lambda}^2} k_\mu \,=\, 
p_\mu + \overline{p}_\mu + \frac{p\cdot \overline{p}}{2\overline{\Lambda}^2} p_\mu\,.
\ee

This leads to
\be
p\cdot \overline{p}^{(1)} \,=\, k\cdot p + \overline{k}\cdot p + \frac{(k\cdot \overline{k})(k\cdot p)}{2\overline{\Lambda}^2}\,,
\ee
and, neglecting terms proportional to $(1/\overline{\Lambda}^4)$, one has
\be
\overline{p}^{(1)}_\mu \,=\, k_\mu + \overline{k}_\mu - p_\mu + \frac{k\cdot \overline{k}}{2\overline{\Lambda}^2} k_\mu - \frac{(k\cdot p+\overline{k}\cdot p)}{2\overline{\Lambda}^2} p_\mu\,,
\ee
and
\begin{align}
\overline{p}^{(1)\,2} &\,=\, 2 k\cdot \overline{k} - 2 k\cdot p - 2 \overline{k}\cdot p + \frac{(k\cdot \overline{k})^2}{\overline{\Lambda}^2} - \frac{(k\cdot \overline{k})(k\cdot p)}{\overline{\Lambda}^2} - \frac{(k\cdot p + \overline{k}\cdot p)^2}{\overline{\Lambda}^2}\,, \\
k\cdot \overline{p}^{(1)} &\,=\, k\cdot \overline{k} - k\cdot p - \frac{(k\cdot p+\overline{k}\cdot p)k\cdot p}{2\overline{\Lambda}^2}\,, \\
\overline{k}\cdot \overline{p}^{(1)} &\,=\, k\cdot \overline{k} - \overline{k}\cdot p + \frac{(k\cdot \overline{k})^2}{2\overline{\Lambda}^2} - \frac{(k\cdot p+\overline{k}\cdot p)\overline{k}\cdot p}{2\overline{\Lambda}^2} \,.
\end{align}

In the reference frame where $k_\mu = E_0 (1, \hat{k})$, $\overline{k}_\mu = E_0 (1, -\hat{k})$, one finds
\begin{align}
\overline{p}^{(1)\,2} &\,=\, 4 E_0^2 - 4 E_0 p_0 + \frac{4E_0^4}{\overline{\Lambda}^2} - \frac{2E_0^3}{\overline{\Lambda}^2} p_0 (1-\cos\theta) - \frac{4E_0^2}{\overline{\Lambda}^2} p_0^2\,, \\
k\cdot \overline{p}^{(1)} &\,=\, 2 E_0^2 - E_0 p_0 (1-\cos\theta) - \frac{E_0^3}{\overline{\Lambda}^2} p_0 (1-\cos\theta)\,, \\
\overline{k}\cdot \overline{p}^{(1)} &\,=\, 2E_0^2 - E_0 p_0 (1+\cos\theta) + \frac{2E_0^4}{\overline{\Lambda}^2} - \frac{E_0^2}{\overline{\Lambda}^2} p_0^2 (1+\cos\theta)\,.
\end{align}

From the expression of $\overline{p}^{(1)\,2}$, one arrives to 
\be
\frac{\partial \overline{p}^{(1)\,2}}{\partial p_0} \,=\, - 4E_0 - \frac{2E_0^3}{\overline{\Lambda}^2} (1-\cos\theta) - \frac{8E_0^2}{\overline{\Lambda}^2} p_0,\quad\quad\quad E^{(1)}\,=\,E_0\left(1 - \frac{E_0^2}{2\overline{\Lambda}^2} (1-\cos\theta)\right)\,.
\ee 
One can do a similar analysis for the other channels.
 
In order to compute the cross section, we consider these four invariant functions
\be
\label{eq:Finv}
F_{\pm}\,=\,t^2 \pm u^2\,=\,(k\cdot p + \overline{k}\cdot \overline{p})^2 \pm (k\cdot \overline{p} + \overline{k}\cdot p)^2\, ,
\ee
\be
\label{eq:F-inv}
\begin{split}
\overline{F}_{\pm}\,=\,\overline{t}^2 \pm \overline{u}^2\,=\,&\left[(k\cdot p + \overline{k}\cdot \overline{p})^2 - (k\cdot p + \overline{k}\cdot \overline{p})\left[(k\cdot p)^2+(\overline{k}\cdot \overline{p})^2\right]/\overline{\Lambda}^2\right] \\
&\pm \left[(k\cdot \overline{p} + \overline{k}\cdot p)^2 - (k\cdot \overline{p} + \overline{k}\cdot p)\left[(k\cdot \overline{p})^2+(\overline{k}\cdot p)^2\right]/\overline{\Lambda}^2\right] \,,
\end{split}
\ee
and the corresponding phase-space integrals $\widehat{F}_{\pm}^{(\alpha)}(E_0)$, $\widehat{\overline{F}}_{\pm}^{(\alpha)}(E_0)$. Then, the two cross sections computed with the two assumptions of the dynamical factor $A$ proposed in Sec.~\ref{sec:dynamical} are
\begingroup\makeatletter\def\f@size{9}\check@mathfonts
\def\maketag@@@#1{\hbox{\m@th\fontsize{10}{10}\selectfont \normalfont#1}}%
\begin{align}
\overline{\sigma}^{(1)} &\,=\, \frac{e^4}{256 \sin^4\theta_W\cos^4\theta_W E_0^2\left(1+\frac{E_0^2}{\Lambda^2}\right)} \,\frac{1}{\left[(\overline{s} - \overline{M}_Z^2)^2 + \overline{\Gamma}_Z^2 \overline{M}_Z^2\right]} \, \left[(C_V^2 + C_A^2)^2 \sum_\alpha \widehat{F}_+^{(\alpha)}(E_0) - 4 C_V^2 C_A^2 \sum_\alpha \widehat{F}_-^{(\alpha)}(E_0)\right]\,, \\
\overline{\sigma}^{(2)} &\,=\, \frac{e^4}{256 \sin^4\theta_W\cos^4\theta_W E_0^2\left(1+\frac{E_0^2}{\Lambda^2}\right)} \,\frac{1}{\left[(\overline{s} - \overline{M}_Z^2)^2 + \overline{\Gamma}_Z^2 \overline{M}_Z^2\right]} \, \left[(C_V^2 + C_A^2)^2 \sum_\alpha \widehat{\overline{F}}_+^{(\alpha)}(E_0) - 4 C_V^2 C_A^2 \sum_\alpha \widehat{\overline{F}}_-^{(\alpha)}(E_0)\right] \,.
\end{align}
\endgroup

Substituting Eqs.~\eqref{eq:Finv} and \eqref{eq:F-inv} in  Eq.~\eqref{eq:Falpha}, one obtains the cross sections of Eqs.~\eqref{eq:cross_section_final_1}-\eqref{eq:cross_section_final_2}.

\chapter{Scalar of curvature of momentum space} 
\label{appendix:cotangent}

We show in this appendix that when one considers a metric in the cotangent bundle starting from a maximally symmetric momentum space with the proposal of Ch.~\ref{ch:cotangent}, the scalar of curvature of the momentum space is also constant. The definition of the momentum curvature tensor for flat spacetime is~\eqref{eq:Riemann_p} 
\begin{equation}
S_{\sigma}^{\mu\nu\rho}(k)\,=\, \frac{\partial C^{\mu\nu}_\sigma(k)}{\partial k_\rho}-\frac{\partial C^{\mu\rho}_\sigma(k)}{\partial k_\nu}+C_\sigma^{\lambda\nu}(k) \,C^{\mu\rho}_\lambda(k)-C_\sigma^{\lambda\rho}(k)\,C^{\mu\nu}_\lambda(k)\,,
\end{equation} 
which can be rewritten using Eq.~\eqref{eq:affine_connection_p}
\begin{equation}
\begin{split}
S^{\sigma\kappa\lambda\mu}(k)\,&=\, \frac{1}{2}\left(\frac{\partial^2 g_k^{\sigma\mu}(k)}{\partial k_\kappa \partial k_\lambda}+\frac{\partial^2 g_k^{\kappa\lambda}(k)}{\partial k_\sigma \partial k_\mu}-\frac{\partial^2 g_k^{\sigma\lambda}(k)}{\partial k_\kappa \partial k_\mu}-\frac{\partial^2 g_k^{\kappa\mu}(k)}{\partial k_\sigma \partial k_\lambda}\right)\\
&+g_k^{\nu\tau}(k)\left(C_\nu^{\kappa\lambda}(k)\,C^{\sigma\mu}_\tau(k)-C_\nu^{\kappa\mu}(k)\,C^{\sigma\lambda}_\tau(k)\right)\,,
\end{split}
\end{equation} 
where we have risen the low index. We have proposed in Ch.~\ref{ch:cotangent} that a possible way to consider a curvature in spacetime in the MRK and in the metric is to replace $k\rightarrow \bar{k}=\bar{e}k$, so the momentum curvature tensor is 
\begin{equation}
\begin{split}
S^{\sigma\kappa\lambda\mu}(\bar{k})\,&=\, \frac{1}{2}\left(\frac{\partial^2 g_{\bar{k}}^{\sigma\mu}(\bar{k})}{\partial \bar{k}_\kappa \partial \bar{k}_\lambda}+\frac{\partial^2 g_{\bar{k}}^{\kappa\lambda}(\bar{k})}{\partial \bar{k}_\sigma \partial \bar{k}_\mu}-\frac{\partial^2 g_{\bar{k}}^{\sigma\lambda}(\bar{k})}{\partial \bar{k}_\kappa \partial \bar{k}_\mu}-\frac{\partial^2 g_{\bar{k}}^{\kappa\mu}(\bar{k})}{\partial \bar{k}_\sigma \partial \bar{k}_\lambda}\right)\\
&+g_{\bar{k}}^{\nu\tau}(\bar{k})\left(C_\nu^{\kappa\lambda}(\bar{k})\,C^{\sigma\mu}_\tau(\bar{k})-C_\nu^{\kappa\mu}(\bar{k})\,C^{\sigma\lambda}_\tau(\bar{k})\right)\,, 
\end{split}
\end{equation}
which contracting gives 
\begin{equation}
S^{\sigma\kappa\lambda\mu}(\bar{k})g^{\bar{k}}_{\sigma\lambda}(\bar{k})g^{\bar{k}}_{\kappa\mu}(\bar{k})\,=\,\text{const}\,,
\end{equation}
due to the fact that the starting momentum space was a maximally symmetric space, and where $g^{\bar{k}}_{\kappa\nu}(\bar{k})$ is the inverse of the metric 
\begin{equation}
g^{\bar{k}}_{\kappa\nu}(\bar{k})g_{\bar{k}}^{\kappa\mu}(\bar{k})\,=\,\delta^\mu_\nu\,.
\end{equation}
From this, one can obtain the scalar of curvature in momentum space from the cotangent bundle metric 
\begin{equation}
g_{\mu\nu}(x,k)\,=\,e^\rho_\mu(x)g^{\bar{k}}_{\rho\sigma}(\bar{k})e^\sigma_\nu(x)\,,
\end{equation}
with the momentum curvature tensor depending on momentum and spacetime coordinates
\begin{equation}
\begin{split}
S^{\sigma\kappa\lambda\mu}(x,k)\,&=\, \frac{1}{2}\left(\frac{\partial^2 g^{\sigma\mu}(x,k)}{\partial k_\kappa \partial k_\lambda}+\frac{\partial^2 g^{\kappa\lambda}(x,k)}{\partial k_\sigma \partial k_\mu}-\frac{\partial^2 g^{\sigma\lambda}(x,k)}{\partial k_\kappa \partial k_\mu}-\frac{\partial^2 g^{\kappa\mu}(x,k)}{\partial k_\sigma \partial k_\lambda}\right)\\
&+g^{\nu\tau}(x,k)\left(C_\nu^{\kappa\lambda}(x,k)\,C^{\sigma\mu}_\tau(x,k)-C_\nu^{\kappa\mu}(x,k)\,C^{\sigma\lambda}_\tau(x,k)\right)\,.
\end{split}
\end{equation} 
After some computations one arrives to 
\begin{equation}
S^{\sigma\kappa\lambda\mu}(x,k)g_{\sigma\lambda}(x,k)g_{\kappa\mu}(x,k)\,=\,S^{\sigma\kappa\lambda\mu}(\bar{k})g^{\bar{k}}_{\sigma\lambda}(\bar{k})g^{\bar{k}}_{\kappa\mu}(\bar{k})\,=\,\text{const}\,.
\end{equation}
Then, through the way we propose here, if the starting momentum space is maximally symmetric, the resulting metric in the cotangent bundle has a constant momentum scalar of curvature. Now we can understand why we have found that there are 10 transformations for the momentum (momentum isometries of the metric) for a fixed point $x$: 4 related with translations and 6 which leave the momentum origin invariant (the phase space point $(x,0)$).

\end{appendices}


\end{document}